\documentclass[twocolumn]{aastex63}

\pdfoutput=1

\usepackage{enumerate}
\usepackage{enumitem}
\usepackage{natbib,ifthen}
\usepackage{graphicx}
\usepackage{fancyhdr}
\usepackage{amssymb,amsmath}
\usepackage{multirow,bigdelim}
\usepackage{wasysym}  
\usepackage{amssymb}

\newcommand*{\ditto}{---\texttt{"}---}
\usepackage[T1]{fontenc}

\shorttitle{Hot Cores in the LMC}
\shortauthors{Sewi{\l}o et al.}

\begin{document}


\title{ALMA Observations of Molecular Complexity in the Large Magellanic Cloud: The N\,105 Star-Forming Region}


\correspondingauthor{Marta Sewi{\l}o}
\email{marta.m.sewilo@nasa.gov}

\author{Marta Sewi{\l}o}
\affiliation{Exoplanets and Stellar Astrophysics Laboratory, NASA Goddard Space Flight Center, Greenbelt, MD 20771, USA}
\affiliation{Department of Astronomy, University of Maryland, College Park, MD 20742, USA}
\affiliation{Center for Research and Exploration in Space Science and Technology, NASA Goddard Space Flight Center, Greenbelt, MD 20771}

\author{Martin Cordiner}
\affiliation{Astrochemistry Laboratory, NASA Goddard Space Flight Center, Greenbelt, MD 20771, USA}
\affiliation{Institute for Astrophysics and Computational Sciences, The Catholic University of America, Washington, DC 20064, USA}

\author{Steven B. Charnley}
\affiliation{Astrochemistry Laboratory, NASA Goddard Space Flight Center, Greenbelt, MD 20771, USA}

\author{Joana M. Oliveira}
\affiliation{Lennard-Jones Laboratories, Keele University, ST5 5BG, UK}

\author{Emmanuel Garcia Berrios}
\affiliation{Astrochemistry Laboratory, NASA Goddard Space Flight Center, Greenbelt, MD 20771, USA}
\affiliation{Institute for Astrophysics and Computational Sciences, The Catholic University of America, Washington, DC 20064, USA}

\author{Peter Schilke}
\affiliation{I. Physikalisches Institut der Universit{\"a}t zu K{\"o}ln, Z{\"u}lpicher Str. 77, 50937, K{\"o}ln, Germany}

\author{Jacob L. Ward}
\affiliation{Astronomisches Rechen-Institut, Zentrum f{\"u}r Astronomie der Universit{\"a}t Heidelberg, M{\"o}nchhofstr. 12-14, 69120 Heidelberg Germany}

\author{Jennifer Wiseman}
\affiliation{NASA Goddard Space Flight Center, 8800 Greenbelt Rd, Greenbelt, MD 20771, USA}

\author{Remy Indebetouw}
\affiliation{Department of Astronomy, University of Virginia, PO Box 400325, Charlottesville, VA 22904, USA}
\affiliation{National Radio Astronomy Observatory, 520 Edgemont Rd, Charlottesville, VA 22903, USA}

\author{Kazuki Tokuda}
\affiliation{Department of Physical Science, Graduate School of Science, Osaka Prefecture University, 1-1 Gakuen-cho, Naka-ku, Sakai, Osaka 599-8531, Japan}
\affiliation{National Astronomical Observatory of Japan, National Institutes of Natural Science, 2-21-1 Osawa, Mitaka, Tokyo 181-8588, Japan}

\author{Jacco Th. van Loon}
\affiliation{Lennard-Jones Laboratories, Keele University, ST5 5BG, UK}

\author{\'{A}lvaro S\'{a}nchez-Monge}
\affiliation{I. Physikalisches Institut der Universit{\"a}t zu K{\"o}ln, Z{\"u}lpicher Str. 77, 50937, K{\"o}ln, Germany}

\author{Veronica Allen}
\affiliation{Kapteyn Astronomical Institute, University of Groningen, P.O. Box 800, 9700 AV Groningen, The Netherlands}

\author{C.-H. Rosie Chen}
\affiliation{Max-Planck-Institut f{\"u}r Radioastronomie, Auf dem H{\"u}gel 69, D-53121 Bonn, Germany}

\author{Roya Hamedani Golshan}
\affiliation{I. Physikalisches Institut der Universit{\"a}t zu K{\"o}ln, Z{\"u}lpicher Str. 77, 50937, K{\"o}ln, Germany}

\author{Agata Karska}
\affiliation{Institute of Astronomy, Faculty of Physics, Astronomy and Informatics, Nicolaus Copernicus University, ul. Grudzi\k{a}dzka 5, 87-100 Toru\'{n}, Poland}

\author{Lars E. Kristensen}
\affiliation{Niels Bohr Institute, Centre for Star \& Planet Formation, University of Copenhagen, {\O}ster Voldgade 5-7, 1350 Copenhagen K, Denmark}

\author{Stan E. Kurtz}
\affiliation{Instituto de Radioastronom\'{i}a y Astrof\'{i}sica, Universidad Nacional Aut\'{o}noma de M\'{e}xico, Apdo. Postal 3-72, 58090 Morelia, Michoac\'{a}n, Mexico}

\author{Toshikazu Onishi}
\affiliation{Department of Physical Science, Graduate School of Science, Osaka Prefecture University, 1-1 Gakuen-cho, Naka-ku, Sakai, Osaka 599-8531, Japan}

\author{Sarolta Zahorecz}
\affiliation{Department of Physical Science, Graduate School of Science, Osaka Prefecture University, 1-1 Gakuen-cho, Naka-ku, Sakai, Osaka 599-8531, Japan}
\affiliation{National Astronomical Observatory of Japan, National Institutes of Natural Science, 2-21-1 Osawa, Mitaka, Tokyo 181-8588, Japan}

\begin{abstract}
The Large Magellanic Cloud (LMC) is the nearest laboratory for detailed studies on the formation and survival of complex organic molecules (COMs), including biologically important ones, in low-metallicity environments---typical for earlier cosmological epochs.  We report the results of 1.2 mm continuum and molecular line observations of three fields in the star-forming region N\,105 with the Atacama Large Millimeter/submillimeter Array (ALMA). N\,105 lies at the western edge of the LMC bar with on-going star formation traced by H$_2$O, OH, and CH$_3$OH masers, ultracompact H\,{\sc ii} regions, and young stellar objects.  Based on the spectral line modeling, we estimated rotational temperatures, column densities, and fractional molecular abundances for twelve 1.2 mm continuum sources.  We identified sources with a range of chemical make-ups, including two bona fide hot cores and four hot core candidates. The CH$_3$OH emission is widespread and associated with all the continuum sources. COMs CH$_3$CN and CH$_3$OCH$_3$ are detected toward two hot cores in N\,105 together with smaller molecules typically found in Galactic hot cores (e.g., SO$_2$, SO, and HNCO) with the molecular abundances roughly scaling with metallicity. We report a tentative detection of the astrobiologically relevant formamide molecule (NH$_2$CHO) toward one of the hot cores; if confirmed, this would be the first detection of NH$_2$CHO in an extragalactic sub-solar metallicity environment.  We suggest that metallicity inhomogeneities resulting from the tidal interactions between the LMC and the Small Magellanic Cloud (SMC) might have led to the observed large variations in COM abundances in LMC hot cores. 
\end{abstract}

\section{Introduction}
\label{s:intro}

The Large Magellanic Cloud (LMC), a gas-rich companion of the Milky Way, is the nearest laboratory for detailed studies on the formation and survival of complex organic molecules (COMs; $\geq$6 atoms, \citealt{herbst2009}), including those of astrobiological importance, in a low-metallicity environment ($Z_{\rm LMC}$ $\sim$ 0.3--0.5 $Z_{\odot}$; \citealt{russell1992}; \citealt{westerlund1997}; \citealt{rolleston2002}).  Both simple and complex molecules are present during each phase of star and planet formation. Following their incorporation into comets, interstellar COMs might have been delivered to early Earth providing important ingredients for the origin of life (e.g., \citealt{ehrenfreund2000}; \citealt{mumma2011}; \citealt{caselli2012}).  The metallicity of the LMC is similar to galaxies around the peak of star formation in the Universe (z$\sim$1.5; e.g., \citealt{pei1999}, \citealt{mehlert2002}; \citealt{madau2014}), making it an ideal template for studying star formation and complex chemistry in low-metallicity systems at earlier cosmological epochs where direct observations are impossible.

\begin{figure*}[ht!]
\centering
\includegraphics[width=\textwidth]{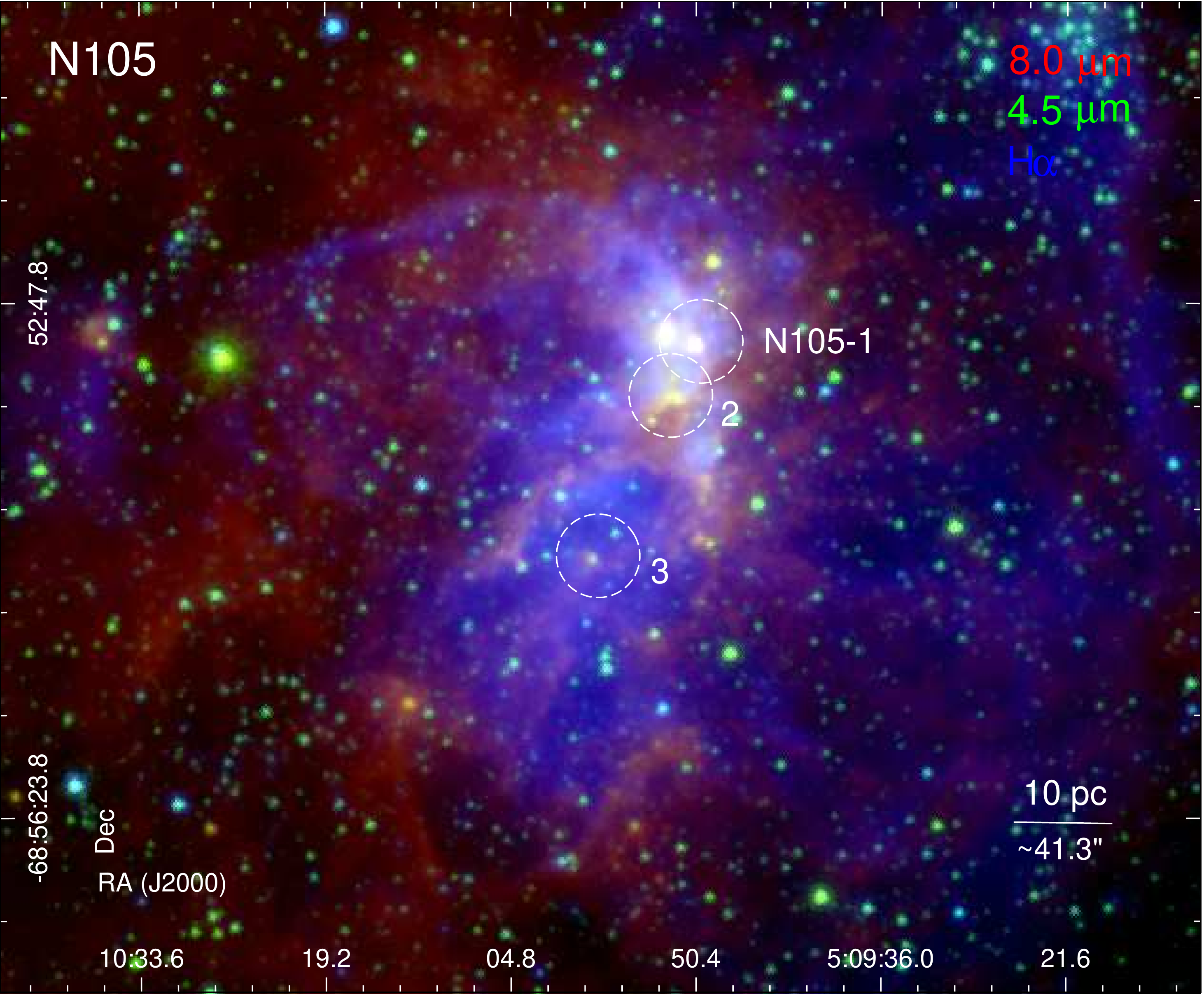}
\caption{Three color-composite image of the N\,105 star-forming region combining the SAGE/IRAC 8.0 $\mu$m ({\it red}), 4.5 $\mu$m ({\it green}; \citealt{meixner2006}), and MCELS H$\alpha$ ({\it blue}; \citealt{smith1998}) images. The ALMA fields are indicated with dashed circles and labeled.  \label{f:n105}}
\end{figure*}

The LMC provides a unique opportunity to study the physics and chemistry of star formation in an environment which is profoundly different than in the Galaxy.  The elemental abundances of gaseous C, O, and N atoms and the dust-to-gas ratio are lower (e.g., \citealt{dufour1975,dufour1984}; \citealt{koornneef1984}; \citealt{duval2014}), and the intensity of the UV radiation field is higher (10--100 times, but with large variations; e.g., \citealt{browning2003}; \citealt{welty2006}) when compared with the Galactic values. The deficiency of dust (and consequently less shielding) and strong UV radiation field lead to warmer dust temperatures in the LMC (e.g., \citealt{vanloon2010b}).  Gamma-ray observations indicate that the cosmic-ray density in the LMC is a factor of four lower than that measured in the solar neighborhood  (e.g., \citealt{abdo2010}; \citealt{knodlseder2013}).  All of these characteristics of the LMC's environment may have direct consequences (with unclear relative importance) on the formation efficiency and survival of COMs. The formation of COMs requires dust surface chemistry on cold grains and cosmic ray processing of grain mantles (e.g., \citealt{herbst2009}; \citealt{oberg2016})

The LMC is sufficiently close ($50.0\pm1.1$ kpc;  \citealt{pietrzynski2013}) to enable detailed studies on individual stars and protostars. The entire star-forming regions can be imaged relatively easily. Not plagued by distance ambiguities, line-of-sight confusion, and extinction that hamper Galactic studies, the LMC has been subject of varied star formation studies (both photometric and spectroscopic) and has been surveyed at a wide wavelength range offering a rich context for interpreting new observations. The LMC has a history of interacting with both its neighbor---the Small Magellanic Cloud (SMC), another dwarf irregular galaxy with an even lower metallicity than the LMC ($Z_{\rm SMC}$ $\sim$ 0.1--0.2 $Z_{\odot}$; \citealt{russell1992}; \citealt{rolleston2002}), and the Milky Way.  The tidal interactions between the LMC and SMC influence the star formation history in each galaxy (e.g., \citealt{fujimoto1990}; \citealt{bekki2007a}; \citealt{fukui2017}; \citealt{tsuge2019}).

\begin{figure*}[ht!]
\centering
\includegraphics[width=0.49\textwidth]{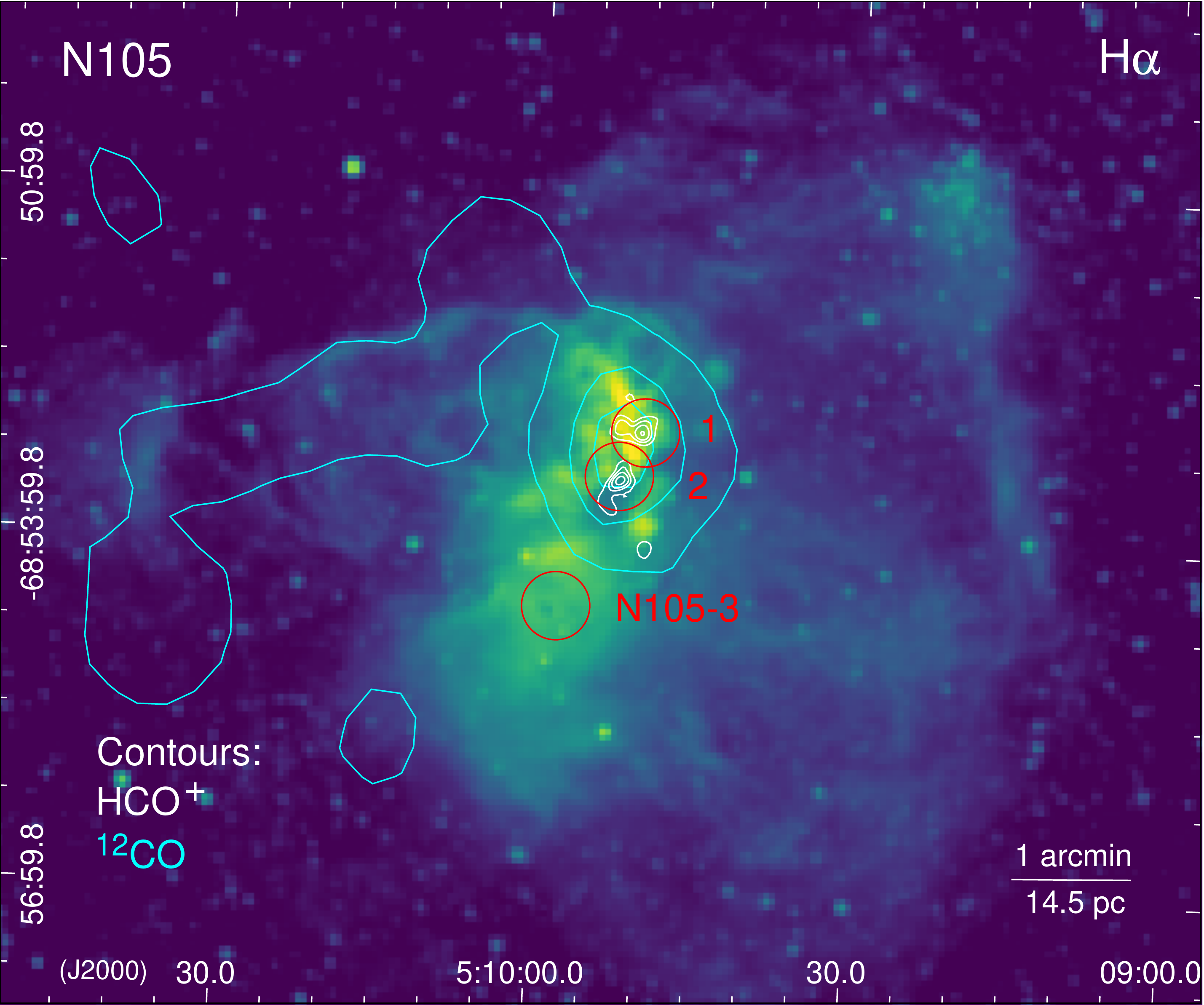}
\includegraphics[width=0.49\textwidth]{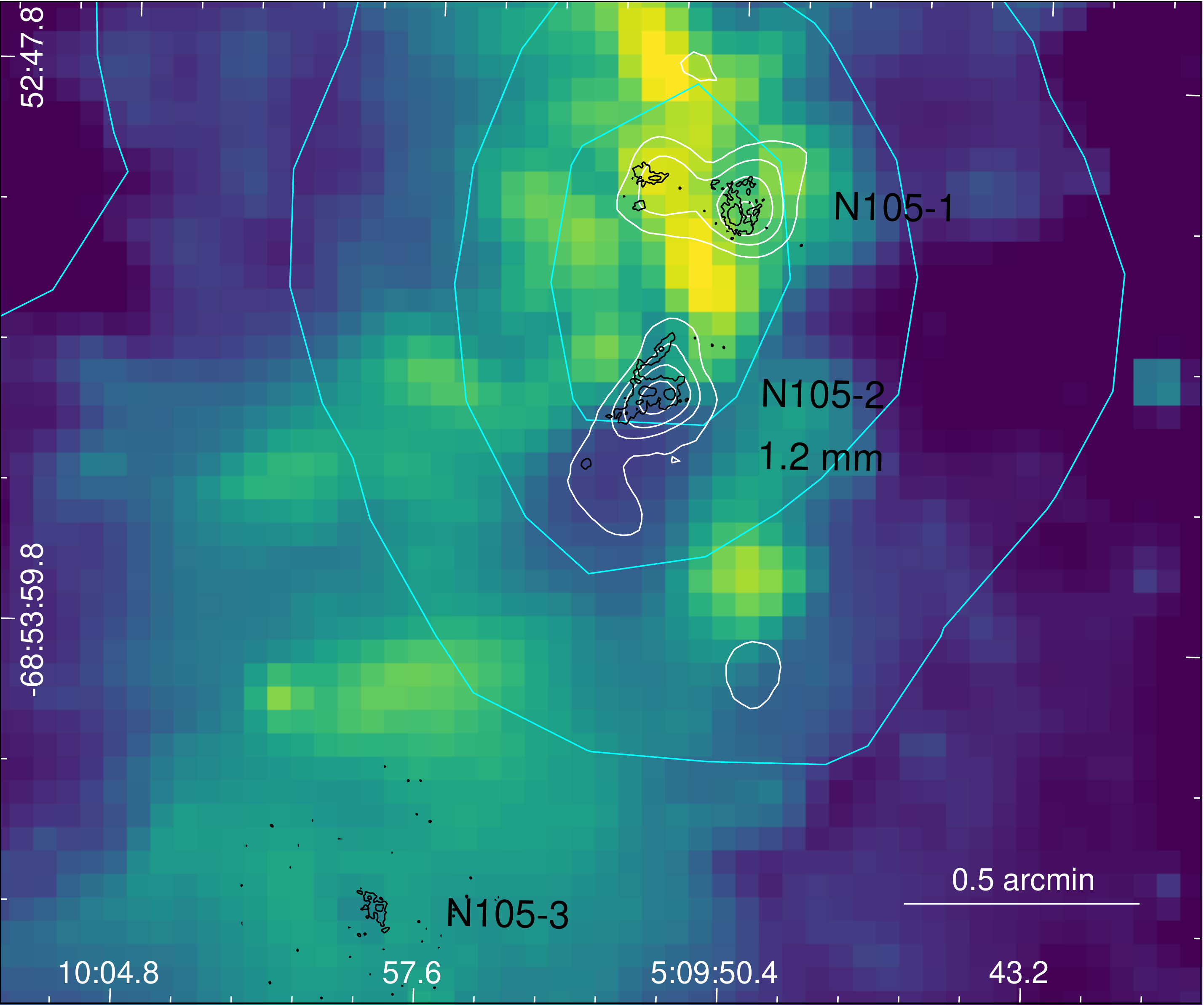}
\caption{The MCELS H$\alpha$ image showing the N\,105 optical nebula ({\it left}) and a zoom-in on the central region of N\,105, referred to in literature as N\,105A ({\it right}). In both images, the cyan contours correspond to the single-dish $^{12}$CO ($J$ = 1--0) emission from the Magellanic Mopra Assessment (MAGMA) survey with an effective angular resolution of $\sim$45$''$, with contour levels of (20, 50, 80)\% of the $^{12}$CO integrated intensity peak of 17.3 K km s$^{-1}$ (Mopra Telescope; \citealt{wong2011}).  The white contours correspond to the HCO$^{+}$ ($J$ = 1--0) emission with contour levels of (10, 30, 50, 80)\% of the HCO$^{+}$ integrated intensity peak of 4.1 Jy beam$^{-1}$ km s$^{-1}$ (ATCA, synthesized beam of $6\rlap.{''}3\times7\rlap.{''}$1; \citealt{seale2012}).  The HCO$^{+}$ observations only covered the molecular cloud traced by the MAGMA survey, i.e., the N\,105--3 field was not observed with ATCA. The three ALMA fields are indicated in red in the left panel, while the ALMA 1.2 mm continuum emission contours are shown in black in the right panel (see Figs.~\ref{f:N105B6cont}).  \label{f:region} }
\end{figure*}

\subsection{Hot Molecular Cores in the LMC} 

Methanol (CH$_3$OH), methyl cyanide (CH$_3$CN), and larger COMs have been found in the LMC toward hot cores (\citealt{sewilo2018}; \citealt{shimonishi2020}):  small ($D\lesssim0.1$ pc), hot ($T_{\rm kin}\gtrsim100$ K), and dense ($n_{\rm H}\gtrsim10^{6-7}$ cm$^{-3}$) regions around forming massive stars where ice mantles have recently been removed from dust grains as a result of thermal evaporation and/or sputtering in shock waves (e.g., \citealt{garay1999}; \citealt{kurtz2000}; \citealt{cesaroni2005}; \citealt{palau2011}).  A typical Galactic hot core has a very rich spectrum at submm wavelengths including lines from many complex organics - the products of interstellar grain-surface chemistry or post-desorption gas chemistry (e.g., \citealt{herbst2009}; \citealt{oberg2016}; \citealt{jorgensen2020}). Methanol has also been detected toward a handful of other locations in the LMC, but outside hot cores (``cold methanol''; see e.g., \citealt{sewilo2019}). 

Observational, theoretical, and laboratory studies indicate that COMs are mainly formed on dust grains through ice chemistry in the young stellar object's (YSO's) envelope which has an inward temperature gradient due to heating from the central protostar (e.g., \citealt{herbst2009}; \citealt{vandishoeck2014}; \citealt{vandishoeck2018}; \citealt{oberg2016}).  During the YSO accretion phase, the composition of the icy grain mantles change as they approach the protostar and eventually sublimate when the grains reach the inner hot core region in its immediate surrounding: (1) Initially, they only contain simple ices that are formed in the molecular cloud phase by a condensation of atoms and molecules from the gas-phase and by subsequent grain surface chemistry (e.g., H$_2$O, CH$_4$, NH$_3$, CO$_2$, H$_2$CO, CH$_3$OH).  First, the ices on the grain surface are formed though hydrogenation (adding H atoms that are the most mobile ice constituents at $T\sim10$~K), and then also through chemical reactions involving CO; (2) When exposed to UV radiation (e.g., cosmic ray interactions with H$_2$), simple ices can partially dissociate into radicals; (3) The radicals become mobile when the temperature increases with decreasing distance from the protostar, and they combine to form new species, including more complex molecules; (4) As the dust grain approaches the central protostar, the temperature becomes high enough for ice mantles to sublimate ($T$ $\sim$ 100--150 K). The molecules released to the gas-phase include simple ices from the original ice mantles, as well as newly formed complex organics.  Gas-phase chemistry following ice sublimation can also contribute to the formation of some COMs \citep{taquet2016}.

YSOs are also associated with jets and outflows at a range of velocities which interact with the envelope and cloud material and produce shocks, enabling shock and hot gas chemistry. Shocks can sublimate or sputter icy grain mantles, releasing ice chemistry products (molecules such as CH$_3$OH and other COMs) into the gas (e.g., \citealt{arce2008}).  At high velocities, shocks can also sputter the grain cores, releasing the Si and S atoms and as a consequence, enhancing the production of Si- and S-bearing species such as SiO, SO$_2$, and SO (e.g., \citealt{schilke1997}; \citealt{gusdorf2008}; \citealt{vandishoeck2018} and references therein).  In addition to jets and outflows, low velocity shocks can be produced in YSOs at the envelope-disk interface where sublimation and sputtering of ices can occur (e.g., \citealt{aota2015}; \citealt{miura2017}).  In summary, the formation of COMs is mainly a result of the chemical processes taking place in icy grain mantles in the protostellar envelope. The ice chemistry products (including COMs) become observable after icy grain mantles are sublimated close to the protostar or sublimated and sputtered in shocks in the jets/outflows or at the envelope-disk interface. Some COMs may be the result of the gas-phase chemistry following ice sublimation.

Prior to the present study, COMs with more than six atoms had only been detected toward two hot cores in the LMC:  A1 and B3 in the star-forming region N\,113 (N\,113 A1 and N\,113 B3; \citealt{sewilo2018,sewilo2019}).  \citet{sewilo2018} reported the detection of methyl formate (HCOOCH$_3$) and dimethyl ether (CH$_3$OCH$_3$), together with their likely parent species CH$_3$OH, with fractional abundances with respect to H$_2$  (corrected for a reduced metallicity in the LMC with respect to the Milky Way) at the lower end, but within the range measured toward Galactic hot cores.  
 
This was a surprising result, because the previous theoretical and observational studies indicated a deficiency of CH$_3$OH in the LMC (e.g., \citealt{acharyya2015};  \citealt{shimonishi2016a, shimonishi2016b};  \citealt{nishimura2016}).  For example, \citet{shimonishi2016b} claimed a hot core detection toward the massive YSO ST11 in the LMC based on the derived physical conditions and the presence of simple molecules connected to the gas chemistry (e.g., SO$_2$), but no CH$_3$OH or other COMs were detected. They concluded that CH$_3$OH is depleted by 2--3 orders of magnitude as compared with Galactic hot cores.  The underabundance of the CH$_3$OH ice and a low detection rate of CH$_3$OH masers were also reported in the LMC (e.g., \citealt{sinclair1992}; \citealt{green2008}; \citealt{shimonishi2016a}). 

The fact that molecules whose formation requires the hydrogenation of CO on grain surfaces were not detected (e.g., CH$_3$OH, HNCO) or underabundant (e.g., H$_2$CO) in ST11 and that the CH$_3$OH ice is underabundant in the LMC YSOs led \citet{shimonishi2016a} to propose a ``warm ice chemistry'' model in which the observed differences between the chemistry of the LMC and Galactic sources are a consequence of the dust being warmer in the LMC due to the strong interstellar radiation field. High dust temperatures in the LMC ($T\gtrsim20$ K)  suppress the hydrogenation of CO on grain surfaces due to the decrease in available hydrogen atoms, leading to inefficient production of CH$_3$OH. At the same time, the model also predicts an enhancement in CO$_2$ production due to the increased mobility of the parent species which explains the increased CO$_2$/H$_2$O ice column density ratio observed toward LMC YSOs (e.g., \citealt{shimonishi2008}; \citealt{oliveira2009,oliveira2011}); alternatively, this increased ratio can be explained by the underabundance of H$_2$O (e.g., \citealt{oliveira2011}). The predictions of the warm ice chemistry model are consistent with astrochemical simulations for the appropriate elemental depletions (\citealt{acharyya2015,acharyya2018}; \citealt{pauly2018}). There is evidence that CH$_3$OH and other complex organics observed toward YSOs (hot cores and outflow shocks) might have formed in cold ($\sim$10 K) molecular cloud phase preceding the onset of star formation (see a discussion in Section~\ref{s:methanol}). 

A picture of a chemically diverse hot core population in the LMC has recently started emerging with the detection of a hot core ST16 exhibiting CH$_3$OH and CH$_3$CN emission, but no larger COMs, and a general underabundance of organic species compared with Galactic hot cores \citep{shimonishi2020}. \citet{shimonishi2020} suggested that LMC hot cores can be divided into ``organic-poor'' and ``organic-rich.'' 
This classification, however, is based on only a handful of objects and needs a verification. 

If confirmed based on a larger sample of LMC hot cores, organic-rich hot cores would be those sources that are associated with larger COMs and have molecular abundances roughly scaled with metallicity (as in N\,113 A1 and B3). In the organic-poor hot cores, the low abundances of organic molecules cannot be explained by the decreased abundance of C and O. No COMs (ST11) or only CH$_3$OH and CH$_3$CN (ST16) are detected. In ST11 and ST16, H$_2$CO, CH$_3$OH, HNCO, CS, H$_2$CS, and SiO are significantly less abundant, while HCO$^{+}$, SO, SO$_2$, and NO are comparable with or more abundant than Galactic hot cores, after being corrected for metallicity.  The organic-poor hot cores are unique to the low-metallicity environment of the LMC. \citet{shimonishi2020} argue that a large chemical diversity of organic molecules seen in the LMC hot cores can be a consequence of the different grain temperature at the initial (ice-forming) stage of star formation. They support their conclusions with astrochemical simulations. The analysis of a larger sample of hot cores in the LMC is required to verify the hot core classification scheme suggested by \citet{shimonishi2020} and get a better understanding of the complex chemistry in the metal-poor environment. 

\begin{deluxetable*}{lccccccc}[ht!]
\centering
\tablecaption{Spectral Cube Parameters\label{t:linedata}}
\tablewidth{0pt}
\tablehead{
\colhead{Field} &
\colhead{RA} &
\colhead{Decl.} &
\colhead{Spectral} &
\colhead{Frequency Range} & 
\colhead{Synth. Beam: ($\Theta_B$, PA)}  &
\multicolumn{2}{c}{Data Cube rms\tablenotemark{\footnotesize a}} \\
\cline{7-8}
\colhead{} & 
\colhead{($^{\rm h}$ $^{\rm m}$ $^{\rm s}$)} &
\colhead{($^{\rm \circ}$ $'$ $''$)} &
\colhead{window} &
\colhead{(GHz)} &
\colhead{($'' \times ''$, $^{\circ}$)} &
\colhead{(mJy beam$^{-1}$)}  &
\colhead{(K)}
}
\startdata
N\,105--1 & 05:09:50.47 & $-$68:53:04.9 & 242 GHz & 241.27653--243.14971 & $0.538\times0.497$, 36.8 & 1.97 & 0.15 \\ 
                                                       & & & 245 GHz & 243.66769--245.54088 & $0.533\times0.492$, 35.0 & 1.88 & 0.15 \\ 
                                                       & & & 258 GHz & 256.71495--258.58814 & $0.511\times0.470$, 38.5 & 2.05 & 0.16 \\ 
                                                       & & & 260 GHz & 258.54806--260.42125 & $0.506\times0.467$, 36.7 & 2.28 & 0.18 \\ 
N\,105--2 & 05:09:52.37 & $-$68:53:26.6 & 242 GHz & 241.27653--243.14971 & $0.539\times0.497$, 37.8 & 1.97 & 0.15 \\ 
                                                       & & & 245 GHz & 243.66769--245.54088 & $0.534\times0.492$, 35.5 & 1.87 & 0.15 \\ 
                                                       & & & 258 GHz & 256.71495--258.58814 & $0.510\times0.471$, 38.7 & 2.05 & 0.16 \\  
                                                       & & & 260 GHz & 258.54806--260.42125 & $0.507\times0.467$, 37.0 & 2.25 & 0.17 \\ 
N\,105--3 & 05:09:58.66 & $-$68:54:34.1 & 242 GHz & 241.27653--243.14971 & $0.540\times0.496$, 39.1 & 1.97 & 0.15 \\ 
                                                       & & & 245 GHz & 243.66769--245.54088 & $0.534\times0.492$, 36.1 & 1.88 & 0.15 \\ 
                                                       & & & 258 GHz & 256.71495--258.58814 & $0.510\times0.471$, 39.0 & 2.05 & 0.16 \\ 
                                                       & & & 260 GHz & 258.54806--260.42125 & $0.507\times0.467$, 36.8 & 2.28 & 0.17 \\ 
\enddata
\tablenotetext{a}{The rms noise per 0.56 km s$^{-1}$ channel estimated with the CASA task {\sc imstat} in line-free channels.}
\end{deluxetable*}

\subsection{The N\,105 Star-Forming Region}

In this paper, we report the results of our observations of three fields in the star-forming region N\,105 in the LMC with ALMA which include a detection of two hot cores that increase a previously known very small sample of four hot cores in the LMC. The LHA~120--N\,105 (hereafter N\,105, \citealt{henize1956}; or DEM\,L86, \citealt{davies1976}) nebula is the star-forming region located at the western edge of the LMC bar (e.g., \citealt{ambrocio1998}).  The H$\alpha$ image of N\,105 reveals a bright central region (referred to in literature as N\,105A) surrounded by a faint extended emission (see Figs.~\ref{f:n105} and \ref{f:region}).  A sparse cluster NGC\,1858 (e.g., \citealt{bica1996}) with age estimates in a range 8--17 Myr (\citealt{vallenari1994}; \citealt{alcaino1986}) and an associated OB association LH\,31 (e.g., \citealt{lucke1970}) are embedded within N\,105A.  LH\,31 contains 18 OB stars and two Wolf-Rayet stars, and coincides with the strongest X-ray emission in the region (e.g., \citealt{vallenari1994}; \citealt{dunne2001}).  Despite the presence of the OB association, the dense cloud N\,105A shows little evidence for feedback from massive stars (e.g., \citealt{ambrocio1998}; \citealt{oliveira2006}).  The N\,105 optical nebula is associated with the thermal radio continuum source MC\,23 or B0510--6857 (e.g., \citealt{mcgee1972}, \citealt{ellingsen1994}, \citealt{filipovic1998}). 

On-going star formation in N\,105A is traced by H$_2$O (e.g., \citealt{scalise1982}; \citealt{whiteoak1983}; \citealt{lazendic2002}; \citealt{oliveira2006}; \citealt{ellingsen2010}), OH (e.g., \citealt{haynes1981}; \citealt{brooks1997}), and CH$_3$OH masers (e.g., \citealt{green2008}; \citealt{ellingsen2010}), ultracompact (UC) H\,{\sc ii} regions (\citealt{indebetouw2004}), and YSOs. About 40 YSOs have been identified based on the {\it Spitzer} Space Telescope (3.6--70 $\mu$m; \citealt{carlson2012} and references therein) and the {\it Herschel} Space Observatory (100--500 $\mu$m; \citealt{sewilo2010}; \citealt{seale2014}) data within the H$\alpha$ nebula.  The {\it Spitzer} images shown in Fig.~\ref{f:n105} reveal a complex structure of the dust and Polycyclic Aromatic Hydrocarbon (PAH) emission in N\,105, with the brightest emission coinciding with the position of the most massive YSOs in N\,105A. 
 
Active star-forming sites in N\,105 coincide with the position of the molecular cloud detected in single-dish observations of $^{12}$CO and $^{13}$CO (1--0), tracing gas densities of $\sim$10$^{2}$--10$^{3}$ cm$^{-3}$ (e.g., \citealt{israel1993}, HPBW$\sim$45$''$ at 115 GHz; \citealt{chin1997}, 45$''$; \citealt{fukui1999} and \citealt{fukui2008}, 2$\rlap.{'}$6;  \citealt{wong2011}, 45$''$ -- see Fig.~\ref{f:region}).  High-resolution ($\sim$6$\rlap.{''}$7) interferometric observations of HCN and HCO$^{+}$ (1--0)  toward the peak of the CO emission with the Australia Telescope Compact Array (ATCA) revealed the densest gas in N\,105 (\citealt{seale2012}; see Fig.~\ref{f:region}).  Two of the three fields we observed with ALMA are located in this region and are associated with H$_2$O and OH masers, while the third field covers a lower density region to the south and is associated with a CH$_3$OH maser. 

The paper is organized as follows: In Section~\ref{s:data}, we describe the observations and the archival data used in the paper. In Section~\ref{s:continuum}--\ref{s:nh2}, we present the analysis of the 1.2 mm continuum and spectral line data. In Section~\ref{s:IR}, we investigate the physical characteristics of the observed fields and chemical properties of selected sources in the N\,105 star-forming region based on the data ranging from the optical to radio wavelengths.  The discussion is presented in Section~\ref{s:discussion}, while in Section~\ref{s:summary}, we provide the summary and conclusions of our study.

\begin{deluxetable*}{lccccc}
\centering
\tablecaption{ALMA Pointings and 242.4 GHz / 1.2 mm continuum Image Parameters\label{t:contdata}}
\tablewidth{0pt}
\tablehead{
\colhead{Field} &
\colhead{RA} &
\colhead{Decl.} &
\colhead{Synth. Beam: ($\Theta_B$, PA)}  &
\multicolumn{2}{c}{Image rms} \\   
\cline{5-6}
\colhead{} & 
\colhead{($^{\rm h}$ $^{\rm m}$ $^{\rm s}$)} &
\colhead{($^{\rm \circ}$ $'$ $''$)} &
\colhead{($'' \times ''$, $^{\circ}$)} &
\colhead{($\mu$Jy beam$^{-1}$)} &
\colhead{(mK)} 
}
\startdata
N\,105--1 & 05:09:50.47 & $-$68:53:04.9 & $0.506\times0.471$, 37.2 & 69 & 6.0 \\   
N\,105--2 & 05:09:52.37 & $-$68:53:26.6 & $0.510\times0.473$, 37.4 & 51 & 4.4 \\   
N\,105--3 & 05:09:58.66 & $-$68:54:34.1 & $0.507\times0.470$, 38.1 & 27 & 2.4 \\   
\enddata
\end{deluxetable*}

\section{The Data}
\label{s:data}
 
The analysis presented in this paper is primarily based on the ALMA Cycle 7 Band 6 observations (Section~\ref{s:ALMAobservations}). However,  we also present the results of near-infrared (near-IR) spectroscopic observations with the {\it Very Large Telescope}/$K$-band Multi-Object Spectrograph ({\it VLT}/KMOS) for three sources located in the ALMA Cycle 7 fields (Section~\ref{s:KMOSobservations}).

\subsection{Source Selection and ALMA Observations}
\label{s:ALMAobservations}

We selected six fields in the LMC for Cycle 7 observations that have common characteristics with those hosting N\,113 A1 and B3, at that time, the only known LMC hot cores with COMs: they are associated with massive {\it Spitzer} YSOs, H$_2$O/OH masers, and SO emission, a well-known hot core and shock tracer (e.g., \citealt{chernin1994}; \citealt{mookerjea2007}).  We also observed an additional field centered on a Stage 0/I protostar (e.g., \citealt{sewilo2010}) associated with one of four 6.67 GHz and the only 12.2 GHz CH$_3$OH maser known in the LMC \citep{sinclair1992}, making it a good hot core candidate. In total, seven fields were observed with the ALMA 12m Array in Band 6 (with a single pointing each) as part of the Cycle 7 project 2019.1.01720.S (PI M. Sewi{\l}o). 

The SO 3$_2$--2$_1$ line emission toward N\,113 A1 and B3 hot cores was serendipitously detected in our ALMA Cycle 3 observations  (2015.1.01388.S, PI M. Sewi{\l}o; see also \citealt{sewilo2018}).  Enhanced SO emission can occur in hot cores following reactions S $+$ OH and O $+$ SH  where the radicals and atoms are produced from the gas-phase destruction of H$_2$O and H$_2$S molecules evaporated/sputtered from ices (e.g., \citealt{charnley1997}).  

A similar Band 3 correlator setup as for N\,113 that covered the SO line was used in an unrelated project targeting massive YSOs in the LMC (2017.1.00093.S, PI T. Onishi), providing us with an opportunity to search for sources with a serendipitous SO detection.  We have identified four Band 3 fields with SO detections and associated with masers (three with H$_2$O masers and one with an OH maser; e.g., \citealt{ellingsen2010}; J. Ott, {\it priv. comm.}), resembling  the A1 and B3 hot cores in N\,113 which are also associated with masers.  All the Band 3 fields were observed with the same setup, resulting in an ALMA synthesized beam of $\sim$2$\rlap.{''}$13 $\times$ 1$\rlap.{''}$57 and a channel width of 2.96 km s$^{-1}$.  These observations also include the (1--0) transitions of  $^{13}$CO and C$^{18}$O, CS (2--1), and the 3 mm continuum; all four Cycle 5 Band 3 fields are associated with dense gas tracers (C$^{18}$O and CS).  Six out of seven fields included in our  Cycle 7 Band 6 observations are centered on regions with SO emission and H$_2$O/OH masers within these four Cycle 5 Band 3 fields and thus are most likely to host hot cores. 

Here, we present the results for three fields observed in Cycle 7, all located in the N\,105 star-forming region. Two of the fields are associated with SO emission and H$_2$O/OH masers; we have dubbed them `N\,105--1' and `N\,105--2'.  The third field is associated with methanol masers and we will refer to it as `N\,105--3'; no prior ALMA observations are available for N\,105--3.   All three ALMA fields in N\,105 hosting hot core candidates are shown in Fig. ~\ref{f:n105} and their positions are listed in Table~\ref{t:contdata}.  

The observations of all fields were executed twice on October 21, 2019 with 43 antennas and baselines from 15 m to 783 m.  The (bandpass, flux, phase) calibrators were (J0519$-$4546, J0519$-$4546, J0440$-$6952) and (J0538$-$4405, J0538$-$4405, J0511$-$6806) for the first and second run, respectively.  The targets were observed again on October 23, 2019 with 43 antennas and baselines from 15 m to 782 m.  The calibrators were the same as for the first run on October 21.  The total on-source integration time for all seven fields was 91.8 min. for all three executions. The maximum recoverable scale calculated from the 5$^{th}$ percentile baseline length for the final data set combining all executions varied between 5$\rlap.{''}$6 and 5$\rlap.{''}$2 for a sky frequency range covered by our observations ($\sim$241.3--260.4 GHz). The spectral setup included four 1875 MHz spectral windows centered on frequencies of 242.4 GHz, 244.8 GHz, 257.85 GHz, and 259.7 GHz, each with 3840 channels, providing a spectral resolution of   1.21--1.13 km~s$^{-1}$. Henceforth, we will refer to the spectral windows as the ``242~GHz / 245~GHz / 258~GHz / 260~GHz spectral window.'' 

The data were calibrated and imaged with version 5.6.1-8 of the ALMA pipeline in CASA (Common Astronomy Software Applications; \citealt{mcmullin2007}).  The continuum in each spectral window was identified and subtracted before cube imaging. The CASA task \texttt{tclean} was used for imaging using the Hogbom deconvolver, standard gridder, Briggs weighting with a robust parameter of 0.5, and masking using the `auto-multithresh' algorithm.   The spectral cubes have a cell size of  $0\rlap.{''}092 \times 0\rlap.{''}092 \times 0.56$ km s$^{-1}$. Additional information on the data cubes is included in Table~\ref{t:linedata}.  The 242.4 GHz (1.2 mm) continuum image parameters are listed in Table~\ref{t:contdata}.  All the images have been corrected for primary beam attenuation.

\subsection{VLT/KMOS Near-Infrared Spectroscopy}
\label{s:KMOSobservations}

Three near-IR sources in the ALMA Cycle 7 fields in N\,105 were observed with the {\it VLT}/KMOS \citep{sharples2013} as part of a survey of  YSO candidates under program 0101.C-0856(A) (PI J. L. Ward) using the $H+K$ grating with a spectral resolving power of 2000 and a spatial pixel scale of 0$\rlap.{''}$2.  The observations took place on the night of August 28--29, 2018 with seeing ranging from 0$\rlap.{''}$55 to 1$\rlap.{''}$66. The measured FWHM of one of the sources in N\,105 (ID 558354728325) in the {\it K}-band is 8.2 pixels, corresponding to approximately 1$\rlap.{''}$6. KMOS is able to perform the Integral Field Spectroscopy in the near-IR bands for 24 targets simultaneously using 24 configurable arms. The KMOS observations were carried out using a standard nod-to-sky procedure with an integration time of 150\,s, four detector integration times (DITs) and three dither positions, yielding a total on-source integration time of 1800\,s. Telluric absorption correction, response curve correction, and absolute flux calibration were carried out using observations of telluric standard stars using three integral-field units (IFUs). The data were reduced with the standard {\it VLT}/KMOS pipeline using the {\sc esoreflex} data reduction package \citep{davies2013}. 

The $K$-band continuum image is produced by integrating over a third-order polynomial fit to the data for every spatial pixel (spaxel) over the spectral range 2.028--2.290\,$\mu$m. The Br$\gamma$ and H$_{2}$ line emission images are produced by fitting a Gaussian profile to the emission lines at every position in the image. Each IFU has a square field of view of 2$\rlap.{''}8\times2\rlap.{''}$8.

The measured KMOS $H$- and $K$-band fluxes are found to be significantly lower than those determined by the near-IR surveys covering this region, 2MASS $JHK_S$ and the VISTA $YJK_S$ survey of the Magellanic Clouds system (VMC; \citealt{cioni2011}), on average by a factor of 24. Thus, the  KMOS fluxes are not reliable enough to be used directly.  Instead, for the subsequent analysis, we have scaled the extracted spectra so that the sum of the spectral region from 2.028--2.295 $\mu$m is consistent with the $K$-band magnitude of the corresponding point source from the VMC survey catalog. The publicly available VMC catalog was queried using the VISTA Science Archive (VSA\footnote{http://horus.roe.ac.uk/vsa}; \citealt{cross2012}) to obtain aperture photometry in $YJK_{\rm S}$ (VMCDR4). For reference, N\,105 is located in VMC tile LMC\,6\_4.

To improve astrometry of the KMOS images,  we computed the cross-correlation functions for all the KMOS fields with the VMC survey and calculated the RA and Dec values that the KMOS data should be shifted by to match the VMC data. First, the KMOS {\it K}-band continuum images were flipped, rotated by 4.9918 degrees, and rescaled to match the orientation and pixel scale of the VMC data using the {\sc WCSTools} package. The {\sc CORREL\_IMAGES} function in IDL was then used to compute the 2D cross-correlation function between the KMOS and VMC images. The 2D Gaussian profiles were fitted to the cross-correlation functions giving the most probable offsets (the centroid position of the Gaussian) in the VMC survey coordinate frame. Flipping and rotating the offsets between the KMOS and VMC images then converts them from the VMC coordinate system into RA and Dec. There is still a small shift between the KMOS and VMC data; however, this shift is sub-pixel and thus not significant and the association between the VMC and KMOS sources can be established reliably. The precision of the KMOS astrometry is limited by the relatively poor resolution of the KMOS data due to seeing. The results of the {\it VLT}/KMOS observations are discussed in Section~\ref{s:KMOSresults}.

\begin{figure}[ht!]
\includegraphics[trim = 0mm 14mm 0mm 20mm, clip, width=0.48\textwidth]{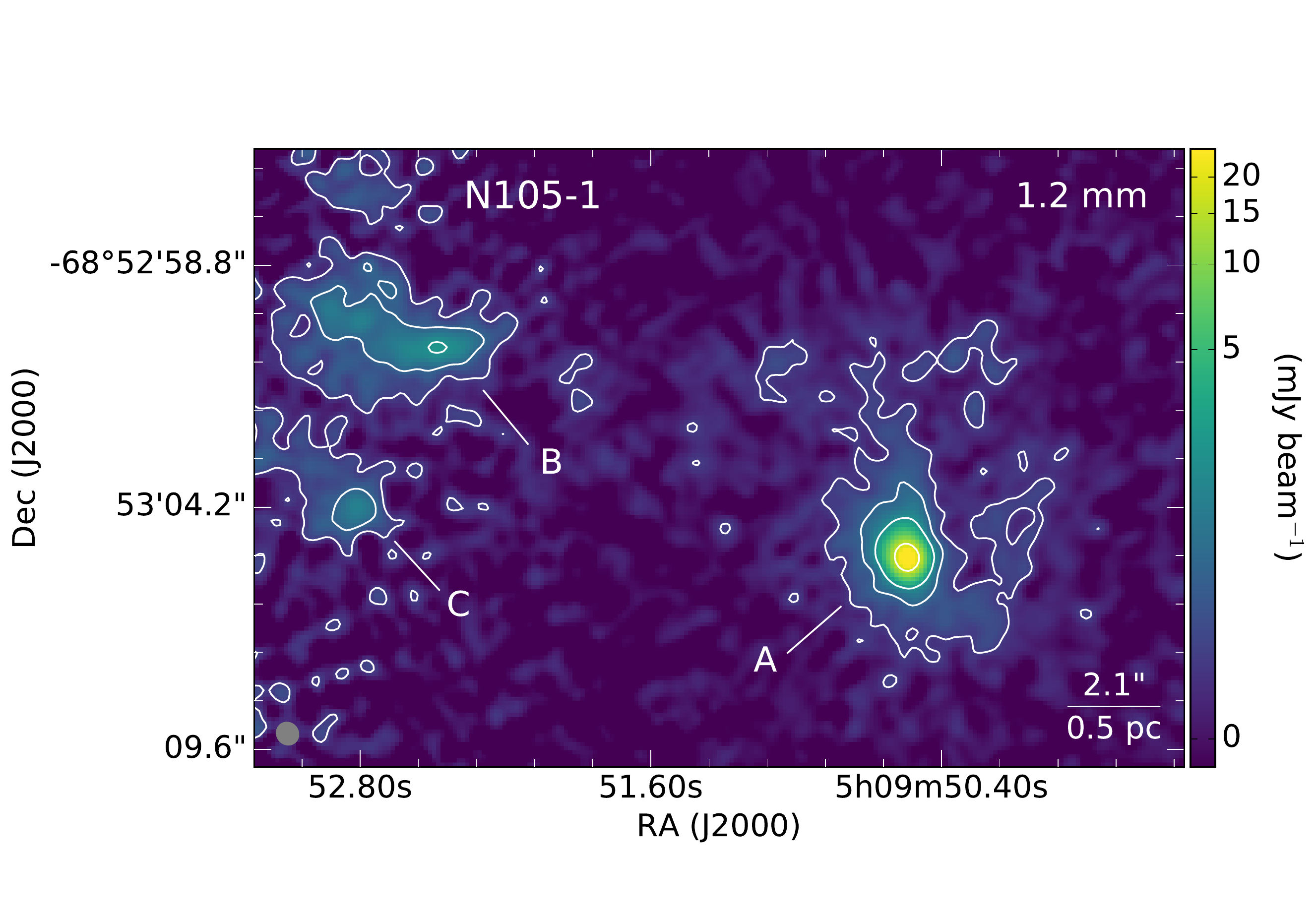}
\includegraphics[width=0.47\textwidth]{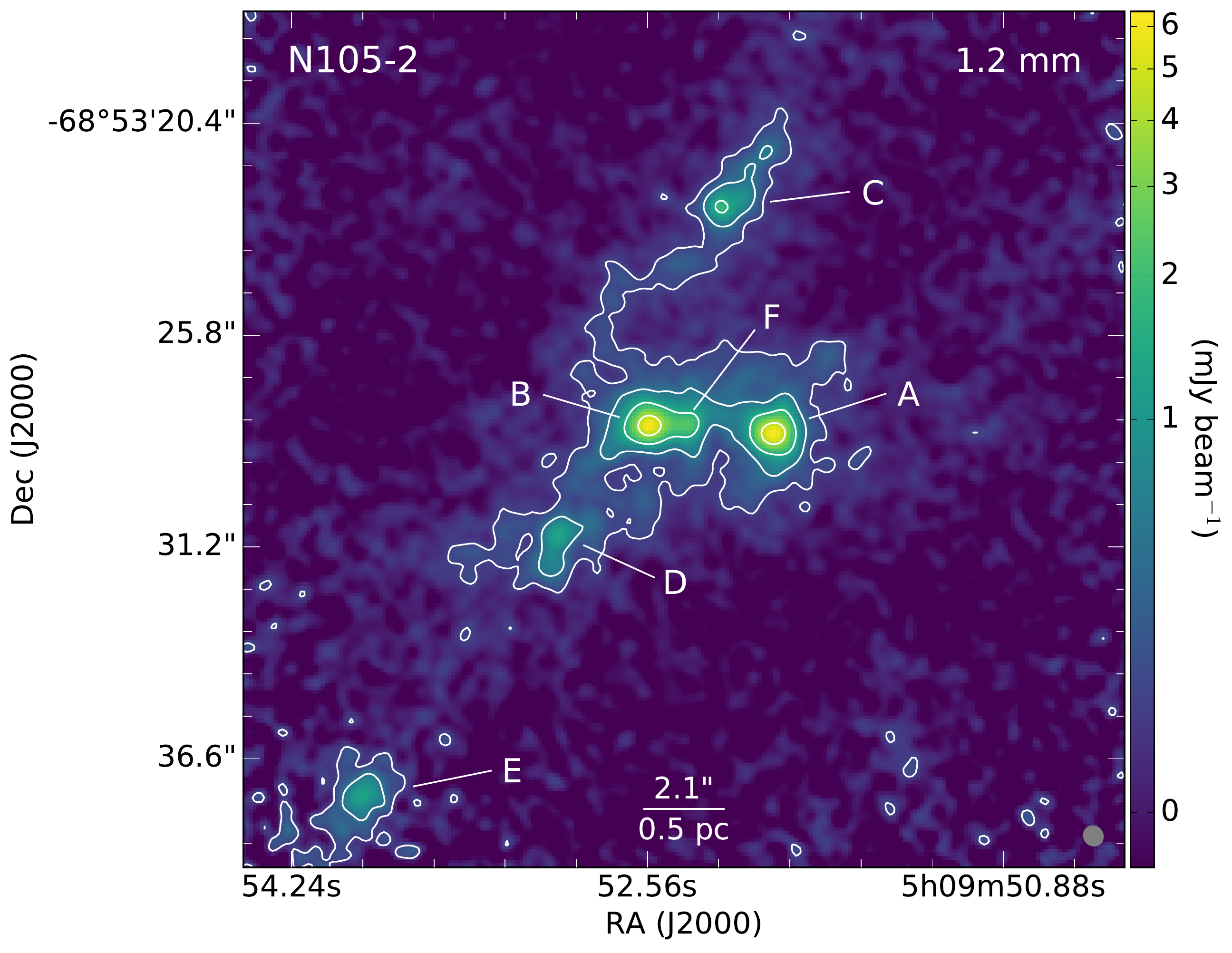}
\includegraphics[width=0.48\textwidth]{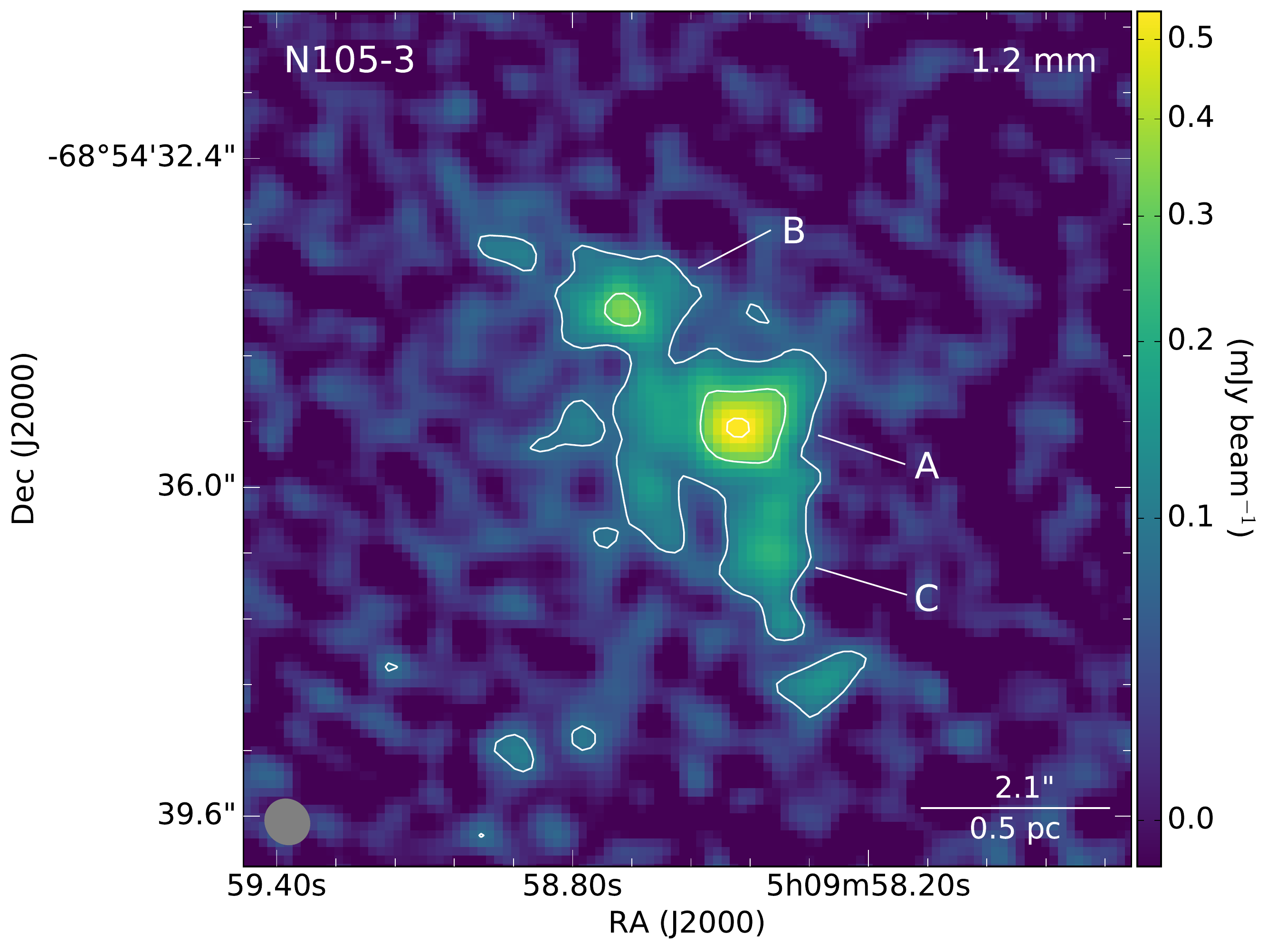}
\caption{The 1.2 mm continuum images of N\,105--1 ({\it top}),  N\,105--2 ({\it center}), and N\,105--3 ({\it bottom}).  The 1.2 mm continuum contour levels are (3, 10, 30, 250) $\times$  the image rms noise ($\sigma$) of 6.9$\times$10$^{-5}$ Jy beam$^{-1}$ for N\,105--1,  (3, 10, 30, 80) $\times$ 5.1$\times$10$^{-5}$ Jy beam$^{-1}$ for N\,105--2, and (3, 10, 20) $\times$ 2.7$\times$10$^{-5}$ Jy beam$^{-1}$ for N\,105--3.   The 1.2 mm continuum sources identified in this paper are labelled.  Sources N\,105--1 B, N\,105--1 C, and N\,105--2 E are located at the edge of the corresponding field (the ALMA field of view).  The size of the ALMA synthesized beam is shown in the lower left ({\it top} and {\it bottom}) or lower right ({\it center}) corner. \label{f:N105B6cont}}
\end{figure}

\section{1.2 mm Continuum Emission and Source Identification}
\label{s:continuum}

Figure~\ref{f:N105B6cont} shows the 1.2 mm continuum images of N\,105--1, N\,105--2,  and N\,105--3.  Each field contains multiple continuum components. We have assigned the identification letters (A, B, C, etc.) to all the 1.2 mm continuum sources associated with the molecular or ionized gas emission peaks in the order of decreasing continuum peak intensity. We will refer to individual sources by providing the field name followed by the letter indicating the source name within this field (e.g., N\,105--2\,A is referred to as 2\,A). We have identified an additional continuum peak which likely is a separate source, but it is blended with N\,105--2\,B in our images; we dubbed it 2\,F. The continuum signal-to-noise ratio is larger than ten for all but one source; 3\,C is an 8$\sigma$ detection.  

We have inspected the ATCA 4.8 GHz (6 cm; synthesized beam: $2\rlap.{''}19\times1\rlap.{''}70$) and 8.6 GHz (3 cm; $1\rlap.{''}82\times1\rlap.{''}24$) images of N\,105 presented in \citet{indebetouw2004} covering N\,105--1 and N\,105--2 to check if any of the ALMA 1.2 mm continuum sources in these fields are associated with the radio emission and thus might need a correction for a contribution from the free-free emission to the mm-wave continuum emission. Three of the four ATCA radio sources detected by \citet{indebetouw2004} in N\,105 are located in regions observed with ALMA (see Fig.~\ref{f:radio}).  B0510$-$6857\,W, the brightest radio source with 4.8 GHz / 6 cm and 8.6 GHz / 3 cm flux densities of $26\pm1$ mJy and $39\pm1$ mJy, respectively, corresponds to N\,105--1\,A \citep{indebetouw2004}. The ATCA source B0510$-$6857\,E lies just to the east of N\,105--1\,B, while B0510$-$6857\,S is located between N\,105--2\,A, 2\,B, and 2\,C (see also Section~\ref{s:IR}).  

N\,105--1\,A requires a correction for the contribution from the free-free emission to its Band 6 continuum emission. We assume that the dust thermal emission and the free-free emission from ionized gas are the dominant sources of the 242.4 GHz continuum emission and estimate their relative contributions in two ways: by extrapolating the 4.8 GHz and 8.6 GHz flux densities to higher frequencies and by analyzing the data from two mm-wave bands following the method described in \citet{brunetti2019}. We have estimated that $\sim$35\% of the 242.4 GHz continuum emission is free-free using the first method under the assumption that the free-free emission becomes optically thin at frequencies higher than 8.6 GHz. Flux densities were measured on the images with common beam and pixel sizes. 

To estimate the relative contributions of the dust and free-free emission to the 242.4 GHz continuum emission using the method outlined in \citet{brunetti2019}, we have utilized the 111.5 GHz continuum image from the Cycle 5 project 2017.1.00093.S (see Section~\ref{s:data}).  The 111.5 GHz continuum image was made using the 12m data only and has a synthesized beam and sensitivity of $2\rlap.{''}27\times1\rlap.{''}66$ and $2.7\times10^{-4}$ Jy beam$^{-1}$, respectively.  We have combined the 111.5 GHz and 242.4 GHz flux densities using Eq. 4 in \citet{brunetti2019} assuming the dust opacity spectral index $\beta$ of 1.7 for N\,105 (\citealt{gordon2014}: the mean value calculated from pixels in the dust opacity spectral index map covering N\,105; we have used the expectation values (`exp') from the Broken Emissivity Law Model, BEMBB) to estimate the dust-only flux density in Band 6. The resulting dust and free-free emission contributions to the 242.4 GHz continuum emission are $\sim$45\% and $\sim$55\%, respectively.  For $\beta=1$/$\beta=2$, the free-free contribution would be $\sim$54\%/$\sim$56\%.

The estimated contribution of the free-free to the 242.4 GHz continuum emission for 1\,A ranges from $\sim$35\% to $\sim$55\%.  The lower value calculated by extrapolating the cm-wave flux densities to higher frequencies may be underestimated if the turnover frequency (the frequency where the free-free emission becomes optically thin) is higher than 8.6 GHz for N\,105--1\,A. N\,105--1\,A is likely at the early UC H\,{\sc ii} stage, if not at an earlier hypercompact (HC) H\,{\sc ii} region stage (e.g., \citealt{kurtz2002}; \citealt{kurtz2005}; \citealt{sewilo2004}), and has a rising spectrum from 4.8 GHz to 8.6 GHz with a spectral index $\alpha=+0.6$ ($S_{\nu}\propto\nu^{\alpha}$, where $S_{\nu}$ is a flux density at a frequency $\nu$).  It would not be unexpected if its spectrum continues to rise to higher frequencies (\citealt{yang2019,yang2021} and references therein).  Considering these uncertainties, we assume that half of the continuum emission at 242.4 GHz is free-free. The correction is applied to the continuum data to calculate H$_2$ column densities and masses as described in Section~\ref{s:nh2}.

While there is no radio emission peak coinciding with 1\,B and 1\,C, the ATCA images reveal the presence of the faint extended emission at the location of these sources, thus a small contamination of the 1.2 mm continuum emission with the free-free emission is possible. 

No high-resolution cm-wave image covering N\,105--3 is available; however, there is no indication of the presence of the significant ionized gas emission (no H  recombination lines have been detected and similarly to N\,105--2, the field lies in the H$\alpha$-dark region). Therefore, we expect the 1.2 mm emission detected toward N\,105--3 to be the thermal emission from dust.

\subsection{Association with YSOs and Masers} 
\label{s:ysos}

Each of our ALMA fields contains high-mass YSO candidates identified based on the {\it Spitzer}'s 3.6--70 $\mu$m data from the LMC-wide ``{\it Spitzer} Surveying the Agents of Galaxy Evolution'' (SAGE, \citealt{meixner2006}; \citealt{sage}) survey (e.g., \citealt{whitney2008}; \citealt{gruendl2009}; \citealt{carlson2012}).  {\it Spitzer} is mostly sensitive to Stage I YSOs with disks and envelopes and some more evolved Stage II YSOs with disks and remnant or no envelopes. Subsets of YSO candidates were followed-up with near- to far-IR spectroscopic observations which confirmed their nature and allowed for investigating their physical and chemical characteristics (e.g., \citealt{seale2009}; \citealt{oliveira2009}; \citealt{sewilo2010}; \citealt{carlson2012};  \citealt{ward2016}; \citealt{jones2017}; and \citealt{oliveira2019}).  

Four out of six YSO candidates in the ALMA fields of view in N\,105 were confirmed spectroscopically as bona fide YSOs by \citet{seale2009} using the {\it Spitzer} Infrared Spectrograph (IRS) observations (5--37 $\mu$m).  Two sources were classified as `Group P' and another two as `Group PE'  YSOs.  Both Group P and PE sources show strong PAH emission features. More evolved Group PE sources also show strong fine-structure lines such as [S\,{\sc IV}] 10.5 $\mu$m, [Ne\,{\sc II}] 12.8 $\mu$m, [Ne\,{\sc III}]  15.5 $\mu$m,  [S\,{\sc III}] 18.7 $\mu$m  and  33.5 $\mu$m,  and  [S\,{\sc III}] 34.8 $\mu$m. The sources from both groups may show some absorption from silicates, particularly at 10 $\mu$m; the silicate absorption features are difficult to identify unambiguously in the presence of strong PAH emission features at 6.2, 7.7, 8.6, and 11.3 $\mu$m.  Group P and PE sources can also exhibit the CO$_2$ 15.2 $\mu$m ice absorption feature in the {\it Spitzer}/IRS spectra (\citealt{seale2011}; see also Section~\ref{s:co2}). 

Below we provide a more detailed discussion on YSO candidates, spectroscopically confirmed YSOs, and masers (H$_2$O, OH, CH$_3$OH) in individual ALMA fields in N\,105.

\begin{figure}
\centering
\includegraphics[width=0.48\textwidth]{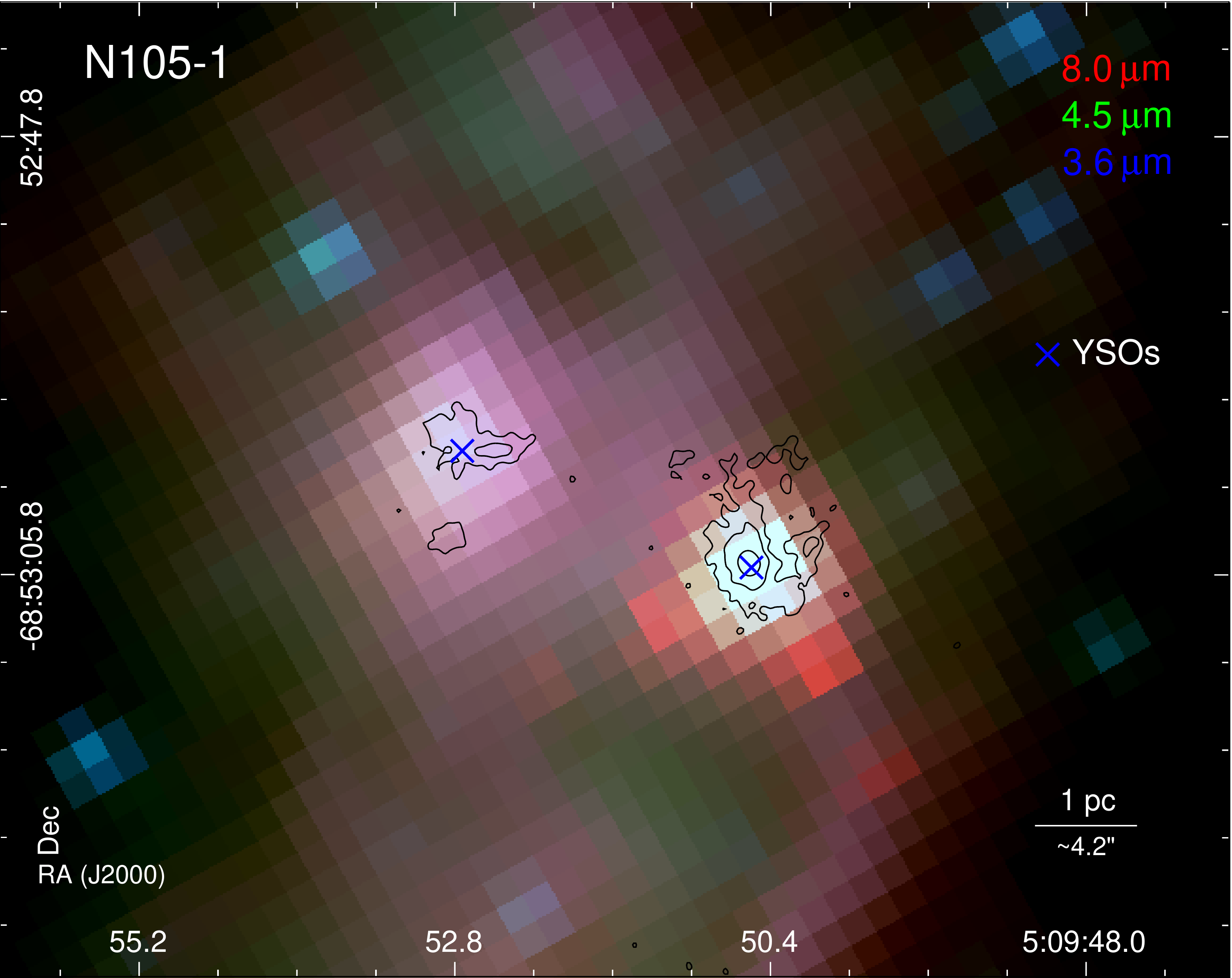}
\includegraphics[width=0.48\textwidth]{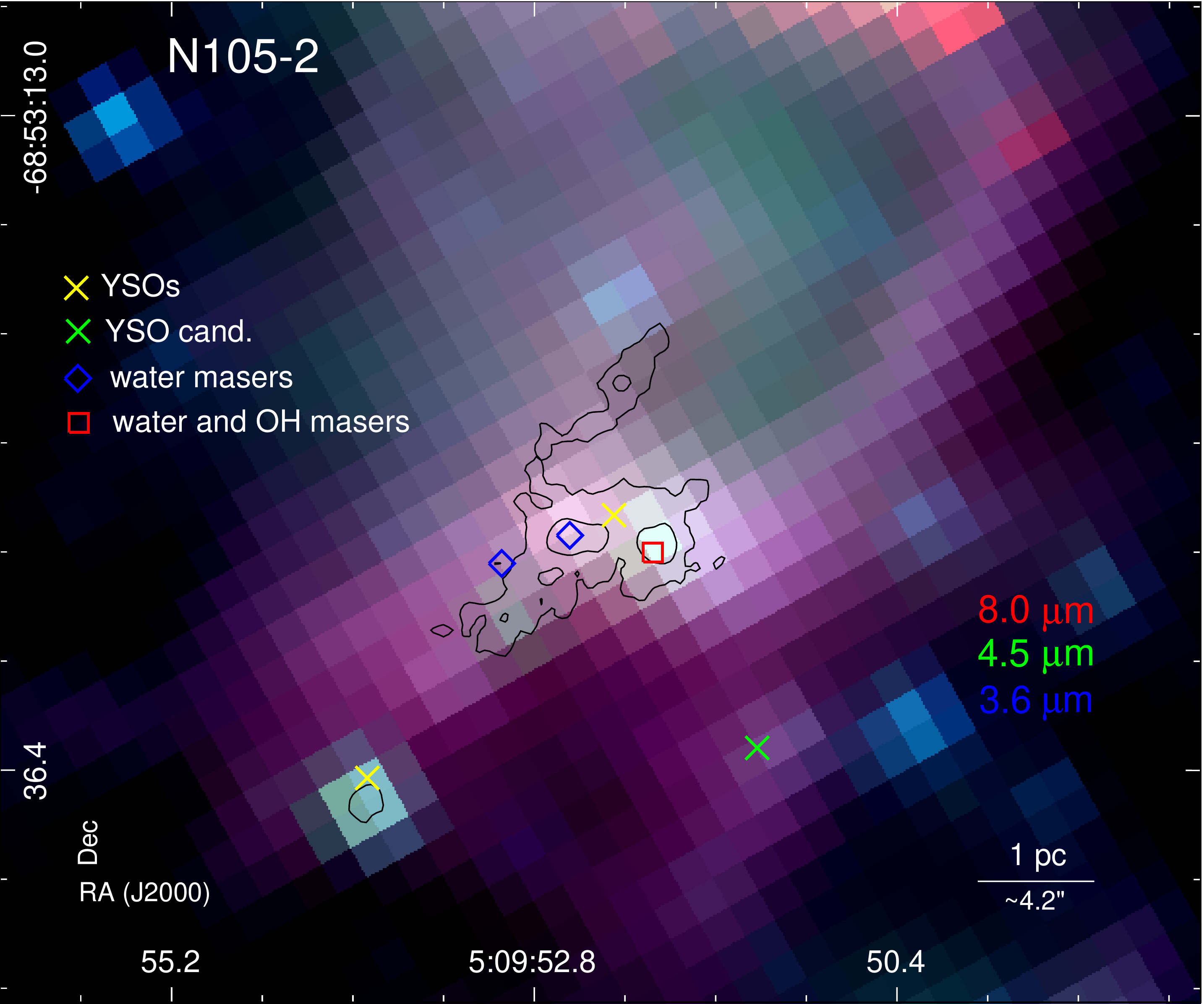}
\includegraphics[width=0.48\textwidth]{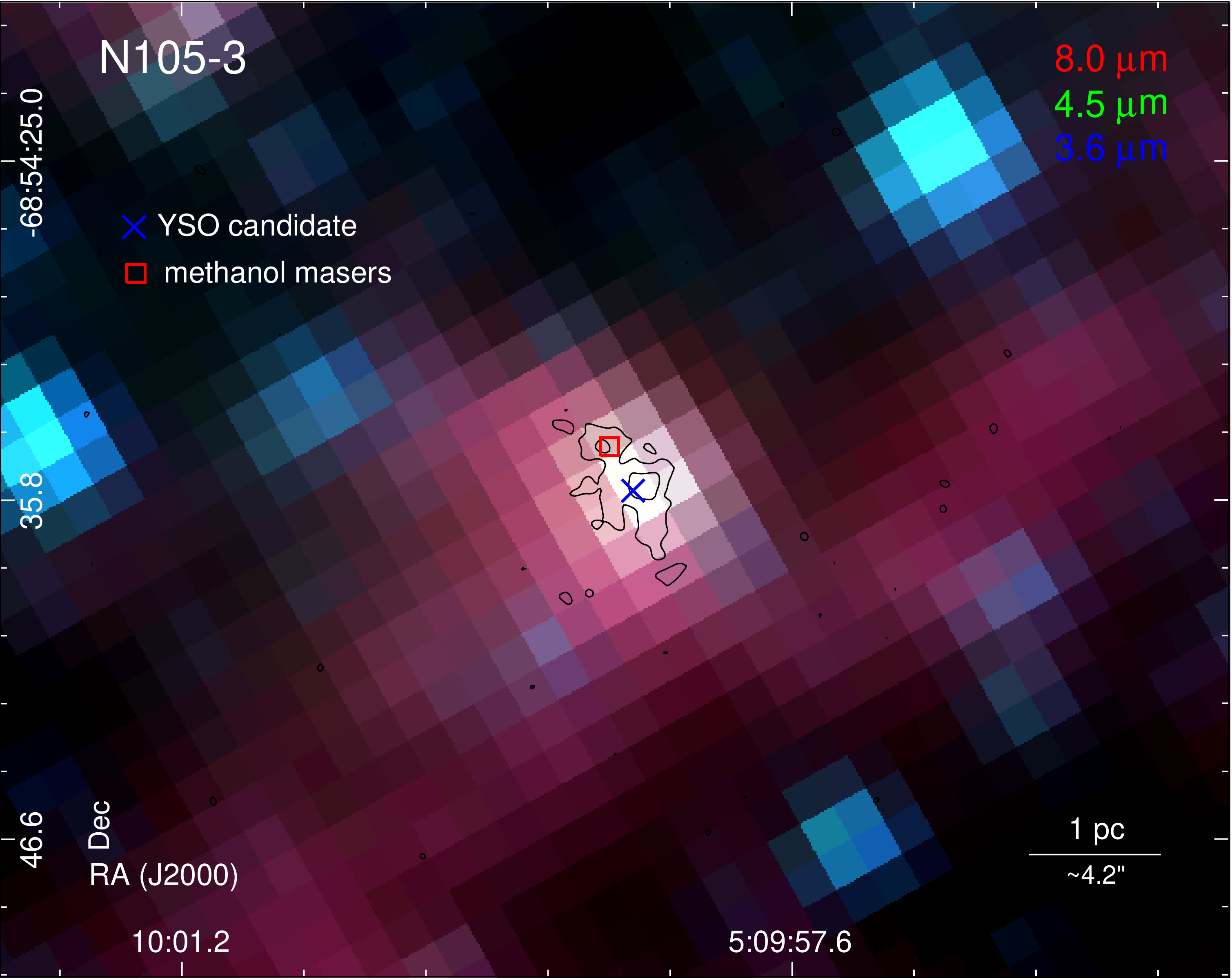}
\caption{{\it Top}: Three-color mosaic of N\,105--1 ({\it top}), N\,105--2 ({\it center}), and N\,105--3 ({\it bottom}) combining the {\it Spitzer}/SAGE IRAC 8.0 $\mu$m ({\it red}), 4.5 $\mu$m ({\it green}), and 3.6 $\mu$m ({\it blue}) images. The positions of YSOs, YSO candidates, and masers are marked as indicated in the legends. The 1.2 mm continuum contour levels correspond to (3, 10, 100)$\sigma_1$ for N\,105--1, (3, 20)$\sigma_2$ for N\,105--2, and  (3, 10)$\sigma_3$ for N\,105--3, where $\sigma_1$, $\sigma_2$, and $\sigma_3$ are rms noise levels in the corresponding 1.2 mm continuum images not corrected for the primary beam attenuation for clarity: (6.8, 5.0, 2.5)$\times$10$^{-5}$ Jy beam$^{-1}$. \label{f:N105spitzer}}
\end{figure}

\noindent {\bf N\,105--1:} N\,105--1 hosts two spectroscopically confirmed YSOs from \citet{seale2009}:  050950.53$-$685305.5 (source \#318 or SSTISAGE1C\,J050950.53$-$685305.4 from \citealt{whitney2008}) and 050952.73$-$685300.7 (see Fig.~\ref{f:N105spitzer}). Source 050950.53$-$685305.5 is associated with the bright 1.2 mm continuum source N\,105--1\,A, while 050952.73$-$685300.7 coincides with N\,105--1\,B and an extended emission to the east (see Section~\ref{s:IR}).  YSO 050950.53$-$685305.5 was classified by  \citet{seale2009} as a Group P and 050952.73$-$685300.7 as a Group PE source. 

No maser detection has been reported in literature toward N\,105--1. 

\noindent {\bf N\,105--2:} Three {\it Spitzer} YSO candidates from \citet{gruendl2009} are in the N\,105--2 field (050952.26$-$685327.3, 050953.89$-$685336.7, and 050951.31$-$685335.6; Fig.~\ref{f:N105spitzer}), two of which were spectroscopically confirmed as YSOs by \citet{seale2009} and are associated with the 1.2 mm continuum emission.  The  YSO 050953.89$-$685336.7 (Group P source in \citealt{seale2009}) coincides with N\,105--2\,E, while the \citet{gruendl2009}'s catalog position of 050952.26$-$685327.3 (Group PE) lies between the 2\,A and 2\,B continuum peaks.   The inspection of the {\it Spitzer} images shows that no source in the image is visible at this position, but there are two {\it Spitzer} sources in the vicinity. The {\it Spitzer}/IRAC resolution is just about resolving these two sources corresponding to the 1.2 mm continuum peaks 2\,A and 2\,B, separated by only $\sim$1.5 pixels. It is likely that the source-finding routine used by \citet{gruendl2009} found only 2\,A (the brighter peak at 4.5 $\mu$m) but not 2\,B in the shorter IRAC bands, and only 2\,B (the brighter peak at 8.0 $\mu$m) but not 2\,A in the longer IRAC bands. The sources are close enough to be identified as a single object during the band-merging process, resulting in a catalog photometry and position (a weighted mean of the positions found in individual bands) being a combination of these two sources. The catalog position roughly in between the two {\it Spitzer} sources supports this interpretation. The {\it Spitzer}/IRS spectrum of 050952.26$-$685327.3 analyzed by \citet{seale2009} most likely includes contributions from both nearby sources as well.  The SAGE IRAC {\it point source} catalog does not include any sources in the central part of the ALMA field. 

No 1.2 mm continuum emission has been detected toward the position of the YSO candidate 050951.31$-$685335.6 in N\,105--2 located to the southwest from 2\,A and 2\,B (Fig.~\ref{f:N105spitzer}). 

Two sources in N\,105--2 are associated with masers. N\,105--2\,A and 2\,B coincide with the 22 GHz H$_2$O masers (\citealt{whiteoak1983}; \citealt{whiteoak1986}; \citealt{lazendic2002};  \citealt{ellingsen2010};  J. Ott, {\it priv. comm.}, see also \citealt{schwarz2012}). Source 2\,A is also associated with the 1665-/1667-MHz OH maser (\citealt{haynes1981}; \citealt{gardner1985}; \citealt{brooks1997}). No methanol masers have been detected toward N\,105--2 (\citealt{green2008} and \citealt{ellingsen2010}).  

The maser positions used for investigating correlations with the ALMA and infrared emission (e.g., in Fig.~\ref{f:N105spitzer}) come from \citet{green2008} who summarize previous observations of different types of masers and provide accurate positions (within sub-arcsec) obtained using the interferometric observations with ATCA.  

One H$_2$O maser spot in N\,105--2 detected with ATCA is offset toward southeast from the 1.2 mm continuum source 2\,B (J. Ott, {\it priv. comm.}, \citealt{schwarz2012}; see e.g., Fig.~\ref{f:N105spitzer}). \citet{imai2013} reported a detection of another H$_2$O maser spot at a distance of $\sim$3$\rlap.{''}$4 from the N\,105--2\,B 1.2 mm continuum peak toward north - northeast.  However, they incorrectly associate this source with that reported in \citet{oliveira2006} who was not able to estimate an accurate position of the maser spot based on their Parkes 64--m telescope observations, but argued that it is likely related to the maser detected by \citet{lazendic2002}; the accurate position of the H$_2$O maser provided by \citet{lazendic2002} indicates that the maser emission originates in the vicinity of N\,105--2\,A. Due to this positional uncertainty, we do not show the position of the H$_2$O maser from \citet{imai2013}, which is not reported in other surveys, in the images.   

\noindent {\bf N\,105--3:} One {\it Spitzer} YSO candidate lies within the N105\,--3 field (050958.52$-$685435.5, \citealt{gruendl2009}; \citealt{carlson2012}; see Fig.~\ref{f:N105spitzer}), with the {\it Spitzer} catalog position corresponding to the 1.2 mm continuum peak of source N\,105--3\,A. No follow-up spectroscopic observations exist for 050958.52$-$685435.5. 

In N\,105--3, the 1.2 mm continuum source 3\,B is associated with CH$_3$OH masers:  6.7 GHz and 12.2 GHz (\citealt{green2008}; \citealt{ellingsen2010}). No H$_2$O masers have been detected toward this field (e.g., \citealt{ellingsen2010}). 

Three of the YSOs in N\,105 have been well-fit with the \citet{robitaille2006} YSO radiation transfer models by \citet{carlson2012}: 050950.53$-$685305.5 (source SSTISAGEMA\,J050950.53$-$685305.4 in \citealt{carlson2012}; 1\,A),  050953.89$-$685336.7 (SSTISAGEMA\,J050953.91$-$685337.1; 2\,E), and 050958.52$-$685435.5 (SSTISAGEMA\,J050958.52$-$685435.2; 3\,A). All sources were found to be massive with stellar masses and luminosities for the best-fit YSO models of $(31.3\pm2.6, 23.0\pm3.2, 17.9\pm1.4)$ M$_{\odot}$ and $(14\pm2, 6.6\pm2.2,3.6\pm0.5)\times10^4$ $L_{\odot}$ for (050950.53$-$685305.5, 050953.89$-$685336.7, 050958.52$-$685435.5).

\begin{figure*}
\centering
\includegraphics[width=\textwidth]{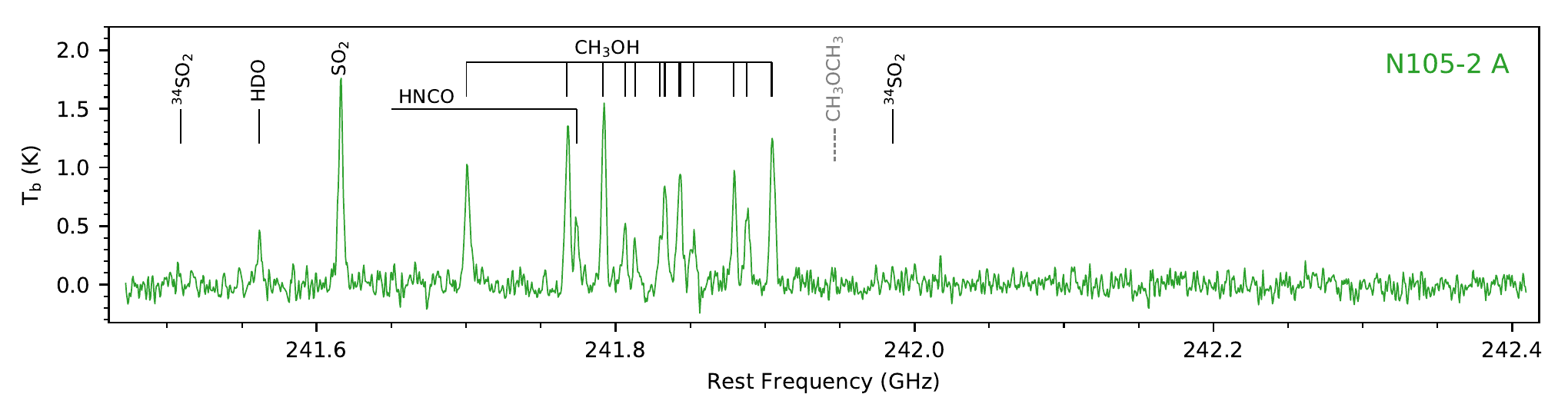}
\includegraphics[width=\textwidth]{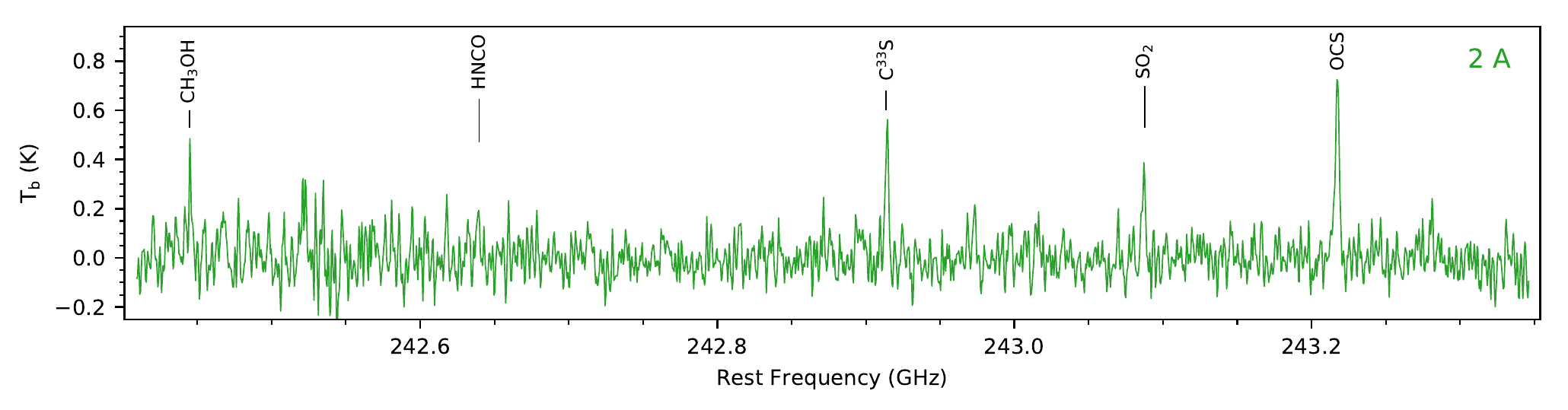}
\includegraphics[width=\textwidth]{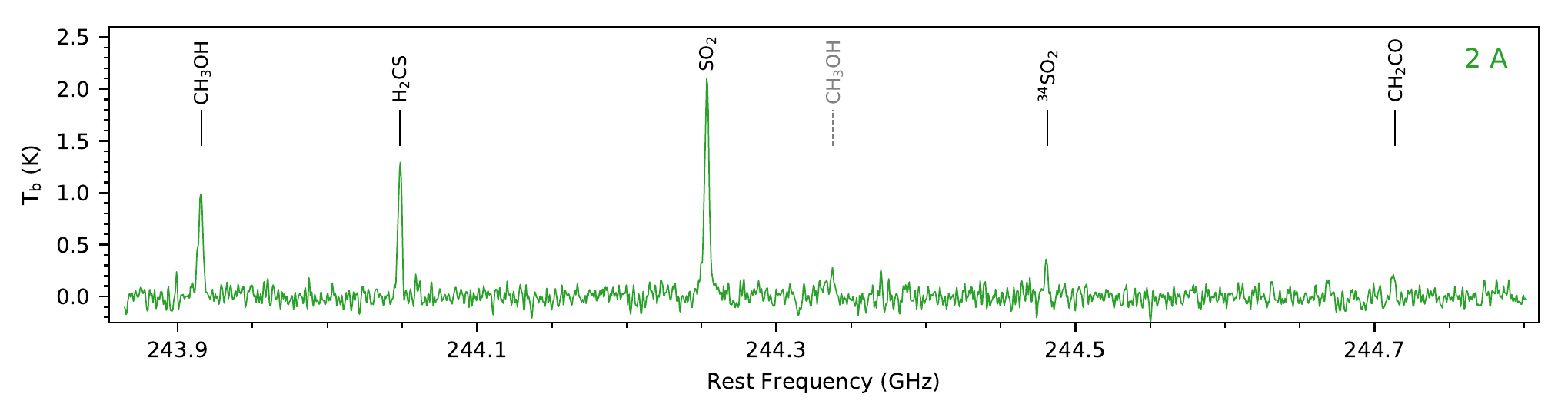}
\includegraphics[width=\textwidth]{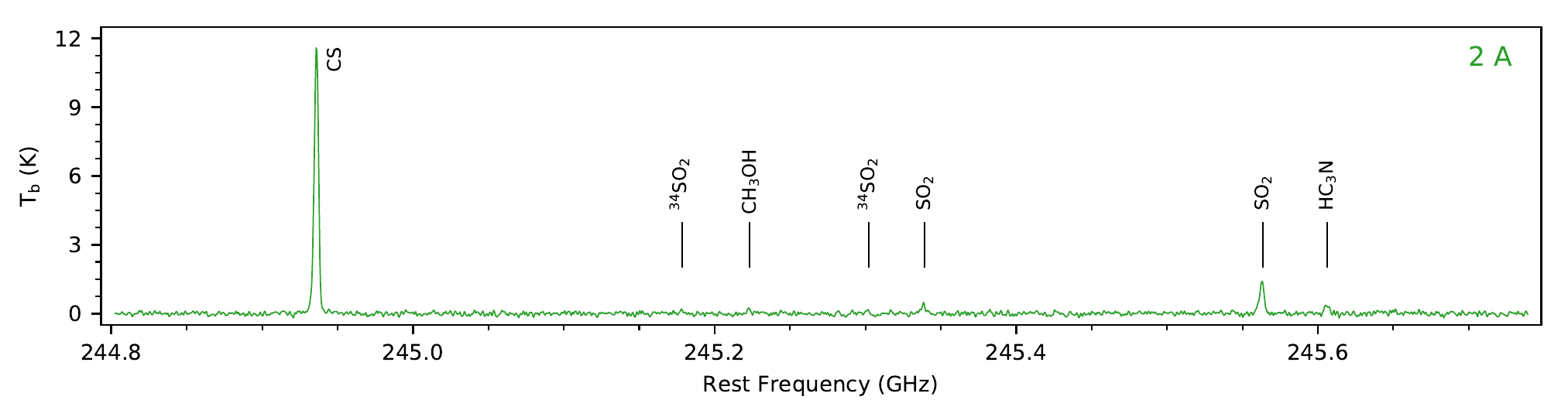}
\caption{ALMA Band 6 spectra of the most chemically-rich source detected in N\,105 (N\,105--2\,A)  in the $\sim$242~GHz ({\it top two panels}) and $\sim$245~GHz ({\it bottom two panels}) spectral windows.  Reliable and tentative detections are indicated and labeled in black and gray, respectively. The spectra were extracted as the mean over the area enclosed by the 50\% of the 1.2 mm continuum peak contour. The $\sim$242~GHz and $\sim$245~GHz spectra of 2\,A with the model spectra overlaid are shown in Fig.~\ref{f:spec2A_1_app} in Appendix~\ref{s:appspectra}. \label{f:spec2A_1}}
\end{figure*}

\begin{figure*}
\centering
\includegraphics[width=\textwidth]{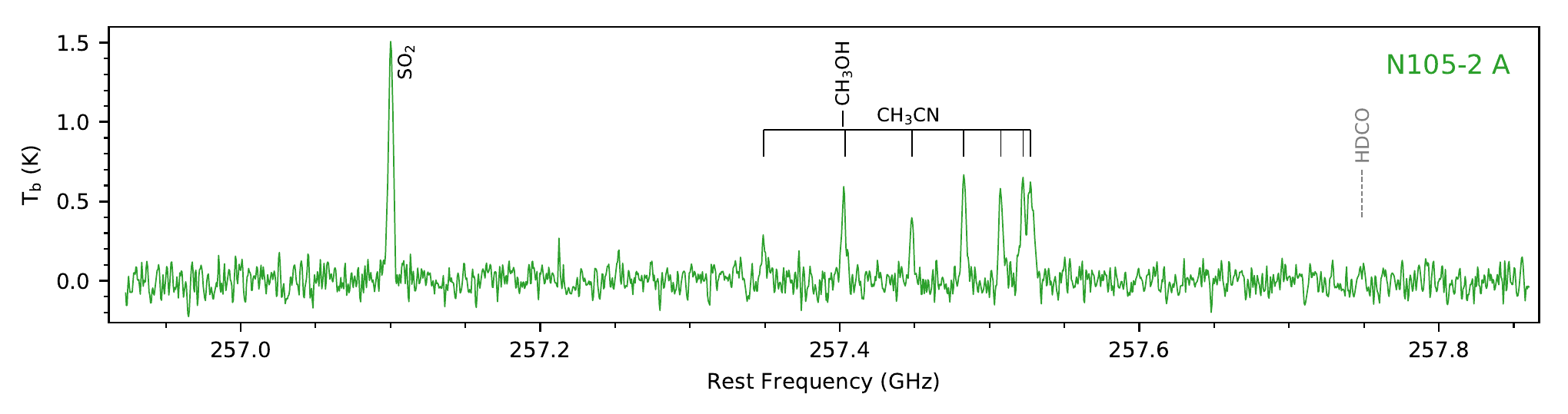}
\includegraphics[width=\textwidth]{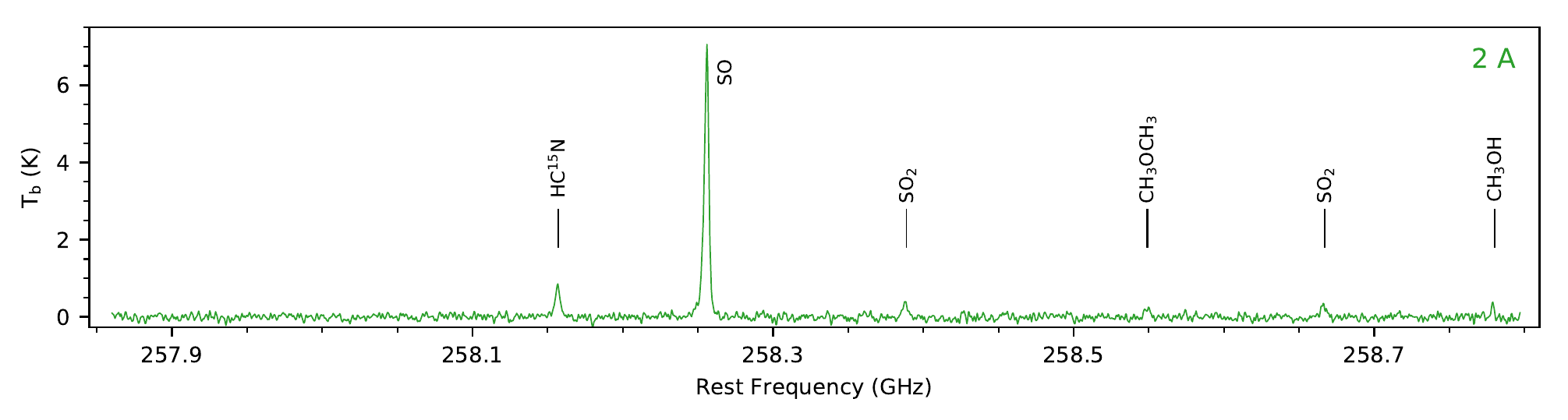}
\includegraphics[width=\textwidth]{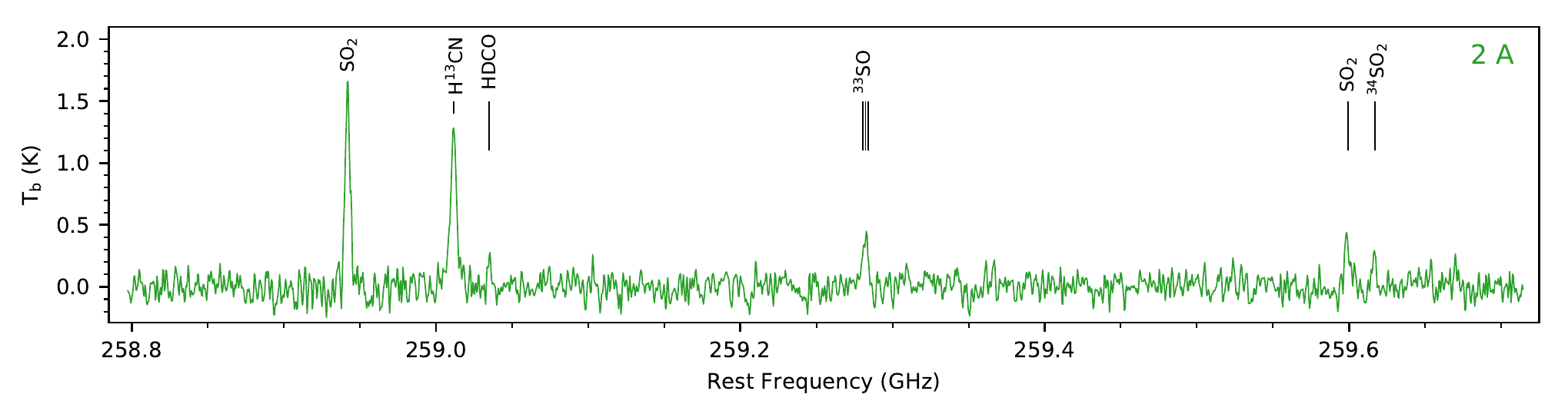}
\includegraphics[width=\textwidth]{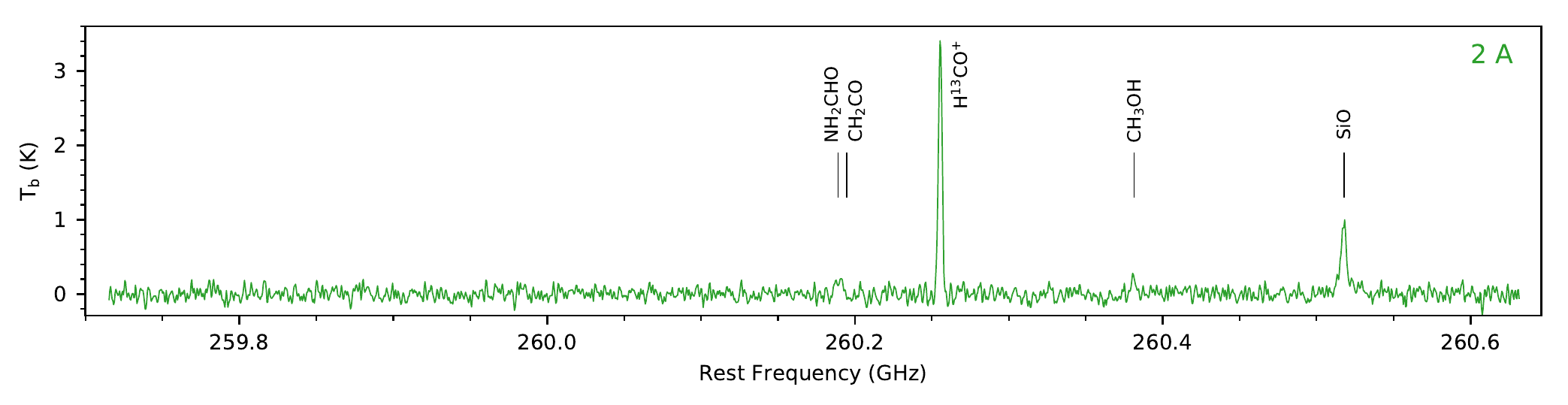}
\caption{ALMA Band 6 spectra of  the most chemically-rich source detected in N\,105 (N\,105--2\,A) in the $\sim$258~GHz ({\it top two panels}) and $\sim$260~GHz ({\it bottom two panels}) spectral windows.  Reliable and tentative detections are indicated and labeled in black and gray, respectively. The spectra were extracted as the mean over the area enclosed by the 50\% of the 1.2 mm continuum peak contour. The $\sim$258~GHz and $\sim$260~GHz spectra of 2\,A with the model spectra overlaid are shown in Fig.~\ref{f:spec2A_2_app} in Appendix~\ref{s:appspectra}.  \label{f:spec2A_2}}
\end{figure*}

\section{Spectral Line Analysis}
\label{s:spectral}

For sources 1\,A--C, 2\,A--E, and 3\,A--B, spectra were extracted as the mean within the area enclosed by the contour corresponding to the 50\% of the source's 1.2 mm continuum emission peak intensity. As a result, for these sources, the physical parameters determined based on spectral modeling provide averages over these spectral extraction areas which are listed in Table~\ref{t:datanh2} (see Section~\ref{s:modeling} for a discussion on spectral modeling). This spectral extraction method could not be applied to sources 2\,F and 3\,C which are faint and associated with an extended continuum emission; the 50\% of the 1.2 mm continuum peak intensity contour encloses other, brighter sources in N\,105--2 and N\,105--3.  For 2\,F and 3\,C, we derive physical parameters at the peak of the continuum emission. Spectra of the chemically-richest source 2\,A, are shown in Figs.~\ref{f:spec2A_1} and \ref{f:spec2A_2} as examples. Spectra for all the sources are presented in Appendix~\ref{s:appspectra}. 

A selection of the spectral extraction method based on a larger area rather than a single pixel associated with a continuum peak was motivated by the fact that the peaks of the molecular line emission are not always coincident with the continuum peaks (see a discussion in Section~\ref{s:mom0}). Moreover, the resulting spectra are less noisy than the single-pixel spectra.

\subsection{Line Identification}

The initial spectral line identification was carried out in the CASA task {\tt Viewer} which uses the NRAO's spectral line database {\it Splatalogue}\footnote{http://www.cv.nrao.edu/php/splat}. The line identification was later verified by comparing the spectra to the predictions of models for the subset of molecules detected in Galactic hot cores, assuming local thermodynamic equilibrium (LTE) conditions. All the detectable lines predicted by the spectral model in the observed frequency ranges must be present in the observed spectrum with relative intensities for different transitions approximately consistent with the model predictions. The spectral analysis and modeling results are described in detail in Section~\ref{s:modeling}. In our analysis we use molecular data from the Cologne Database for Molecular Spectroscopy (CDMS\footnote{http://www.astro.uni-koeln.de/cdms}; \citealt{muller2005}) where available; otherwise, we use the Jet Propulsion Laboratory (JPL) Millimeter and Submillimeter Spectral Line Catalog\footnote{http://spec.jpl.nasa.gov/} (\citealt{pickett1998}; see Section~\ref{s:modeling} for details). We use the CDMS quantum number notation from the Splatalogue throughout the paper. 

Table~\ref{t:detections} lists all the molecular lines detected toward the continuum sources in N\,105. We detected S-bearing species: SO, $^{33}$SO, SO$_2$, $^{34}$SO$_2$, CS, C$^{33}$S, OCS, H$_2$CS; N-bearing species: HNCO, HC$_3$N, HC$^{15}$N, H$^{13}$CN; three deuterated molecules: HDO, HDCO, and HDS, as well as SiO, H$^{13}$CO$^{+}$, and CH$_2$CO.

We detected COMs CH$_3$OH, CH$_3$CN, and CH$_3$OCH$_3$ in N\,105.  All three COMs are observed toward sources 2\,A and 2\,B. CH$_3$CN is also detected toward 2\,C. CH$_3$OH is identified in the spectra of all the continuum sources, making it the most widespread COM in our observations.  

We also report a tentative detection of formamide (NH$_2$CHO) toward N\,105--2\,A.  We detected a single NH$_2$CHO transition (260.18984820 GHz): this is a $\sim$3.2$\sigma$ detection of the brightest NH$_2$CHO transition within the frequency range covered by our ALMA observations.  The NH$_2$CHO line is blended with a ketene (CH$_2$CO) line (260.19198200 GHz); the lines are separated by $\sim$2.13 MHz which corresponds to $\sim$2.45 km~s$^{-1}$ or 2.19 channel widths. The significance of this detection in the low-metallicity environment is discussed in Section~\ref{s:formamide}. 

Extragalactic detection of deuterated species were first reported in the LMC star-forming regions by \citet{chin1996} who detected DCO$^{+}$ toward three (N\,113, N\,44\,BC, N\,159\,HW) and DCN toward one star-forming region (N\,113; see also \citealt{heikkila1997} for N\,159, and \citealt{sewilo2018} and \citealt{wang2009} for N\,113). \citet{martin2006} reported a tentative detection of DNC and N$_2$D$^{+}$ in the nucleus of the starburst galaxy NGC\,253. Most recently, \citet{muller2020} reported the detection of ND, NH$_2$D, and HDO with ALMA at redshift $z=0.89$ in the spiral galaxy intercepting the line of sight to the quasar PKS\,1830$-$211. We detected deuterated formaldehyde (HDCO), deuterated hydrogen sulfide (HDS), and deuterated water (HDO) toward hot cores 2\,A (HDCO and HDO) and 2\,B (HDO), and a hot core candidate 2\,C (HDS). These are the first extragalactic detections of HDCO and HDS, and the first detection of HDO in an extragalactic star-forming region.  A detailed discussion on the detection of HDO in the LMC will be included in a separate paper. Our observations did not cover any H$_2$CO, H$_2$O, or H$_2$S transitions, preventing us from calculating the deuterium fractionation (the abundance ratio of deuterated over hydrogenated isotopologues, D/H) toward N\,105--2. 

Several hydrogen recombination lines are observed toward source 1\,A: H49$\epsilon$ (241.86116 GHz), H54$\eta$ (243.94239 GHz), H53$\eta$ (257.19399 GHz), H41$\gamma$ (257.63549 GHz), and H36$\beta$ (260.03278 GHz). This is the first extragalactic detection of the $\epsilon$, $\eta$, and $\gamma$ transitions of the hydrogen recombination lines and will be reported elsewhere.


\startlongtable
\begin{deluxetable*}{lccccccccccccccc}
\centering
\tablecaption{Spectral Lines Detected Toward Continuum Sources in N\,105--1, N\,105--2, and N\,105--3\tablenotemark{\footnotesize a}  \label{t:detections}}
\tablewidth{0pt}
\tablehead{
\multicolumn{1}{c}{Species} &
\multicolumn{1}{c}{Transition} &
\colhead{Frequency} &
\colhead{$E_{\rm U}$}&
\colhead{1A} & 
\colhead{1B} & 
\colhead{1C} & 
\colhead{2A} & 
\colhead{2B} &
\colhead{2C} &
\colhead{2D} &
\colhead{2E} &
\colhead{2F} &
\colhead{3A} &
\colhead{3B} &
\colhead{3C} \\
\colhead{} &
\colhead{} &
\colhead{(MHz)} &
\colhead{(K)} &
\colhead{} &
\colhead{} &
\colhead{} &
\colhead{} &
\colhead{} &
\colhead{} &
\colhead{} &
\colhead{} &
\colhead{} &
\colhead{} &
\colhead{} &
\colhead{} 
}
\startdata
\multicolumn{16}{c}{COMs} \\
\hline
CH$_3$OH  &  5$_{-0, 5}$--4$_{-0, 4}$ E, v$_t$=0 & 241700.159& 47.94& $\checkmark?$& $\checkmark$& $\checkmark?$& $\checkmark$& $\checkmark$& $\checkmark$& $\checkmark$& $\checkmark?$& $\checkmark$& $\checkmark?$& $\checkmark$? \tablenotemark{\footnotesize d}& $\checkmark?$\\
CH$_3$OH  &  5$_{1, 5}$--4$_{1, 4}$ E, v$_t$=0 & 241767.234& 40.39& $\checkmark$& $\checkmark$& $\checkmark$& $\checkmark$& $\checkmark$& $\checkmark$& $\checkmark$& $\checkmark$& $\checkmark$& $\checkmark$& $\checkmark$? \tablenotemark{\footnotesize d}& $\checkmark$\\
CH$_3$OH  &  5$_{0, 5}$--4$_{0, 4}$ A, v$_t$=0 & 241791.352& 34.82& $\checkmark$& $\checkmark$& $\checkmark$& $\checkmark$& $\checkmark$& $\checkmark$& $\checkmark$& $\checkmark$& $\checkmark$& $\checkmark$& $\checkmark$? \tablenotemark{\footnotesize d}& $\checkmark$\\
CH$_3$OH  &  5$_{4, 2}$--4$_{4, 1}$ A, v$_t$=0 & 241806.524& 115.17& $-$& $-$& $-$& $\checkmark$& $\checkmark$& $\checkmark?$& $-$& $-$& $\checkmark?$& $-$& $\checkmark$? \tablenotemark{\footnotesize d}& $-$\\
CH$_3$OH  &  5$_{4, 1}$--4$_{4, 0}$ A, v$_t$=0 & 241806.525& 115.17& $-$& $-$& $-$& $\checkmark$& $\checkmark$& $\checkmark?$& $-$& $-$& $\checkmark?$& $-$& $\checkmark$? \tablenotemark{\footnotesize d}& $-$\\
CH$_3$OH  &  5$_{4, 2}$--4$_{4, 1}$ E, v$_t$=0 & 241813.255& 122.73& $-$& $-$& $-$& $\checkmark$& $\checkmark$& $\checkmark?$& $-$& $-$& $\checkmark?$& $-$& $\checkmark$? \tablenotemark{\footnotesize d}& $-$\\
CH$_3$OH  &  5$_{-4, 1}$--4$_{-4, 0}$ E, v$_t$=0 & 241829.629& 130.82& $-$& $-$& $-$& $\checkmark$& $\checkmark$& $\checkmark$& $-$& $-$& $-$& $-$& $\checkmark$? \tablenotemark{\footnotesize d}& $-$\\
CH$_3$OH  &  5$_{3, 3}$--4$_{3, 2}$ A, v$_t$=0 & 241832.718& 84.62& $-$& $-$& $-$& $\checkmark$& $\checkmark$& $\checkmark$& $\checkmark?$& $-$& $\checkmark?$& $-$& $\checkmark$? \tablenotemark{\footnotesize d}& $-$\\
CH$_3$OH  &  5$_{3, 2}$--4$_{3, 1}$ A, v$_t$=0 & 241833.106& 84.62& $-$& $\checkmark$& $-$& $\checkmark$& $\checkmark$& $\checkmark$& $\checkmark?$& $-$& $\checkmark?$& $-$& $\checkmark$? \tablenotemark{\footnotesize d}& $-$\\
CH$_3$OH  &  5$_{2, 4}$--4$_{2, 3}$ A, v$_t$=0 & 241842.284& 72.53& $-$& $\checkmark$& $\checkmark?$& $\checkmark$& $\checkmark$& $\checkmark$& $\checkmark$& $-$& $\checkmark?$& $-$& $\checkmark$? \tablenotemark{\footnotesize d}& $-$\\
CH$_3$OH  &  5$_{-3, 3}$--4$_{-3, 2}$ E, v$_t$=0 & 241843.604& 82.53& $-$& $-$& $-$& $\checkmark$& $\checkmark$& $\checkmark$& $\checkmark$& $-$& $\checkmark?$& $-$& $\checkmark$? \tablenotemark{\footnotesize d}& $-$\\
CH$_3$OH  &  5$_{3, 2}$--4$_{3, 1}$ E, v$_t$=0 & 241852.299& 97.53& $-$& $-$& $-$& $\checkmark$& $\checkmark$& $-$& $-$& $-$& $\checkmark?$& $-$& $\checkmark$? \tablenotemark{\footnotesize d}& $-$\\
CH$_3$OH  &  5$_{-1, 4}$--4$_{-1, 3}$ E, v$_t$=0 & 241879.025& 55.87& $\checkmark?$ \tablenotemark{\footnotesize b}& $\checkmark$& $\checkmark?$& $\checkmark$& $\checkmark$& $\checkmark$& $\checkmark$& $\checkmark?$& $\checkmark$& $\checkmark?$& $\checkmark$? \tablenotemark{\footnotesize d}& $-$\\
CH$_3$OH  &  5$_{2, 3}$--4$_{2, 2}$ A, v$_t$=0 & 241887.674& 72.54& $-$& $-$& $\checkmark?$& $\checkmark$& $\checkmark$& $\checkmark$& $\checkmark?$& $-$& $-$& $-$& $\checkmark$? \tablenotemark{\footnotesize d}& $-$\\
CH$_3$OH  &  5$_{2, 3}$--4$_{2, 2}$ E, v$_t$=0 & 241904.147& 60.73& $\checkmark?$ \tablenotemark{\footnotesize b}& $\checkmark$& $\checkmark$& $\checkmark$& $\checkmark$& $\checkmark$& $\checkmark$& $\checkmark?$& $\checkmark$& $-$& $\checkmark$? \tablenotemark{\footnotesize d}& $-$\\
CH$_3$OH  &  5$_{-2, 4}$--4$_{-2, 3}$ E, v$_t$=0 & 241904.643& 57.07& $\checkmark?$ \tablenotemark{\footnotesize b}& $\checkmark$& $\checkmark$& $\checkmark$& $\checkmark$& $\checkmark$& $\checkmark$& $\checkmark?$& $\checkmark$& $-$& $\checkmark$? \tablenotemark{\footnotesize d}& $-$\\
CH$_3$OH  &  14$_{1, 14}$--13$_{2, 11}$ E, v$_t$=0 & 242446.084& 248.94& $-$& $-$& $-$& $\checkmark$& $\checkmark?$& $-$& $-$& $-$& $-$& $-$& $-$& $-$\\
CH$_3$OH  &  5$_{1, 4}$--4$_{1, 3}$ A, v$_t$=0 & 243915.788& 49.66& $-$& $\checkmark$& $-$& $\checkmark$& $\checkmark$& $\checkmark$& $\checkmark$& $-$& $\checkmark$& $-$& $-$& $-$\\
CH$_3$OH  &  22$_{3, 19}$--22$_{2, 20}$ A, v$_t$=0 & 244330.372& 636.78& $\checkmark?$& $-$& $-$& $-$& $-$& $-$& $-$& $-$& $-$& $-$& $-$& $-$\\
CH$_3$OH  &  9$_{-1, 9}$--8$_{-0, 8}$ E, v$_t$=1 & 244337.983& 395.66& $\checkmark?$& $-$& $-$& $\checkmark?$& $-$& $-$& $-$& $-$& $-$& $-$& $-$& $-$\\
CH$_3$OH  &  21$_{3, 18}$--21$_{2, 19}$ A, v$_t$=0 & 245223.019& 585.76& $-$& $-$& $-$& $\checkmark$& $-$& $-$& $-$& $-$& $-$& $-$& $-$& $-$\\
CH$_3$OH  &  18$_{3, 16}$--18$_{2, 17}$ A, v$_t$=0 & 257402.086& 446.55& $-$& $-$& $-$& $\checkmark$& $\checkmark?$ \tablenotemark{\footnotesize c}& $-$& $-$& $-$& $-$& $-$& $-$& $-$\\
CH$_3$OH  &  19$_{3, 17}$--19$_{2, 18}$ A, v$_t$=0 & 258780.248& 490.58 & $-$& $-$& $-$& $\checkmark$& $-$& $-$& $-$& $-$& $-$& $-$& $-$& $-$\\
CH$_3$OH  &  20$_{3, 18}$--20$_{2, 19}$ A, v$_t$=0 & 260381.463& 536.97& $-$& $-$& $-$& $\checkmark$& $-$& $-$& $-$& $-$& $-$& $-$& $-$& $-$\\
CH$_3$OCH$_3$ &  13$_{1, 13}$--12$_{0, 12}$ EA & 241946.249& 81.13& $-$& $-$& $-$& $\checkmark?$& $-$& $-$& $-$& $-$& $-$& $-$& $-$& $-$\\
CH$_3$OCH$_3$ &  13$_{1, 13}$--12$_{0, 12}$ AE& 241946.249& 81.13& $-$& $-$& $-$& $\checkmark?$& $-$& $-$& $-$& $-$& $-$& $-$& $-$& $-$\\
CH$_3$OCH$_3$ &  13$_{1, 13}$--12$_{0, 12}$ EE& 241946.542& 81.13& $-$& $-$& $-$& $\checkmark?$& $-$& $-$& $-$& $-$& $-$& $-$& $-$& $-$\\
CH$_3$OCH$_3$ &  13$_{1, 13}$--12$_{0, 12}$ AA& 241946.835& 81.13& $-$& $-$& $-$& $\checkmark?$& $-$& $-$& $-$& $-$& $-$& $-$& $-$& $-$\\
CH$_3$OCH$_3$ &  14$_{1, 14}$--13$_{0, 13}$ EA & 258548.819& 93.33& $-$& $-$& $-$& $\checkmark$& $\checkmark$& $-$& $-$& $-$& $-$& $-$& $-$& $-$\\
CH$_3$OCH$_3$ &  14$_{1, 14}$--13$_{0, 13}$ AE& 258548.819& 93.33& $-$& $-$& $-$& $\checkmark$& $\checkmark$& $-$& $-$& $-$& $-$& $-$& $-$& $-$\\
CH$_3$OCH$_3$ &  14$_{1, 14}$--13$_{0, 13}$ EE& 258549.063& 93.33& $-$& $-$& $-$& $\checkmark$& $\checkmark$& $-$& $-$& $-$& $-$& $-$& $-$& $-$\\
CH$_3$OCH$_3$ &  14$_{1, 14}$--13$_{0, 13}$ AA& 258549.308& 93.33& $-$& $-$& $-$& $\checkmark$& $\checkmark$& $-$& $-$& $-$& $-$& $-$& $-$& $-$\\
CH$_3$CN  &  14$_6$--13$_6$ & 257349.180 & 349.73& $-$& $-$& $-$& $\checkmark$& $-$\tablenotemark{\footnotesize e} & $-$& $-$& $-$& $-$& $-$& $-$& $-$\\
CH$_3$CN  &  14$_5$--13$_5$ & 257403.585 & 271.23& $-$& $-$& $-$& $\checkmark$& $\checkmark?$ \tablenotemark{\footnotesize c} & $-$& $-$& $-$& $-$& $-$& $-$& $-$\\
CH$_3$CN &  14$_4$--13$_4$ & 257448.128 & 206.98& $-$& $-$& $-$& $\checkmark$& $\checkmark?$& $-$& $-$& $-$& $-$& $-$& $-$& $-$\\
CH$_3$CN &  14$_3$--13$_3$ & 257482.792 & 156.99 & $-$& $-$& $-$& $\checkmark$& $\checkmark$& $-$& $-$& $-$& $-$& $-$& $-$& $-$\\
CH$_3$CN  &  14$_2$--13$_2$ & 257507.562 & 121.28& $-$& $-$& $-$& $\checkmark$& $\checkmark$& $\checkmark?$& $-$& $-$& $-$& $-$& $-$& $-$\\
CH$_3$CN  &  14$_1$--13$_1$ & 257522.428 & 99.85& $-$& $-$& $-$& $\checkmark$& $\checkmark$& $\checkmark?$& $-$& $-$& $-$& $-$& $-$& $-$\\
CH$_3$CN  &  14$_0$--13$_0$ & 257527.384 & 92.71& $-$& $-$& $-$& $\checkmark$& $\checkmark$& $\checkmark$& $-$& $-$& $-$& $-$& $-$& $-$\\
NH$_2$CHO &  12$_{2, 10}$--11$_{2, 9}$ & 260189.090 & 92.36& $-$& $-$& $-$& $\checkmark$& $-$& $-$& $-$& $-$& $-$& $-$& $-$& $-$\\
\hline
\multicolumn{16}{c}{Other Molecules} \\
\hline
HNCO  &  11$_{0, 11}$--10$_{0, 10}$  & 241774.032& 69.63& $-$& $\checkmark$& $-$& $\checkmark$& $\checkmark$& $\checkmark$& $-$& $-$& $-$& $-$& $-$& $-$\\
HNCO  &  11$_{1, 10}$--10$_{1, 9}$  & 242639.705& 113.15& $-$& $-$& $-$& $\checkmark$& $\checkmark$& $-$& $-$& $-$& $-$& $-$& $-$& $-$\\
HC$_3$N  &  27--26 & 245606.320& 165.04& $-$& $-$& $-$& $\checkmark$& $\checkmark$& $-$& $-$& $-$& $-$& $-$& $-$& $-$\\
HC$^{15}$N &  3--2 & 258156.996 & 24.78& $-$& $\checkmark$& $\checkmark$& $\checkmark$& $\checkmark$& $\checkmark$& $\checkmark$& $-$& $\checkmark?$& $-$& $-$& $-$\\
H$^{13}$CN &  3--2 & 259011.798 & 24.86& $\checkmark$& $\checkmark$& $\checkmark$& $\checkmark$& $\checkmark$& $\checkmark$& $\checkmark$& $-$& $\checkmark$& $-$& $-$& $-$\\
H$^{13}$CO$^{+}$ &  3--2 & 260255.339 & 24.98& $\checkmark$& $\checkmark$& $\checkmark$& $\checkmark$& $\checkmark$& $\checkmark$& $\checkmark$& $\checkmark$& $\checkmark$& $\checkmark$& $\checkmark$& $\checkmark?$\\
CH$_2$CO   &  12$_{1, 11}$--11$_{1, 10}$ & 244712.269  & 89.40 & $-$& $-$& $-$& $\checkmark$& $-$& $-$& $-$& $-$& $-$& $-$& $-$& $-$\\
CH$_2$CO   &  13$_{1, 13}$--12$_{1, 12}$ & 260191.982  & 100.47 & $-$& $-$& $-$& $\checkmark$& $-$& $-$& $-$& $-$& $-$& $-$& $-$& $-$\\
SO $^{3}\Sigma$  &  6$_6$--5$_5$ & 258255.826 & 56.50 & $\checkmark$& $\checkmark$& $\checkmark$& $\checkmark$& $\checkmark$& $\checkmark$& $\checkmark$& $\checkmark$& $\checkmark$& $\checkmark$& $\checkmark$& $\checkmark$\\
OCS  &  20--19 & 243218.036 & 122.58& $-$& $-$& $-$& $\checkmark$& $\checkmark$& $\checkmark$& $-$& $-$& $-$& $-$& $-$& $-$\\
H$_2$CS&  7$_{1, 6}$--6$_{1, 5}$ & 244048.504 & 60.03& $\checkmark?$& $\checkmark$& $\checkmark$& $\checkmark$& $\checkmark$& $\checkmark$& $\checkmark$& $-$& $\checkmark$& $-$& $-$& $-$\\
CS  &  5--4 & 244935.557 & 35.27& $\checkmark$& $\checkmark$& $\checkmark$& $\checkmark$& $\checkmark$& $\checkmark$& $\checkmark$& $\checkmark$& $\checkmark$& $\checkmark$& $\checkmark$ & $\checkmark$\\
C$^{33}$S  &  5--4 & 242913.610 & 34.98& $-$& $\checkmark$& $\checkmark?$& $\checkmark$& $\checkmark$& $\checkmark$& $\checkmark$& $-$& $\checkmark$& $-$& $-$& $-$\\
SO$_2$ &  5$_{2, 4}$--4$_{1, 3}$ & 241615.797 & 23.59& $\checkmark?$& $\checkmark$& $-$& $\checkmark$& $\checkmark$& $\checkmark$& $-$& $-$& $\checkmark$& $\checkmark$& $-$& $-$\\
SO$_2$  &  5$_{4, 2}$--6$_{3, 3}$ & 243087.647 & 53.07& $-$& $-$& $-$& $\checkmark$& $\checkmark$& $-$& $-$& $-$& $-$& $-$& $-$& $-$\\
SO$_2$  &  26$_{8, 18}$--27$_{7, 21}$ & 243245.435 & 479.58& $-$& $-$& $-$& $-$& $\checkmark$& $-$& $-$& $-$& $-$& $-$& $-$& $-$\\
SO$_2$\tablenotemark{\footnotesize f}  &  14$_{0, 14}$--13$_{1, 13}$ & 244254.218 & 93.90 & $\checkmark$& $\checkmark?$& $-$& $\checkmark$& $\checkmark$& $\checkmark$& $-$& $-$& $\checkmark$& $-$& $-$& $-$\\
SO$_2$  &  26$_{3, 23}$--25$_{4, 22}$ & 245339.233 & 350.79& $-$& $-$& $-$& $\checkmark$& $\checkmark$& $-$& $-$& $-$& $-$& $-$& $-$& $-$\\
SO$_2$\tablenotemark{\footnotesize f}  &  10$_{3, 7}$--10$_{2, 8}$ & 245563.422 & 72.72& $\checkmark$& $\checkmark$& $-$& $\checkmark$& $\checkmark$& $\checkmark?$& $-$& $-$& $-$& $\checkmark$& $-$& $-$\\
SO$_2$  &  7$_{3, 5}$--7$_{2, 6}$ & 257099.966 & 47.84& $\checkmark$& $\checkmark?$& $-$& $\checkmark$& $\checkmark$& $\checkmark$& $-$& $-$& $-$& $-$& $-$& $-$\\
SO$_2$  &  32$_{4, 28}$--32$_{3, 29}$ & 258388.716 & 531.12& $-$& $-$& $-$& $\checkmark$& $\checkmark$& $-$& $-$& $-$& $-$& $-$& $-$& $-$\\
SO$_2$  &  20$_{7, 13}$--21$_{6, 16}$ & 258666.969 & 313.19& $-$& $-$& $-$& $\checkmark$& $\checkmark$& $-$& $-$& $-$& $-$& $-$& $-$& $-$\\
SO$_2$\tablenotemark{\footnotesize f}  &  9$_{3, 7}$--9$_{2, 8}$ & 258942.199 & 63.47& $\checkmark$& $\checkmark$& $-$& $\checkmark$& $\checkmark$& $-$& $-$& $-$& $-$& $-$& $-$& $-$\\
SO$_2$  &  30$_{4, 26}$--30$_{3, 27}$ & 259599.448 & 471.52& $-$& $-$& $-$& $\checkmark$& $\checkmark$& $-$& $-$& $-$& $-$& $-$& $-$& $-$\\
$^{34}$SO$_2$  &  16$_{1, 15}$--15$_{2, 14}$ & 241509.046 & 130.31& $-$& $-$& $-$& $\checkmark$& $\checkmark$ & $-$& $-$& $-$& $-$& $-$& $-$& $-$\\
$^{34}$SO$_2$  &  8$_{3, 5}$--8$_{2, 6}$ & 241985.449 & 54.38& $-$& $-$& $-$& $\checkmark$& $\checkmark$& $-$& $-$& $-$& $-$& $-$& $-$& $-$\\
$^{34}$SO$_2$  &  18$_{1, 17}$--18$_{0, 18}$ & 243936.052 & 162.59& $-$& $-$& $-$& $-$& $\checkmark$& $-$& $-$& $-$& $-$& $-$& $-$& $-$\\
$^{34}$SO$_2$  &  14$_{0, 14}$--13$_{1, 13}$ & 244481.517 & 93.54& $-$& $-$& $-$& $\checkmark$& $\checkmark$& $-$& $-$& $-$& $\checkmark?$& $-$& $-$& $-$\\
$^{34}$SO$_2$  &  15$_{2, 14}$--15$_{1, 15}$ & 245178.587 & 118.72& $-$& $-$& $-$& $\checkmark$& $\checkmark?$& $-$& $-$& $-$& $-$& $-$& $-$& $-$\\
$^{34}$SO$_2$  &  6$_{3, 3}$--6$_{2, 4}$ & 245302.239 & 40.66 & $-$& $-$& $-$& $\checkmark$& $\checkmark?$& $-$& $-$& $-$& $\checkmark$& $-$& $-$& $-$\\
$^{34}$SO$_2$  &  13$_{3, 11}$--13$_{2, 12}$ & 259617.203 & 104.91& $-$& $-$& $-$& $\checkmark$& $\checkmark$& $-$& $-$& $-$& $-$& $-$& $-$& $-$\\
$^{33}$SO &  6$_{7, 6}$--5$_{6, 5}$ & 259280.331 & 47.12& $-$& $-$& $-$& $\checkmark$& $\checkmark$& $-$& $-$& $-$& $-$& $-$& $-$& $-$\\
$^{33}$SO &  6$_{7, 7}$--5$_{6, 6}$ & 259282.276 & 47.12& $-$& $-$& $-$& $\checkmark$& $\checkmark$& $-$& $-$& $-$& $-$& $-$& $-$& $-$\\
$^{33}$SO &  6$_{7, 8}$--5$_{6, 7}$ & 259284.027 & 47.12& $-$& $-$& $-$& $\checkmark$& $\checkmark$& $-$& $-$& $-$& $-$& $-$& $-$& $-$\\
$^{33}$SO &  6$_{7, 9}$--5$_{6, 8}$ & 259284.027 & 47.12& $-$& $-$& $-$& $\checkmark$& $\checkmark$& $-$& $-$& $-$& $-$& $-$& $-$& $-$\\
SiO  &  6--5 & 260518.009 & 43.76& $-$& $\checkmark?$& $-$& $\checkmark$& $\checkmark$& $\checkmark$& $\checkmark$& $-$& $\checkmark$& $-$& $-$& $-$\\
HDO           &  2$_{1, 1}$--2$_{1, 2}$ & 241561.550 & 95.22 & $-$& $-$& $-$& $\checkmark$& $\checkmark$& $-$& $-$& $-$& $-$& $-$& $-$& $-$\\
HDCO           &  4$_{2, 3}$--3$_{2, 2}$ & 257748.701 & 62.78& $-$& $-$& $-$& $\checkmark?$& $-$& $-$& $-$& $-$& $-$& $-$& $-$& $-$\\
HDCO           &  4$_{2, 2}$--3$_{2, 1}$ & 259034.910 & 62.87& $-$& $-$& $-$& $\checkmark$& $-$& $-$& $-$& $-$& $-$& $-$& $-$& $-$\\
HDS&  1$_{0, 1}$--0$_{0, 0}$ & 244555.580 & 11.74& $-$& $-$& $-$& $-$& $-$& $\checkmark$ & $-$& $-$& $-$& $-$& $-$& $-$\\
HDS&  2$_{1, 1}$--2$_{0, 2}$& 257781.410 & 47.04& $-$& $-$& $-$& $-$& $-$& $\checkmark?$ & $-$& $-$& $-$& $-$& $-$& $-$\\
\enddata
\tablenotetext{a}{Spectroscopic parameters were taken from the CDMS catalog for all species except HDO for which the data were taken from the JPL database (see Section~\ref{s:modeling}). The symbols `$\checkmark$', `$\checkmark?$', and `$-$' indicate, respectively, a detection, a tentative detection, and a non-detection of a given transition.}
\tablenotetext{b}{Methanol transitions likely blended with the H49$\epsilon$ recombination line toward 1\,A.}
\tablenotetext{c}{The  257402.086 MHz CH$_3$OH and  257403.5848 MHz CH$_3$CN transitions are blended in the spectrum of 2\,B.}
\tablenotetext{d}{In the spectrum of 3\,B, the $J = 5-4$ CH$_3$OH Q-branch as a whole is securely detected, even though the individual lines are not clearly visible.}
\tablenotetext{e}{Even though this is a $<$3$\sigma$ detection for 2\,B, this CH$_3$CN transition is used in the rotational diagram analysis as it still contributes useful information  (see Section~\ref{s:modeling} and Fig.~\ref{f:rotdiag}).}
\tablenotetext{f}{The SO$_2$ transitions likely suffering from significant opacity effects (as defined by $E_{\rm U}<100$ K and ${\rm log}(S \mu^2)>1.0$) and thus excluded from the fitting in the rotational diagram for 2\,A and 2\,B (see Section~\ref{s:modeling} and Fig.~\ref{f:rotdiag}).}
\end{deluxetable*}

\subsection{Spatial Distribution of the Molecular Line Emission}
\label{s:mom0}

Figures~\ref{f:N105int1a}--\ref{f:N105int3} show the integrated intensity (moment 0) images for molecular species detected toward N\,105 with ALMA. Methanol is detected toward all the continuum sources with the faintest emission associated with those in N\,105--3.  Other species detected toward all the sources are SO, CS, and H$^{13}$CO$^{+}$. 

\subsubsection{N\,105--1}
\begin{figure*}[h!]
\includegraphics[width=0.485\textwidth]{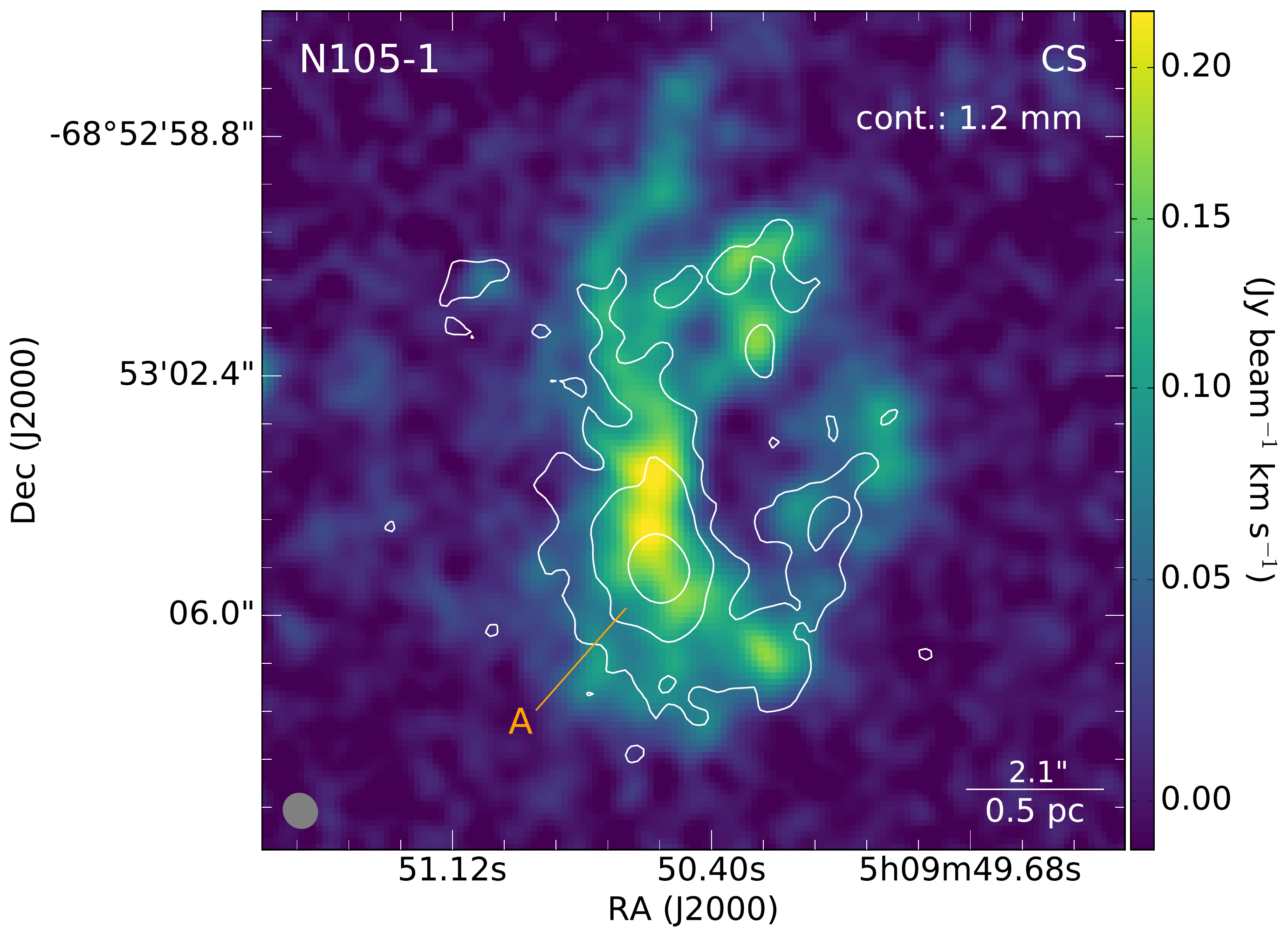}
\hfill
\includegraphics[width=0.485\textwidth]{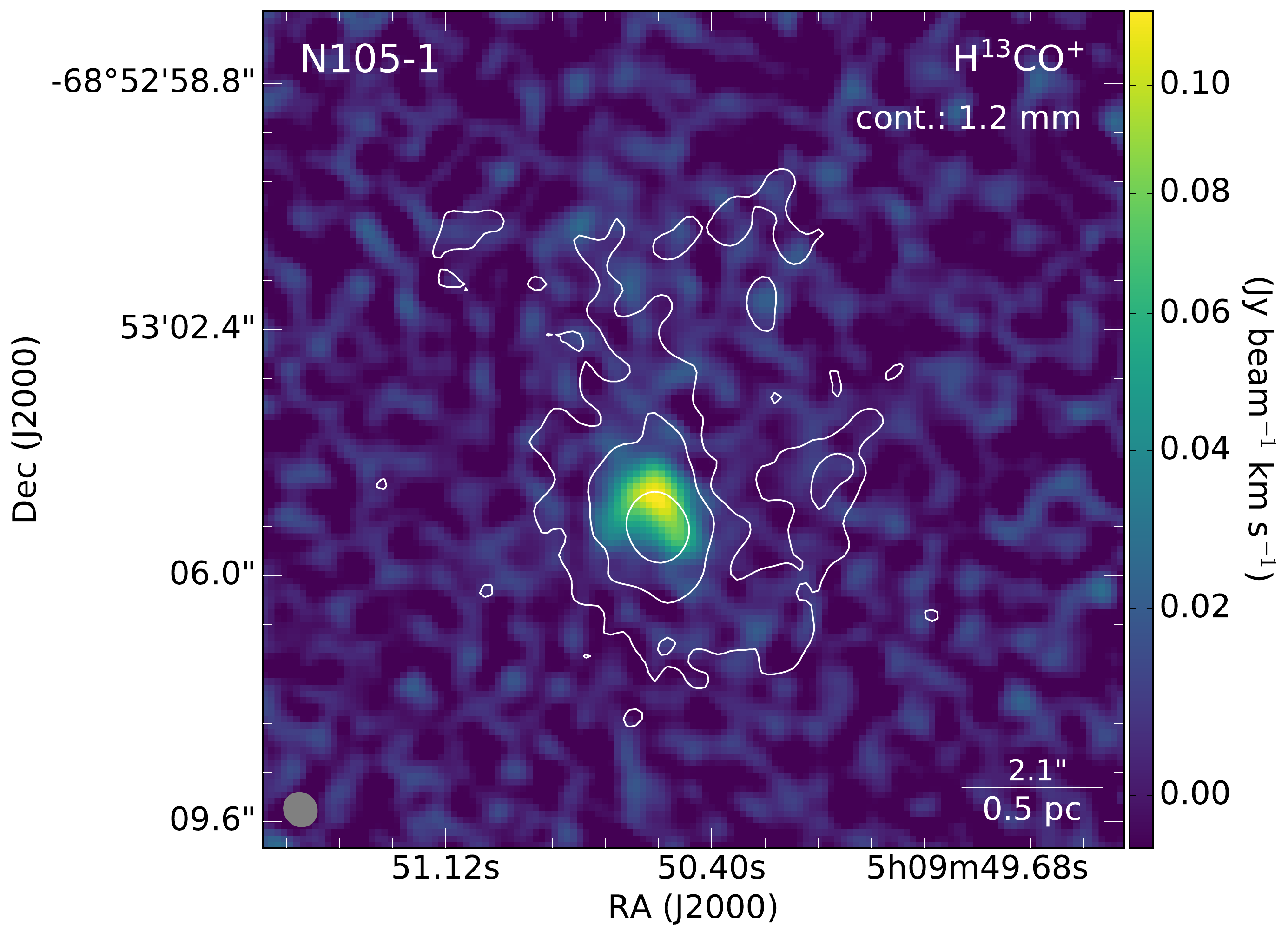}
\hfill
\includegraphics[width=0.485\textwidth]{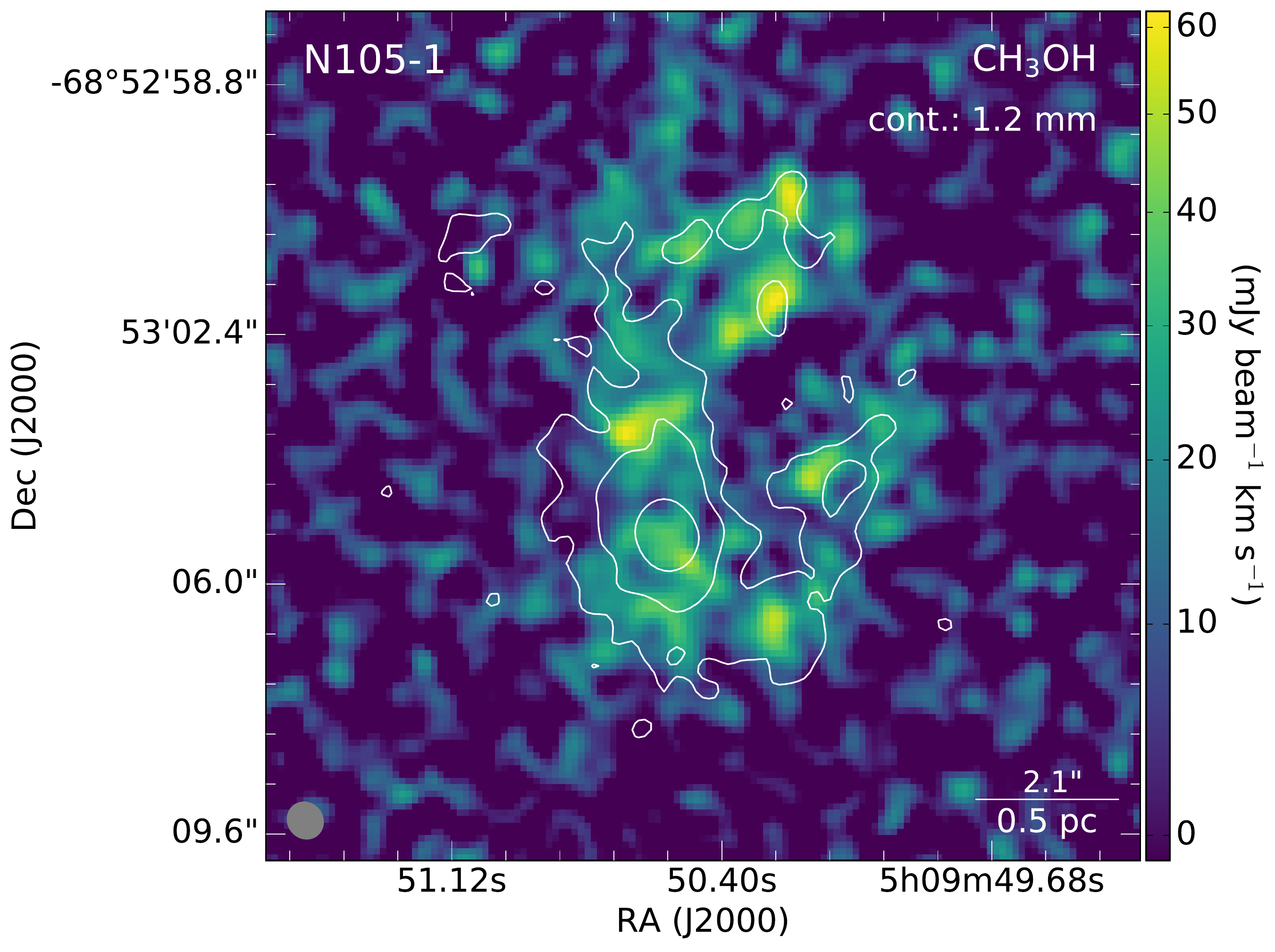}
\hfill
\includegraphics[width=0.485\textwidth]{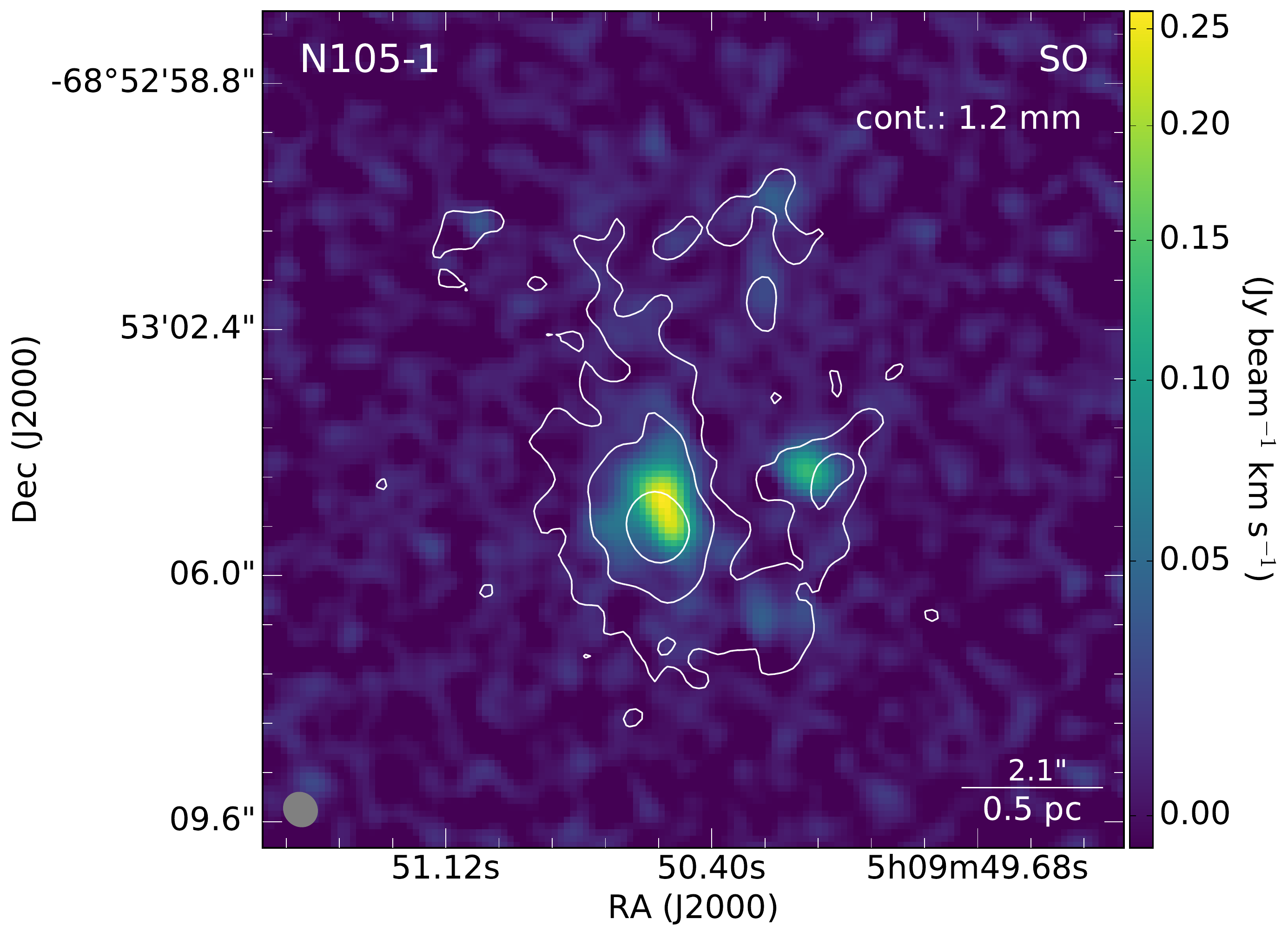}
\caption{{\it From upper left to lower right}: The CS (5--4),  H$^{13}$CO$^{+}$ (3--2), CH$_3$OH (combined 5$_{-1, 5}-4_{-1, 4}$ and 5$_{0, 5}-4_{0, 4}{\rm ++}$ transitions), and SO $6_6-5_5$ integrated intensity images of N\,105--1 around the continuum source 1\,A (color maps).  The white contours in each image correspond to the 1.2 mm continuum emission with contour levels of (3, 10, 100)$\sigma$.  The grey ellipse shown in the lower left corner of each image corresponds to the size of the ALMA synthesized beam (see Table~\ref{t:linedata}). \label{f:N105int1a}}
\end{figure*}

\begin{figure*}[h!]
\includegraphics[width=0.32\textwidth]{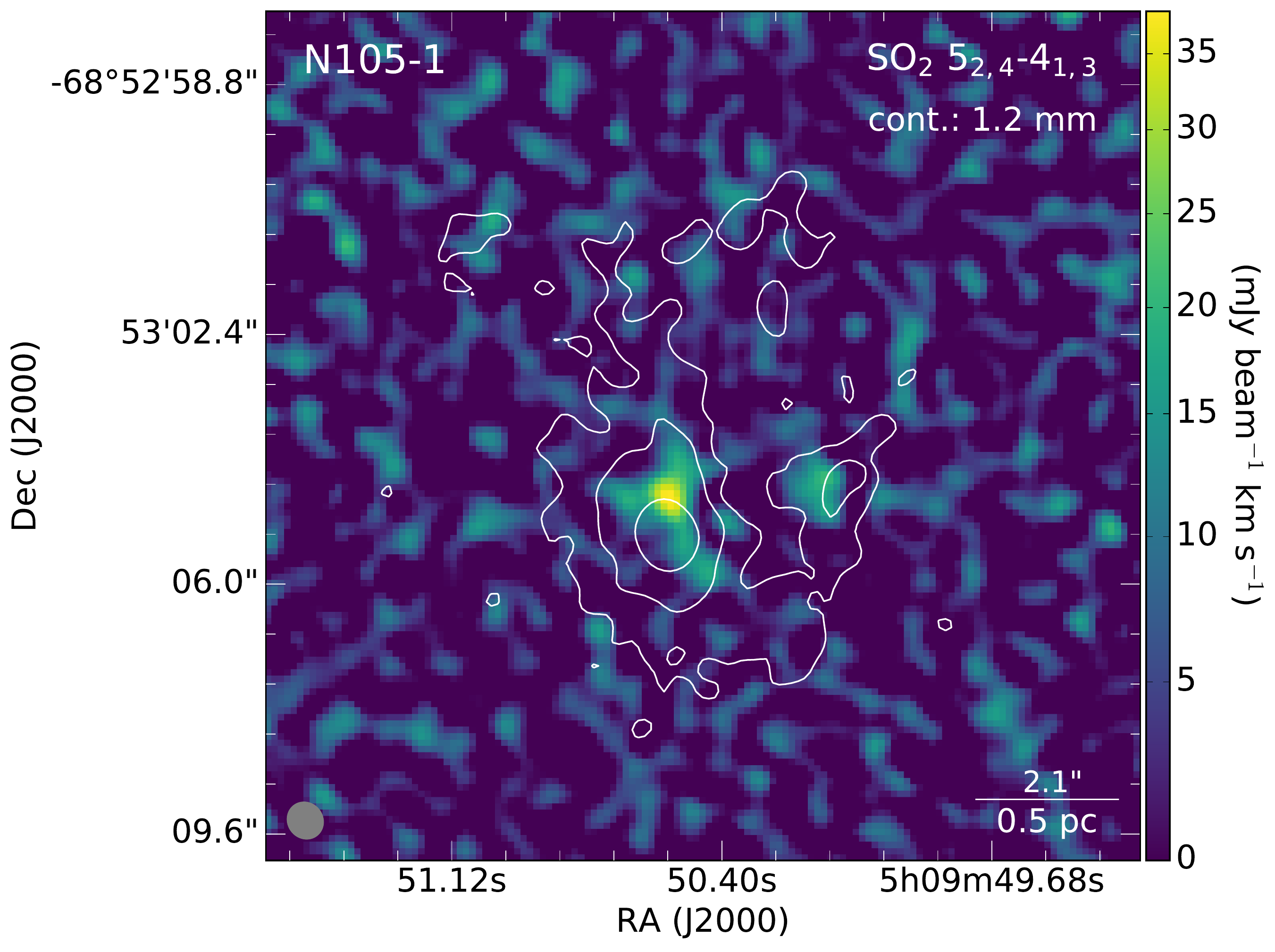}
\hfill
\includegraphics[width=0.32\textwidth]{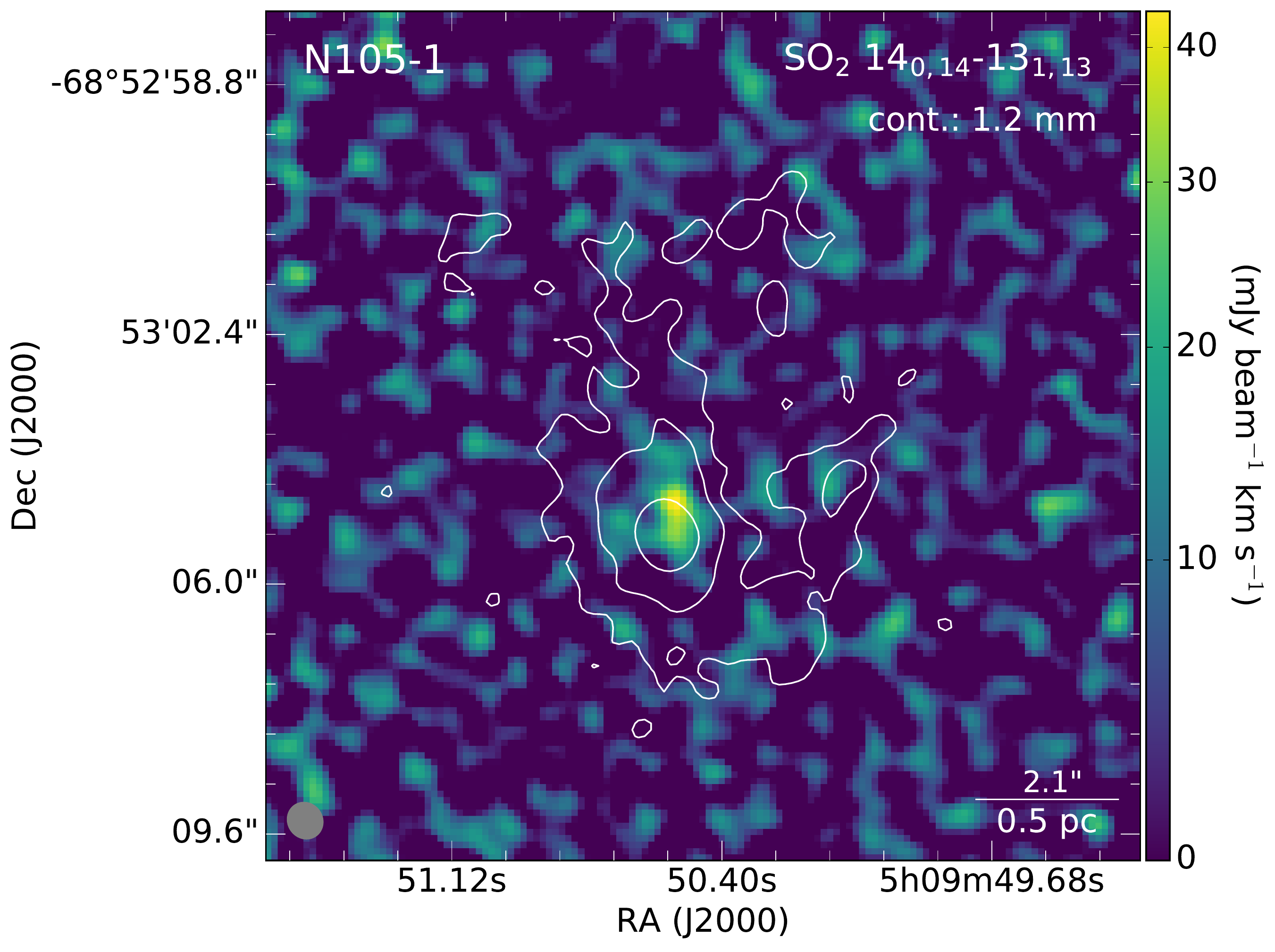}
\hfill
\includegraphics[width=0.32\textwidth]{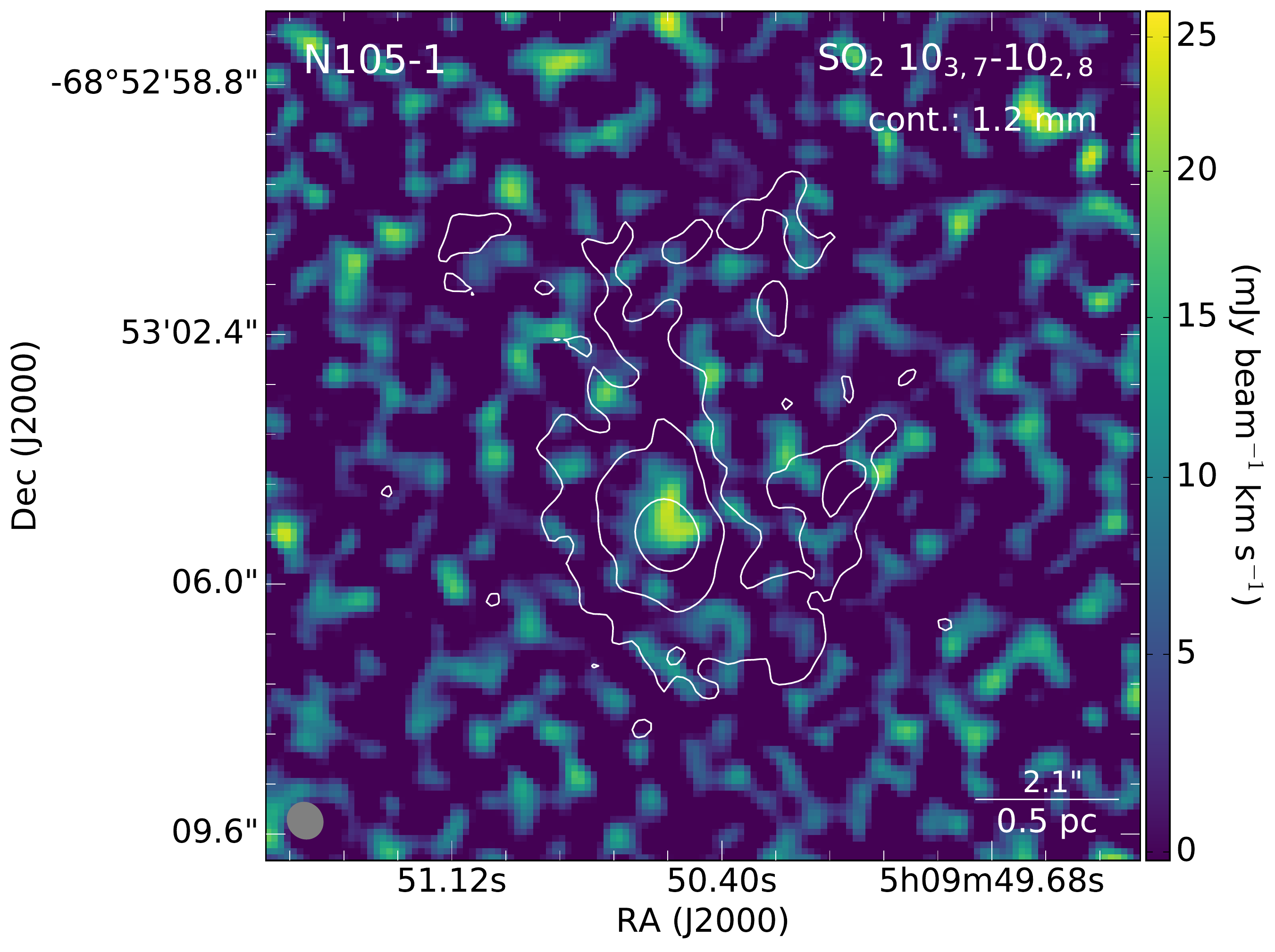}
\caption{The SO$_2$ 5$_{2, 4}-4_{1, 3}$ ({\it left}), 14$_{0, 14}-13_{1, 13}$ ({\it center}), and  10$_{3, 7}-10_{2, 8}$ ({\it right}) integrated intensity images of N\,105--1\,A. The white contours in each image correspond to the 1.2 mm continuum emission with contour levels of (3, 10, 100)$\sigma$.  \label{f:N105int1b}}
\end{figure*}

\indent Figure~\ref{f:N105int1a} shows the CS, H$^{13}$CO$^{+}$, CH$_3$OH, and SO integrated intensity images of the N\,105--1 field centered on the continuum source A (1\,A), while the distribution of the SO$_2$ emission for three detected transitions toward 1\,A are shown in Fig.~\ref{f:N105int1b}. 

\begin{figure*}[h!]
\centering
\includegraphics[width=0.55\textwidth]{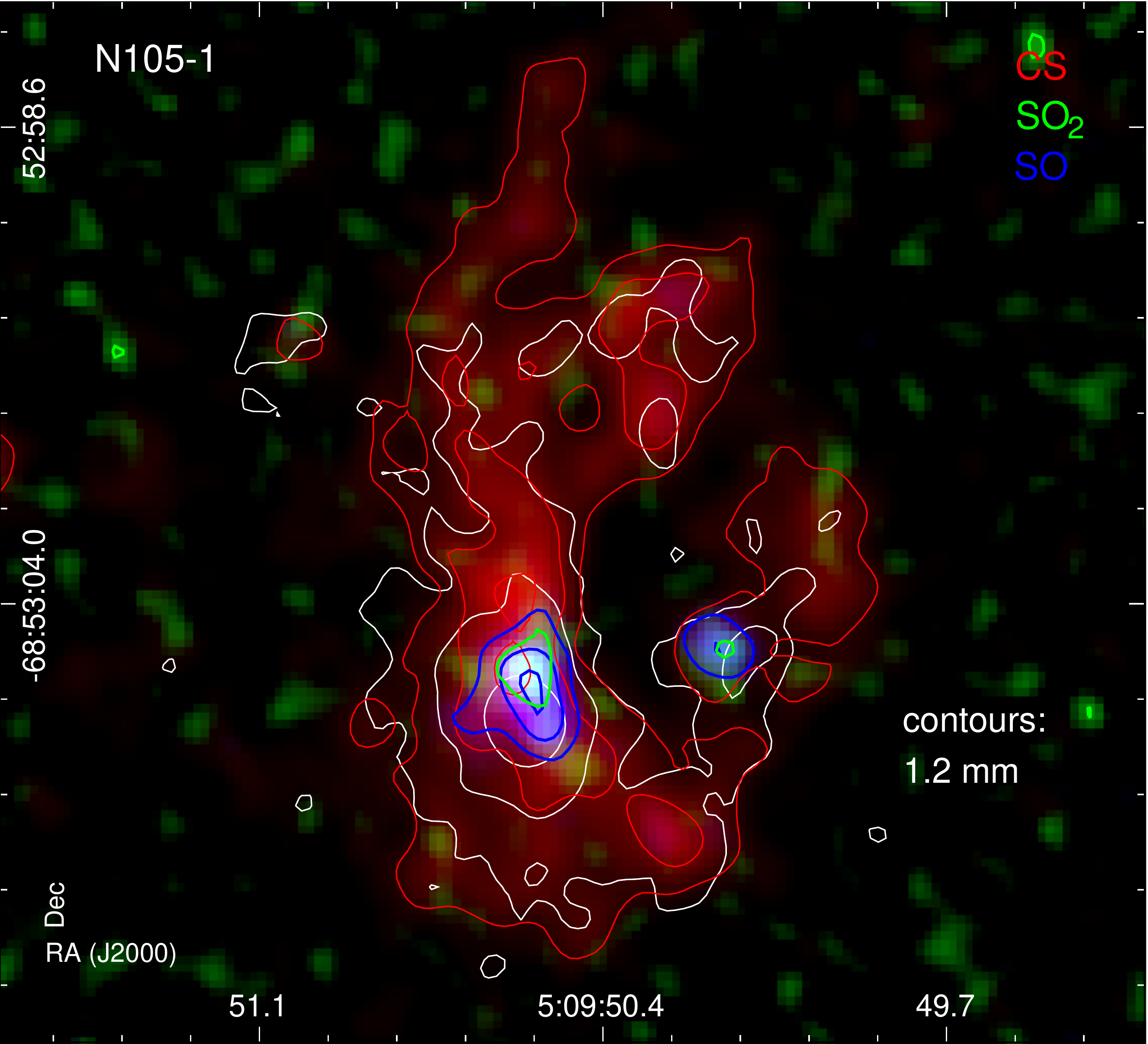}
\caption{Three-color mosaic of the N\,105--1\,A region, combining the CS (5--4) ({\it red}), SO$_2$ 5$_{2, 4}-4_{1, 3}$ ({\it green}), and SO $6_6-5_5$ ({\it blue}) integrated intensity images with the corresponding contours overlaid. The CS contour levels are (20, 50, 90)\% of the CS emission peak of 230.8 mJy beam$^{-1}$ km s$^{-1}$. The SO$_2$ contour level corresponds to 50\% of the SO$_2$ emission peak of 39.4 mJy beam$^{-1}$ km s$^{-1}$ and the SO contour levels to the (20, 50, 90)\% of the SO emission peak of 262.8 mJy beam$^{-1}$ km s$^{-1}$.  The white contours correspond to the 1.2 mm continuum with contour levels of (3, 10, 100)$\sigma$. \label{f:N105int1d}}
\end{figure*}

Toward 1\,A, none of the molecular line peaks coincide with the 1.2 mm continuum peak. 
The CH$_3$OH and CS emissions are extended with multiple peaks throughout the region.  Two of the brightest CS peaks are offset to the north from the 1\,A continuum peak, with the closer one roughly coinciding with the SO$_2$ and SO peaks as illustrated in the three-color image in Fig.~\ref{f:N105int1d}; only the faint extended CH$_3$OH emission has been detected at the position of the two CS peaks.  The H recombination line emission (all transitions) tracing the ionized gas coincides with the 1.2 mm continuum peak. 

\begin{figure*}[h!]
\includegraphics[width=0.482\textwidth]{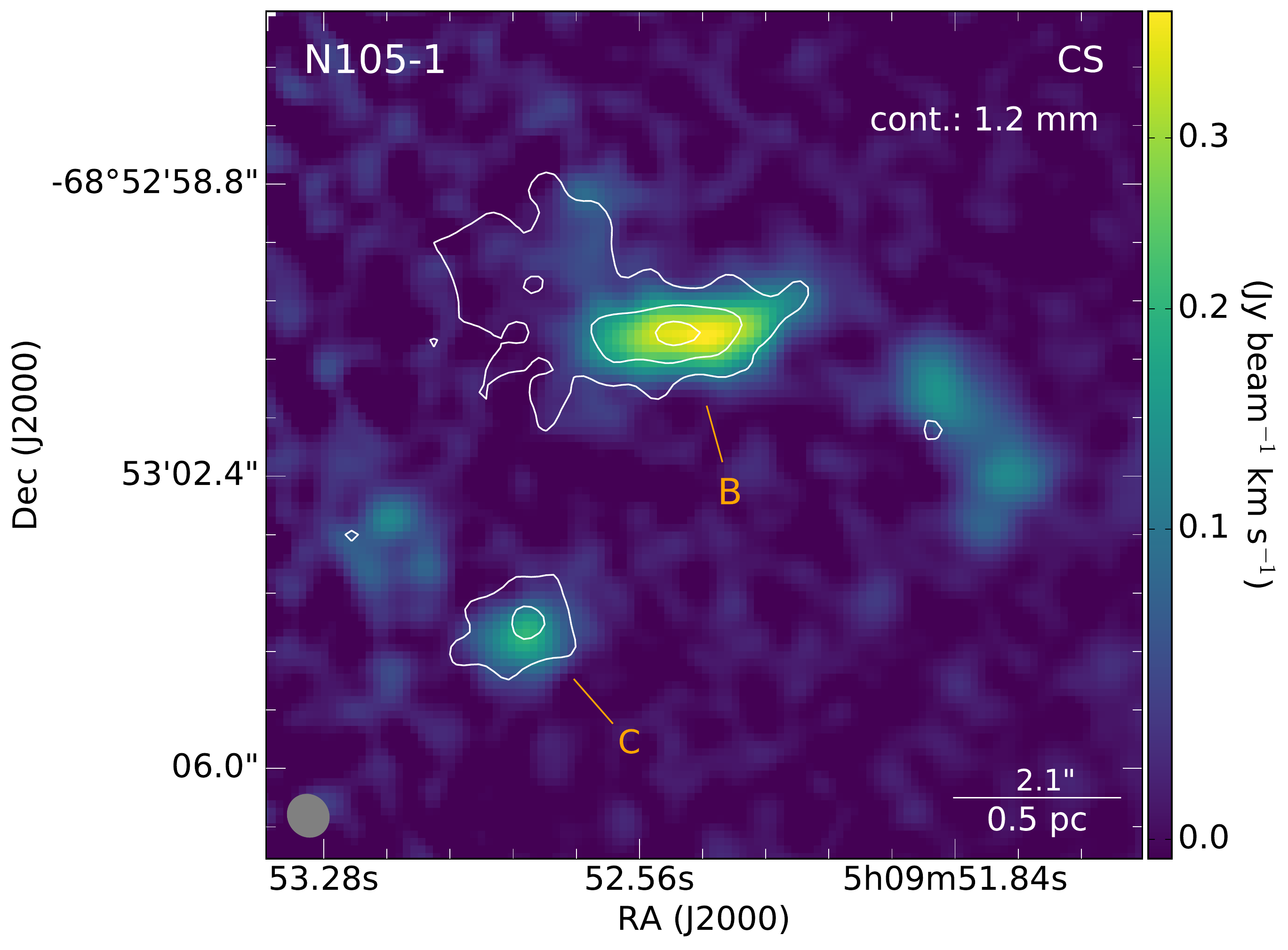}
\hfill
\includegraphics[width=0.485\textwidth]{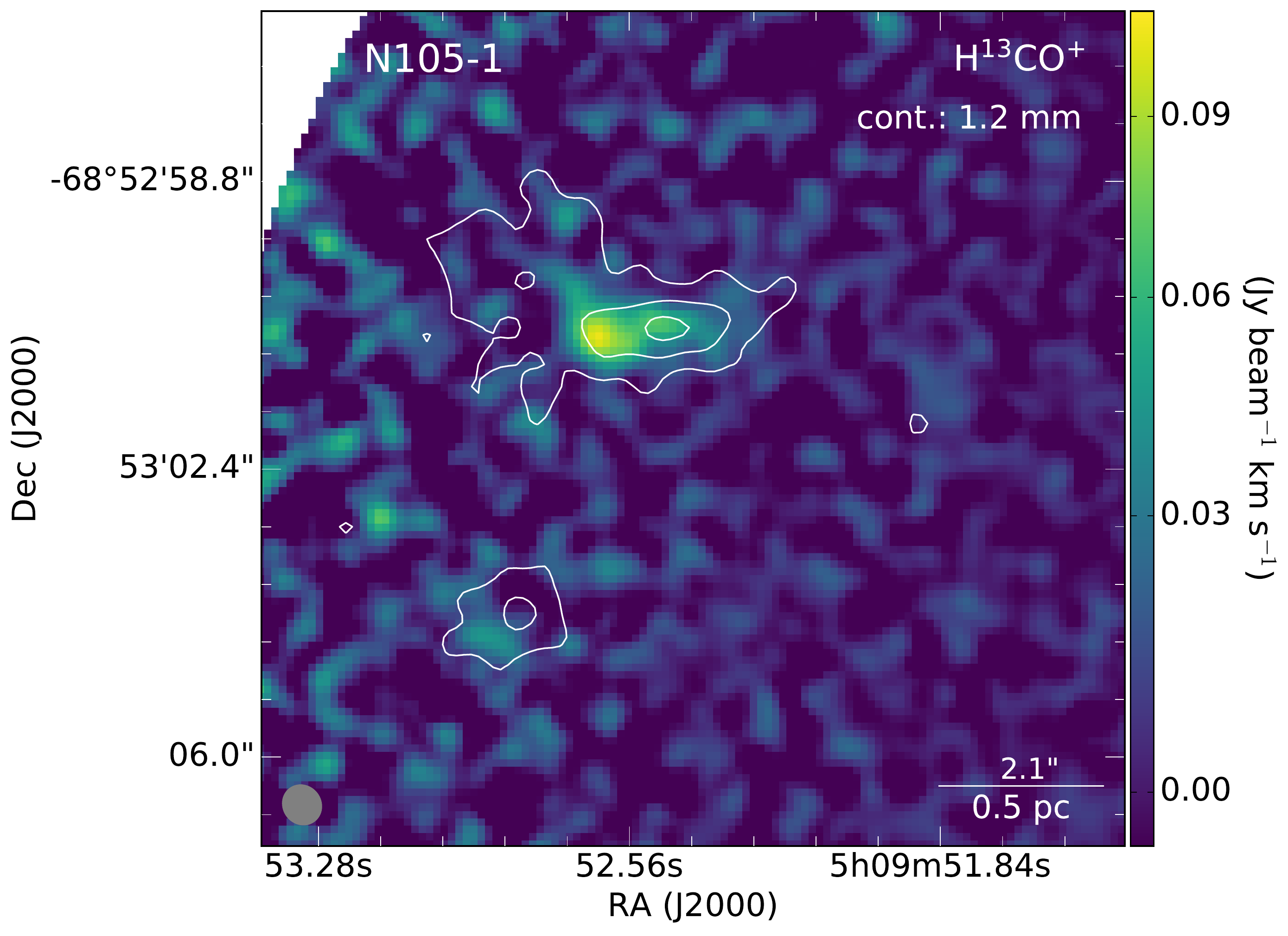}
\hfill
\includegraphics[width=0.485\textwidth]{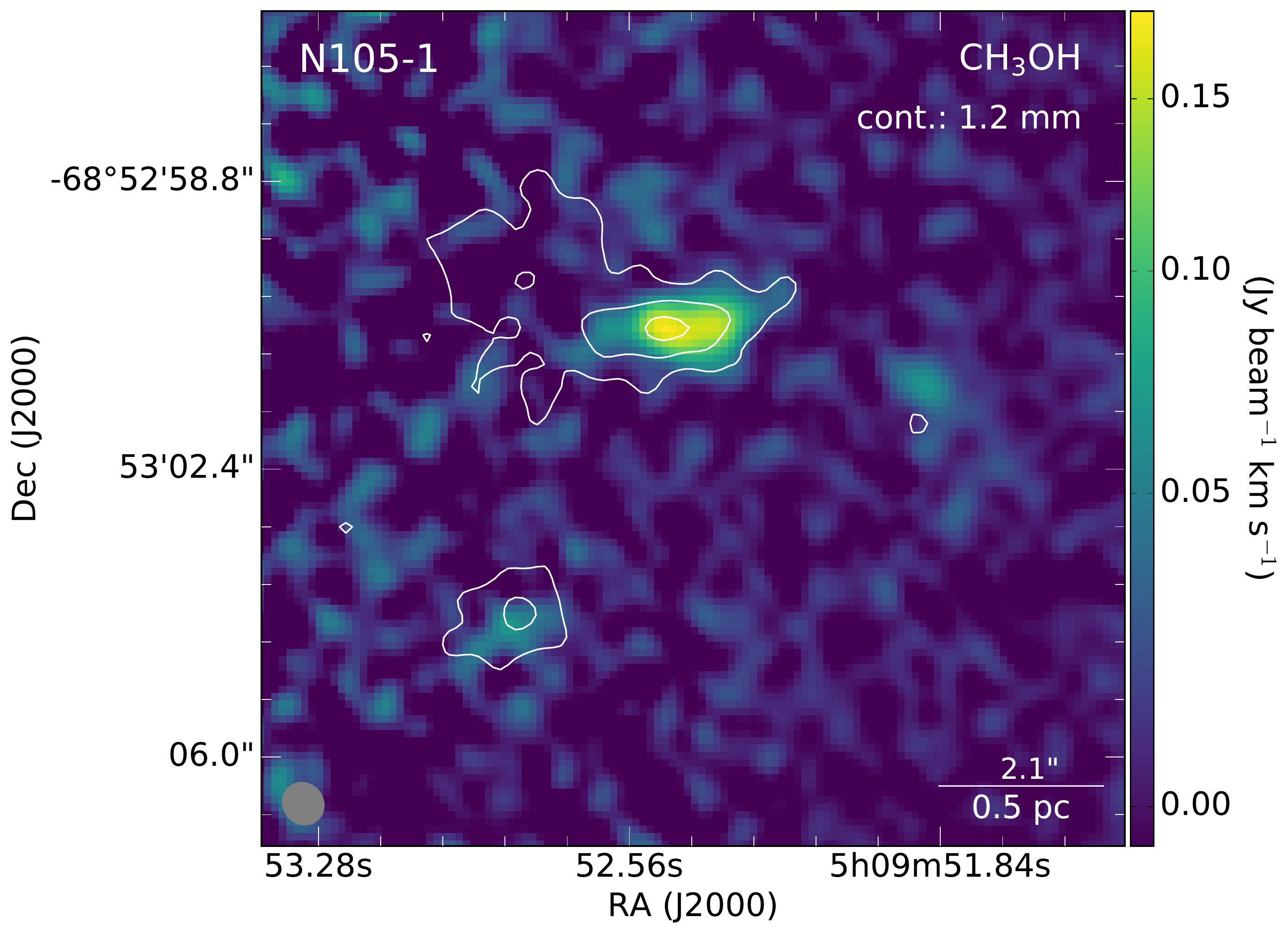}
\hfill
\includegraphics[width=0.485\textwidth]{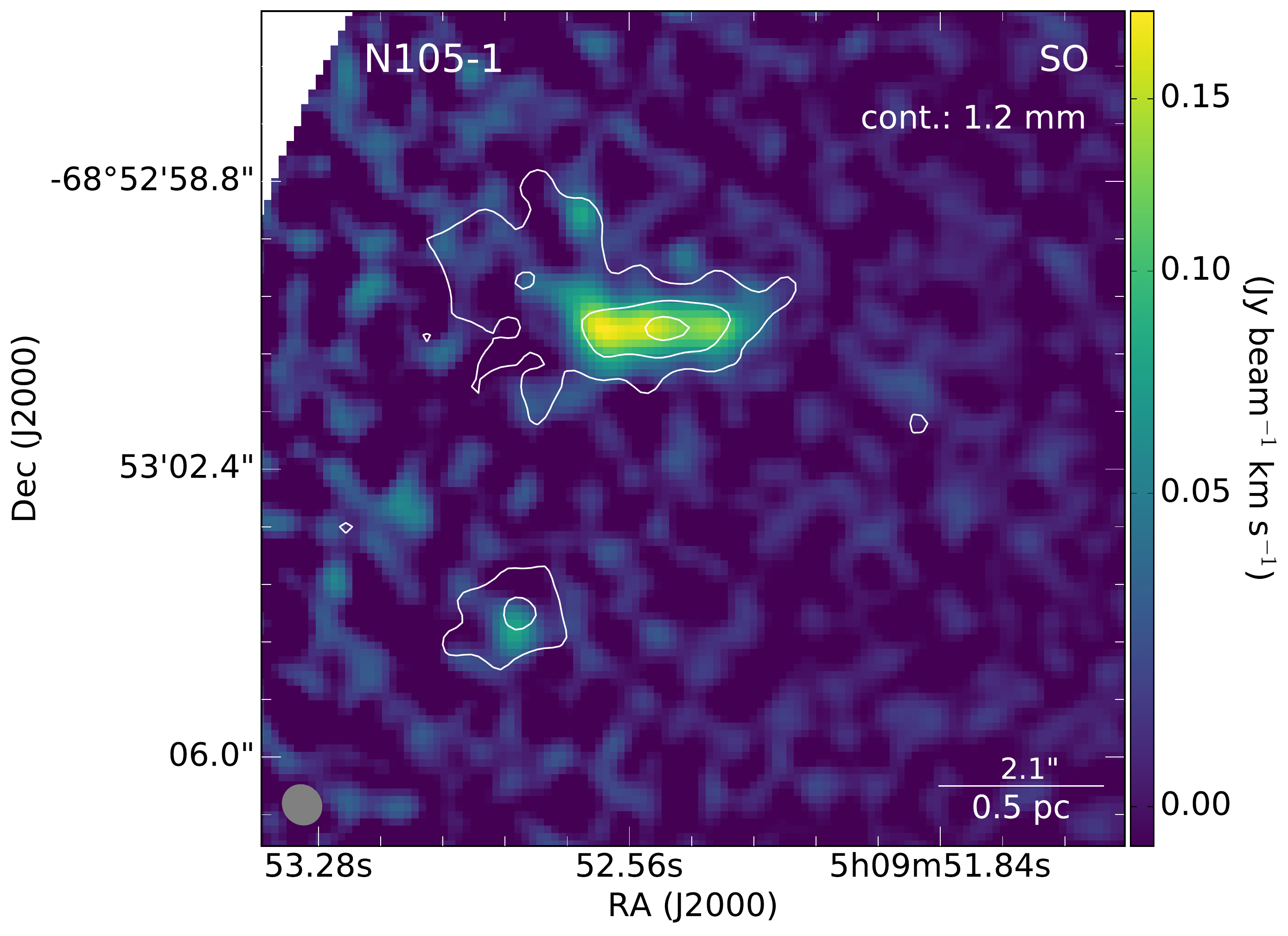}
\caption{{\it From upper left to lower right}: The CS (5--4),  H$^{13}$CO$^{+}$ (3--2), CH$_3$OH (combined 5$_{-1, 5}-4_{-1, 4}$ and 5$_{0, 5}-4_{0, 4}{\rm ++}$ transitions), and SO $6_6-5_5$ integrated intensity images of N\,105--1 around the continuum sources B and C (1\,B and 1\,C).  The white contours in each image correspond to the 1.2 mm continuum emission with contour levels of (3, 8, 15)$\sigma$.  \label{f:N105int1e}}
\end{figure*}

\begin{figure*}[h!]
\centering
\includegraphics[width=0.55\textwidth]{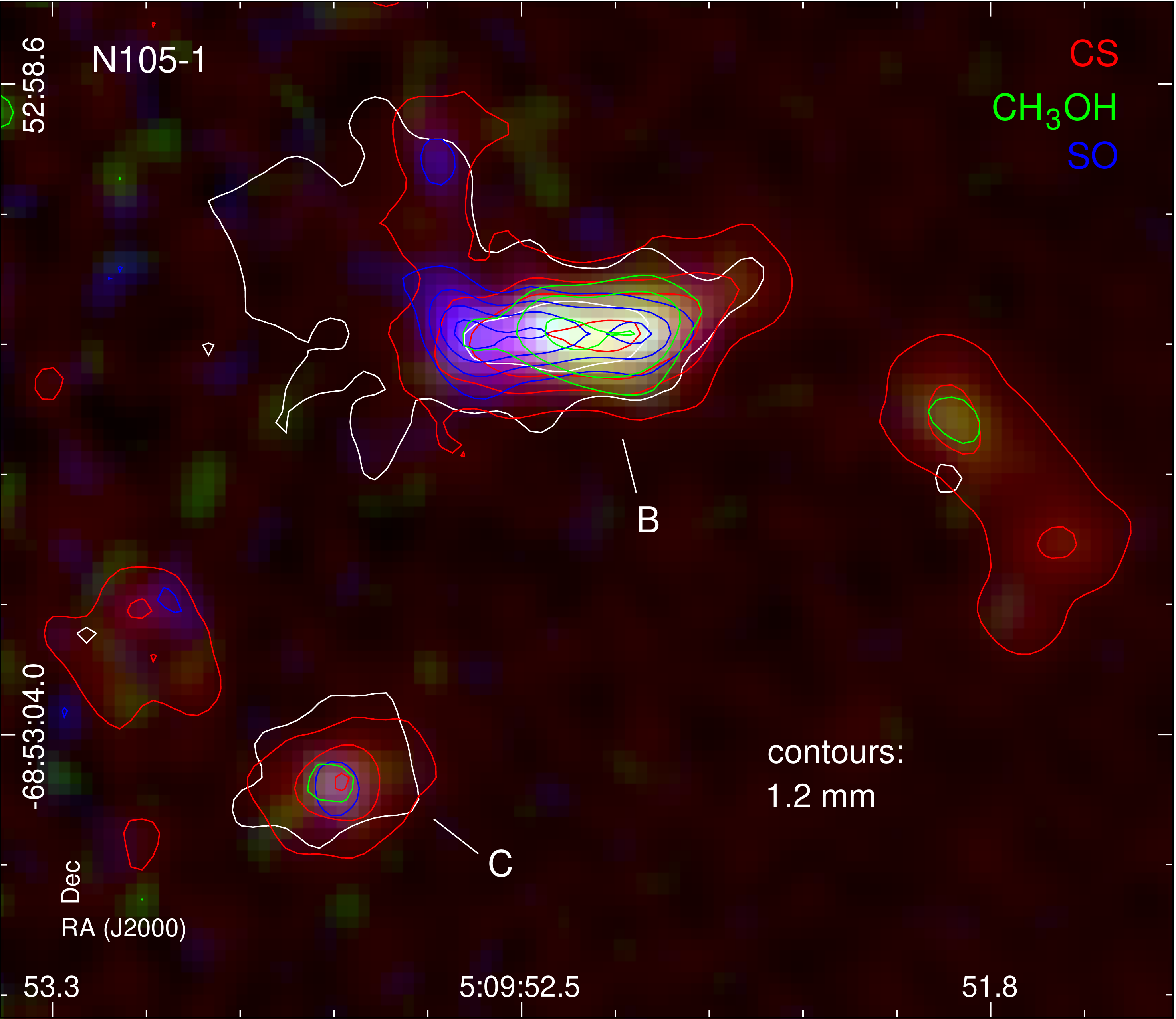}
\caption{Three-color mosaic of the N\,105--1\,B/1\,C region, combining the CS (5--4) ({\it red}), CH$_3$OH  (combined 5$_{-1, 5}-4_{-1, 4}$ and 5$_{0, 5}-4_{0, 4}^{\rm ++}$ transitions) ({\it green}), and SO $6_6-5_5$ ({\it blue}) integrated intensity images with the corresponding contours overlaid. The CS contour levels are (10, 30, 50, 90)\% of the CS emission peak of 400.5 mJy beam$^{-1}$ km s$^{-1}$. The CH$_3$OH contour levels correspond to (30, 50, 90)\% of the CH$_3$OH emission peak of 179.2 mJy beam$^{-1}$ km s$^{-1}$, and the SO contour levels to the (30, 50, 70, 90)\% of the SO emission peak of 182.2 mJy beam$^{-1}$ km s$^{-1}$.  The white contours correspond to the 1.2 mm continuum with contour levels of (3, 10)$\sigma$. 
 \label{f:N105int1f}}
\end{figure*}

The integrated intensity images for N\,105--1 centered on continuum sources B and C (1\,B and 1\,C) are shown in Fig.~\ref{f:N105int1e}. Both 1\,B and 1\,C were detected at the edge of the field-of-view, i.e., in an area of significantly reduced sensitivity, and as a result, their spectra are noisy. In Fig.~\ref{f:N105int1e} we show the integrated intensity images for four species detected toward these sources with the highest signal-to-noise ratio: CS, H$^{13}$CO$^{+}$, CH$_3$OH, and SO. In the three-color image in Fig.~\ref{f:N105int1f}, we compare the distribution of the CH$_3$OH, CS, SO, and  continuum emission.  Figures~\ref{f:N105int1e} and \ref{f:N105int1f} show that toward 1\,B, the brightest CH$_3$OH peak coincides with the 1.2 mm continuum peak. The CH$_3$OH emission extends to the west with another, fainter peak associated with the SO emission. Some faint SO emission is associated with the continuum peak, but two SO peaks are offset to the east and the emission gets brighter with distance from the continuum peak. The CS emission peak is located between the CH$_3$OH/SO peak to the west of the continuum peak. The fainter H$^{13}$CO$^{+}$, H$^{13}$CN, and HNCO emission peaks are offset  from the continuum peak. Both the continuum and molecular line emission (SO, CS, and CH$_3$OH in particular) are elongated in the east-west direction with multiple peaks, indicating that more than one source may be present. The molecular line emission is slightly offset from the 1.2 mm continuum peak toward 1\,C.

\subsubsection{N\,105--2}
\begin{figure*}[h!]
\centering
\includegraphics[width=0.48\textwidth]{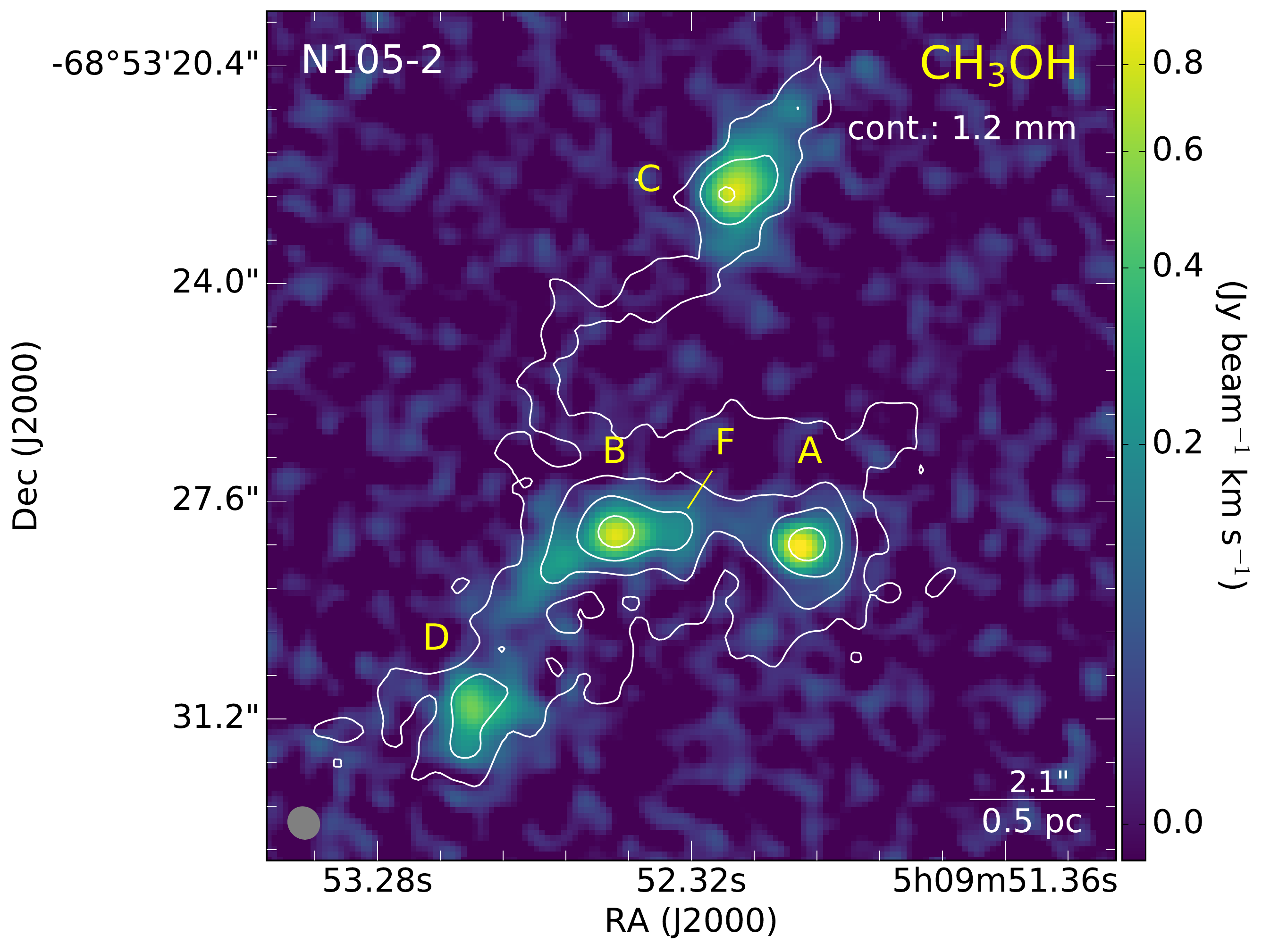} 
\includegraphics[width=0.48\textwidth]{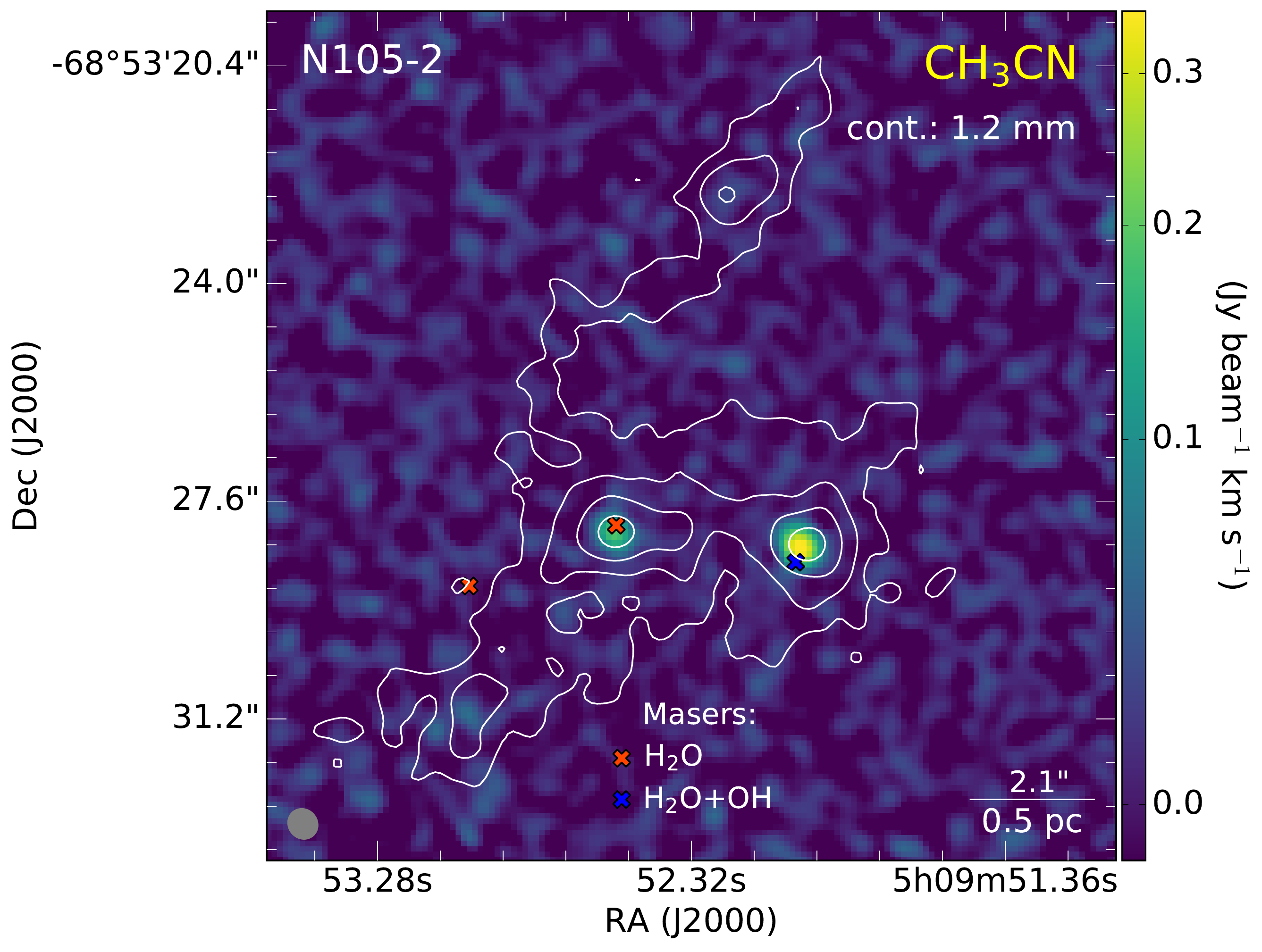}
\hfill
\includegraphics[width=0.329\textwidth]{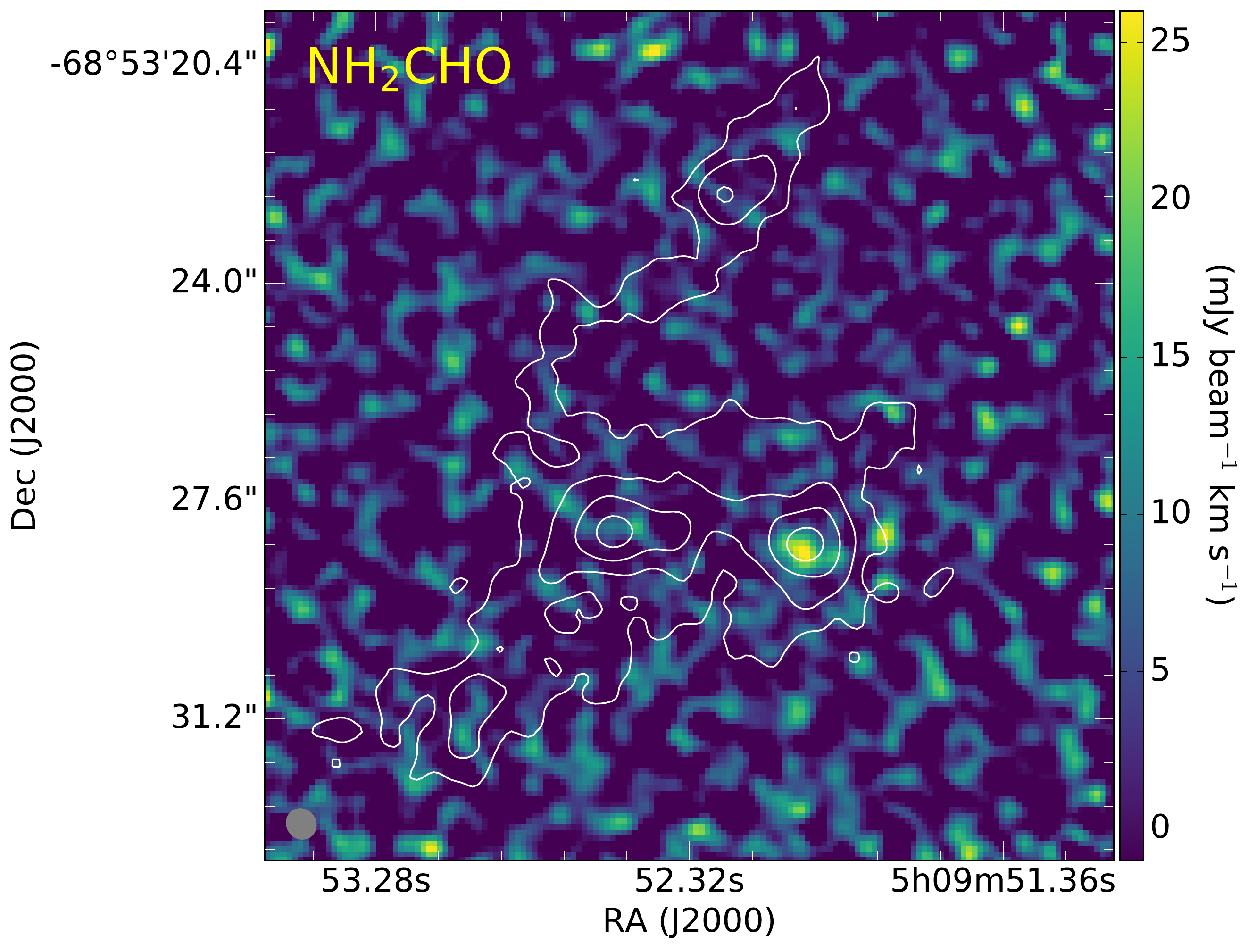} 
\hfill
\includegraphics[width=0.329\textwidth]{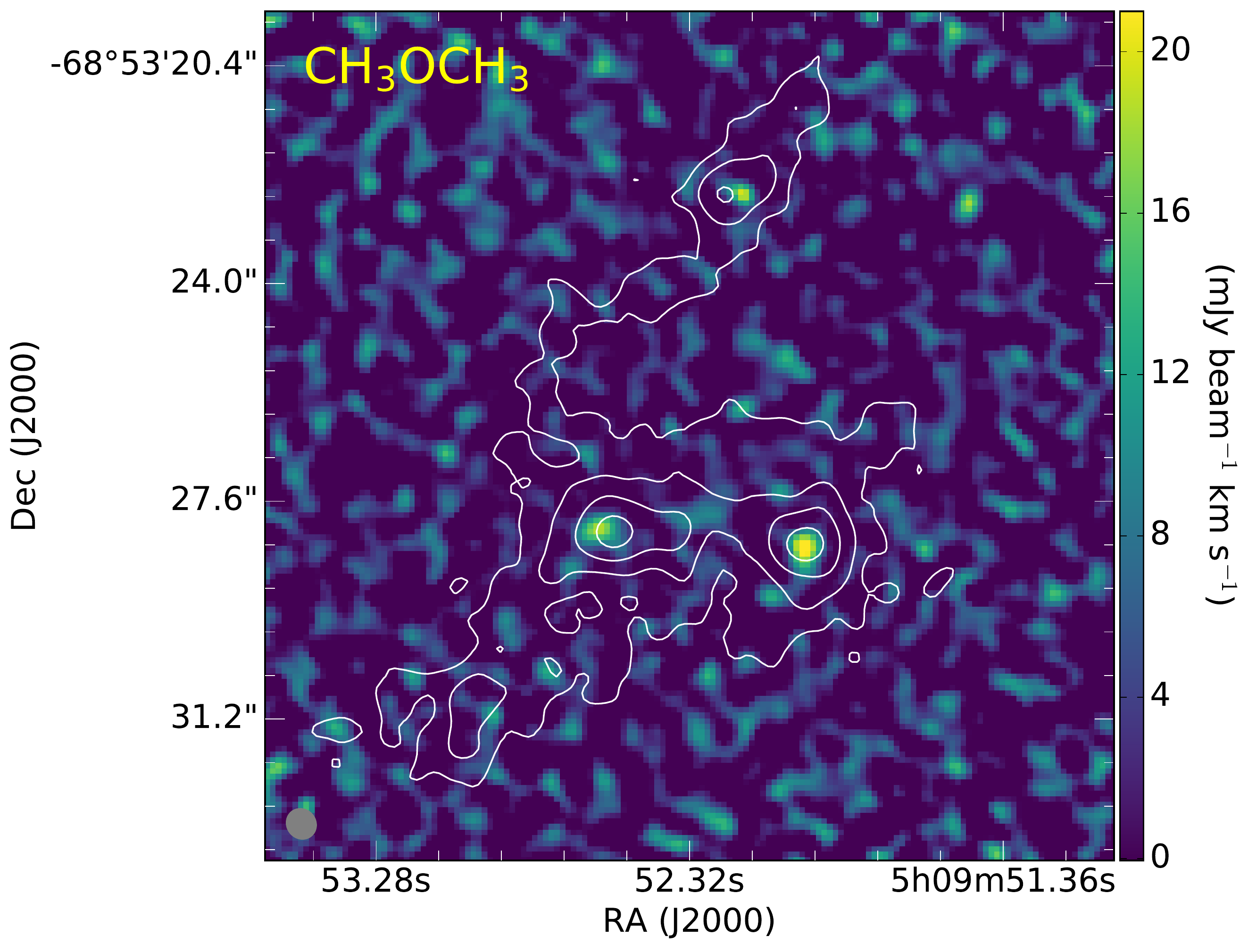} 
\hfill
\includegraphics[width=0.329\textwidth]{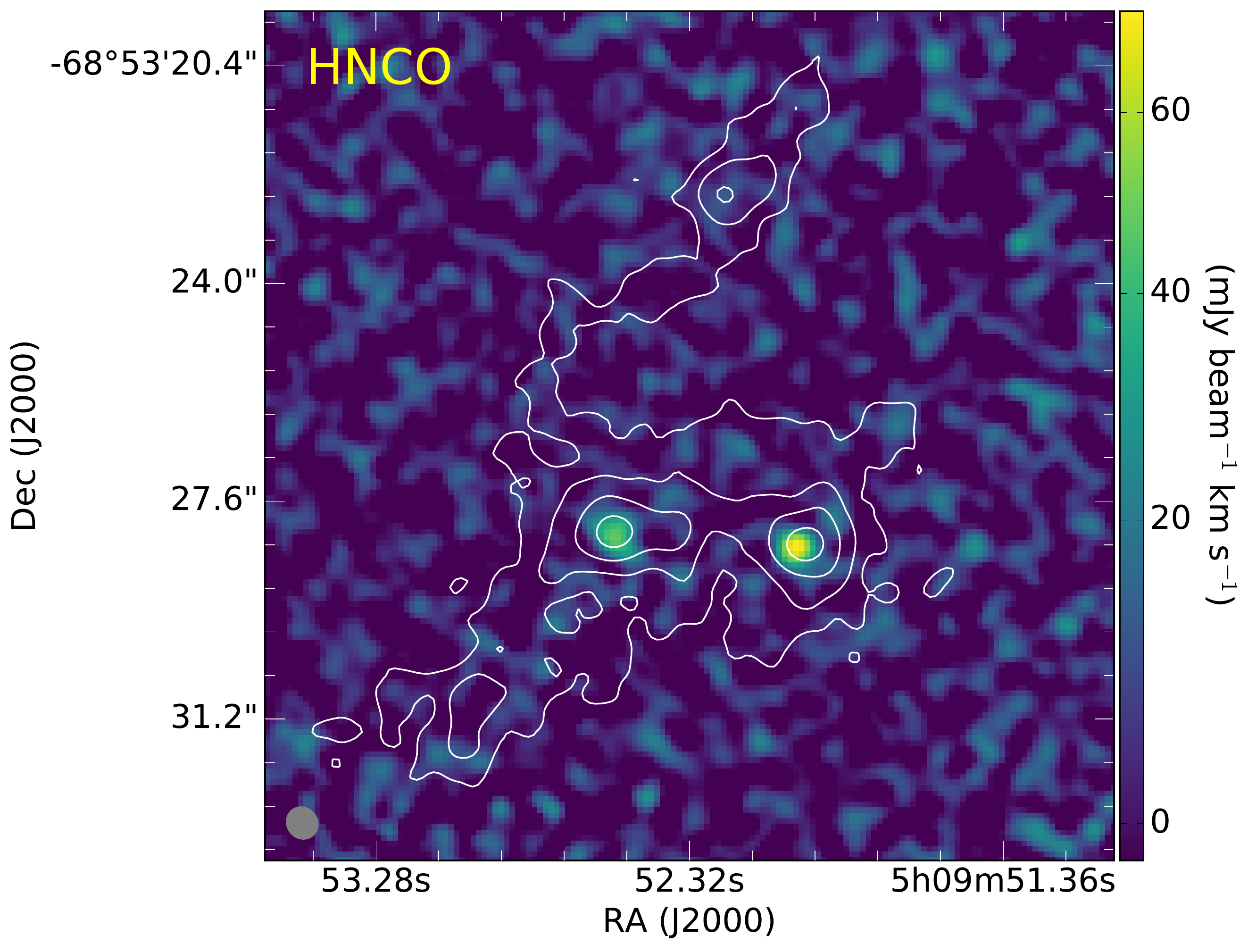} 
\hfill
\includegraphics[width=0.329\textwidth]{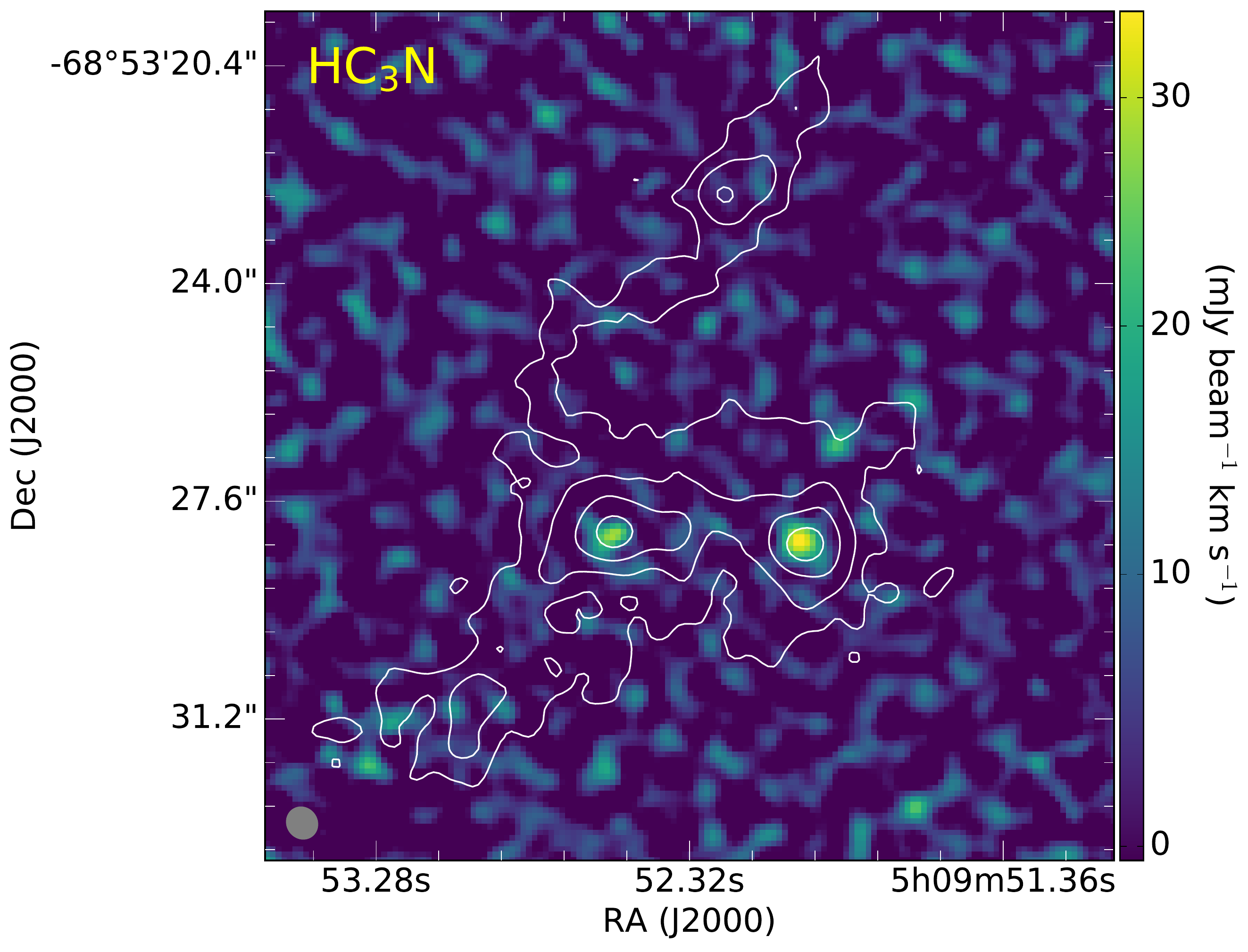} 
\hfill
\includegraphics[width=0.329\textwidth]{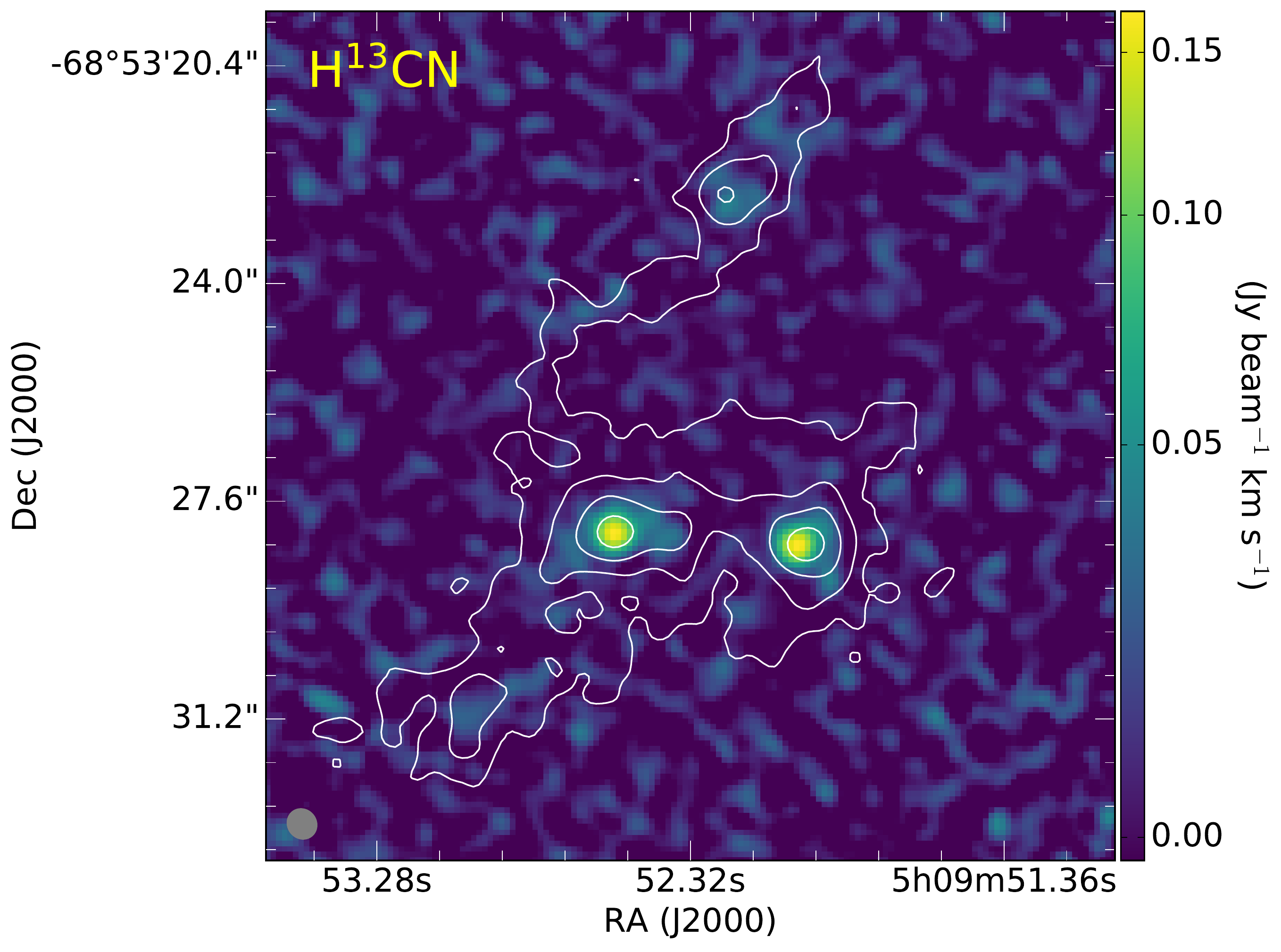} 
\hfill
\includegraphics[width=0.329\textwidth]{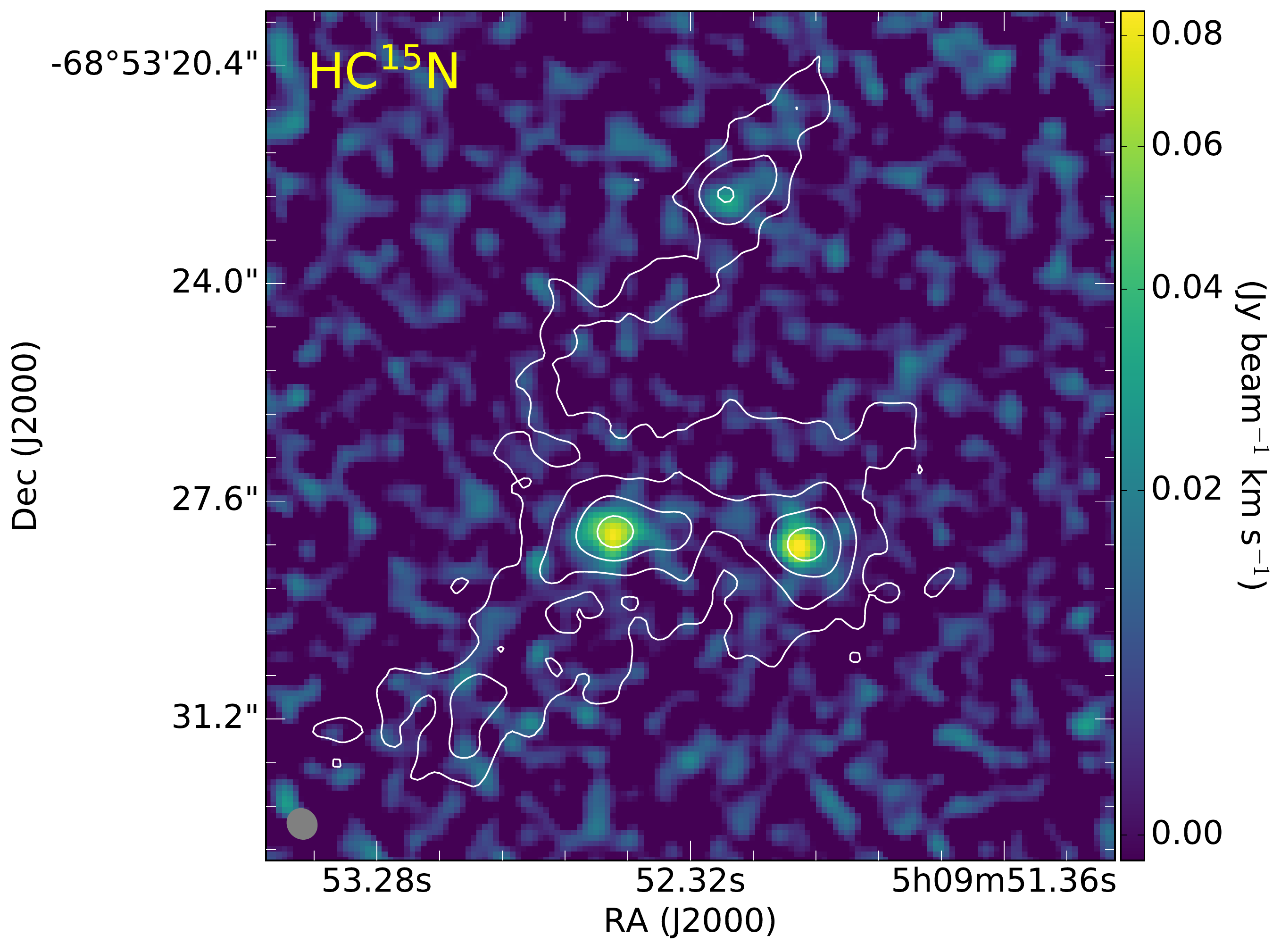} 
\hfill
\includegraphics[width=0.329\textwidth]{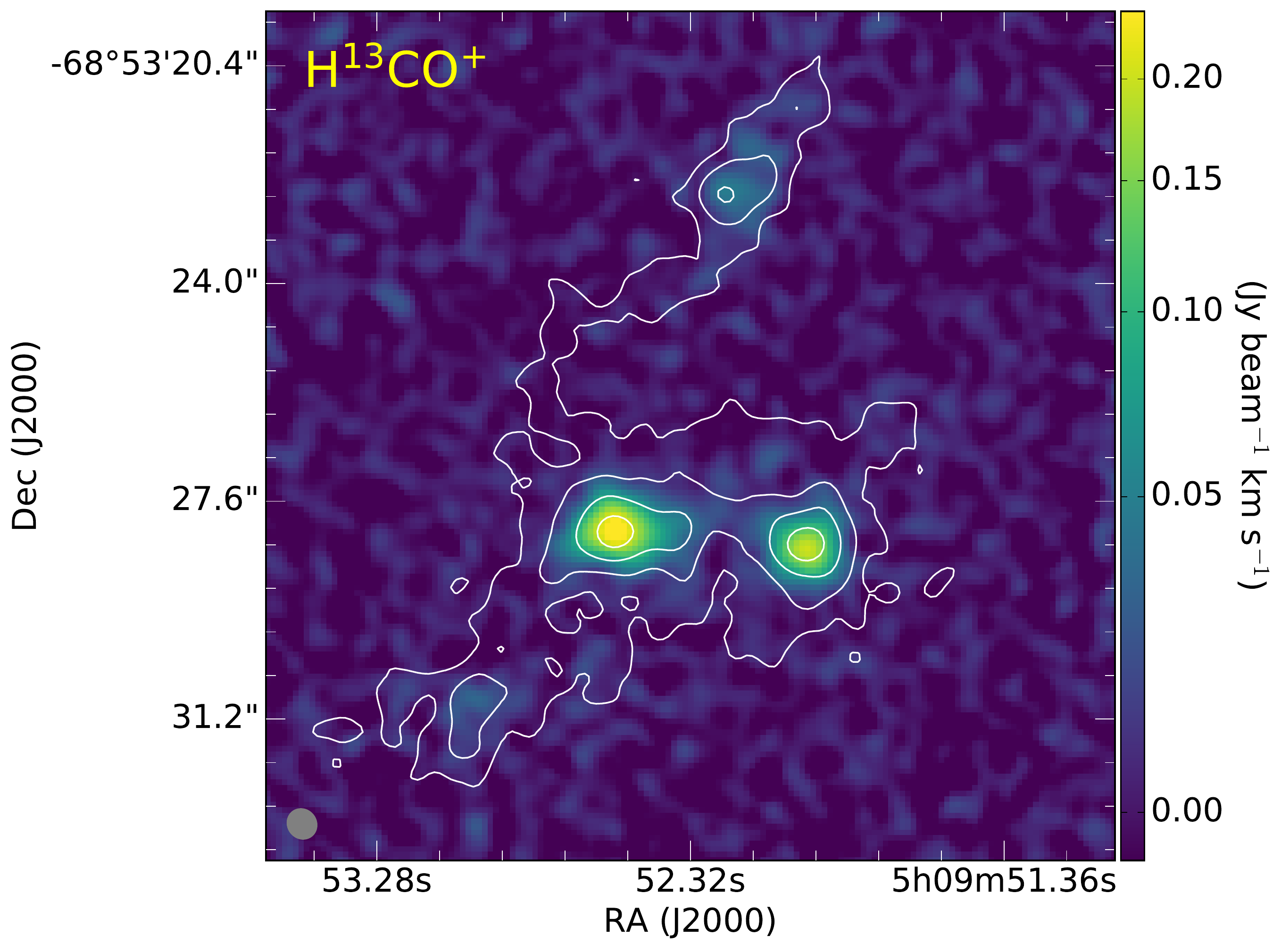} 
\hfill
\includegraphics[width=0.329\textwidth]{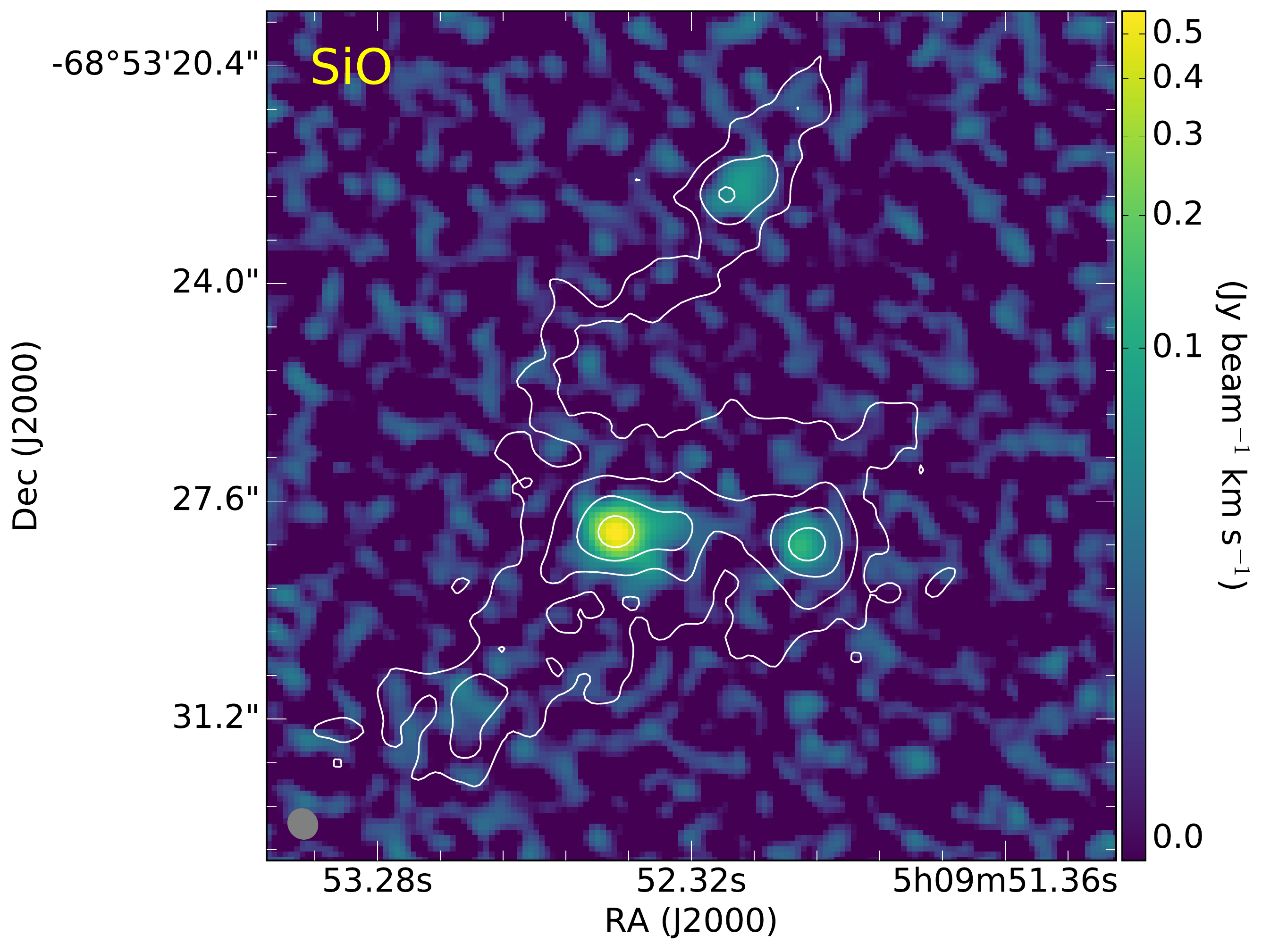} 
\hfill
\includegraphics[width=0.329\textwidth]{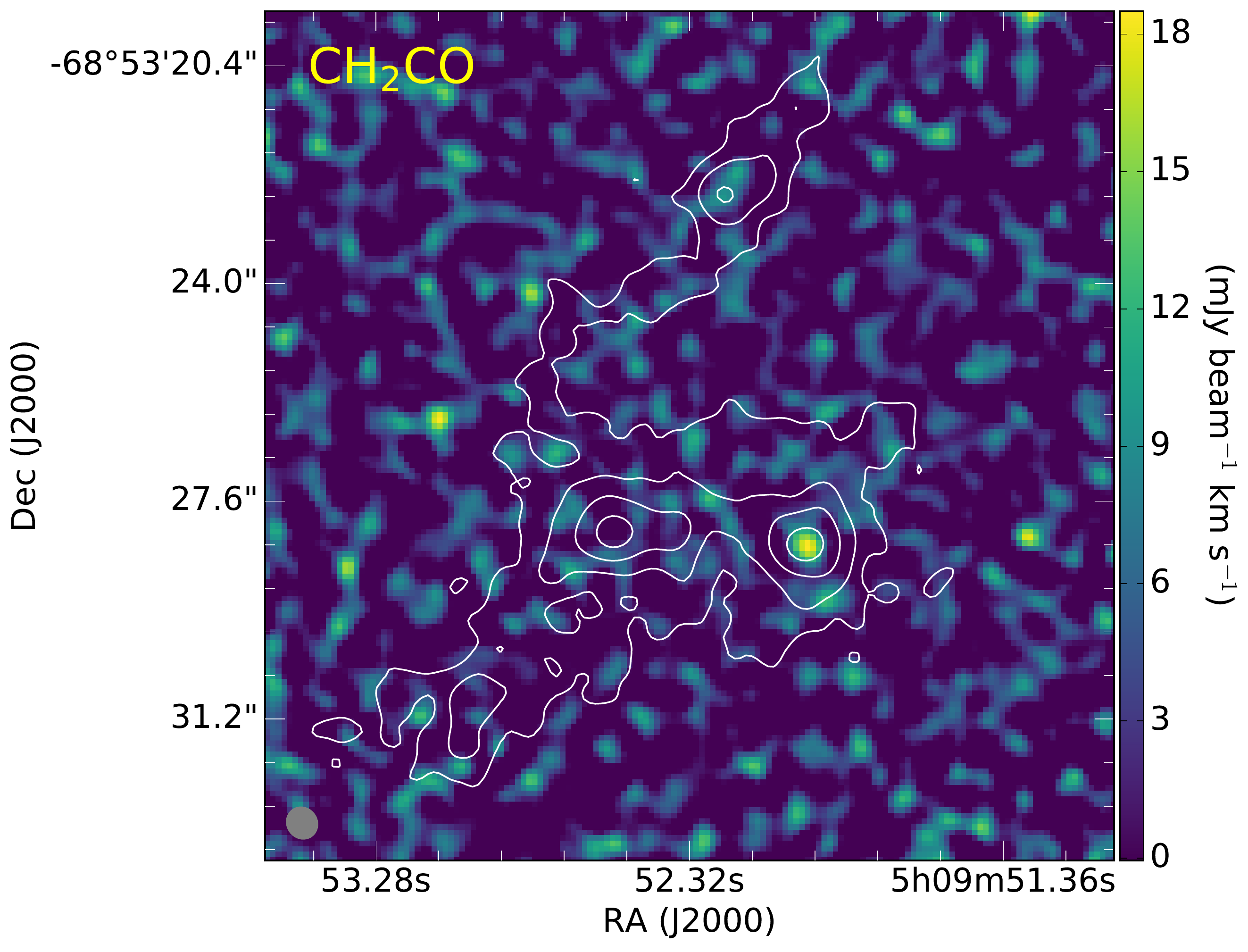} 
\hfill
\caption{The integrated intensity images of the molecular species detected toward N\,105--2\,A--2\,D and 2\,F ({\it from upper left to lower right}): 
CH$_3$OH (integrated over all CH$_3$OH transitions detected in the $\sim$242~GHz spectral window),  
CH$_3$CN (integrated over the $K$=0--6 components of the CH$_3$CN 14$_K$--13$_K$ ladder), 
NH$_2$CHO 12$_{2, 10}$--11$_{2, 9}$, 
CH$_3$OCH$_3$ 14$_{1,14}$--13$_{0,13}$, 
HNCO (integrated over the 11$_{0, 11}$--10$_{0, 10}$ and  11$_{1, 10}$--10$_{1, 9}$ transitions), 
HC$_3$N 27--26, 
H$^{13}$CN 3--2,
HC$^{15}$N 3--2,
H$^{13}$CO$^{+}$ 3--2,
SiO 6--5, and
CH$_2$CO 12$_{1, 11}$--11$_{1, 10}$.
The 1.2 mm continuum contours with contour levels of (3, 10, 30, 80)$\sigma$ are overlaid in each image for reference. The positions of H$_2$O and OH masers are shown in the CH$_3$CN image.  The size of the ALMA synthesized beam is indicated in the lower left corner of each image. 
\label{f:N105int2a}}
\end{figure*}

\begin{figure*}[h!]
\centering 
\includegraphics[width=0.329\textwidth]{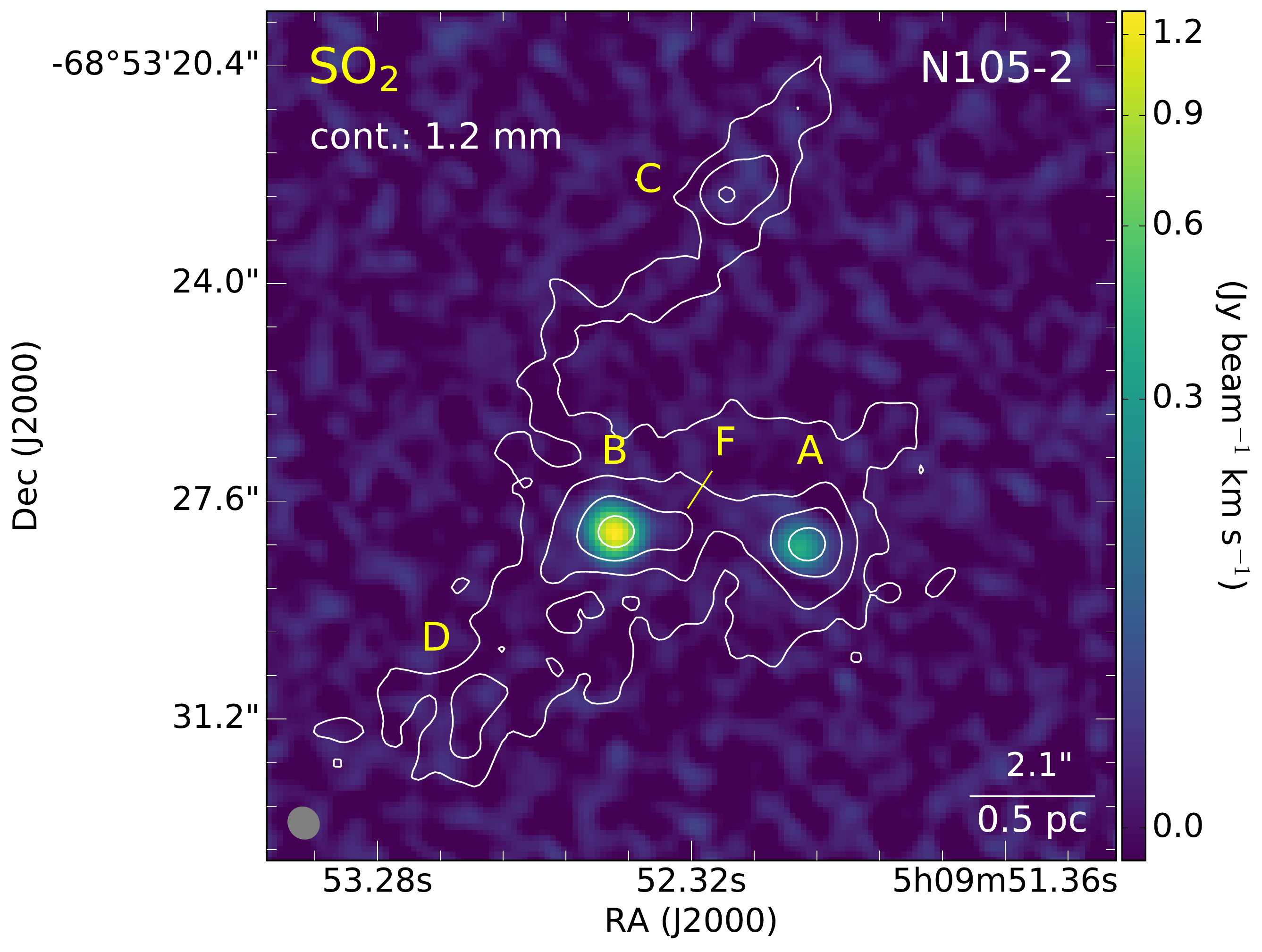} 
\hfill
\includegraphics[width=0.329\textwidth]{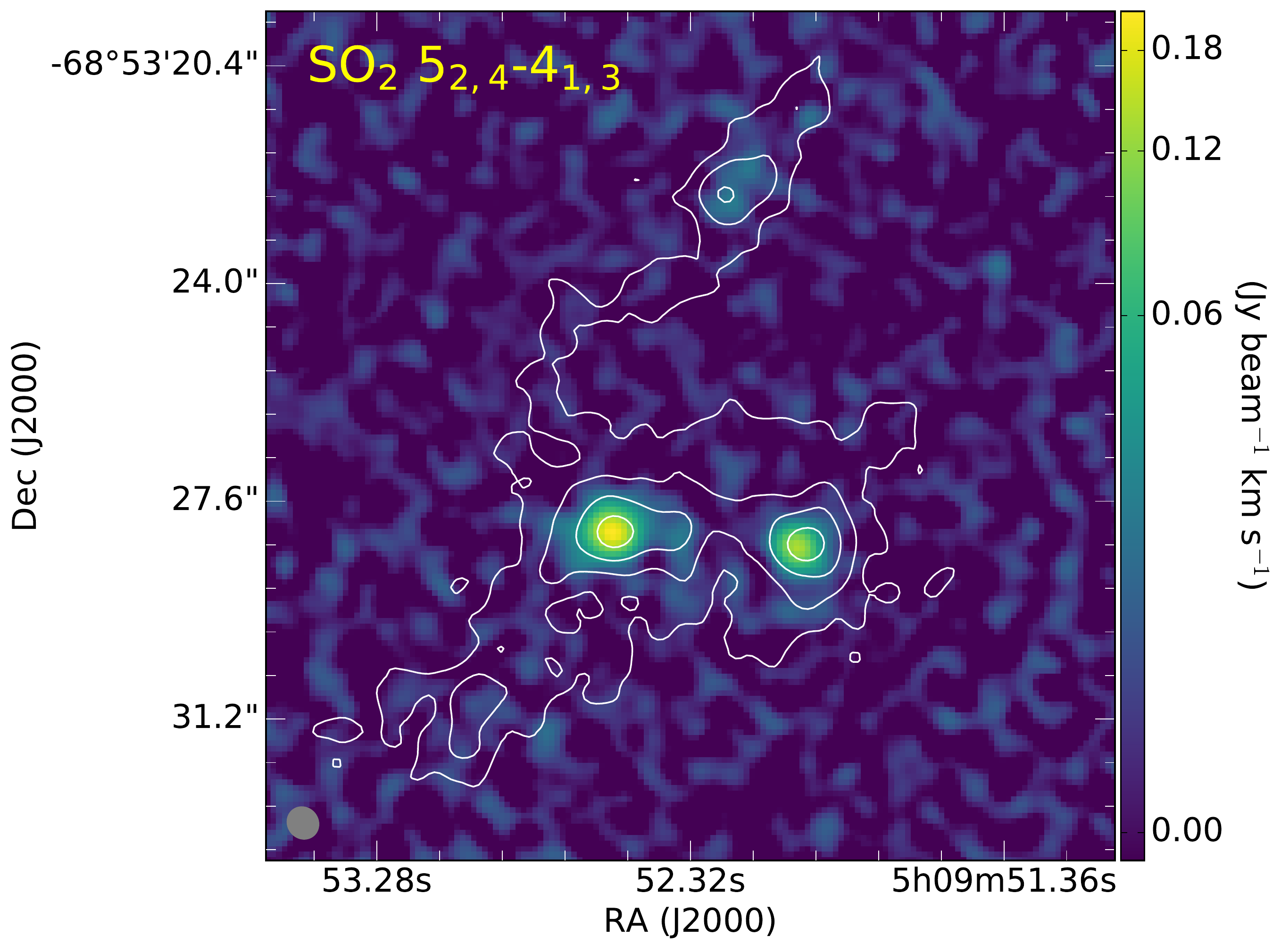} 
\hfill
\includegraphics[width=0.329\textwidth]{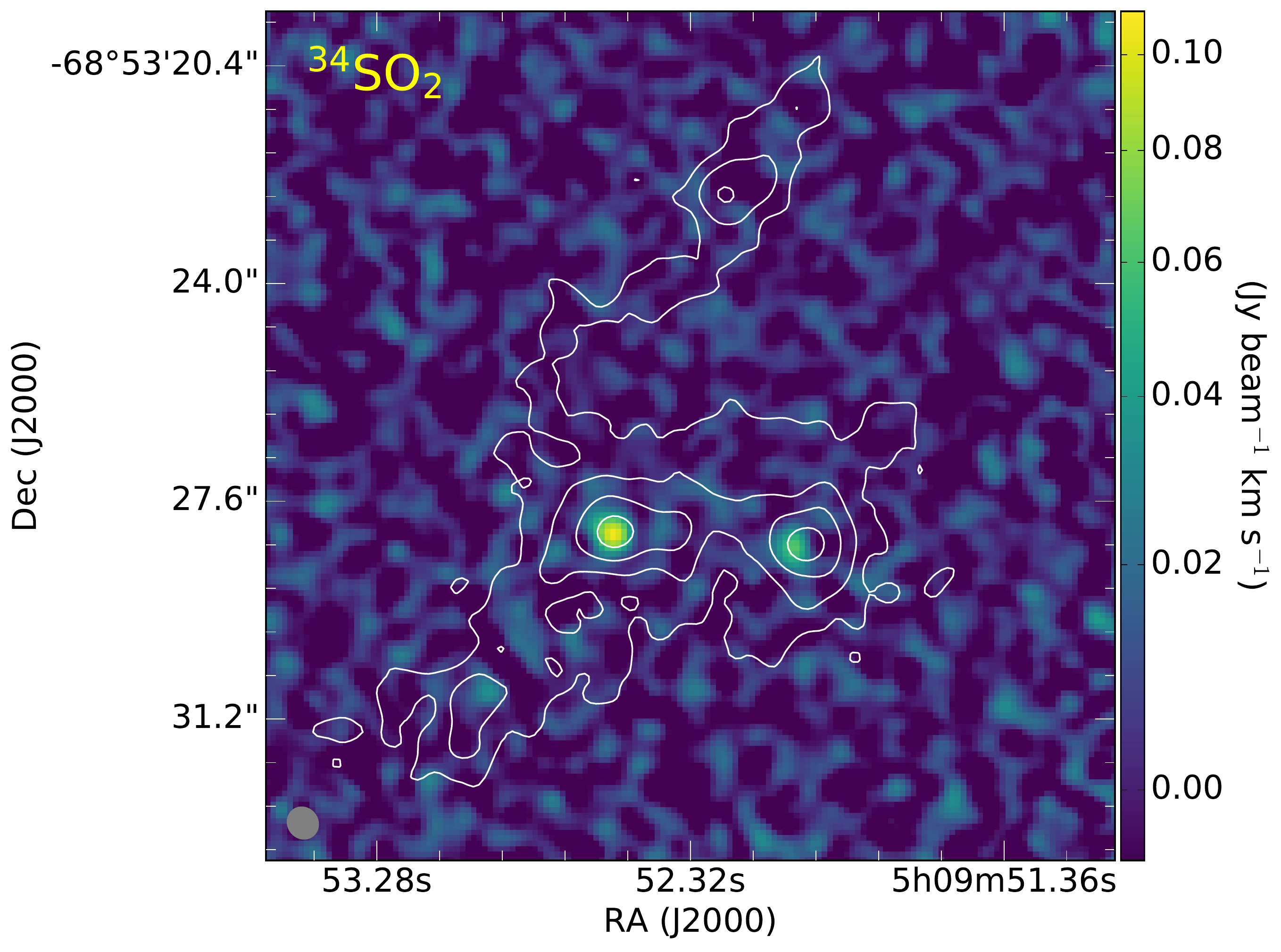} 
\hfill
\includegraphics[width=0.329\textwidth]{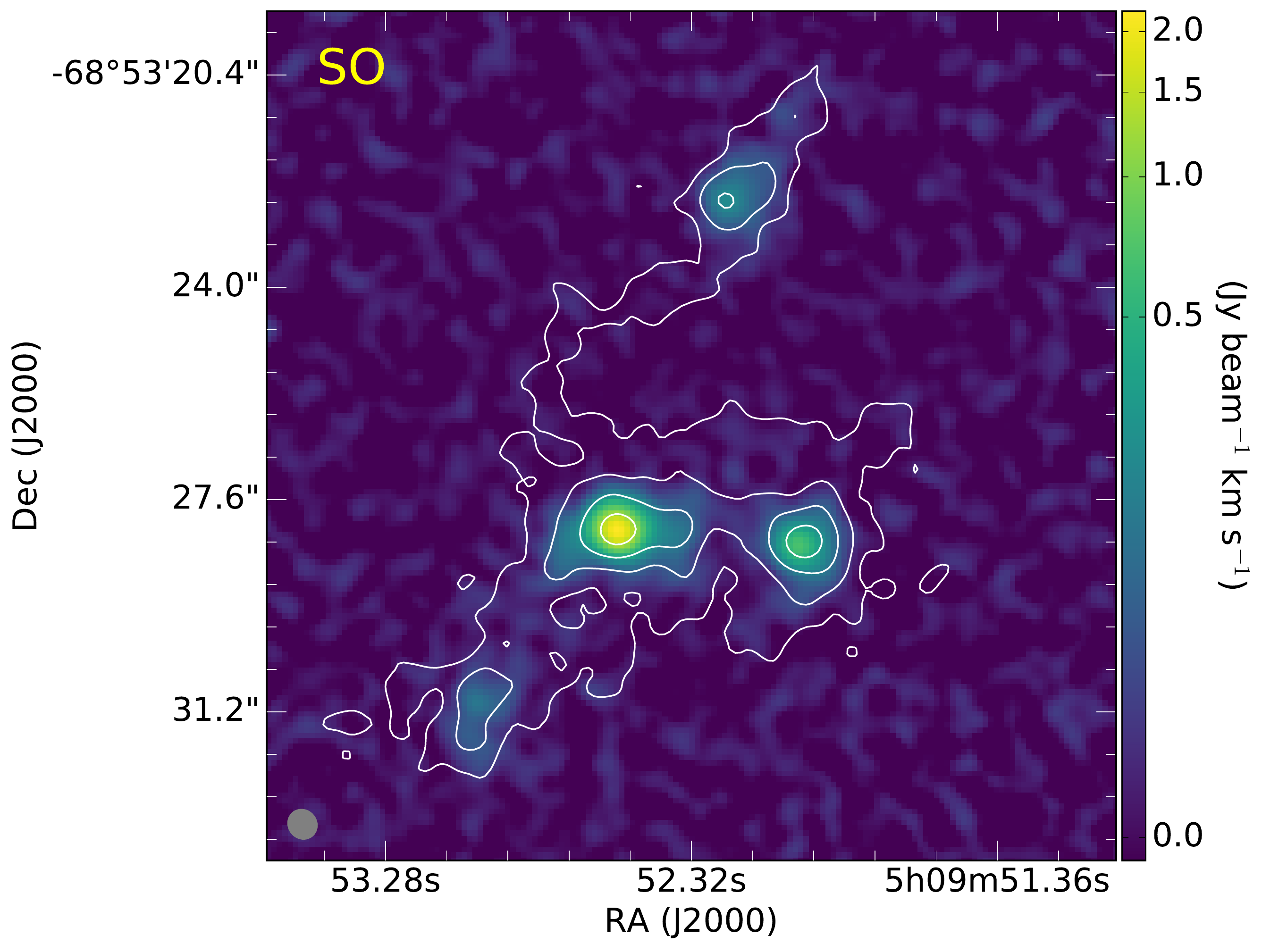} 
\hfill
\includegraphics[width=0.329\textwidth]{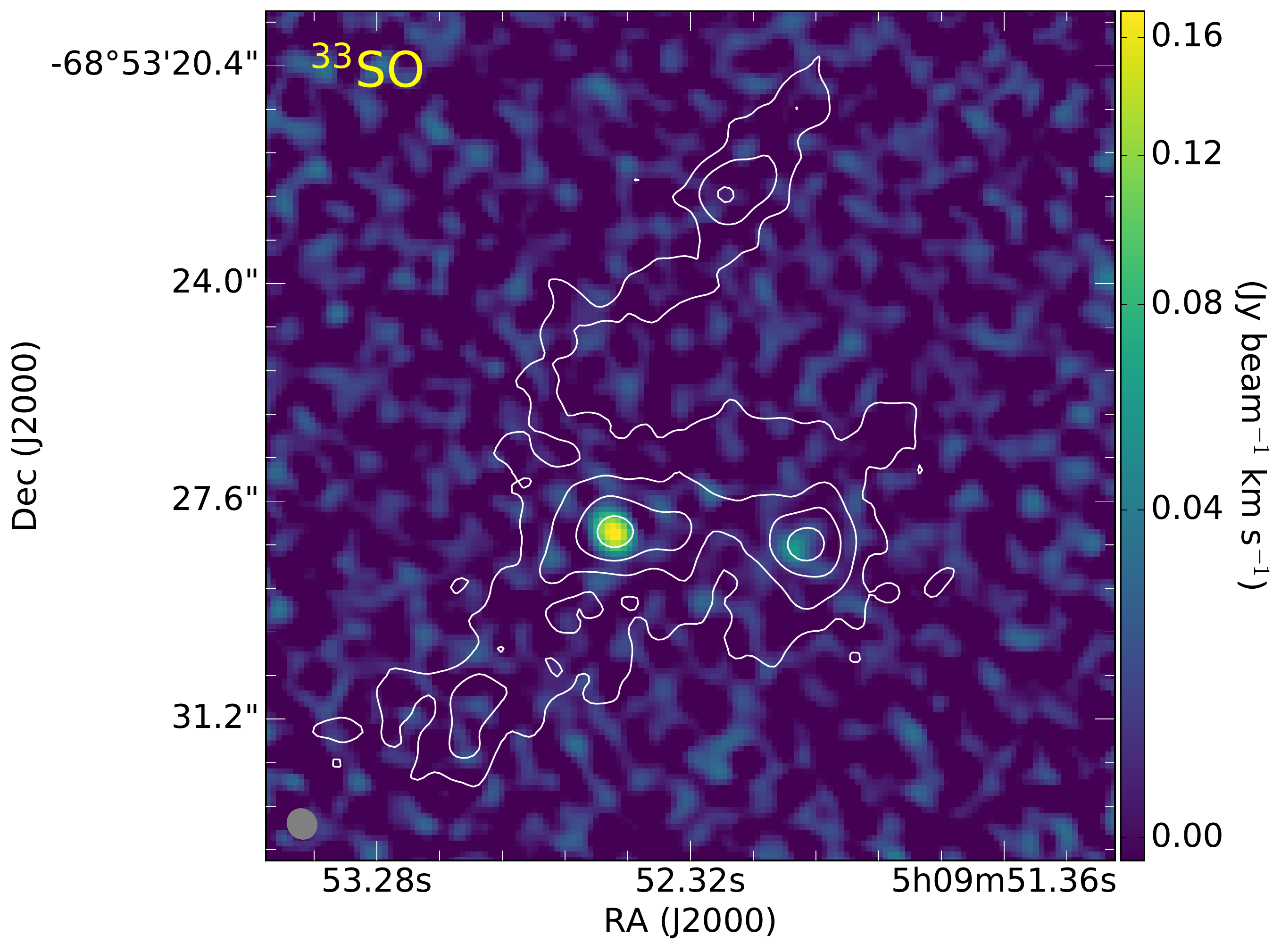} 
\hfill
\includegraphics[width=0.329\textwidth]{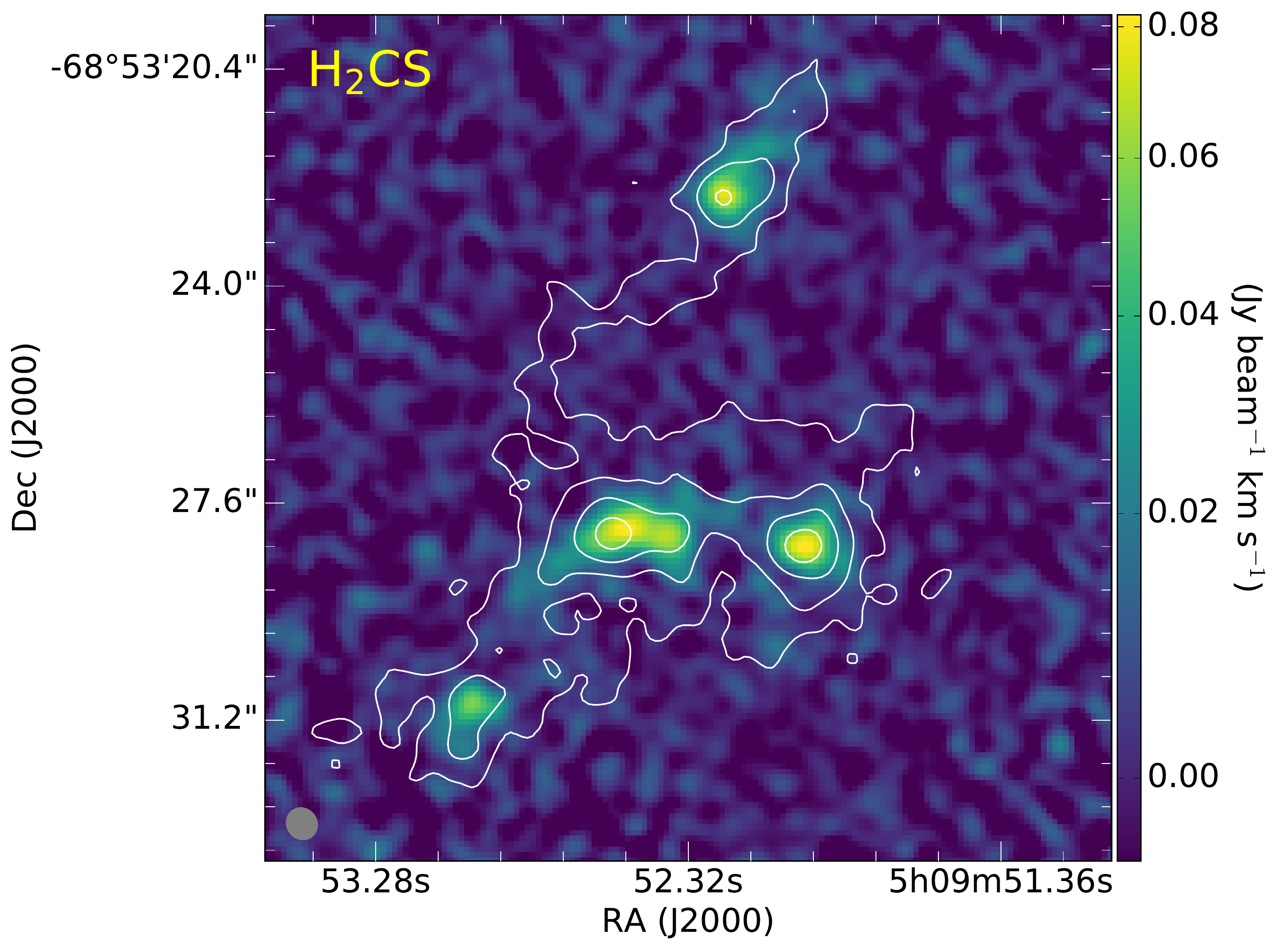} 
\hfill
\includegraphics[width=0.329\textwidth]{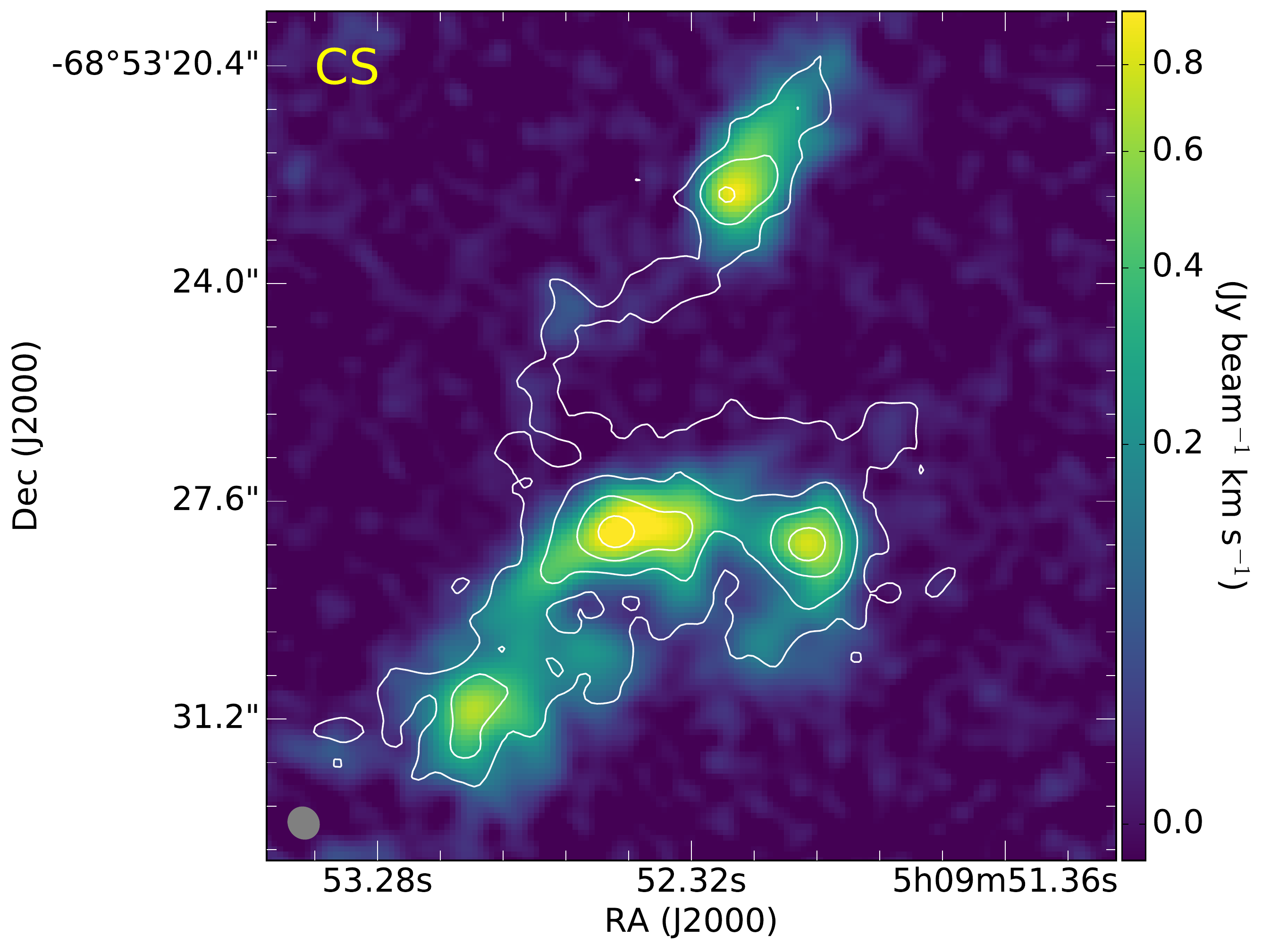} 
\hfill
\includegraphics[width=0.329\textwidth]{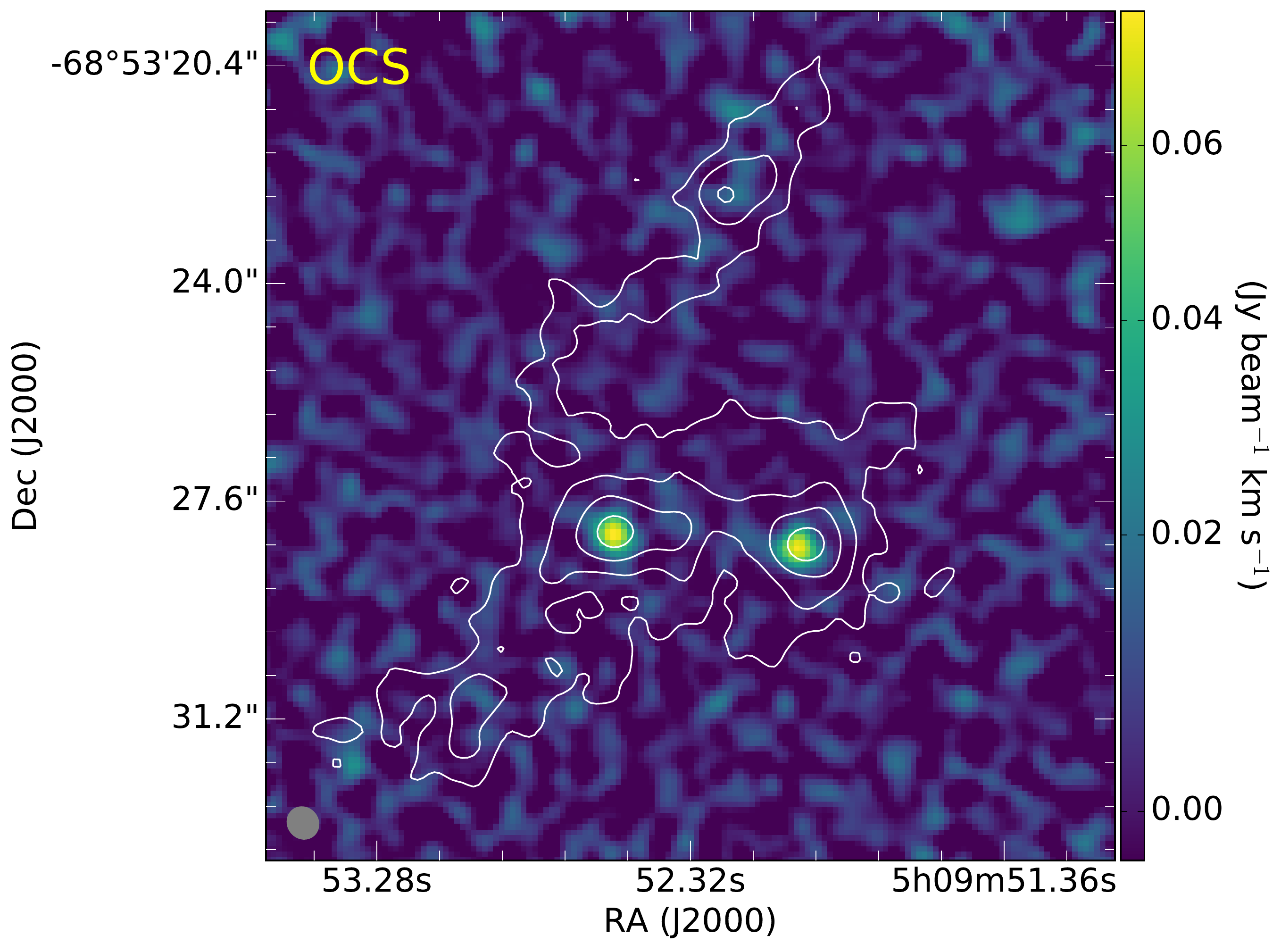} 
\hfill
\includegraphics[width=0.329\textwidth]{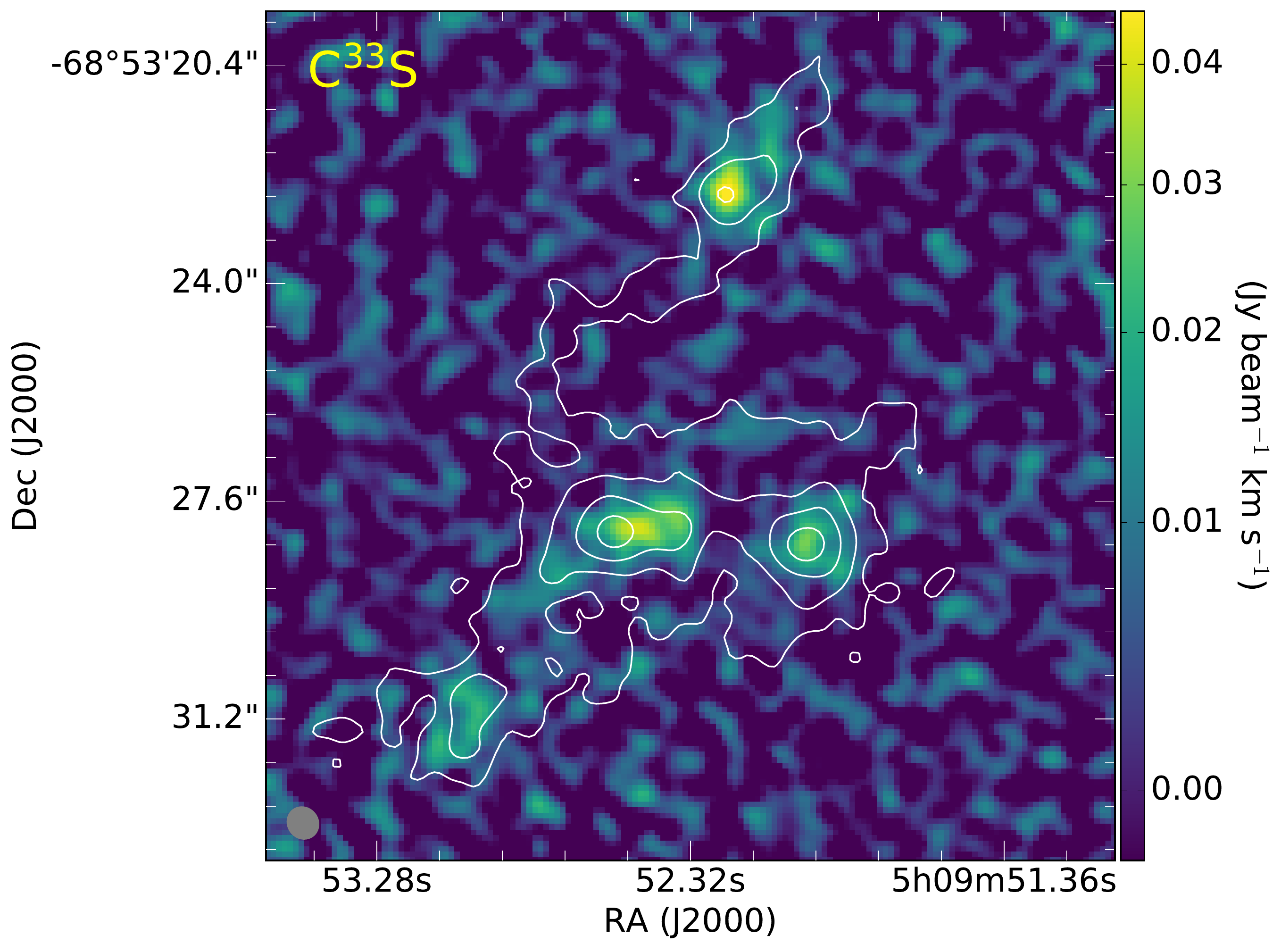} 
\hfill
\includegraphics[width=0.329\textwidth]{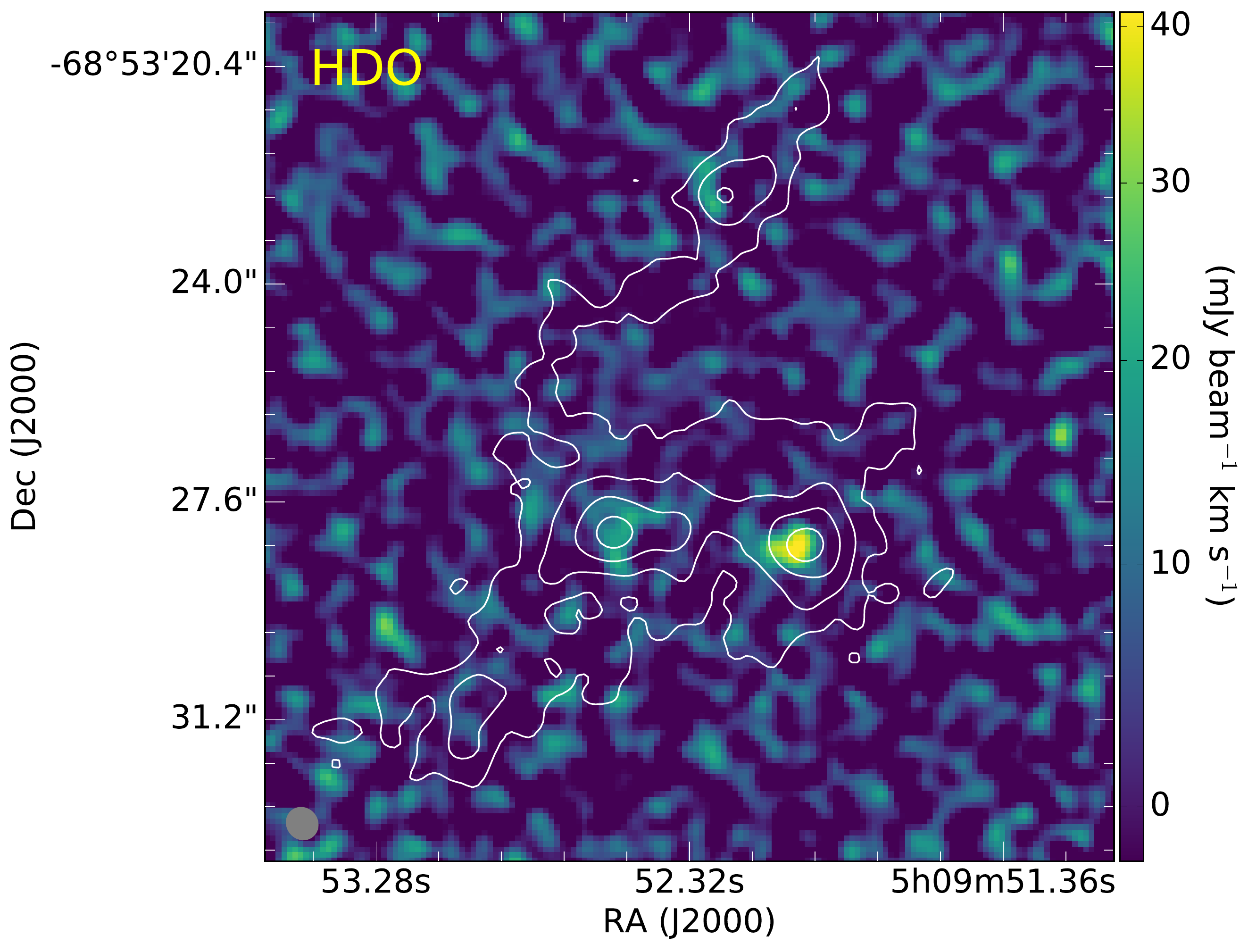} 
\hfill \hfill \hfill \hfill
\includegraphics[width=0.327\textwidth]{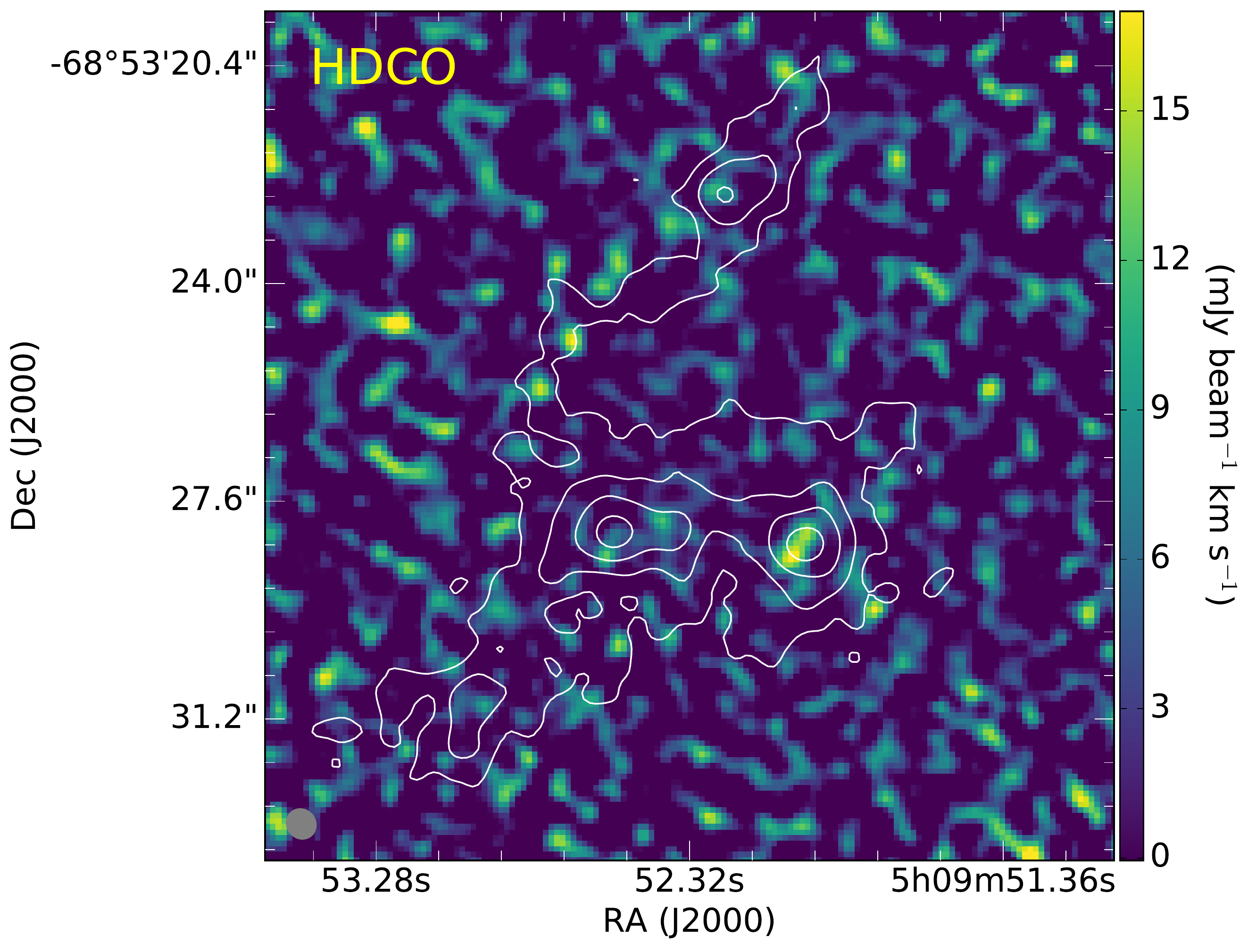} 
\hfill 
\includegraphics[width=0.327\textwidth]{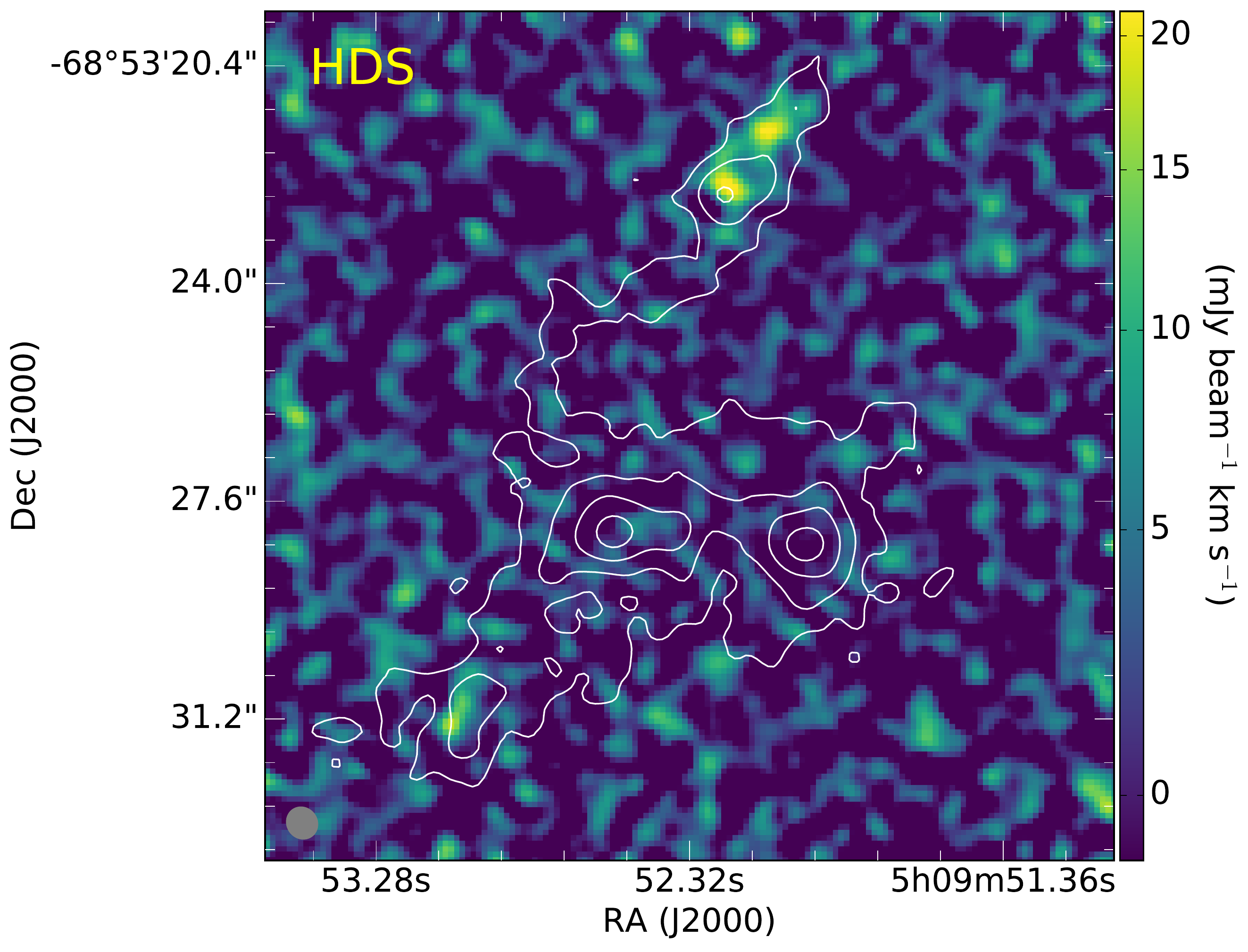} 
\hfill
\caption{Same as Fig.~\ref{f:N105int2a}, but for ({\it from upper left to lower right}): 
SO$_2$ (integrated over all the  transitions detected in the $\sim$245~GHz spectral window), 
SO$_2$ 5$_{2, 4}$--4$_{1, 3}$  (integrated over the channels containing the line emission from component C), 
$^{34}$SO$_2$ (integrated over all the  transitions detected in the $\sim$245~GHz spectral window),
SO 6$_6$--5$_5$,  
$^{33}$SO 7$_6$--6$_5$,
H$_2$CS 7$_{1, 6}$--6$_{1, 5}$,
CS 5--4, 
OCS 20--19, 
C$^{33}$S 7$_6$--6$_5$,
HDO 2$_{1, 1}$--2$_{1, 2}$,  
HDCO 4$_{2, 2}$--3$_{2, 1}$, and 
HDS 1$_{0, 1}$--0$_{0, 0}$.
\label{f:N105int2b}}
\end{figure*}

\begin{figure*}[h!]
\includegraphics[width=0.485\textwidth]{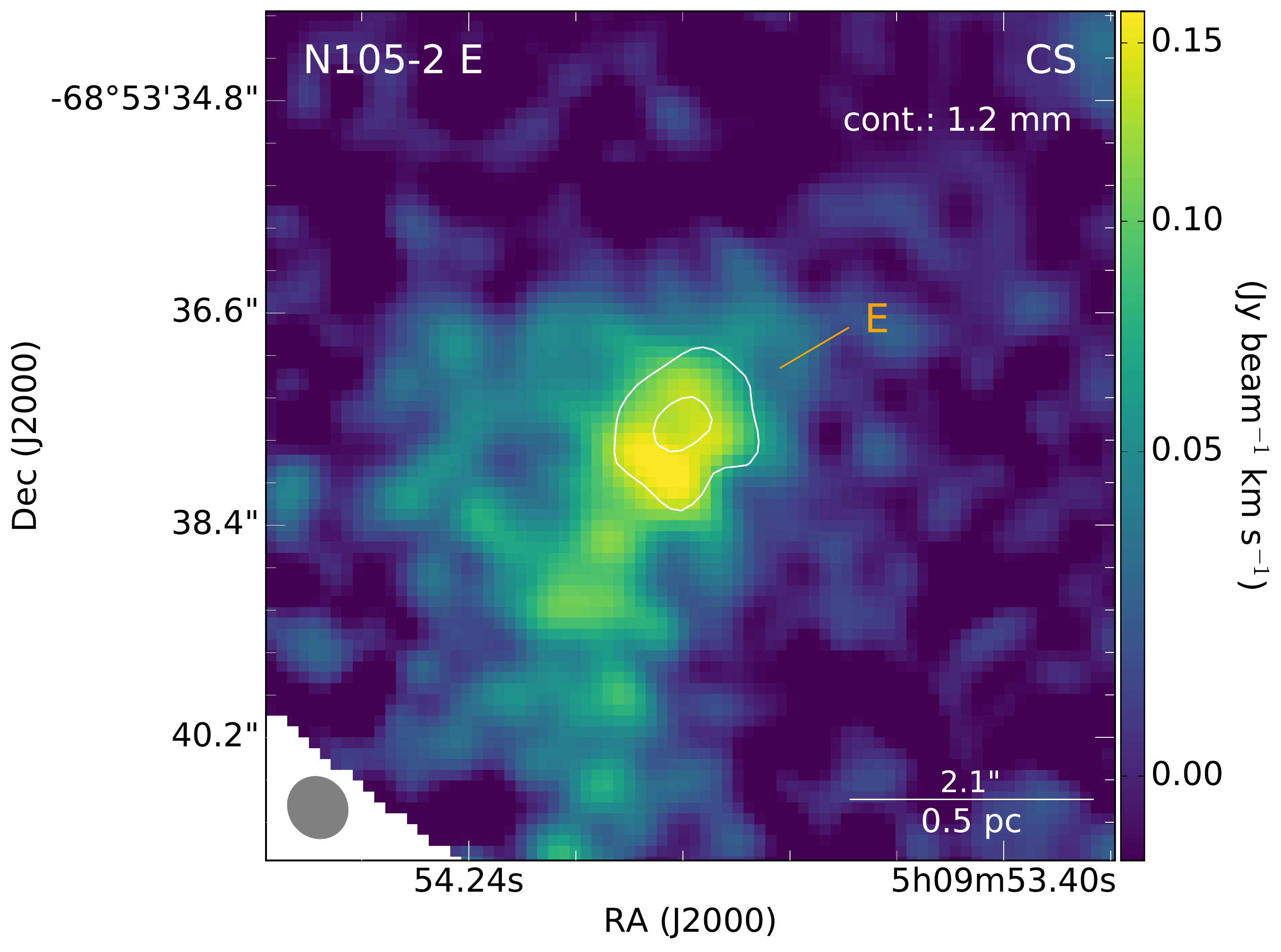}
\hfill
\includegraphics[width=0.485\textwidth]{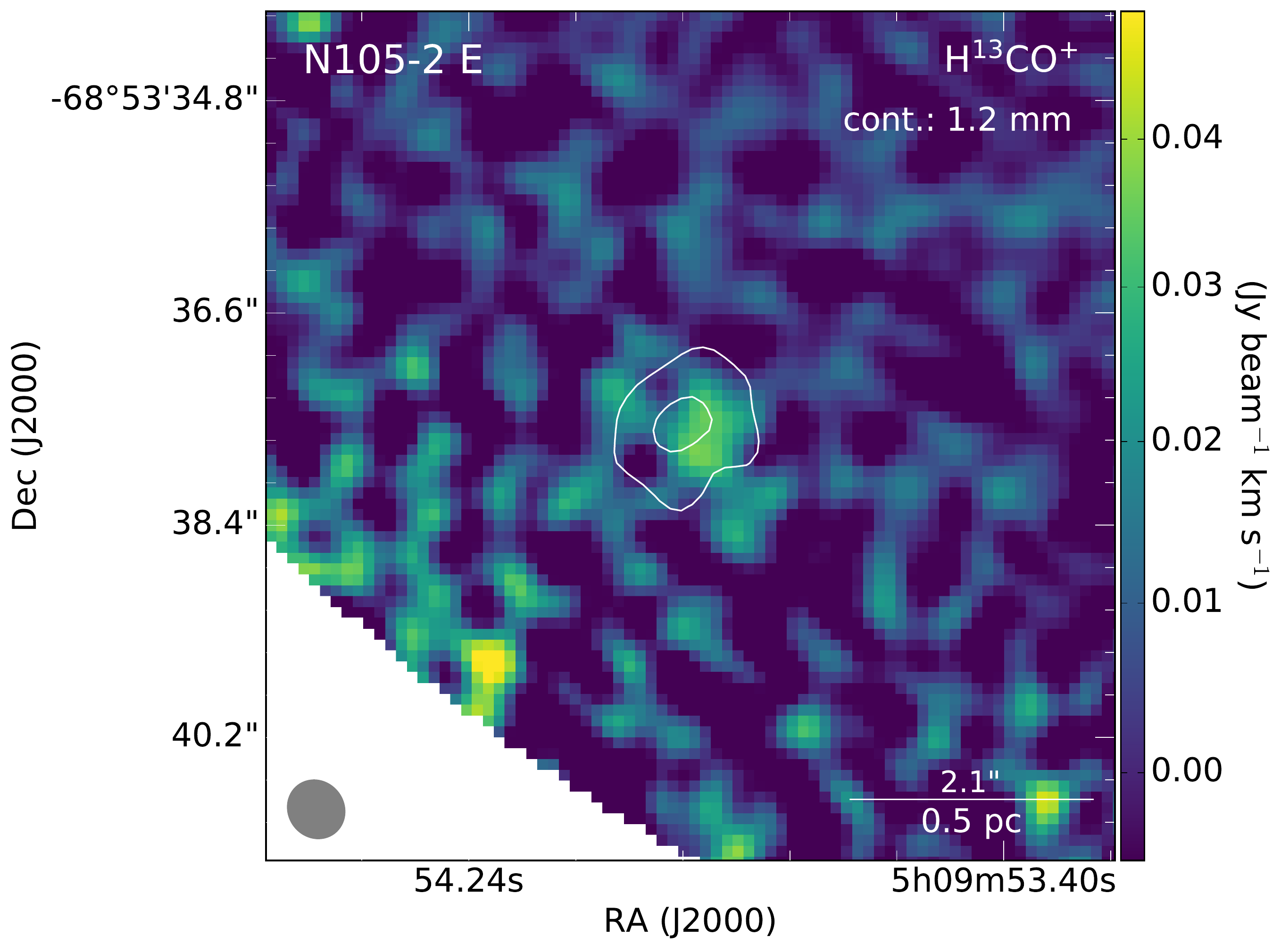}
\hfill
\includegraphics[width=0.485\textwidth]{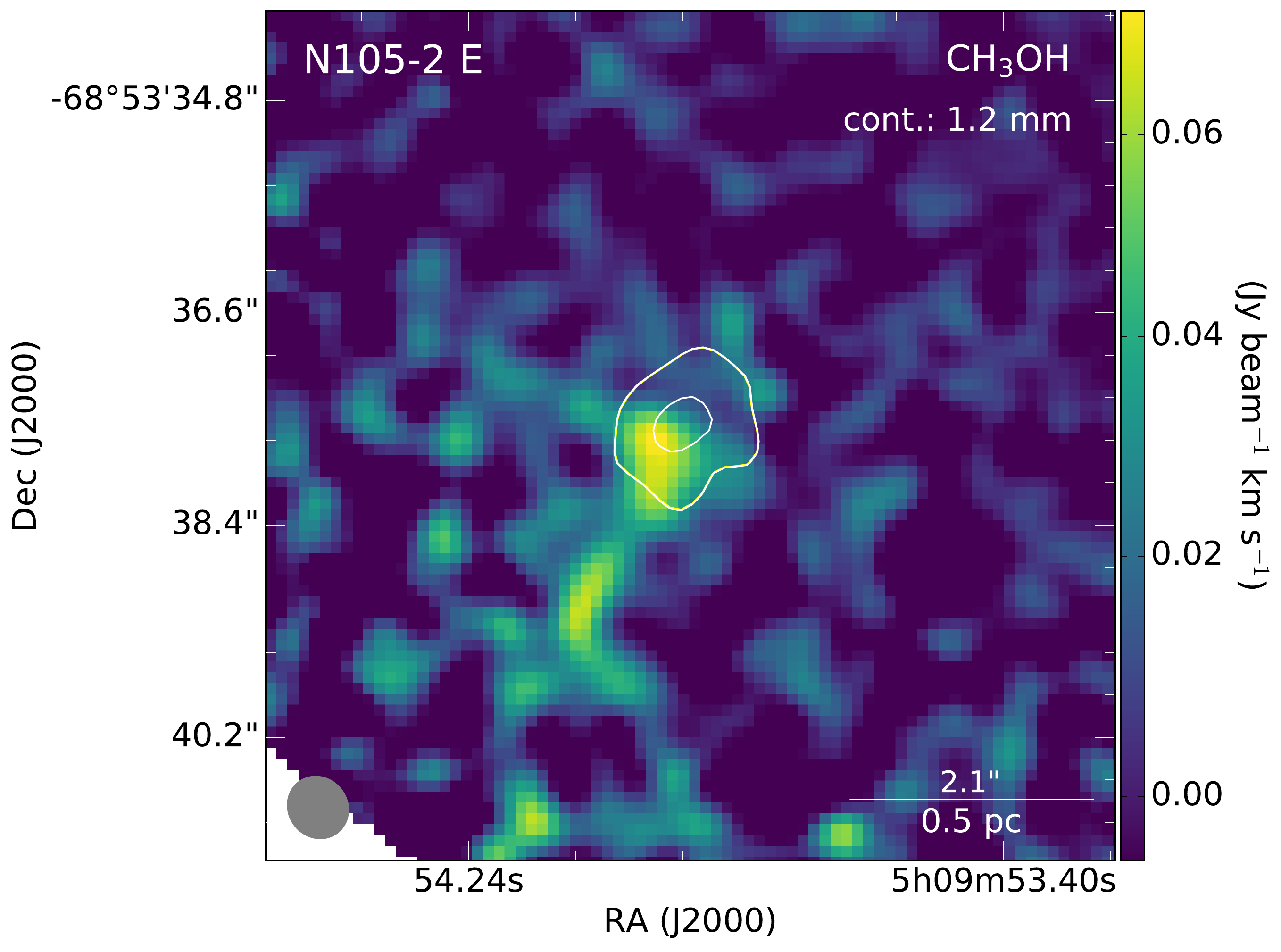}
\hfill
\includegraphics[width=0.485\textwidth]{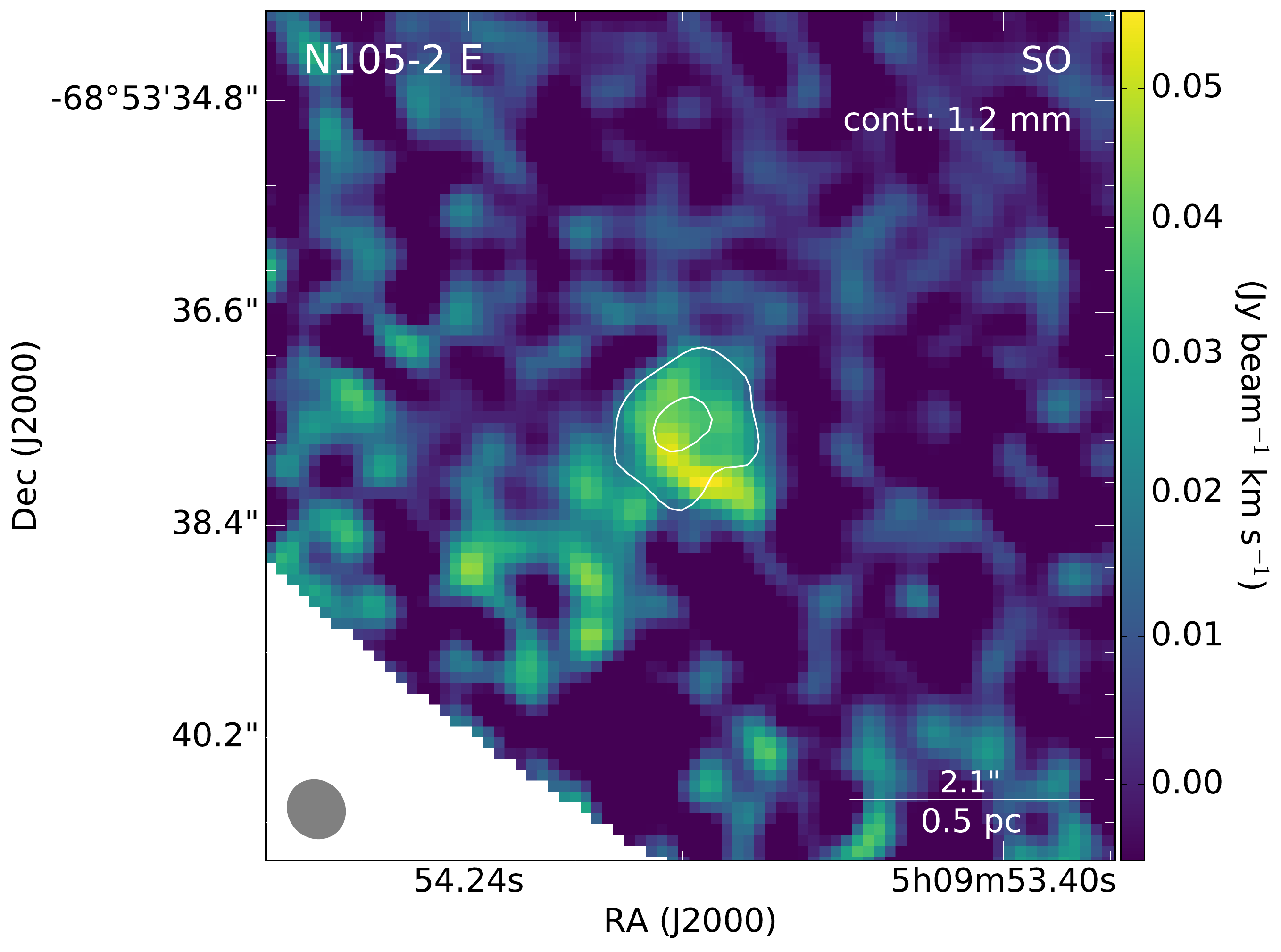}
\caption{{\it From upper left to lower right}: The CS (5--4),  H$^{13}$CO$^{+}$ (3--2),  CH$_3$OH (combined 5$_{-1, 5}-4_{-1, 4}$ and 5$_{0, 5}-4_{0, 4}{\rm ++}$ transitions),  and SO $6_6-5_5$ integrated intensity images of N\,105--2\,E.  The white contours in each image correspond to the 1.2 mm continuum emission with contour levels of (3, 8)$\sigma$.  \label{f:N105int2c}}
\end{figure*}

\indent The integrated intensity images for N\,105--2 are shown in Figs.~\ref{f:N105int2a}--\ref{f:N105int2b} (2\,A--2\,D and 2\,F) and in Fig.~\ref{f:N105int2c} (2\,E). CH$_3$OH is widespread across the N\,105--2 field with both the compact emission associated with continuum sources and the extended emission throughout the region. Similar spatial distributions are seen for CS, H$_2$CS, SO, and H$^{13}$CO. CS has its brightest, most extended component away from the continuum peaks. COMs other than CH$_3$OH have compact morphology and are located toward 2\,A and 2\,B only except CH$_3$CN;  faint CH$_3$CN emission is also detected toward 2\,C. 2\,C is the only source in N\,105 covered by our observations with a detection of the HDS emission.  

The most chemically rich continuum sources in N\,105--2 are 2\,A and 2\,B. In general, the molecular line emission peaks coincide with the continuum peak toward 2\,B. In 2\,A, the CH$_3$CN peak is offset from the continuum peak by $\sim$0$\rlap.{''}$15--0$\rlap.{''}$2 and coincides with the emission peaks of other molecules such as CH$_3$OH, HNCO, HDO, HC$^{15}$N, SO$_2$, OCS, and $^{33}$SO. The emission from other species is slightly offset from both the CH$_3$CN/CH$_3$OH peak and the continuum peak, but within 1--1.5 pixels (or within $\sim$0$\rlap.{''}$2). Such offsets are also observed toward other continuum sources in N\,105--2.

\subsubsection{N\,105--3}

\begin{figure*}[h!]
\includegraphics[width=0.485\textwidth]{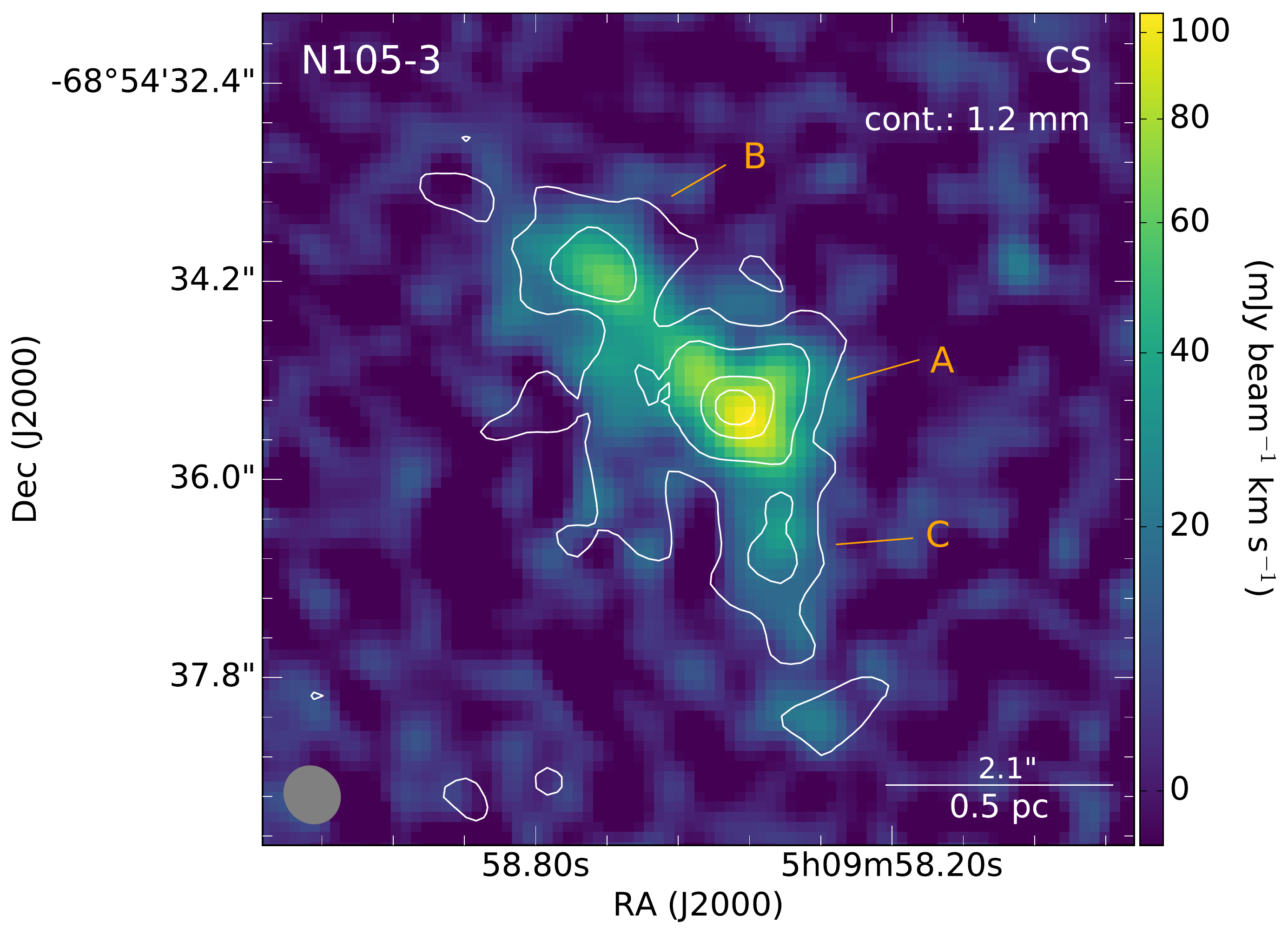}
\hfill
\includegraphics[width=0.485\textwidth]{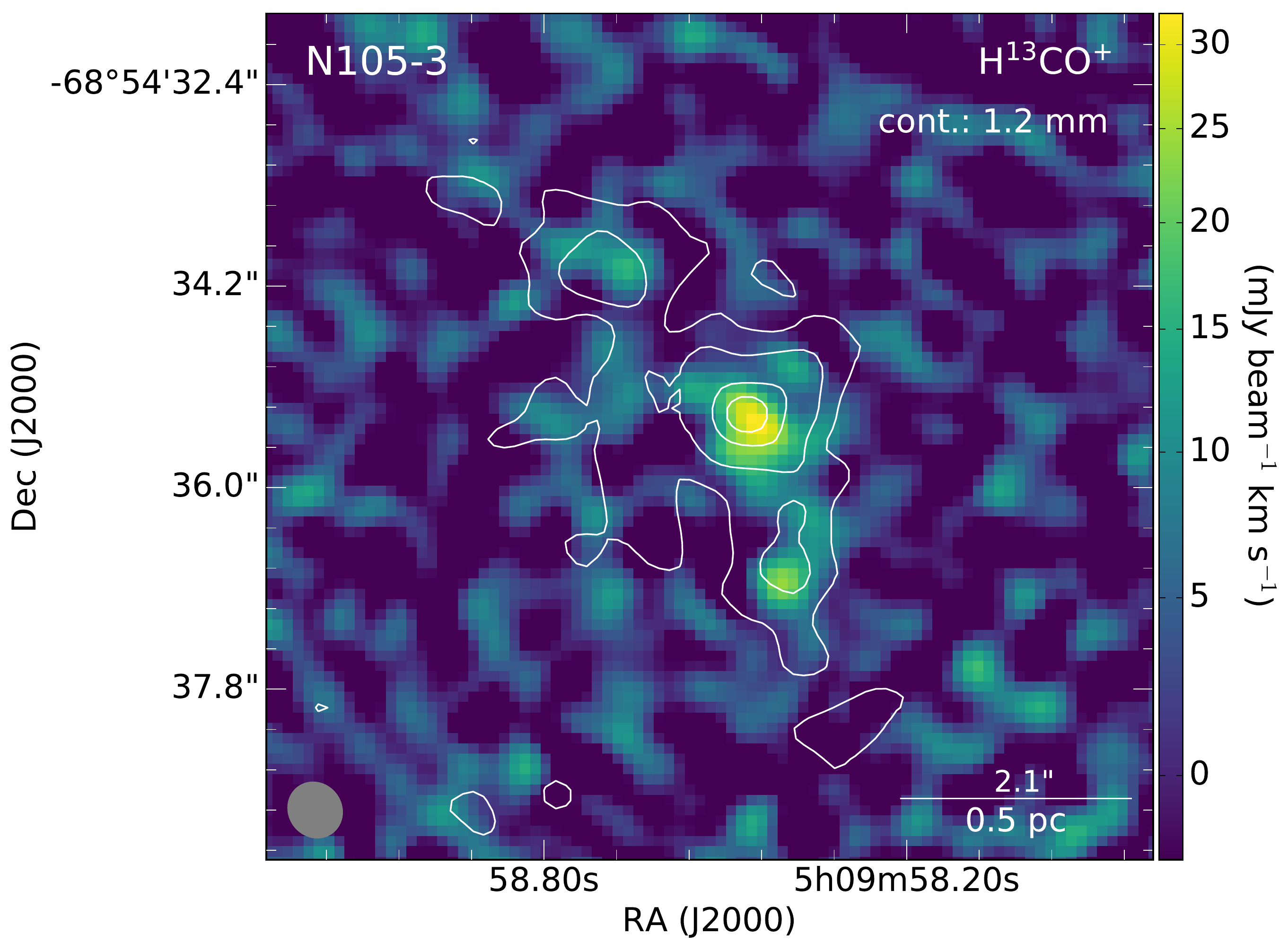}
\hfill
\includegraphics[width=0.485\textwidth]{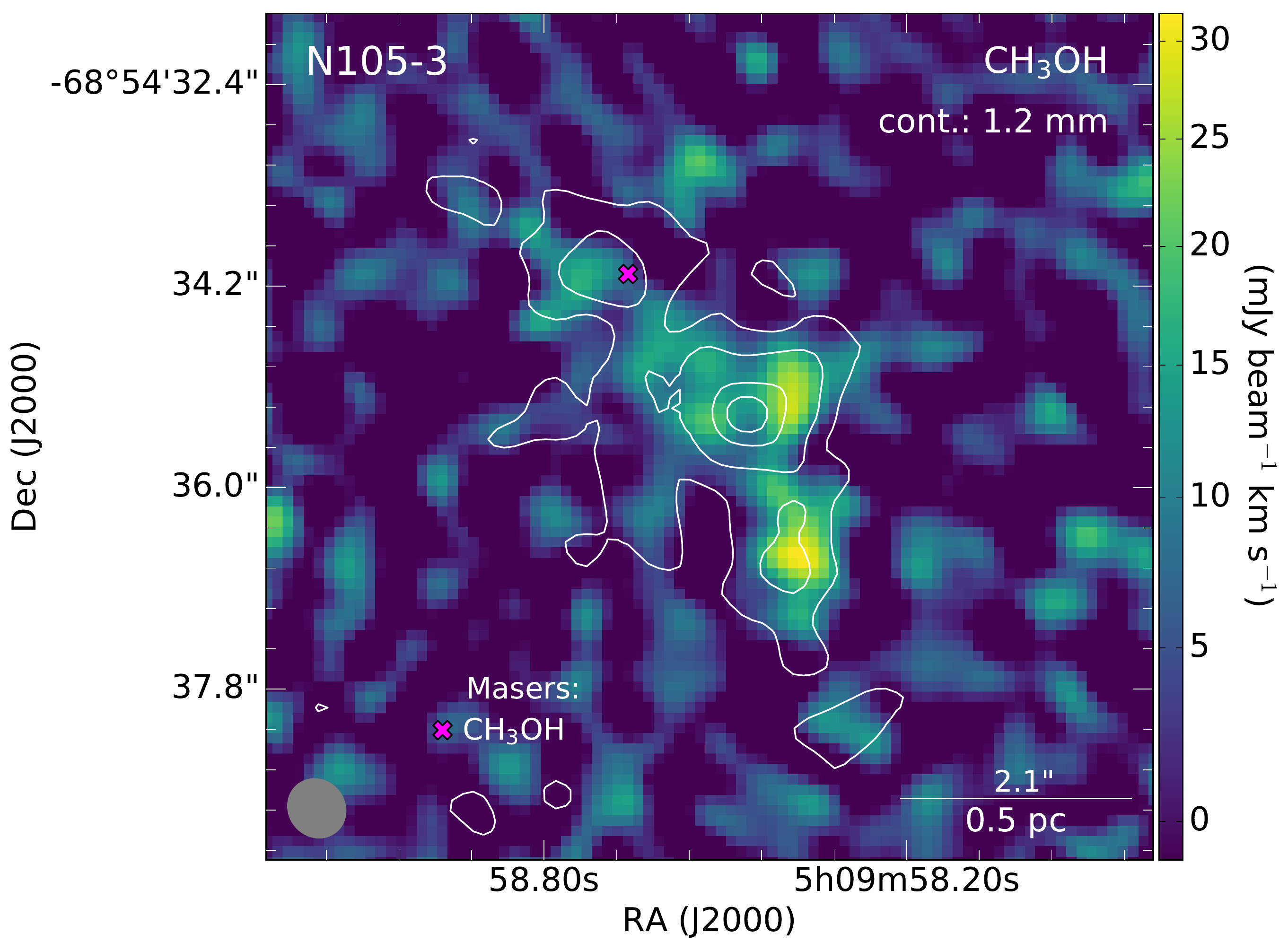}
\hfill
\includegraphics[width=0.485\textwidth]{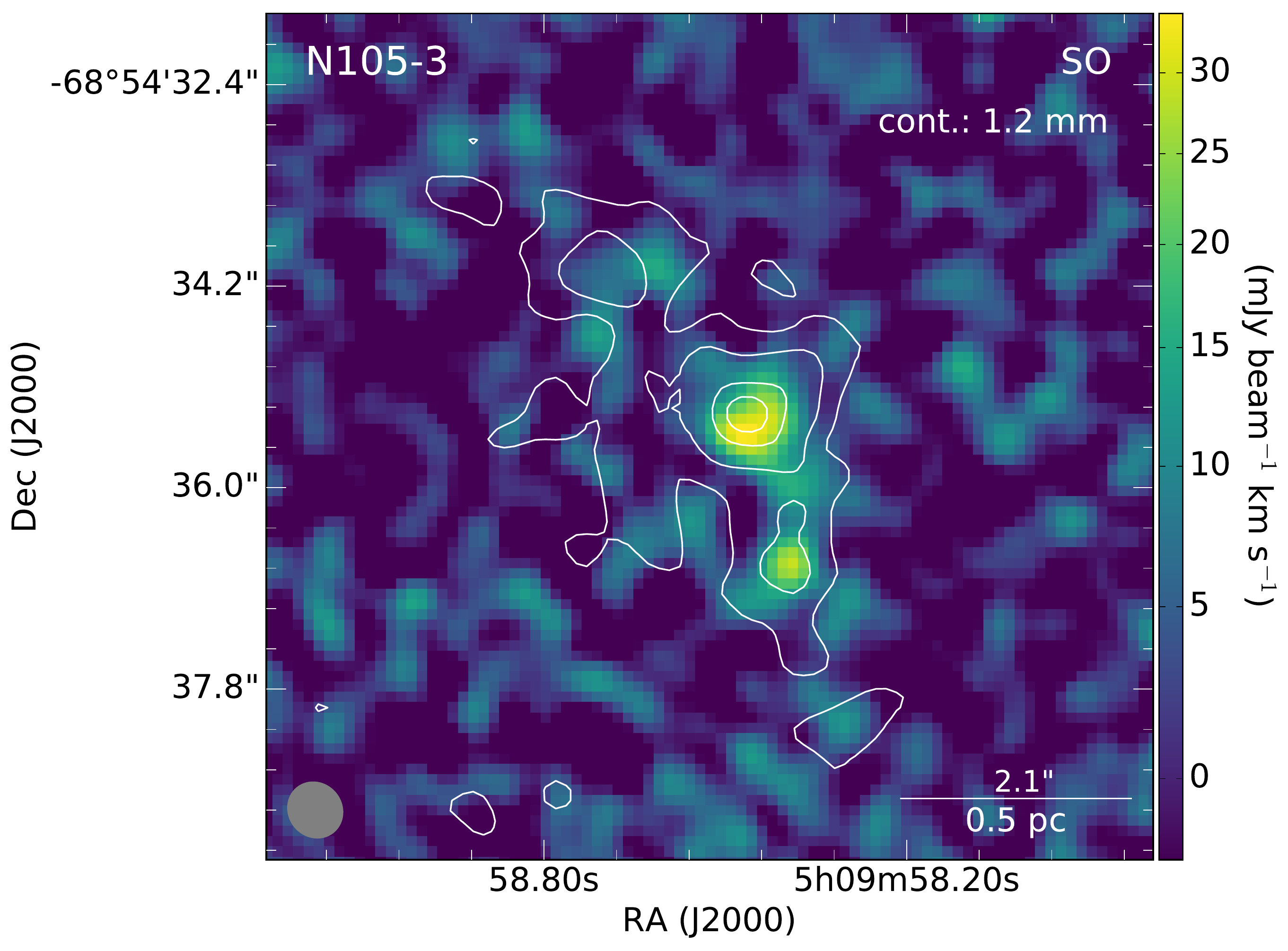}
\caption{{\it From upper left to lower right}: The CS (5--4), H$^{13}$CO$^{+}$ (3--2),  CH$_3$OH (combined 5$_{-1, 5}-4_{-1, 4}$ and 5$_{0, 5}-4_{0, 4}{\rm ++}$ transitions), and SO $6_6-5_5$ integrated intensity images of N\,105--3.  The white contours in each image correspond to the 1.2 mm continuum emission with contour levels of (3, 7, 10, 20)$\sigma$. The position of the CH$_3$OH masers (see Section~\ref{s:ysos}) is indicated in the image at the upper left. \label{f:N105int3}}
\end{figure*}

\indent Figure~\ref{f:N105int3} shows the integrated intensity images of the four species detected toward all the continuum components in N\,105--3: CH$_3$OH, H$^{13}$CO$^{+}$, CS, and SO. The CH$_3$OH emission detected toward 3\,B is the faintest out of all the continuum sources we analyzed in N\,105. None of the molecular line emission peaks (including SO$_2$ and HC$^{15}$N detected toward 3\,A only) are right on the 1.2 mm continuum peaks in N\,105--3, but within 1--2 pixels ($\sim$0$\rlap.{''}$1--0$\rlap.{''}$2).

\clearpage

\subsection{Spectral Modeling}
\label{s:modeling}

\begin{figure*}[ht!]
\centering
\includegraphics[width=0.2\textwidth,angle=-90.]{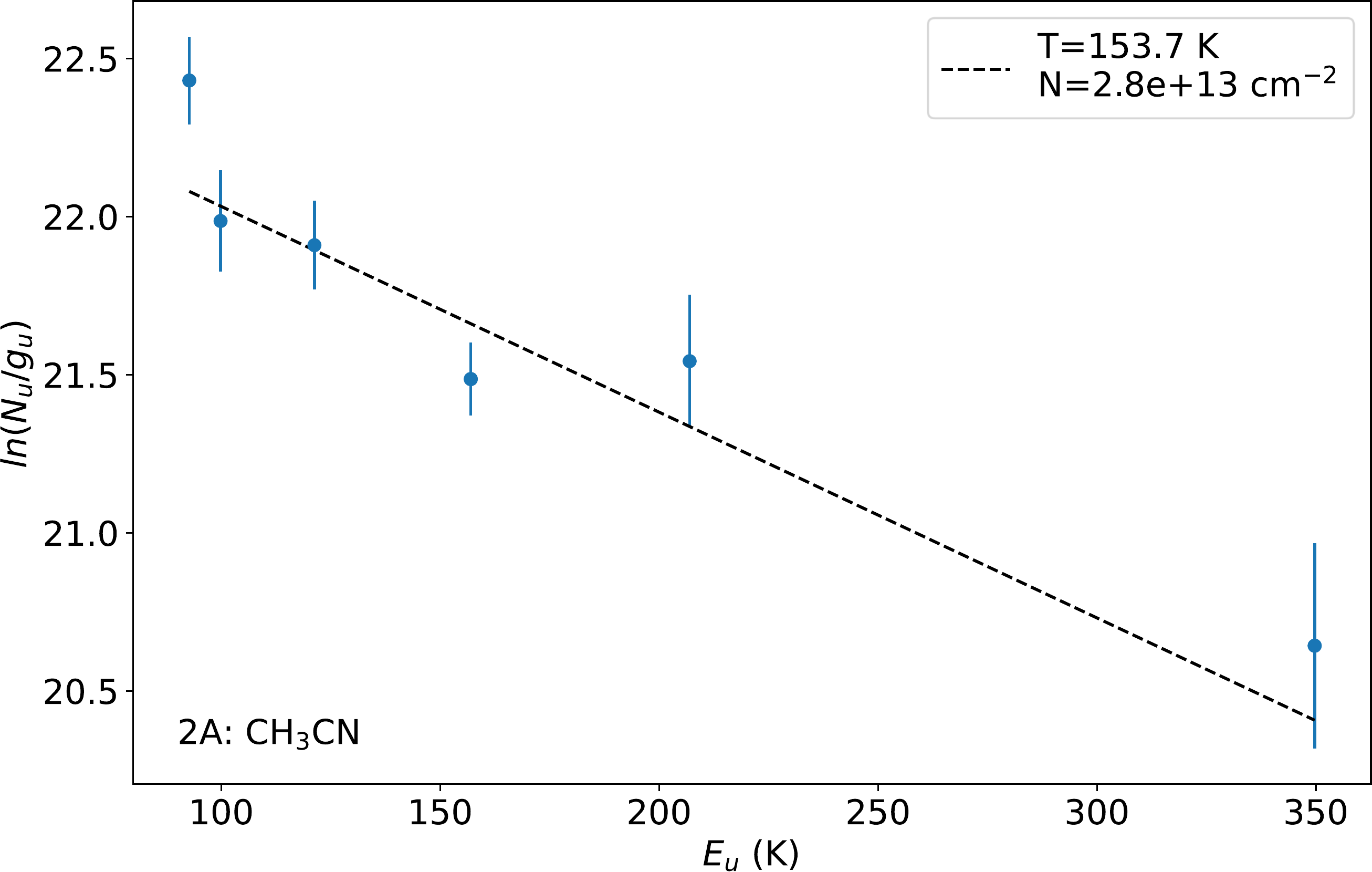} 
\hfill
\includegraphics[width=0.2\textwidth,angle=-90.]{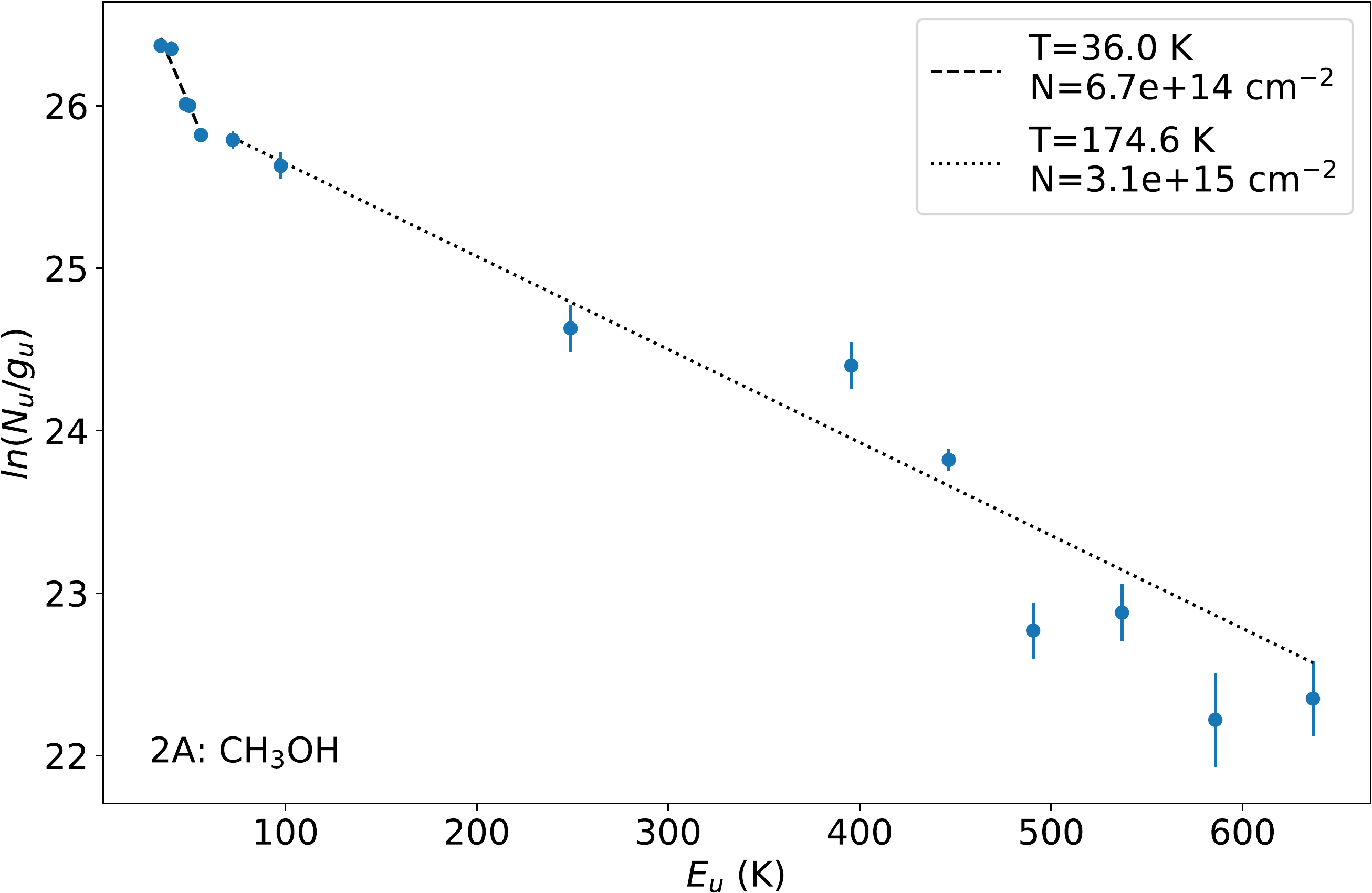}
\hfill
\includegraphics[width=0.2\textwidth,angle=-90.]{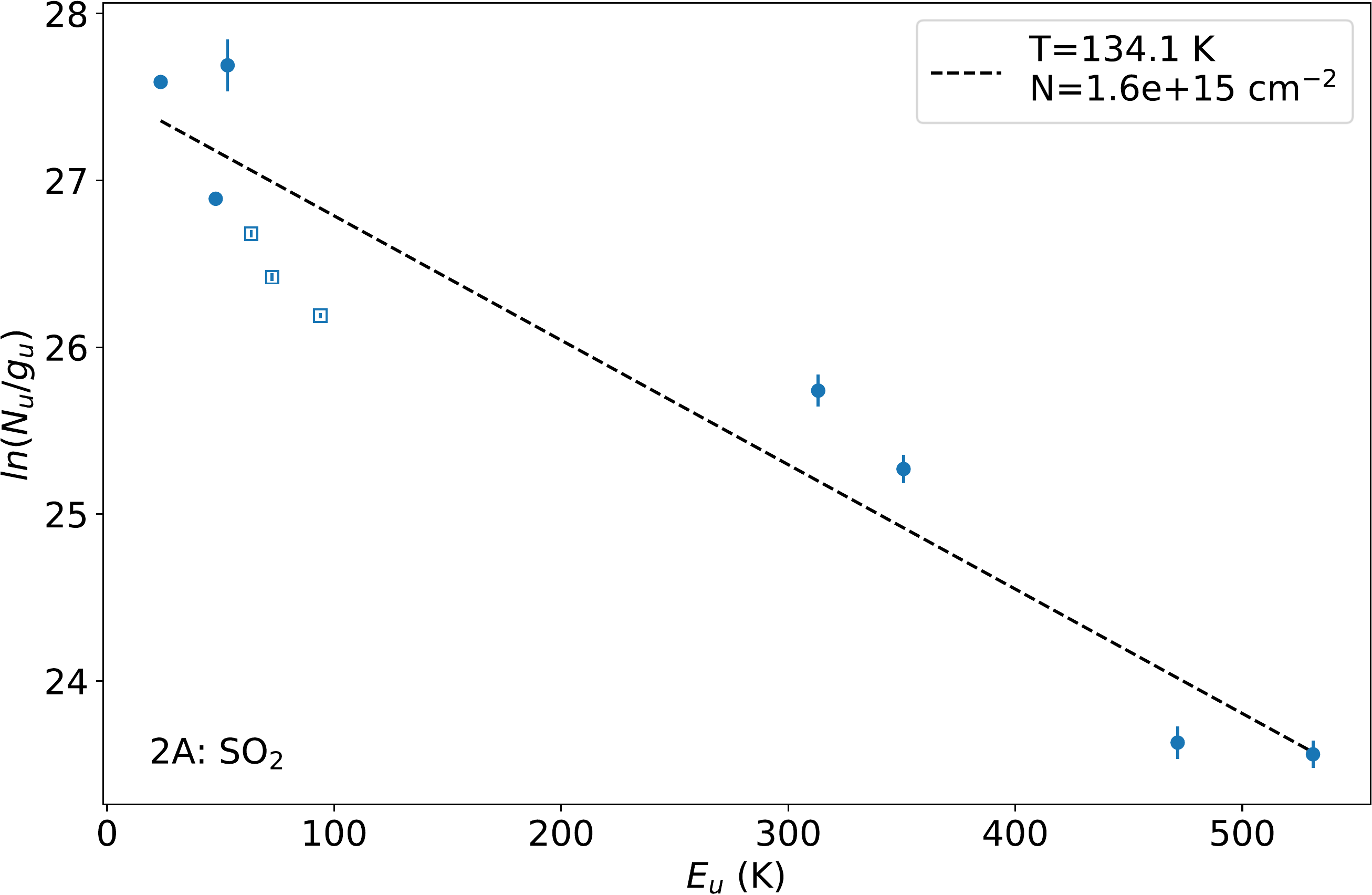}
\hfill
\includegraphics[width=0.2\textwidth,angle=-90.]{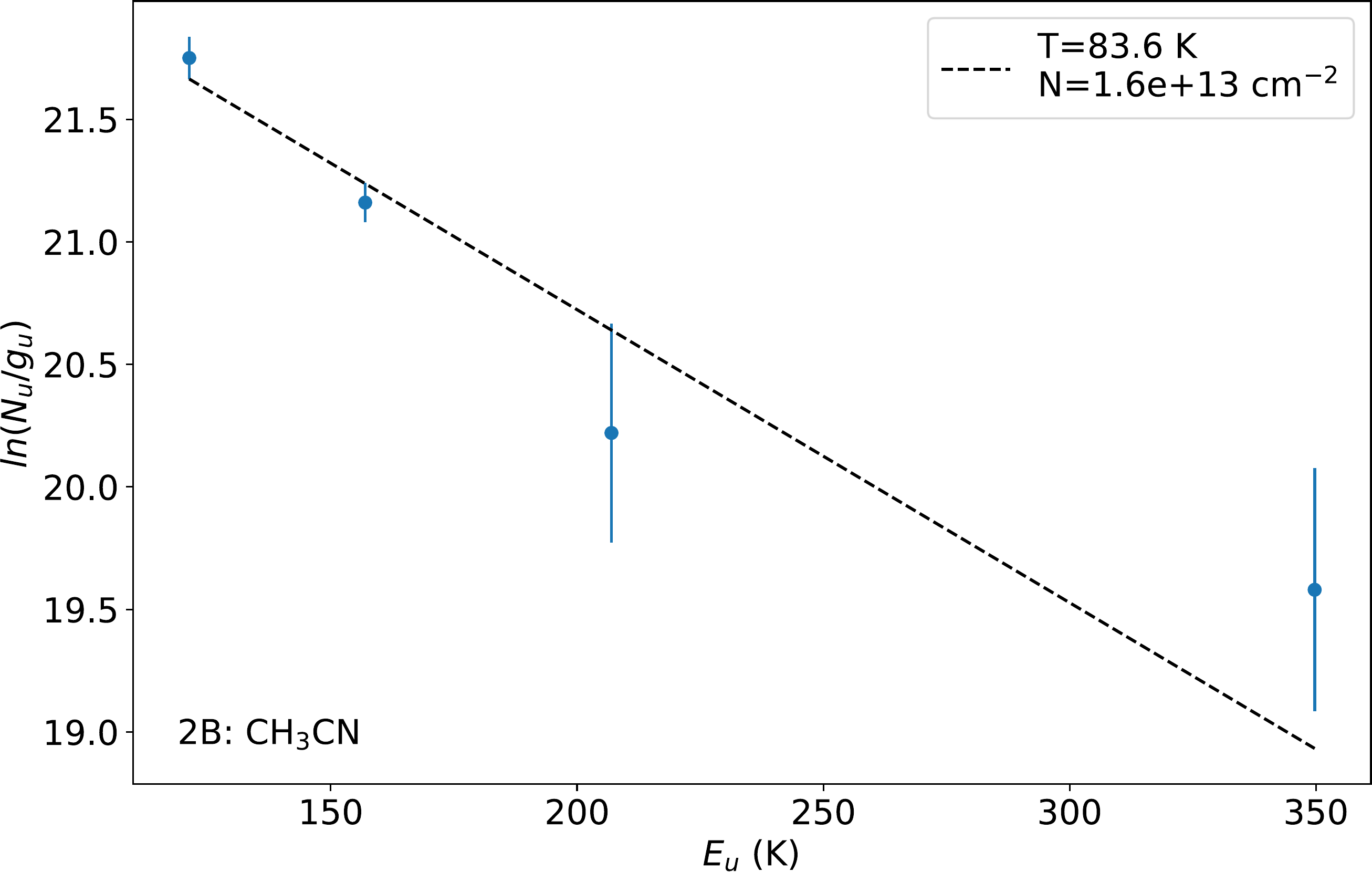}
\hfill
\includegraphics[width=0.2\textwidth,angle=-90.]{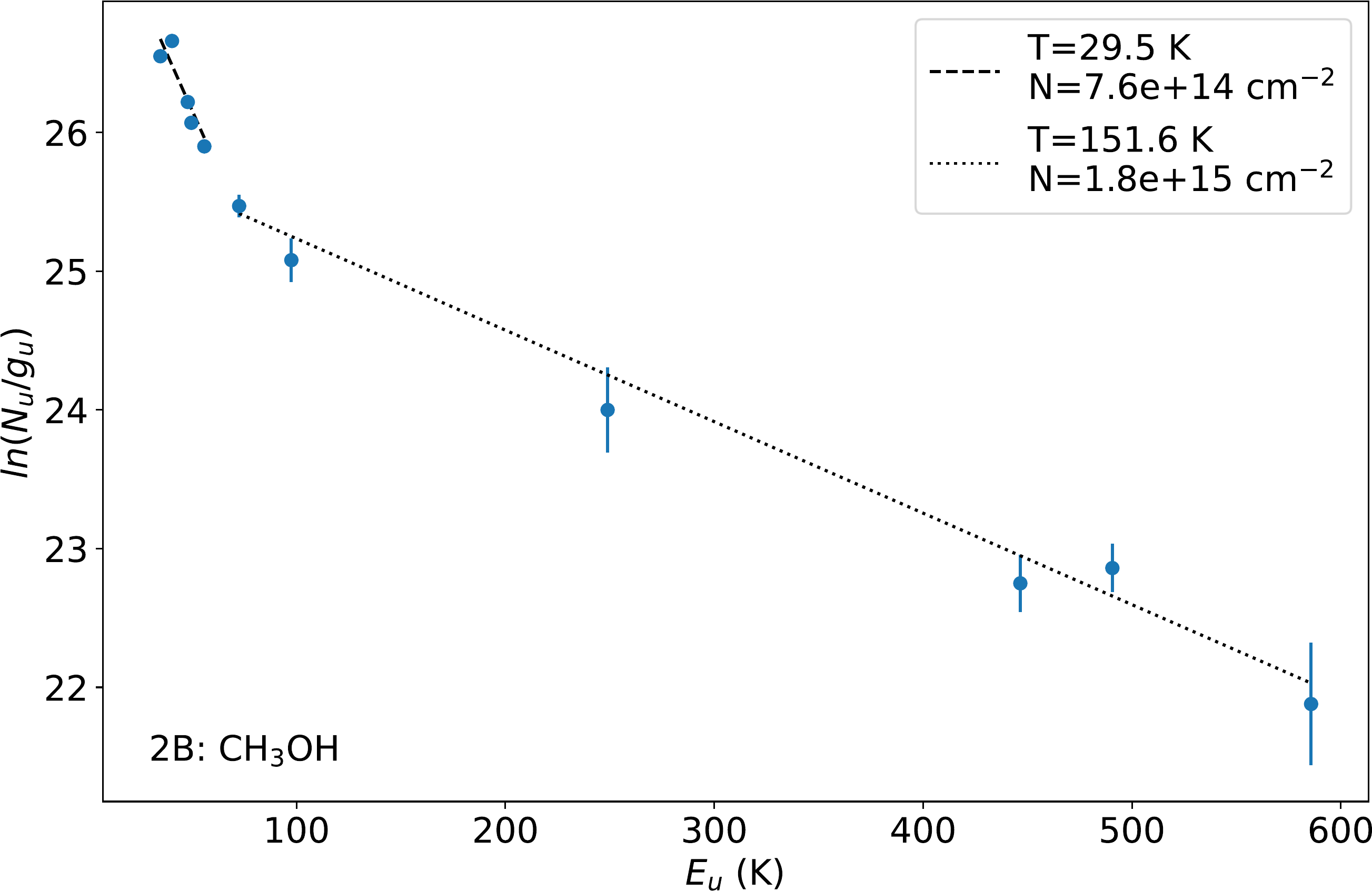}
\hfill
\includegraphics[width=0.2\textwidth,angle=-90.]{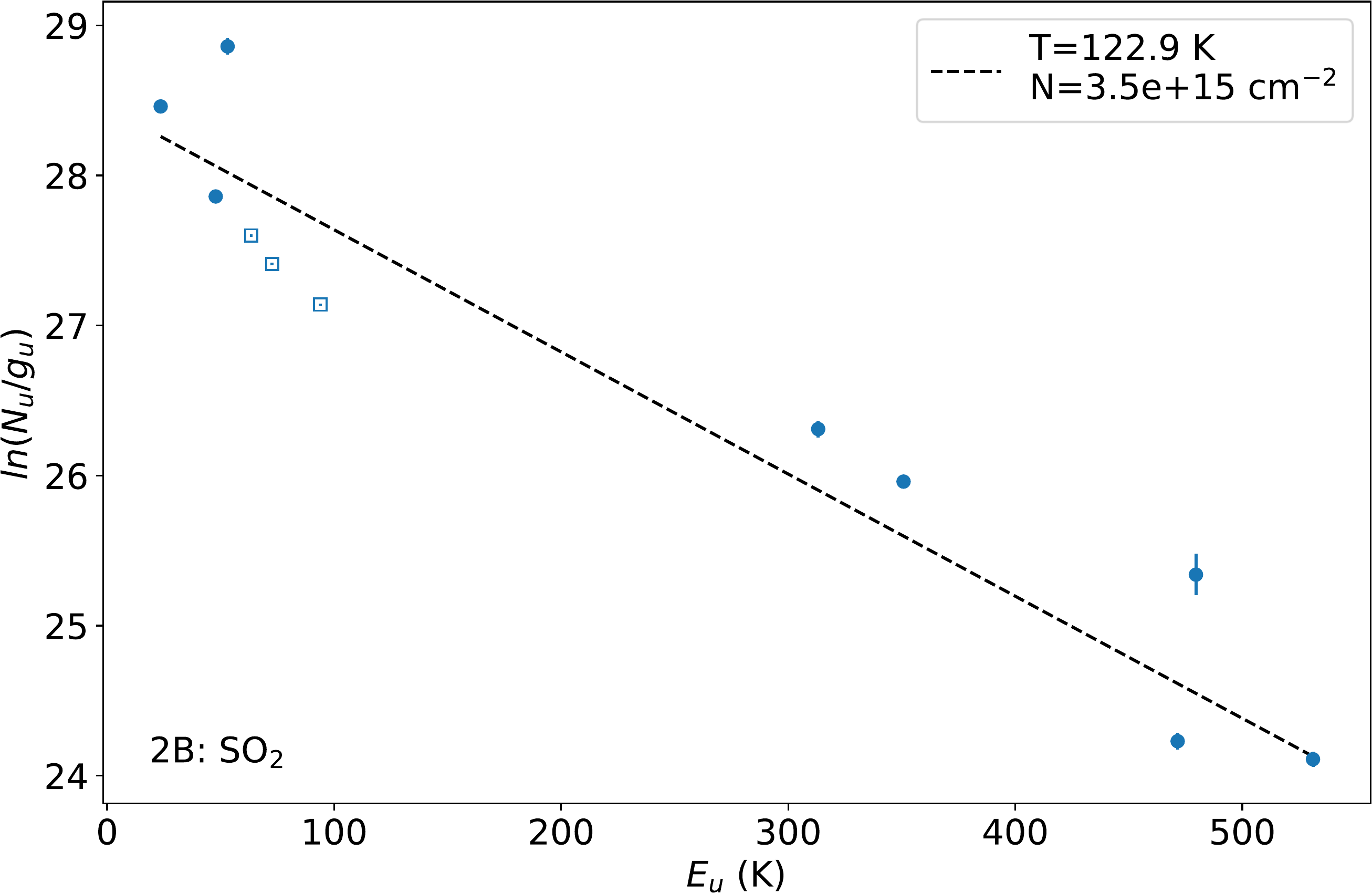} 
\caption{Rotational diagrams for sources N\,105--2\,A ({\it top panel}) and 2\,B ({\it bottom panel}) for  ({\it from left to right)}: CH$_3$CN, CH$_3$OH, and SO$_2$.  Only the transitions with the integrated flux above 2$\sigma$ are included in the diagrams. For CH$_3$CN, two strongest lines had to be excluded for 2\,B because they are blended and another CH$_3$CN transition was excluded for both 2\,A and 2\,B because it is blended with the CH$_3$OH line. Open box symbols indicate the SO$_2$ transitions suffering from significant opacity effects and thus excluded from the fit (see Section~\ref{s:modeling} for details). The rotational temperatures and column densities derived based on the rotational diagram analysis are indicated in the upper right corner in each plot.  \label{f:rotdiag}}
\end{figure*}

We performed an initial assessment of the physical conditions in N\,105 by using a rotational diagram analysis for CH$_3$CN, CH$_3$OH, and SO$_2$ for sources with multiple CH$_3$CN, CH$_3$OH, and SO$_2$ line detections with a range of upper state energies ($E_{\rm U}$). This analysis assumes the gas is in LTE and the lines are optically thin \citep{goldsmith1999}, and not blended with lines from other species.  The rotational diagrams are shown in Fig.~\ref{f:rotdiag} for the most chemically rich sources 2\,A and 2\,B. For both sources, two temperature components for CH$_3$OH are clearly visible in the rotational diagrams, while the SO$_2$ rotational diagrams shows  possible non-LTE effects, i.e., an apparent discontinuity in the distribution of the low- and high-$E_{\rm U}$ data points.  For the fitting in the rotational diagram, we excluded the SO$_2$ lines likely suffering from significant opacity effects, as defined by $E_{\rm U}<100$ K and ${\rm log}(S \mu^2)>1.0$ (also see \citealt{shimonishi2021}).

Spectral line modeling was then performed for all the continuum sources under the assumption of LTE and taking into account line blending and opacity effects, using a least-squares approach similar to \citet{sewilo2018} to simultaneously retrieve best-fitting column densities, rotational temperatures, Doppler shifts and spectral line widths for each species ($i$): [$N^i, T_{\rm rot}^i, v_i, dv^i$]. Simultaneously modeling all lines of every detectable species in our observed frequency range is crucial in order to resolve line blending issues and optical depth effects that might otherwise bias the retrieved parameters.

Spectral line models were generated for each source using a custom Python routine (based on the code used by \citealt{cordiner2017}). Spectroscopic parameters were taken from the Cologne Database for Molecular Spectroscopy (CDMS, \citealt{muller2001}), where available, and additional data for HDO were taken from the Jet Propulsion Laboratory (JPL) Molecular Spectroscopy Database \citep{pickett1998}. Gaussian spectral line opacity profiles were assumed, and the source was assumed to fill the aperture (unity beam-filling factor). The model sums the radiative source terms (in the equation of radiative transfer) in each spectral channel for emission from overlapping lines of both the same and different species. The peak opacity of each spectral line was calculated using equation A2 of \citet{turner1991}, and the final, synthetic $T_B$ spectrum was generated by combining the emission from the full set of lines in our frequency range based on their individual contributions to the radiation source function.

Optimization of the individual [$N^i, T_{\rm rot}^i, v_i, dv^i$] parameters for each species ($i$) was performed using the LMFIT nonlinear least-squares package \citep{newville2014}. Goodness of fit between the observed and synthetic spectra was monitored via the reduced chi-square statistic ($\chi^2_{\rm R}$). A good fit to the observed spectra ($\chi^2_{\rm R}\approx1.0$) was obtained for all species in all sources using a single set of [$N^i, T_{\rm rot}^i, v_i, dv^i$] parameters (cloud components) for each species, apart from CH$_3$OH toward N105--2\,A--2\,D and 2\,F and SO$_2$ for 2\,A, 2\,B, and 2\,F which required two cloud temperature components in order to obtain $\chi^2_{\rm R}\approx1.0$ (the two components are hereafter designated as  "hot" and "cold" due to their significantly different best-fitting temperatures).

Multiple lines of CH$_3$OH were detected in each source, with differing upper-state energies (in the range from 35~K up to $\sim$600~K for 2\,A; see Table~\ref{t:detections}), enabling robust derivations of the CH$_3$OH rotational temperatures. Towards sources 2\,A and 2\,B, multiple lines of the hot-core tracer CH$_3$CN were also detected, resulting in a more reliable estimate of the hot core gas temperature in those sources, since the hot core CH$_3$OH lines are more likely to be contaminated with the ambient interstellar medium (ISM) since CH$_3$OH is more widespread. Temperature information was also available in some cases for SO$_2$ and $^{34}$SO$_2$. For the remaining species with single-line detections (H$^{13}$CO$^{+}$, HCN, HC$^{15}$N, HC$_3$N, CS, C$^{33}$S, H$_2$CS, OCS, SO, $^{33}$SO, SiO, NH$_2$CHO, HNCO, HDCO, CH$_2$CO, HDO, HDS), or multi-line detections of insufficient strength for robust temperature determinations (e.g., CH$_3$OCH$_3$), we fixed their rotational temperatures to the best-fitting CH$_3$CN rotational temperature. When CH$_3$CN was not detected, the other molecules were assumed to follow the temperature of the (hot) CH$_3$OH component. In addition, when multiple lines of SO$_2$ were detected, the temperature of this species was obtained independently, and the temperatures of SO and $^{33}$SO were tied to SO$_2$, due to the chemical similarities between these species.

Error estimates on each free parameter were generated via Monte Carlo noise resampling. This involved the generation of 300 synthetic spectra for each source, obtained by adding random (Gaussian) noise to the best-fitting model spectra (with RMS equivalent to the noise in nearby line-free spectral regions), which were subsequently re-fit to determine the distribution of possible model parameters. $1\sigma$ errors were determined from the $\pm68$\% ranges of the resulting parameter distributions, under the assumption of Gaussian statistics.

In the spectral fitting process, we used the CDMS/JPL partition functions ($Q(T_{\rm rot})$); the CDMS data were used where available, i.e., for all species except HDO.  In our experience, the CDMS catalog tends to have more complete partition functions than the JPL catalog, including information from higher-excitation states where available. In each case, we used the appropriate corresponding partition function from the respective catalog for each species, thus ensuring a consistent statistical weight scheme to that used in that catalog. Partition functions were tabulated at 0, 9.375, 18.75, 37.5, 75, 150, 225 and 300 K, and interpolated using a cubic spline.

For some molecules with noisy or tentative line detections, reliable parameter error estimates could not be obtained using the Monte Carlo resampling method due to the tendency of the radial velocity parameter to drift into spectral regions affected by emission from nearby species, or with zero emission,  leading to erroneous, or insufficient constraints on the model. In these cases, the radial velocities were held fixed at the value given by the initial least-squares fit, with the other parameters allowed to vary freely.

Our spectral modeling procedure implicitly accounts for line opacity effects in the derivation of molecular column densities and rotational temperatures. In general, the deconvolved source sizes are larger or similar in size to the ALMA beam size (see Table~\ref{t:datanh2}). However, there is a possibility for additional, unresolved, high-opacity interstellar cloud components within the beam of our ALMA observations, that could not be distinguished at the resolution and signal-to-noise of our data. In that case, the spectral line opacities could have been underestimated, leading to a corresponding underestimate of the column densities.   

The resulting rotational temperatures ($T_{\rm rot}$), column densities  ($N$), velocities ($v_{\rm LSR}$), and line widths ($\Delta v_{\rm FWHM}$) are listed in Table~\ref{t:tempdens}, along with the estimated abundances with respect to H$_2$ ($N({\rm X})/N(\rm H_2)$) and CH$_3$OH ($N({\rm X})/N(\rm CH_{3}OH)$), where ${\rm X}$ represents a given species. $N({\rm H_2})$ was estimated from the 1.2 mm continuum as described in Section~\ref{s:nh2}.  

Observed spectra with overlaid model fits are presented in Appendix~\ref{s:appspectra}.

\subsubsection{SO$_2$ Excitation in N\,105--2\,A and 2\,B}
\label{s:xclass}

Since the SO$_2$ rotational diagrams for N\,105--2\,A and 2\,B (Fig.~\ref{f:rotdiag}) indicate a problem with the excitation of SO$_2$ under the assumptions specified above, we have performed an additional analysis to investigate it.  In an attempt to improve the spectral model fits, we explored the scenario with the relaxed assumption that the emission is beam filling, while still allowing for two components.  These LTE fits were performed using XCLASS \citep{moller2017}; the results are shown in Table~\ref{t:SO2fits}. XCLASS LTE model spectra for SO$_2$ are overlaid on the observed spectra of 2\,A and 2\,B in Figs.~\ref{f:spec2A_1_app}--\ref{f:spec2B_2_app} in Appendix~\ref{s:appspectra}.

For 2\,A, one large (approximately beam filling) and one very compact component with a size of 0$\rlap.{''}$05 (corresponding to 2500 au at the distance of N\,105) are necessary to fit the data.  The compact component is somewhat warmer than the extended component, representing an approximation of a centrally heated source with the temperature and steep density gradients.  The very compact component could point at a disk at the center. The extended component is needed to adjust the line shapes of the otherwise flat-topped lines. Most of the line emission seems to be due to the compact component which produces optically thick emission; this result is further constrained by the fit to three $^{34}$SO$_2$ lines that cannot be achieved with the optically thin SO$_2$ emission. The corner plot of the Markov Chain Monte Carlo (MCMC) error estimate for the XCLASS LTE fit is shown in Fig.~\ref{f:corner1} in Appendix~\ref{s:fittingresultstab}.  Given the narrowness of the distribution peak for this case in Fig.~\ref{f:corner1}, this seems to be a robust result.  The exact parameters cannot be constrained further with the current data set, and would require higher sensitivity and spatial resolution data for SO$_2$ and its isotopologues. 

For 2\,B, the results are similar to those obtained for 2\,A: the spectra are best fitted by a warmer, high column density and a colder, more extended components.  However, as can be seen in the MCMC corner plot in Fig.~\ref{f:corner2}, there is a large uncertainty regarding source sizes, temperatures, and column densities.  Again, we cannot better constrain the parameters with the current data.


\section{H$_2$ Column Densities, Masses, and Source Sizes}
\label{s:nh2}

Assuming that the dust and gas are well-coupled ($T_{\rm dust} \sim T_{\rm gas} \sim T$), the H$_2$ column density can be estimated from the observed millimeter continuum flux using the formula (e.g., \citealt{hildebrand1983}; \citealt{kauffmann2008}):
\begin{equation}
N({\rm H_2}) = \frac{S_{\rm \nu}^{\rm beam}\,R_{\rm gd}}{\Omega_{\rm A}\,\mu_{\rm H_2}\,m_{\rm H}\,\kappa_{{\rm \nu, d}}\,B_{\nu}(\rm T)},
\end{equation}
where 
$S_{\nu}^{\rm beam}$ is the flux per synthesized beam;
$R_{\rm gd}$ is the gas-to-dust mass ratio;
$\Omega_{\rm A}$ is the beam solid angle: $\Omega_{\rm A}$=$\frac{\pi}{4\, {\rm ln2}}\, \theta_{\rm maj}\,\theta_{\rm min}$ where $\theta_{\rm maj}$ and $\theta_{\rm min}$ are the major and minor axes of the synthesized beam, respectively;
$\mu_{\rm H_2}$ is the mean molecular weight per hydrogen molecule (e.g., \citealt{kauffmann2008}; \citealt{cox2000}; $\mu_{\rm H_2}$ $\approx$ 2.76 for the LMC, \citealt{remyruyer2014});
$m_{\rm H}$ is the mass of the hydrogen atom;
$\kappa_{\rm {\nu, d}}$ is the dust opacity per unit mass (e.g., \citealt{hildebrand1983}; \citealt{shirley2000});
and $B_{\nu}(\rm T)$ is the Planck function.  The assumption of thermal equilibrium between the gas and dust holds for high-density regions ($n_{\rm H_2}\gtrsim10^5$ cm$^{-3}$), including hot cores where the temperature exceeds 100 K (e.g.,  \citealt{goldsmith1978}; \citealt{ceccarelli1996}; \citealt{kaufman1998}). 

Dust opacity for the LMC was derived by \citet{galliano2011} for a large area of the LMC and thus is primarily relevant to the diffuse ISM. For our analysis focused on $\sim$0.1~pc scales,  we adopt a Galactic dust opacity from \citet{ossenkopf1994} for the model with the initial MRN distribution \citep{mathis1977} with thin ice mantles after 10$^{5}$ years of coagulation at a hydrogen gas density of 10$^{6}$ cm$^{-3}$.  For 1.24 mm (242.4 GHz), we adopt the opacity per unit dust mass of $\kappa_{\rm 1.24\,mm, d}$ of 0.993 cm$^{2}$ g$^{-1}$ (Table 1 in \citealt{ossenkopf1994}). 

It has been shown that the gas-to-dust mass ratio ($R_{\rm gd}$) strongly depends on metallicity, but shows a significant scatter (e.g., \citealt{duval2014}; \citealt{remyruyer2014}, \citealt{remyruyer2015}) attributed to differences in star formation histories of the galaxies (see e.g., \citealt{galliano2018}). We determined the LMC gas-to-dust mass ratio by scaling the Galactic value using the empirical broken power-law relationship between gas-to-dust mass ratio and metallicity (defined as $Z_{\rm gal}/Z_{\odot}=[{\rm O/H}]_{\rm gal}/[{\rm O/H}]_{\odot}$, where $[{\rm O/H}]_{\odot}=4.9\times10^{-4}$; \citealt{asplund2009}) from \citet{remyruyer2014}, which adopts the solar gas-to-dust mass ratio of 162 \citep{zubko2004}.  For the ${\rm log}[{\rm O/H}]$ abundance ratio in the LMC H\,{\sc ii} regions of $-$3.6  (or $Z_{\rm LMC}\sim0.5\,Z_{\rm \odot}$;  e.g., \citealt{pagel2003}), we estimate the gas-to-dust mass ratio of 316 for the LMC. This value is in agreement (within the uncertainties) with $R_{\rm gd}$ found by \citet{duval2014} for the LMC based on the {\it Herschel} data. 

\begin{deluxetable*}{rcccccccccccc}
\centering
\rotate
\tablecaption{Continuum Intensities and H$_2$ Column Densities \label{t:datanh2}}
\tablewidth{0pt}
\tablehead{
\multicolumn{1}{c}{Source} &
\colhead{RA (J2000)\tablenotemark{\footnotesize a}} &
\colhead{Dec (J2000)\tablenotemark{\footnotesize a}} &
\colhead{$I_{\rm peak}$\tablenotemark{\footnotesize b}} &
\colhead{$I_{\rm mean}$\tablenotemark{\footnotesize c}} &
\colhead{$Area$\tablenotemark{\footnotesize c}}  &
\colhead{FWHM$_{\rm eff}$}  &
\colhead{FWHM$_{\rm eff, deconv}$}  &
\colhead{$F_{\rm 50}$\tablenotemark{\footnotesize e}}  &
\colhead{$M_{\rm 50, gas}$\tablenotemark{\footnotesize e}}  &
\colhead{$N({\rm H_2})$\tablenotemark{\footnotesize f}} &
\colhead{$N({\rm H_2})$\tablenotemark{\footnotesize g}} &
\colhead{$N({\rm H_2})$\tablenotemark{\footnotesize h}} \\
\cline{11-13}
\colhead{} &
\colhead{($^{\rm h}$~$^{\rm m}$~$^{\rm s}$)} &
\colhead{($^{\rm \circ}$ $'$ $''$)} &
\colhead{(mJy beam$^{-1}$)} &
\colhead{(mJy beam$^{-1}$)} &
\colhead{(arcsec$^{2}$)} &
\colhead{($''$/pc)} &
\colhead{($''$/pc)} &
\colhead{(mJy)} &
\colhead{($M_{\odot}$)} &
\multicolumn{3}{c}{(10$^{23}$ cm$^{-2}$)} 
}
\startdata
N\,105--1\,A & 05:09:50.54 & $-$68:53:05.4 & 31.003\tablenotemark{\footnotesize d} & 22.370\tablenotemark{\footnotesize d} & 0.305 & 0.62/0.15 & 0.39/0.10 & 25.2\tablenotemark{\footnotesize d} & 3706$^{+597}_{-475}$ & 93$^{+15}_{-12}$  & \nodata & 7.4$^{+1.8}_{-1.8}$ \\
1\,B & 05:09:52.48 & $-$68:53:00.7 & 2.329 &   1.575 & 0.923 & 1.08/0.26  & 0.97/0.23 & 5.4 & 1038$^{+109}_{-108}$ & 8.6$^{+1.0}_{-1.0}$ &  \nodata & 4.8$^{+1.0}_{-1.0}$ \\
1\,C & 05:09:52.82 & $-$68:53:04.3 &  1.437 &   1.008 & 0.601 & 0.87/0.21  & 0.73/0.18 & 2.2 & 390$^{+95}_{-87}$ & 5.0$^{+1.2}_{-1.2}$ & \nodata & \nodata \\
N\,105--2\,A & 05:09:51.96 & $-$68:53:28.3 & 6.362 & 4.560 & 0.423& 0.73/0.18  & 0.55/0.13 & 7.1 & 102$^{+12}_{-13}$ & 1.6$^{+0.3}_{-0.2}$ & 1.8$^{+0.2}_{-0.2}$  & 1.7$^{+0.2}_{-0.2}$ \\ 
2\,B & 05:09:52.56 & $-$68:53:28.1 & 6.181 & 4.382 & 0.415 & 0.73/0.18 & 0.54/0.13 & 6.7 & 170$^{+27}_{-26}$ & 2.0$^{+0.2}_{-0.2}$  & 3.1$^{+0.5}_{-0.5}$ & 1.7$^{+0.2}_{-0.2}$ \\            
2\,C & 05:09:52.22 & $-$68:53:22.6 & 1.799 & 1.217 & 0.474 & 0.78/0.19 &  0.60/0.15 & 2.1 & 50$^{+16}_{-12}$ & 0.8$^{+0.3}_{-0.2}$ & \nodata & 2.7$^{+0.5}_{-0.4}$ \\
2\,D & 05:09:52.99 & $-$68:53:31.0 & 1.344 & 0.910 & 0.592 & 0.87/0.21 &  0.72/0.17 & 2.0 & 161$^{+19}_{-18}$ & 2.1$^{+0.3}_{-0.3}$ & \nodata & \nodata \\
2\,E & 05:09:53.90 & $-$68:53:37.6 &  1.269 &  0.896 & 0.618 & 0.89/0.22 & 0.74/0.18 & 2.0 & 501$^{+95}_{-79}$ & 6.2$^{+1.2}_{-1.0}$ & \nodata & \nodata \\
2\,F & 05:09:52.39 & $-$68:53:28.1 & 2.397 & \nodata & \nodata & \nodata & \nodata & \nodata& \nodata & 1.2$^{+0.4}_{-0.3}$ & \nodata & \nodata \\
N\,105--3\,A & 05:09:58.48 & $-$68:54:35.4 & 0.573 & 0.409 & 0.559 & 0.84/0.20 & 0.69/0.17 & 0.85 & 307$^{+69}_{-57}$ & 4.2$^{+1.0}_{-0.8}$  & \nodata &  1.0$^{+0.3}_{-0.3}$\\
3\,B & 05:09:58.70 & $-$68:54:34.1 & 0.342 & 0.241 & 0.381 & 0.70/0.17 &  0.50/0.12 & 0.33 & 5$^{+1}_{-2}$ & 0.09$^{+0.03}_{-0.04}$ & \nodata & \nodata\\
3\,C & 05:09:58.41 & $-$68:54:36.7 & 0.221 &  \nodata   & \nodata & \nodata & \nodata& \nodata & \nodata & 2.2$^{+1.1}_{-0.8}$ & \nodata & \nodata \\
N\,113\,A1 & 05:13:25.17 & $-$69:22:45.5 & 13.136 & 9.399 & 0.606 & 0.88/0.21 &  0.55/0.13 & 10.7 & 214$^{+24}_{-24}$ & 2.7$^{+0.3}_{-0.3}$ & \nodata & \nodata \\
B3 & 05:13:17.18 & $-$69:22:21.5 &   6.306 & 4.332 & 1.128 & 1.2/0.29 & 0.98/0.24 & 9.2 & 184$^{+29}_{-30}$ & 1.2$^{+0.2}_{-0.2}$&  \nodata & \nodata \\
\enddata
\tablenotetext{a}{The continuum peak positions:  at $\sim$242.2 GHz for sources in N\,105 (this paper) and $\sim$224.3 GHz for N\,113 A1 and B3 \citep{sewilo2018}.} 
\tablenotetext{b}{$I_{\rm peak}$ is the observed 1.2 mm continuum intensity peak.}
\tablenotetext{c}{$I_{\rm mean}$ is the 1.2 mm continuum intensity averaged over the area ($Area$) within the contour corresponding to the 50\% of the 1.2  mm continuum peak.  This is the same area used to extract spectra for the analysis (see Section~\ref{s:spectral}). The beam areas for the (N\,105--1, N\,105--2, N\,105--3, N113) observations are (0.270, 0.273, 0.270, 0.534) arcsec$^{2}$.}
\tablenotetext{d}{The observed values, i.e., not corrected for the contribution from the free-free emission (see Section~\ref{s:continuum}).}
\tablenotetext{e}{$F_{\rm 50}$ and $M_{\rm 50, gas}$ are flux densities and masses, respectively, calculated for the area above the 50\% of the peak intensity.}
\tablenotetext{f}{$N({\rm H_2})$  calculated assuming $T = T_{\rm rot} ({\rm CH_{3}OH})$ (see Section~\ref{s:nh2} for details).}
\tablenotetext{g}{$N({\rm H_2})$  calculated assuming $T = T_{\rm rot} ({\rm CH_{3}CN})$.}
\tablenotetext{h}{$N({\rm H_2})$  calculated assuming $T = T_{\rm rot} ({\rm SO_{2}})$.}
\end{deluxetable*}

We calculate $N({\rm H_2})$ using Eq. 1 in the form presented in \citet{kauffmann2008}:
\begin{eqnarray}
N({\rm H_2}) = 2.02\cdot10^{20} (e^{1.439(\lambda/{\rm mm})^{-1}(T/{\rm 10\,K})^{-1}}-1) \\ \nonumber
\cdot \left(\frac{\kappa_{{\rm \nu, d}}/R_{\rm gd}}{0.01\,{\rm cm^2\,g^{-1}}}\right)^{-1}\,\left(\frac{I_{\rm \nu}^{\rm beam}}{{\rm mJy\,beam^{-1}}}\right) \\ 
\cdot \left(\frac{\theta_{\rm HPBW}}{{\rm 10\,arcsec}}\right)^{-2}\,\left(\frac{\lambda}{{\rm mm}}\right)^3 {\rm cm^{-2}}, \nonumber 
\end{eqnarray}
where $\theta_{\rm HPBW}= \sqrt{\theta_{\rm maj}\cdot\theta_{\rm min}}$.  

For $T$, we adopt the temperatures determined based on CH$_3$CN for 2\,A and 2\,B and those based on CH$_3$OH for all other sources in N\,105. If both the hot and cold CH$_3$OH components are present for a given source, we used the temperature of the hot component. 

$I_{\rm \nu}^{\rm beam}$ was measured as a mean intensity within the region used to extract spectra ($I_{\rm mean}$), i.e., the area enclosed by the contour corresponding to 50\% of the 1.2 mm continuum peak (see Section~\ref{s:modeling}).  We used the same areas to measure flux densities ($F$) and determined corresponding masses ($M$) for each source using the formula \citep{kauffmann2008}: 
\begin{eqnarray}
M = 0.12\, M_{\odot} \cdot (e^{1.439(\lambda/{\rm mm})^{-1}(T/{\rm 10\,K})^{-1}}-1) \\
\cdot \left(\frac{\kappa_{{\rm \nu, d}}/R_{\rm gd}}{0.01\,{\rm cm^2\,g^{-1}}}\right)^{-1}\,\left(\frac{F_{\rm \nu}}{{\rm Jy}}\right) \nonumber 
\cdot \left(\frac{D}{{\rm 100\,pc}}\right)^{2} \,\left(\frac{\lambda}{{\rm mm}}\right)^3,  \nonumber
\end{eqnarray}
where $D$ is the distance to the LMC and other parameters are the same as in Eqs. 1 and 2. 

We determined the source sizes utilizing a common definition of an ``effective'' radius: $R_{\rm eff} = 2 \sqrt{A/\pi}$, where $A$ is the source area. We adopted the area contained within the 50\% of the continuum peak intensity contour as $A$ and calculated the source size at the half-peak as FWHM$_{\rm eff} = 2 \cdot R_{\rm eff}$. Assuming the sources can be represented by Gaussian profiles, we calculated the deconvolved sizes FWHM$_{\rm {eff, deconv}} = \sqrt{{\rm FWHM}_{\rm eff}^2 - {\rm HPBW}^2}$, where HPBW (the half-power beam width) is the geometric mean of the minor and major axes of the synthesized beam. 

The continuum peak intensities ($I_{\rm peak}$), mean intensities ($I_{\rm mean}$), observed and deconvolved source sizes (FWHM$_{\rm eff}$ and FWHM$_{\rm {eff, deconv}}$), fluxes and masses calculated for the area above the 50\% of the peak intensity ($F_{50}$ and $M_{\rm {50, gas}}$), and H$_2$ column densities ($N({\rm H_2})$) are listed in Table~\ref{t:datanh2}. 

In addition to $N({\rm H_2})$ calculated based on the CH$_3$OH temperature, Table~\ref{t:datanh2} also lists $N({\rm H_2})$ determined based on CH$_3$CN  and/or SO$_2$ where available (CH$_3$CN: 2\,A and 2\,B; SO$_2$: 1\,A, 1\,B, 2\,A, 2\,B, 2\,C, and 3\,A).  For 2\,A, all values of $N({\rm H_2})$ agree within the uncertainties with $N({\rm H_2})$ calculated using $T(\rm {CH_3CN})$ being $\sim$6\% higher than the average $N({\rm H_2})$. For 2\,B, the differences between $N({\rm H_2})$ calculated using different temperatures is larger with $N({\rm H_2})$ based on $T(\rm {CH_3CN})$ being $\sim$37\% higher than the average $N({\rm H_2})$. 

The largest discrepancy between $N({\rm H_2})$ derived based on different species exists for 1\,A with $N({\rm H_2})$ derived using $T({\rm CH_3OH})$$\sim$12 K being an order of magnitude higher than that derived using $T({\rm SO_2})$$\sim$96 K.  The former $N({\rm H_2})$ value is one to two orders of magnitude larger than $N({\rm H_2})$ for all the other continuum sources in N\,105 (see Table~\ref{t:datanh2}). Such a large difference in temperature between CH$_3$OH and SO$_2$ can be the result of CH$_3$OH and SO$_2$ tracing different physical components (an extended cold CH$_3$OH emission and SO$_2$ produced in outflow shocks) or non-LTE effects (see Section~\ref{s:hotcores}). Considering the large uncertainties in the determination of $N({\rm H_2})$ based on the ALMA data, we have decided to use an independent measurement of $N({\rm H_2})$. 

We follow the procedure described in \citet{shimonishi2020} to estimate $N({\rm H_2})$ based on the value of $A_{\rm V}$ we obtained using the KMOS observations ($42\pm4$ mag; see Section~\ref{s:IR}) which is consistent with that previously reported in literature ($\sim$40 mag, \citealt{oliveira2006}; see also Sections~\ref{s:IR}). We use the relation $N({\rm H_2})/A_{\rm V} = 2.8 \times 10^{21}$ cm$^{-2}$ mag$^{-1}$ for the LMC; the value of $A_{\rm V}$ is doubled before it is used in this formula to obtain the total column density along the line of sight (see  \citealt{shimonishi2020} and references therein).  The resulting $N({\rm H_2})$ for 1\,A is $(2.4\pm0.2)\times10^{23}$ cm$^{-2}$.

We used $N({\rm H_2})$ calculated using the CH$_3$CN temperature (2\,A and 2\,B), CH$_3$OH temperature (the remaining sources except 1\,A), and derived from $A_{\rm V}$ (1\,A) to determine molecular abundances with respect to H$_2$ listed in Table~\ref{t:tempdens}.  

We have estimated the H$_2$ number density, $n_{\rm H_2}$, using the relation $n_{\rm H_2} = N({\rm H_2})/{\rm FWHM}_{\rm eff, deconv}$. All the sources except 3\,B have $n_{\rm H_2}$ of at least a few times 10$^5$ cm$^{-3}$; $n_{\rm H_2}$ of hot cores 2\,A and 2\,B is $\sim$$4.6\times10^5$ cm$^{-3}$ and $\sim$$7.8\times10^5$ cm$^{-3}$, respectively.

\section{ALMA Fields in N\,105 from Optical to Radio Wavelengths}
\label{s:IR}

N\,105A (overlapping with N\,105--1 and N\,105--2) was observed by \citet{indebetouw2004} with ATCA at 8.6 GHz (3 cm) and 4.8 GHz (6 cm) with a resolution of $\sim$1$\rlap.{''}$5 and $\sim$2$''$. They detected four radio continuum sources, three of which are in the ALMA fields: B0510$-$6857\,W coincides with N\,105--1\,A, B0510$-$6857\,E is associated with the 1.2 mm continuum emission extending toward east from N\,105--1\,B, and B0510$-$6857\,S lies between N\,105--2\,A, 2\,B, and 2\,C. \citet{indebetouw2004} determined spectral types of ionizing stars of O6.5 V, O7.5 V, and O8.5 V for radio continuum components W, E, and S, respectively.  Source B0510$-$6857\,S is the faintest out of the three radio components at both wavelengths, while sources W and E are the brightest at 3 cm and 6 cm, respectively.  \citet{indebetouw2004} derived spectral indices of $+$0.6, 0, and $-$0.2 for B0510$-$6857\,W, E, and S, respectively.  In the lower-resolution ($\sim$10$''$) 6.6 GHz image presented in \citet{ellingsen1994}, all the continuum sources from \citet{indebetouw2004} remain unresolved and associated with an extended ionized gas emission. 

Radio source B0510$-$6857\,W / ALMA N\,105--1\,A is associated with the infrared source N\,105A IRS1 from \citet{oliveira2006}, a candidate protostar first identified by  \citet{epchtein1984}.  The 3--4 $\mu$m spectrum from the Infrared Spectrometer And Array Camera (ISAAC) on the ESO-VLT presented by \citet{oliveira2006} displays a very red continuum and strong hydrogen recombination line emission:  Br$\alpha$ and Pf$\gamma$.   \citet{oliveira2006} argue that a non-detection of the Pf$\delta$ line indicates a high dust column density in front of the region of the line emission; they estimate a total visual extinction $A_{\rm V}$ of $\sim$40 mag. The Br$\alpha$ line detected toward IRS1 shows broad wings and is asymmetric, providing a strong evidence for the bipolar outflow.  N\,105A IRS1 is very bright in $L'$-band and it is extremely red ($K_{\rm S}$-$L'$ = 3.9 mag). Based on the analysis of the spectral energy distribution (SED) and IR colors of IRS1, and the presence of strong recombination lines and the outflow, \citet{oliveira2006} concluded that IRS1 is likely an embedded massive YSO ionizing its immediate surroundings.  

Radio source B0510$-$6857\,E coincides with the extended 1.2 mm continuum emission east of N\,105--1\,B and is associated with the infrared source N\,105A `blob A' from \citet{oliveira2006}, while N\,105--1\,B corresponds to `blob B.'  Blob A is a bright core, while blob B is part of the extended and patchy $L'$-band emission surrounding it.  Similarly to IRS1, the spectra of blobs A and B display the H recombination lines; however, the Pf$\delta$ line is detected toward both of them, indicating that the extinction by dust is significantly lower than toward IRS1.  Blob A is also not as red as IRS1 ($K_{\rm S}$-$L'$ = 1.51 mag) and the continuum emission is extremely faint.  The broad $\sim$3.3 $\mu$m PAH emission is underlying the Pf$\delta$ emission in both blob A and B spectra.  The spectral lines toward blob A are broader than those detected toward IRS1 and they are double-peaked, suggesting the presence of the outflow. No signs of an outflow are present in the blob B's spectrum.  Blobs A and B correspond to the VMC sources 558354728344 and 558354728373, respectively; they are unresolved in {\it Spitzer}'s images and both contribute to the emission from the {\it Spitzer} YSO 050952.73$-$685300.7 (see Section~\ref{s:ysos}). 

The area where B0510$-$6857\,E, N\,105--1\,B, and N\,105--1\,C are located is coincident with the peak of the 888 MHz emission in N\,105. The 888 MHz image with the spatial resolution of 13$\rlap.{''}9\times12\rlap.{''}$1 (FWHM) presented in \citet{pennock2021} was obtained with Australian Square Kilometre Array Pathfinder (ASKAP) telescope as part of the Evolutionary Map of the Universe survey. 

The catalog position of the 3 cm / 6 cm source B0510$-$6857\,S is concentrated between N\,105--2 A/2\,B (offset by $\sim$4$''$ north ) and 2\,C; the physical association between this radio emission and  N\,105--2 A--C is unclear. 

The source IRS2 from \citet{oliveira2006} corresponds to ALMA N\,105--2\,A. IRS2 is very faint in $L'$-band/3.78 $\mu$m ($14.30\pm0.34$ mag); in comparison, the $L'$ magnitude of IRS1 is $9.88\pm0.01$. \citet{oliveira2006} concluded that IRS2 could be an embedded YSO based on a suggestive association with water masers.

\begin{figure}
\centering
\includegraphics[width=0.48\textwidth]{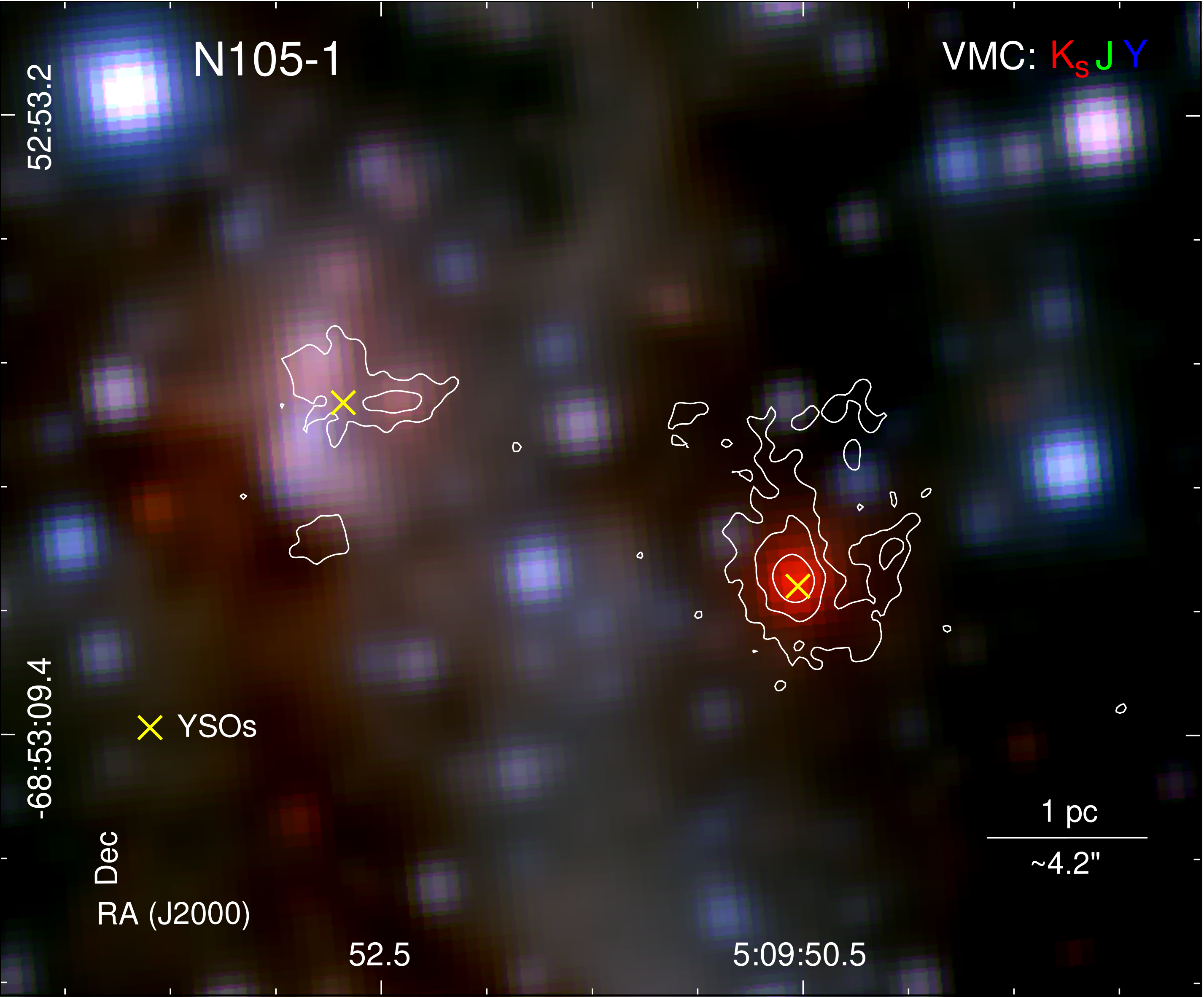} 
\includegraphics[width=0.48\textwidth]{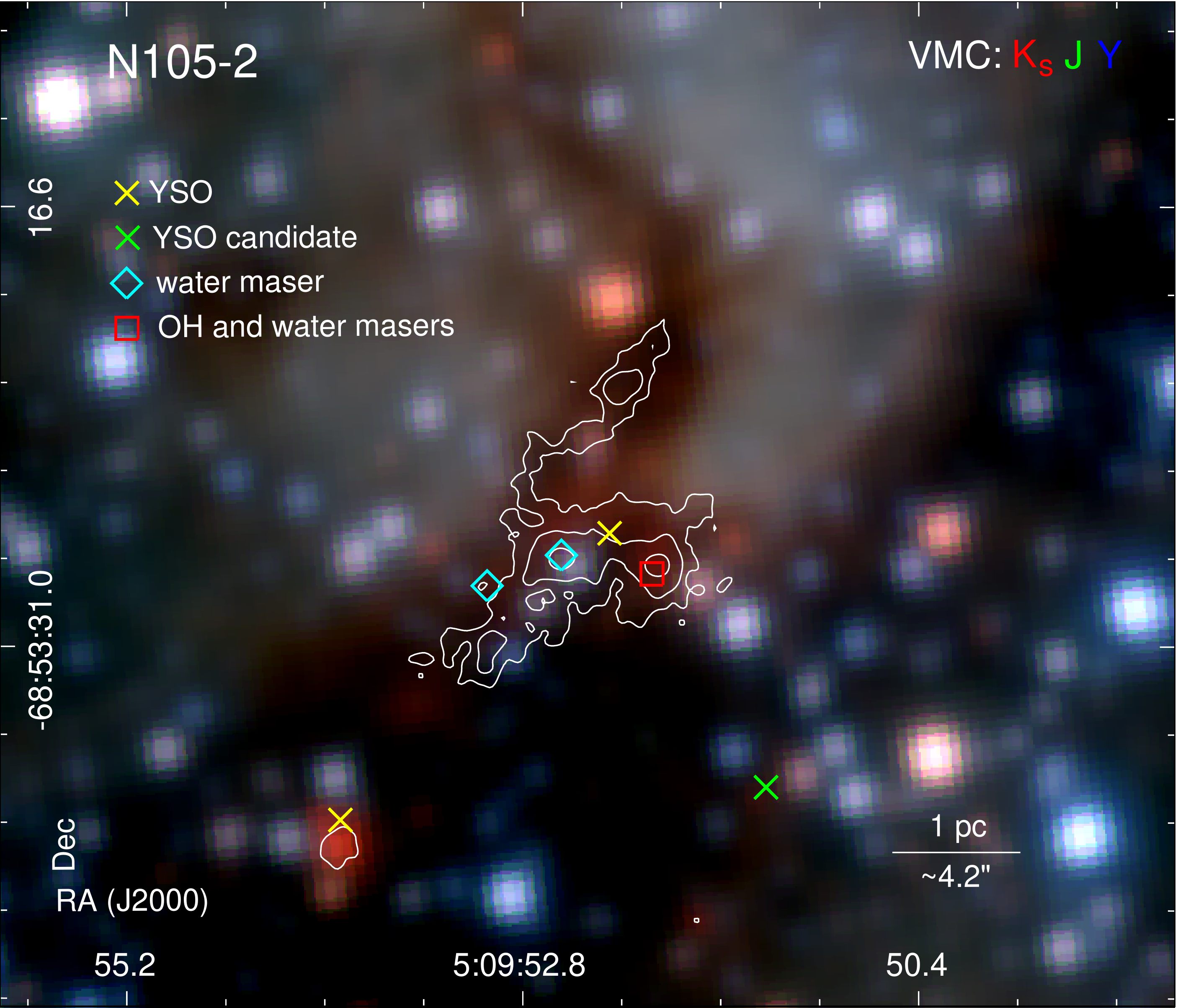}
\includegraphics[width=0.48\textwidth]{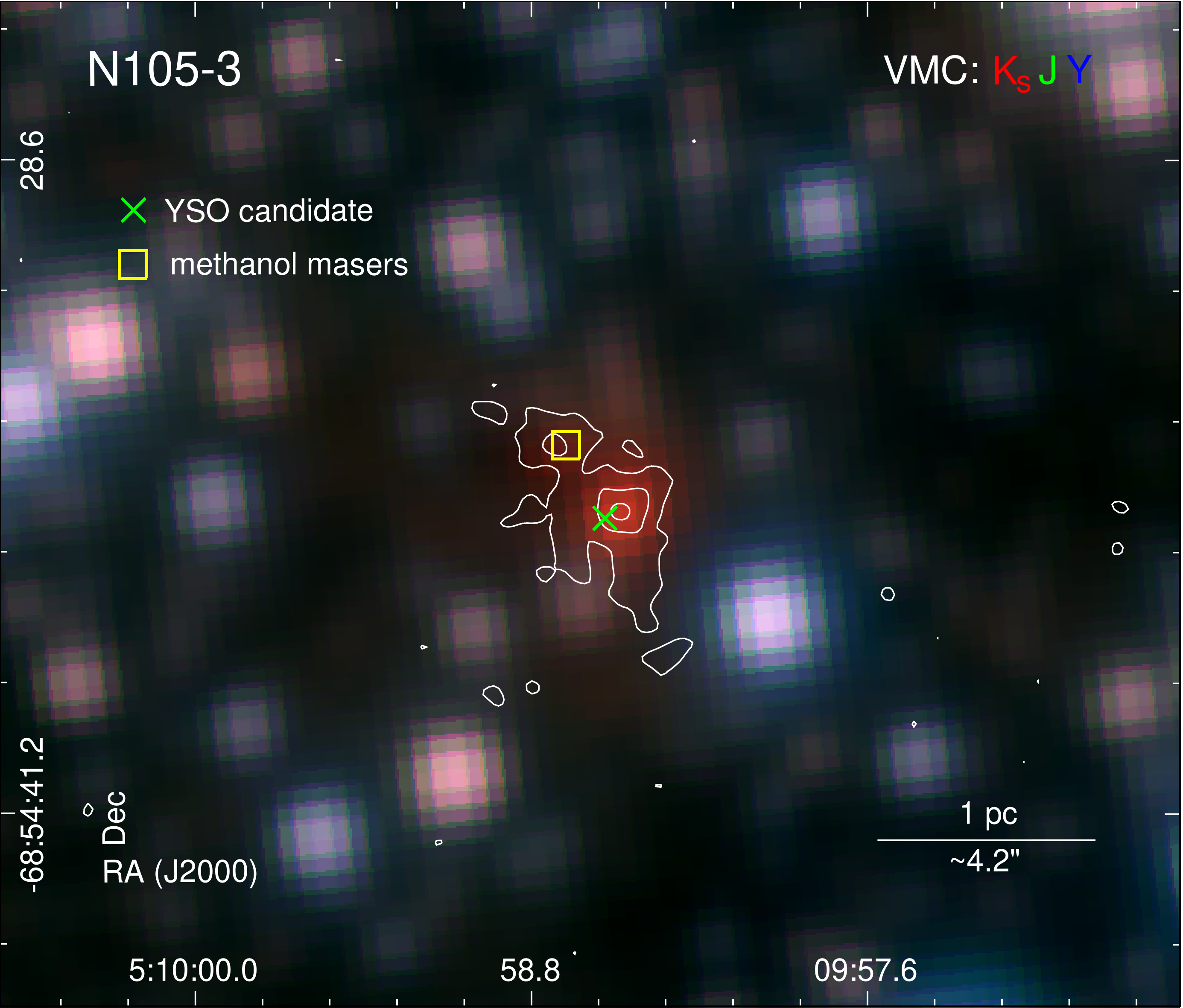}
\caption{Three-color mosaics of N\,105--1 ({\it top}), N\,105--2 ({\it center}), and N\,105--3 ({\it bottom}) combining the VMC $K_{\rm s}$ ({\it red}), $J$ ({\it green}), and $Y$ ({\it blue}) images.  The positions of YSOs, YSO candidates, and masers are marked as indicated in the legends.  The 1.2 mm continuum contours are (3, 10, 60)$\sigma_1$ for N\,105--1, (3, 10, 60)$\sigma_2$ for N\,105--2, and (3, 10, 20)$\sigma_3$ for N\,105--3; $\sigma_1$, $\sigma_2$, and $\sigma_3$ are the same as in Fig.~\ref{f:N105spitzer}.   \label{f:N105VMC}}
\end{figure}

\citet{ambrocio1998} identified and characterized several H$\alpha$ features in N\,105, including two in the brightest part of the region that they dubbed ``bright entities'' (BE).  The boundary between these features (north and south BE) is at the location of an apparent optically dark region that can be seen in the right panel of Fig.~\ref{f:region}. The northern BE is coincident with N\,105A (around the peak of the $^{12}$CO emission) and overlaps with the N\,105--1 and N\,105--2 ALMA fields.  N\,105--3 lies in the southern BE associated with the fainter H$\alpha$ emission.  

\begin{figure*}[ht!]
\centering
\includegraphics[width=0.8\textwidth]{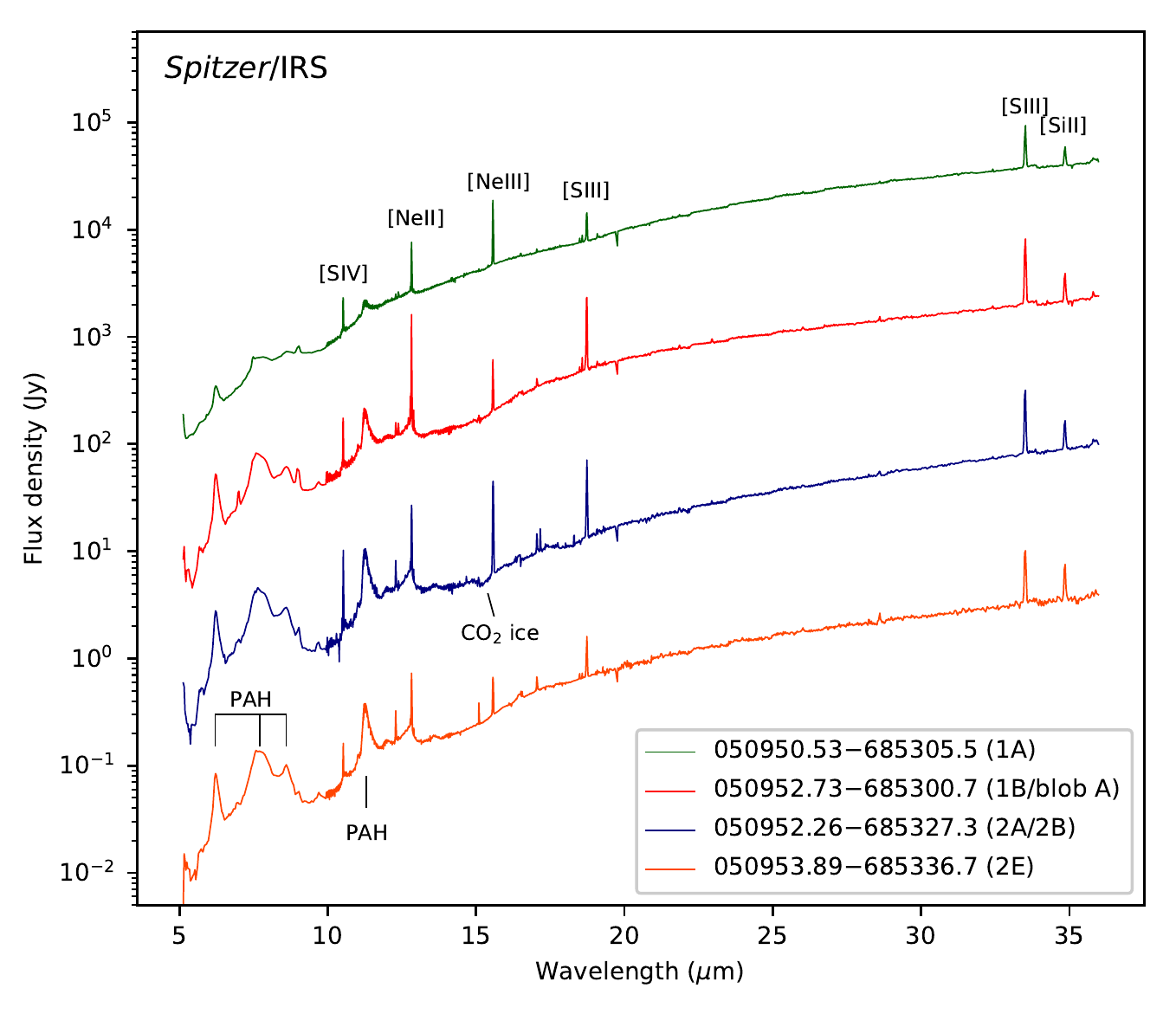}
\caption{The {\it Spitzer}/IRS spectra of YSOs associated with ALMA continuum sources color-coded as indicated in the legend. PAH and CO$_2$ ice features and fine-structure lines are labeled.  The spectra have been scaled by the following multiplicative factors for clarity (from bottom to top): 1, 12, 140, and 1000.  All spectra were analyzed and classified in \citet{seale2009} and \citet{jones2017}.  \label{f:N105spitzerIRS}}
\end{figure*}

The MCELS H$\alpha$ image shows a hint of a filamentary H$\alpha$-dark feature extending from the northern boundary of the larger optically dark region between the two BEs, first toward northwest and then northeast up to the region east of N\,105--1\,B/1\,C and farther toward northeast-east roughly to the edge of the bright H$\alpha$ emission. This dark lane is also visible in the near-IR images of N\,105--2 and N\,105--1 where the extended emission in all VMC bands has been detected. The ALMA 1.2 mm continuum emission in N\,105--2 coincides with the optically dark regions. Sources 2\,A, 2\,B, 2\,D, and 2\,F are located at the northern edge of the larger optically dark region, while the extended emission connecting sources 2\,B and 2\,C, as well as source 2\,C lie in the dark lane. Several near-IR sources are detected throughout the region. Two very faint $K_{\rm S}$-band  sources are associated with the 2\,A and 2\,B continuum peaks; however, no near-IR or mid-IR source appear to coincide with 2\,C, indicating that this source may be the youngest object in N\,105--2. 

\begin{figure*}
\centering
\includegraphics[width=0.6\textwidth]{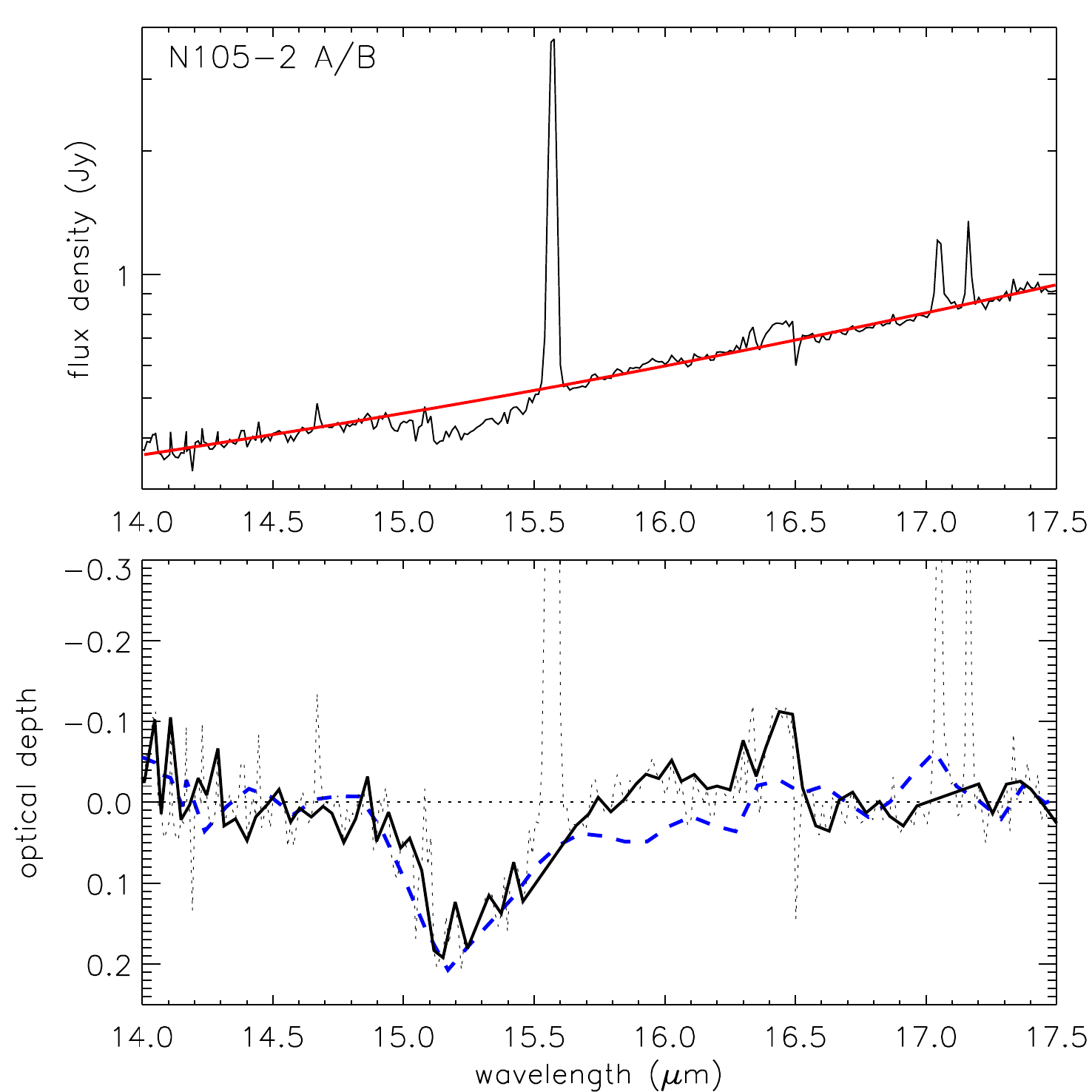}
\caption{The {\it top} panel shows a zoom-in on the {\it Spitzer}/IRS spectrum for the YSO corresponding to the  N\,105--2\,A continuum source, covering the CO$_2$ ice band at 15.2 $\mu$m.  The red solid line shows the continuum fit (a low-degree polynomial). The spectrum and the fitted continuum  were used to calculate the optical depth.  The dotted black line in the {\it bottom} panel shows the resulting optical depth spectrum for 2\,A; the solid black line shows the same optical depth spectrum smoothed for ease of visibility. The blue dashed line shows the spectrum for a well-studied SMC YSO from \citet{oliveira2013} roughly scaled to the same optical depth for guidance of the shape. \label{f:2A_CO2}}
\end{figure*}

The brightest mid- and far-IR emission in N\,105 is associated with the northern BE / N\,105A (see Figs.~\ref{f:n105}, \ref{f:N105spitzer}, and \ref{f:N105herschel3col}--\ref{f:N105herschel3im}). The structure of the {\it Spitzer} 8 $\mu$m emission tracing hot gas and PAHs is relatively complex with filaments and shell-like structures, the latter particularly evident in the southern BE.  The ALMA 1.2 mm continuum sources in N\,105--3 lie at the rim of the bubble outlined by the mid-IR emission.  Similarly, N\,105--2\,A/2\,B appear to be located at the southern rim of the smaller bubble filled with the extended 4.5 $\mu$m emission (likely dominated by H$_2$ emission from outflow shocks; \citealt{cyganowski2011}). The extended 4.5 $\mu$m emission is detected throughout the northern BE / N\,105A.

BEs and the bubble-shaped H$\alpha$ nebulae extending toward east and west are highly excited, displaying a bright [O\,{\sc iii}] 5007 \AA\, emission (e.g., \citealt{ambrocio1998}).  N\,105 is photoionized by massive stars from the LH\,31 OB association and two Wolf-Rayet (WR) stars (Brey\,16 and Brey\,16a; \citealt{breysacher1981}).  \citet{ambrocio1998} found that the excitation level of the northern BE nebula (traced by the [O\,{\sc iii}]/H$\beta$ ratio) is twice as large as the excitation level measured in the southern BE nebula. The WR star Brey\,16a located $\sim$10$''$ northeast from the ALMA continuum source N\,105--1\,B (see Fig.~\ref{f:N105VMC}) is most likely responsible for ionizing the northern BE with a possible contribution from the second WR star (Brey 16) located $\sim$1$'$ east from N\,105--2\,A. 

WR stars are hot, high-luminosity evolved stars with powerful, fast, and dense stellar winds (e.g., \citealt{crowther2007}). WR stars' winds with the highest mechanical luminosities out of all massive stars, sweep up the ambient medium forming shells structures.   Surprisingly, no bubble associated with the progenitor of Brey\,16a is observed in N\,105A.    \citet{ambrocio1998} suggested that the observed morphology of the gas around WR stars is a combined effect of the powerful stellar winds propagating in an inhomogeneous medium and the fact that stars formed deep in their natal molecular cloud; such a scenario would lead to the blister H\,{\sc ii} regions observed in N\,105.  The natal molecular cloud has not been completely disrupted by massive stars yet and it is still being photoevaporated and ionized. The molecular material is likely confined by stellar winds of the WR stars and OB stars in LH\,31, which is consistent with the location of the protostars and masers at the rim of the northern BE.  The on-going star formation in N\,105 might have been triggered by the winds of the progenitors of the WR stars in the northern BE, and likely by the winds of the OB stars in LH\,31 in the southern BE.

\subsection{The Detection of the CO$_2$ Ice Band Toward N\,105--2\,A}
\label{s:co2}

The {\it Spitzer}/IRS spectra of four point sources in N\,105 were first described in \citet{seale2009}. They classified these sources using a principal component analysis, resulting in P (i.e., PAH-dominated) classifications for 1\,A and 2\,E and PE (PAH-dominated with significant fine-structure emission) classifications for 1\,B/blob\,A (see Section~\ref{s:IR}) and 2\,A/2\,B. As it can be seen from Fig.\ref{f:N105spitzerIRS}, the spectra are very similar and the PAH and fine structure emission are very conspicuous. Silicate in absorption was reported by \citet{seale2009} for 2\,A/2\,B and 2\,E, but this is not clear at all from the spectra. \citet{jones2017} reclassified all IRS spectra obtained for LMC sources; they find no evidence of silicate absorption in any of the spectra in N\,105. They classify 1\,A, 1\,B/blob\,A, 2\,A/2\,B, and 2\,E as YSO3/H\,{\sc ii} (1\,A, 1\,B/blob\,A, 2\,A/2\,B) and H\,{\sc ii} (2\,E) types, i.e. relatively evolved YSOs with emerging compact H\,{\sc ii} regions. As pointed out by \citet{jones2017}, the IRS spectra alone are in fact not sufficient to unambiguously classify more evolved YSOs. Furthermore, given the large and variable slit width across the IRS range (3$\rlap.{''}$6/$\sim$0.9 pc at shorter to 11$\rlap.{''}$1/$\sim$2.7 pc at longer wavelengths), point-source spectra can be contaminated by the ambient emission in the wider H\,{\sc ii} region, likely the case in N\,105 (see Figs.~\ref{f:n105}--\ref{f:region}).

Given the new ALMA data we have obtained for N\,105, we have revisited the analysis of these IRS spectra. In Fig.~\ref{f:N105spitzerIRS}, there is a hint of a broad absorption feature at $\sim$15\,$\mu$m  in the spectrum of 2\,A/2\,B. Broad absorption features in the IRS spectra of YSOs are commonly attributed to solid-state (i.e., ice) features of abundant molecules like H$_2$O, CO$_2$, etc. \citep{oliveira2009,oliveira2011,seale2011,oliveira2013}; the feature at $15.2$\,$\mu$m is due to CO$_2$ ice.

Figure~\ref{f:2A_CO2} shows the results of this new analysis. It closely follows that described in \citet{oliveira2009,oliveira2011,oliveira2013}. We first define a pseudo-continuum over the range 14\,$-$\,17\,$\mu$m (using spectral regions free of absorption and emission lines) by fitting a low-degree polynomial (top). The spectrum and the fitted continuum are used to calculate the optical depth (bottom).  The optical depth  for N105--2\,A/2\,B is compared to that of a well studied SMC YSO analyzed in \citet{oliveira2013}; the shape of these absorption features are very similar, except that that of N\,105--2\,A/2\,B lacks the red wing. While the shape of the CO$_2$ ice profile reflects the composition and environmental conditions like temperature \citep[see the extensive discussion in][]{oliveira2009}, in this case the presence of the strong emission line at 15.55 $\mu$m due to [Ne\,{\sc iii}]  precludes any further analysis of the profile shape. For completeness, we analyzed the spectra of the other three sources in the same way; no features attributable to CO$_2$ ice are  detected. 

The column density has been computed using the line strength $A = 1.1\times10^{-17}$~cm~molecule$^{-1}$ \citep{gerakines1995}. Column density calculations are  dependent on the exact determination of the pseudo-continuum and can vary by as much as 25\% \citep{oliveira2009}; furthermore, for this particular spectrum, the relatively low signal-to-noise ratio and  the aforementioned emission line make this determination difficult. We compute a CO$_2$ ice column density of $(2.63\pm0.14)\times10^{17}$ \,cm$^{-2}$; the uncertainty is statistical only and it does not take into account any uncertainties in the continuum determination. This column density is at the lower end of the ranges reported in \citet{oliveira2009} and \citet[$\sim$(1\,$-$\,17) $\times$ 10$^{17}$\,cm$^{-2}$]{seale2011}, and this is perhaps the reason why neither \citet{seale2011}  nor \citet{jones2017} identified CO$_2$ ice in the spectrum of N\,105--2\,A/2\,B.

\begin{figure}[ht!]
\centering
\includegraphics[width=0.47\textwidth]{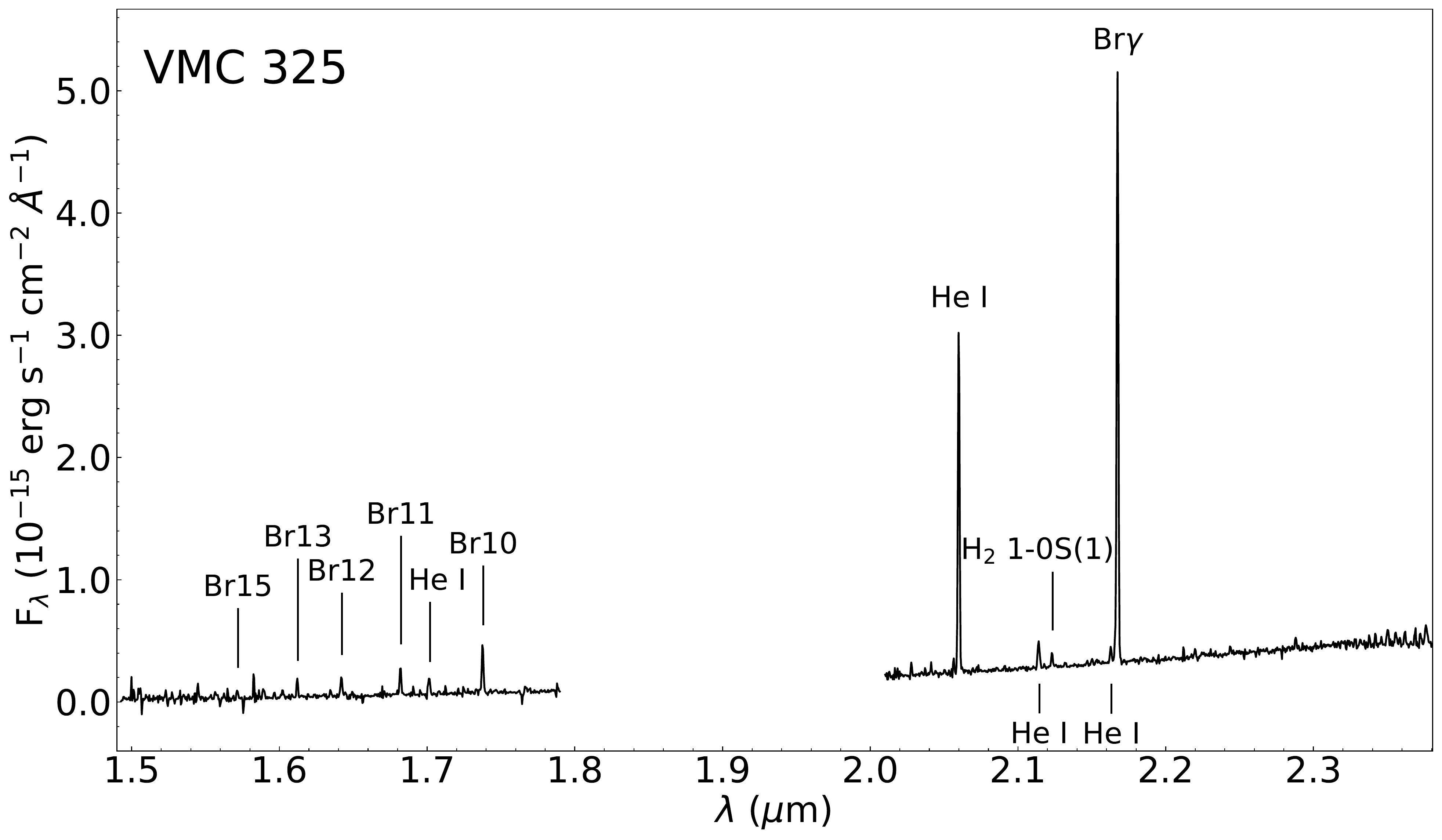}
\includegraphics[width=0.47\textwidth]{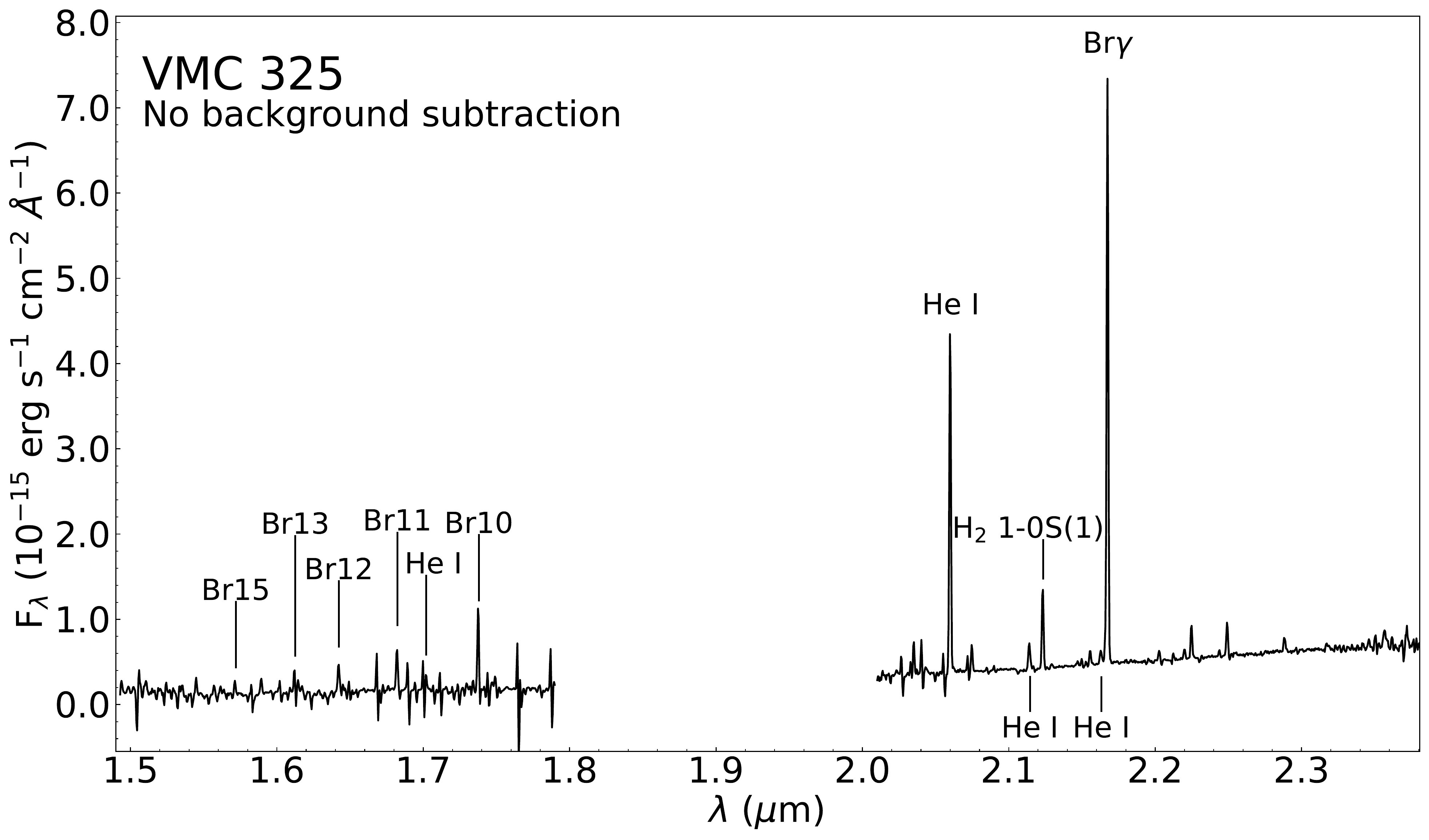}
\includegraphics[width=0.472\textwidth]{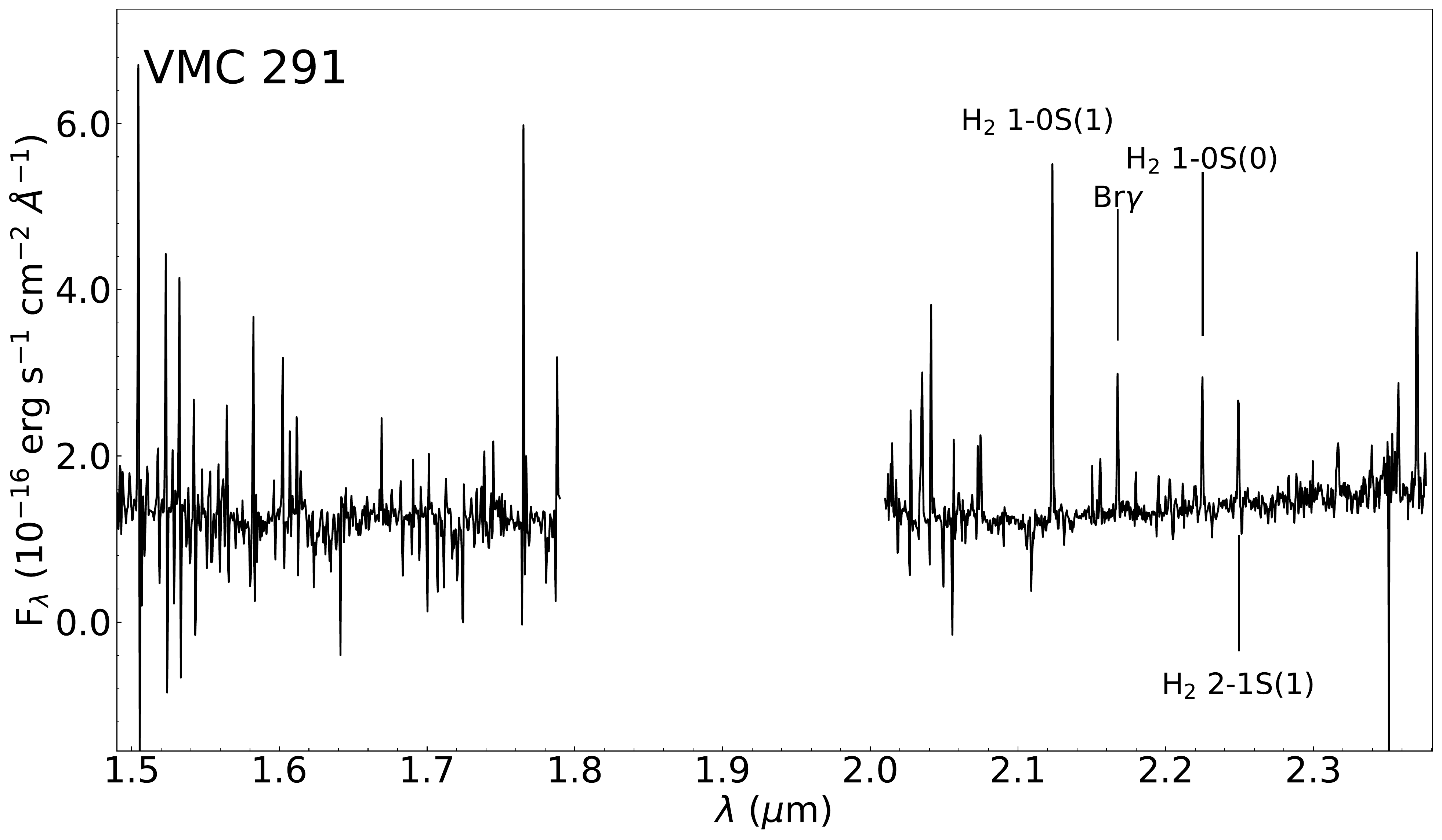}
\includegraphics[width=0.475\textwidth]{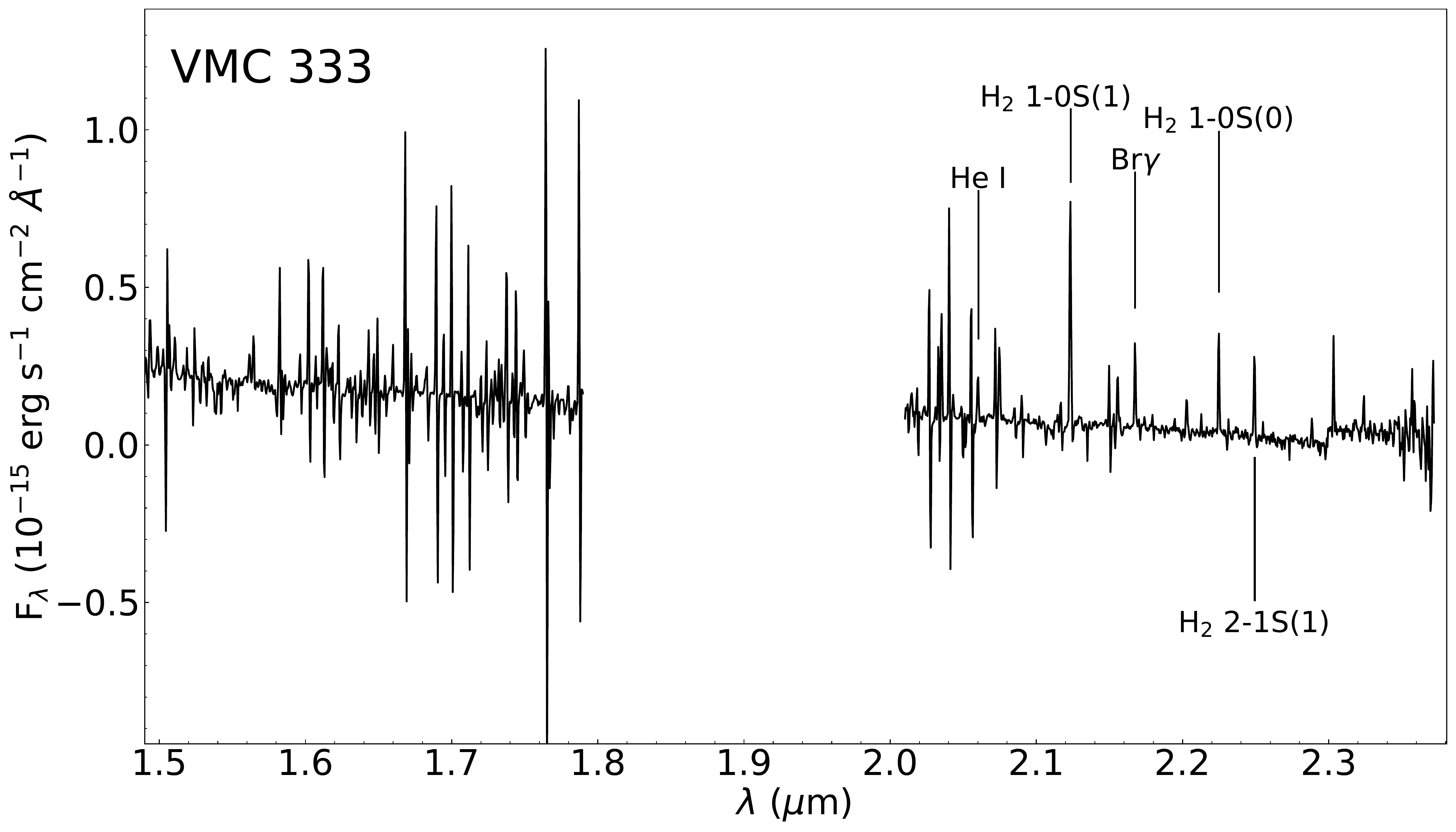} 
\caption{{\it From top to bottom} {\it VLT}/KMOS {\it H}$+${\it K} spectra of the near-IR source VMC\,325 (corresponding to the ALMA continuum source N\,105--1\,A) with and without background subtraction, and the background-subtracted spectra of VMC\,333 (in the vicinity of 2\,B) and VMC\,291 (2\,E). 
 \label{f:KMOSspec}}
\end{figure}

\begin{figure*}[ht!]
\centering 
\includegraphics[width=\textwidth]{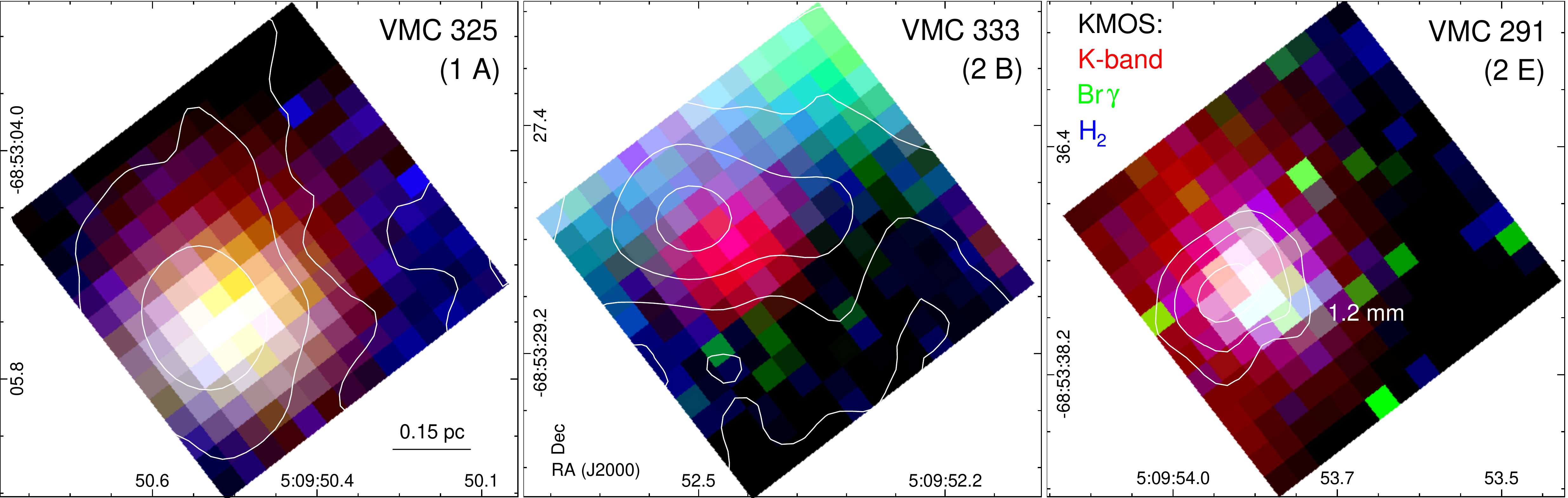}
\caption{Three-color KMOS mosaics of VMC\,325  ({\it left}), VMC\,333 ({\it center}), and VMC\,291 ({\it right}), combining the {\it K}--band ({\it red}), Br$\gamma$ ({\it green}), and H$_2$ ({\it blue}) images.  The 1.2 mm continuum contours are overlaid for reference and the name of the closest ALMA continuum source is indicated in each image; the contour levels are (3, 10, 80)$\sigma_1$, (3, 10, 30, 80)$\sigma_2$, and (3, 5 8)$\sigma_2$ for 1\,A, 2\,B, and 2\,E, respectively, where $\sigma_1$ ($\sigma_2$) is the rms noise in the N\,105--1 (N\,105--2) 1.2 mm continuum image. The size of the images is $2\rlap.{''}8\times2\rlap.{''}$8. \label{f:KMOS3color}}
\end{figure*}

\subsection{KMOS Spectroscopic Results} 
\label{s:KMOSresults} 

The KMOS {\it H}$+${\it K}-band spectra are available for three VMC sources in N\,105 in the area covered by our ALMA observations. Their source catalog IDs are 558354728325, 558354728333, and 558354728291 (hereafter VMC\,325, VMC\,333, and VMC\,291) and are located  nearby ALMA 1.2 mm continuum sources N\,105--1\,A, N\,105--2\,B, and N\,105--2\,E, respectively.  The KMOS spectra are shown in Fig.~\ref{f:KMOSspec}; the detected spectral lines include the H\,{\sc i} (with the brightest Br$\gamma$ at 2.166 $\mu$m), He\,{\sc i} (2.058 $\mu$m), and H$_2$ (2.122 $\mu$m) lines. No CO bandhead (2.3 $\mu$m) and fluorescent Fe\,{\sc ii} (1.688 $\mu$m) emission tracing disks or forbidden [Fe\,{\sc ii}] (1.644 $\mu$m) emission tracing outflows were detected. Figure~\ref{f:KMOS3color} shows three color images of VMC\,325, VMC\,333, and VMC\,291, combining the KMOS {\it K}-band continuum, Br$\gamma$, and H$_2$ images, with the ALMA 1.2 mm continuum contours overlaid.

To analyze the data, we followed the analysis outlined in \citet{ward2016}. Here, we include a brief summary of the data analysis methods, followed by a discussion of the results for individual sources. The extinction ($A_{\rm V}$) values are calculated using the KMOS $H$- and $K$-band continuum measurements and are provided in Table~\ref{t:kmosLacc}. These were measured by fitting a third order polynomial to the continuum spectrum and the integrated flux of the polynomial was measured for the wavelength ranges 1.5365--1.7875 $\mu$m ($H$-band) and 2.028--2.290 $\mu$m ($K$-band; Eq. 1 in \citealt{ward2016}) and assuming an intrinsic $H-K$ color of $-$0.05 mag, corresponding to a B0\,V star. We were unable to estimate $A_{\rm V}$ for VMC\,333 (nearby 2\,B) which has a very blue spectrum, indicating that it is not embedded. The value of $A_{\rm V}$ for VMC\,325 (1\,A; $42\pm4 $ mag, Table~\ref{t:kmosLacc}) is in an excellent agreement with that estimated by \citet{oliveira2006} using the $VLT$/ISAAC spectrum ($\sim$40 mag; see Section~\ref{s:IR}). We applied the extinction correction to our measurements for VMC\,325 (1\,A) and VMC\,291 (2\,E) described below to constrain the physical properties of the observed sources.  

The accretion luminosity ($L_{\rm acc}$)  was estimated from the Br$\gamma$ luminosity ($L_{{\rm Br}{\gamma}}$) assuming the relation of \citet{calvet2004} for intermediate-mass YSOs holds for high-mass YSOs in the LMC: ${\rm log} (L_{\rm acc}) = -0.7 + 0.9({\rm log}(L_{{\rm Br}{\gamma}})+4)$. The accretion luminosities for VMC\,325, VMC\,333, and VMC\,291 are listed in Table~\ref{t:kmosLacc}.

\begin{deluxetable*}{lcccccccc}
\centering
\tablecaption{VMC Photometry, Visual Extinction, and Accretion Luminosity for KMOS targets  \label{t:kmosLacc}}
\tablewidth{0pt}
\tablehead{
\colhead{Object} &
\colhead{VMC} &
\colhead{RA (J2000)} & 
\colhead{Dec (J2000)} &
\colhead{$K_{\rm S}$} & 
\colhead{$J$} & 
\colhead{$Y$} & 
\colhead{$A_{\rm V}$} &
\colhead{$L_{\rm acc}$} \\
\colhead{} & 
\colhead{Source ID} &
\colhead{(deg)} &
\colhead{(deg)} &
\colhead{(mag)} &
\colhead{(mag)} &
\colhead{(mag)} &
\colhead{(mag)} &
\colhead{($L_{\odot}$)}
}
\startdata
VMC\,325 & 558354728325	& 77.4605865	& $-$68.8848206 & 13.837 (0.004) & \nodata             & \nodata& $42\pm4$ &   $(2.9^{+0.9}_{-1.5})\times10^{5}$ \\   
VMC\,333 & 558354728333	& 77.4686862	& $-$68.8912009 & 16.113 (0.014) &  17.415 (0.026)	 & 18.070 (0.036)& \nodata & $(2.0^{+0.1}_{-0.1})\times10^{2}$  \tablenotemark{\footnotesize a}\\   VMC\,291 & 558354728291	& 77.4746443	& $-$68.8937331 & 15.357 (0.009) & \nodata             & \nodata &  $12\pm2$ & $(2.3^{+0.5}_{-0.6})\times10^{2}$ \\   
\enddata
\tablenotetext{\footnotesize a}{Not corrected for extinction.}
\end{deluxetable*}

\noindent {\bf VMC\,325}: Figure~\ref{f:KMOS3color} shows a good positional correlation between the VMC source VMC\,325 and the ALMA N\,105--1\,A 1.2 mm continuum peak, indicating that the near-IR source observed with KMOS is the central protostar in 1\,A.  With $A_{\rm V}$ of $42\pm4$ mag, VMC\,325 is the most embedded source of the three KMOS sources in the ALMA fields, with by far the reddest spectrum.  The spectrum of VMC\,325 / 1\,A exhibits a full Brackett series emission in the {\it H}$+${\it K} bands, a strong He\,{\sc i} and some considerably weaker H$_2$ emission. 

The H$_2$ emission tracing shocks extends toward west-northwest from the continuum source (Fig.~\ref{f:KMOS3color}), in agreement with the direction of the outflow reported in \citet{oliveira2006} (see Section~\ref{s:IR}). The accretion luminosity of $(2.9^{+0.9}_{-1.5})\times10^{5}$ $L_{\odot}$ (see Table~\ref{t:kmosLacc}) is 2--3 orders of magnitude larger than those measured toward similar objects in the Magellanic Clouds (\citealt{ward2016,ward2017}; \citealt{vangelder2020}).  This high value of $L_{\rm acc}$ can be explained by a contribution from the bright UC H\,{\sc ii} region to the Br$\gamma$ emission (e.g., \citealt{armand1996}). This interpretation is supported by the fact that the Br$\gamma$ emission is extended toward VMC\,325 (see Fig.~\ref{f:KMOS3color}).

\noindent {\bf VMC\,333}:  The $K$-band continuum position of VMC\,333 is offset by $\sim$0$\rlap.{''}$4 ($\sim$0.1 pc or $\sim$20,000 AU) from the 1.2 mm continuum source N\,105--2\,B.   VMC\,333 has quite a blue spectrum and is unlikely to be deeply embedded.  Figure~\ref{f:KMOS3color} shows an extended Br$\gamma$ and H$_2$ emission in the region. Following background subtraction most of this emission is removed, but there is still Br$\gamma$ emission coincident with the source; thus, it seems likely that the source does contribute to the emission, although it is unclear whether it is the dominant ionizing source. Assuming it is the dominant ionizing source, the accretion luminosity of VMC\,333 is $(2.0\pm0.1)\times10^{2}$ $L_{\odot}$. 

The position of the closest {\it Spitzer} source to 2\,B is offset by $\sim$0$\rlap.{''}$8, i.e., the distance two times larger than that between 2\,B and VMC\,333 (see Figs.~\ref{f:N105spitzer} and \ref{f:N105VMC}). The properties of this {\it Spitzer} source are uncertain; the {\it Spitzer}/IRS spectrum analyzed in \citet{seale2009} and \citet{jones2017} was obtained at a position between this source and another {\it Spitzer} source nearby 2\,A seen in the images and likely both sources contribute to the emission (see Sections~\ref{s:ysos} and \ref{s:co2}).  At longer {\it Spitzer} and {\it Herschel} wavelengths, these two sources remain unresolved with offsets between the emission peaks in different bands. 

No $K$-band source has been detected with KMOS at the position of the 1.2 mm continuum source 2\,F; the central source may be below a detection limit of the KMOS observations at $K$-band if it is very young. Mid- and far-IR observations matching the ALMA spatial resolution are needed to learn about the central sources in 2\,B and 2\,F. 

The extended H$_2$ emission throughout the KMOS field is consistent with the ALMA data showing evidence for strong shocks toward 2\,B (see Section~\ref{s:chemdiff2AB}).  The brightest H$_2$ emission is located toward east and northeast from 2\,B, but it is also present in the north and southeast. The distribution of the Br$\gamma$ emission is similar to H$_2$, but it is the brightest in the north.

\noindent {\bf VMC\,291}:  There is a very good positional correlation between VMC\,291 and the 1.2 mm continuum source N\,105--2\,E, indicating that VMC\,291 is the central protostar in 2\,E (see Fig.~\ref{f:KMOS3color}).  VMC\,291 has quite a red spectrum (not quite as red as VMC\,325/1\,A). The source is embedded with $A_{\rm V}$ of $12\pm2$ mag. The signal-to-noise ratio in the {\it H}-band spectrum of VMC\,291/2\,E is very poor; however, there is a clear detection of the Br$\gamma$ and H$_2$ emission lines in the $K$-band spectrum (see Fig.~\ref{f:KMOSspec}).  We have estimated the accretion luminosity for VMC\,291 of $(2.3^{+0.5}_{-0.6})\times10^{2}$  $L_{\odot}$.

\begin{figure*}[ht!]
\centering
\includegraphics[width=0.495\textwidth]{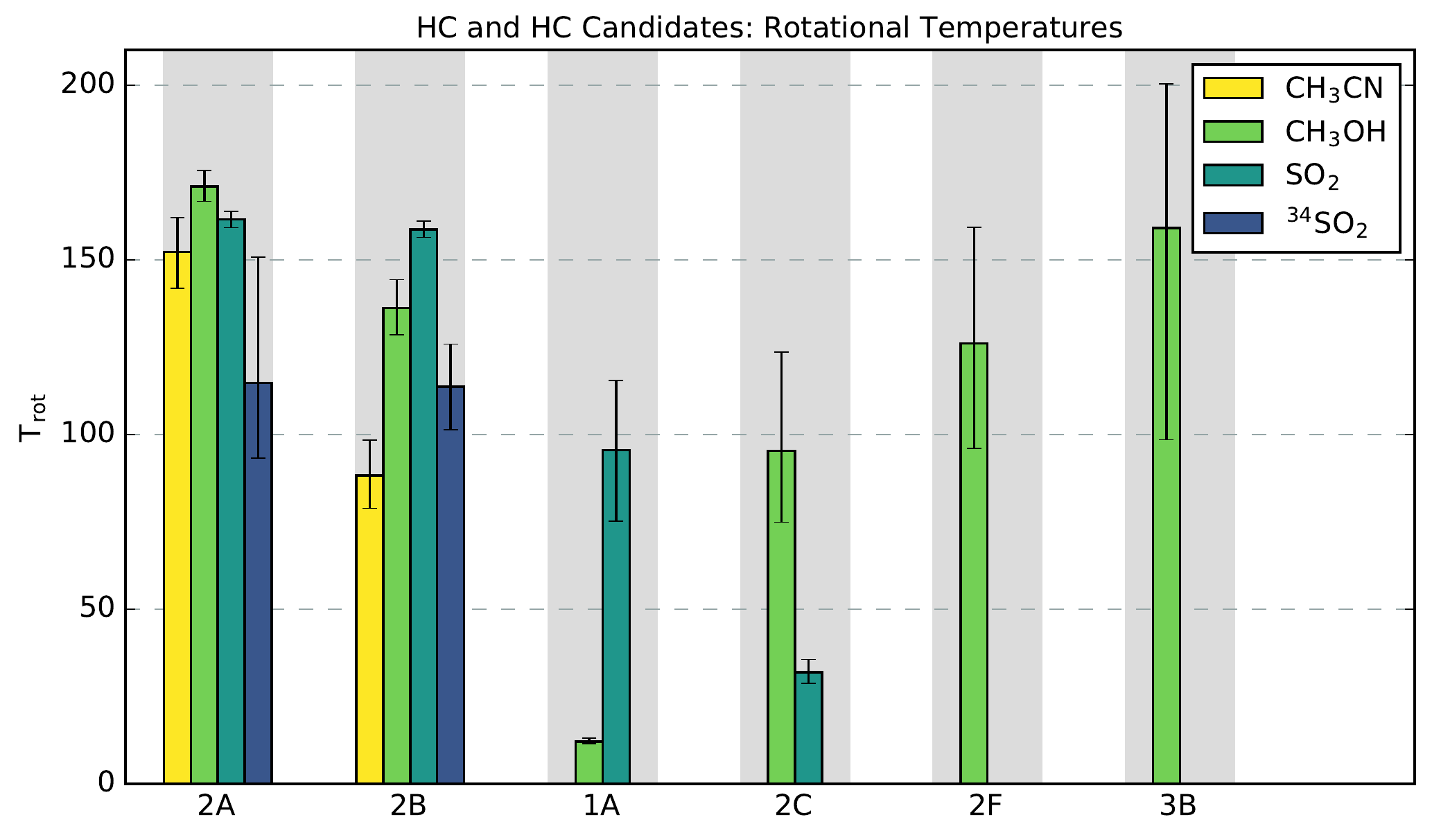}
\includegraphics[width=0.495\textwidth]{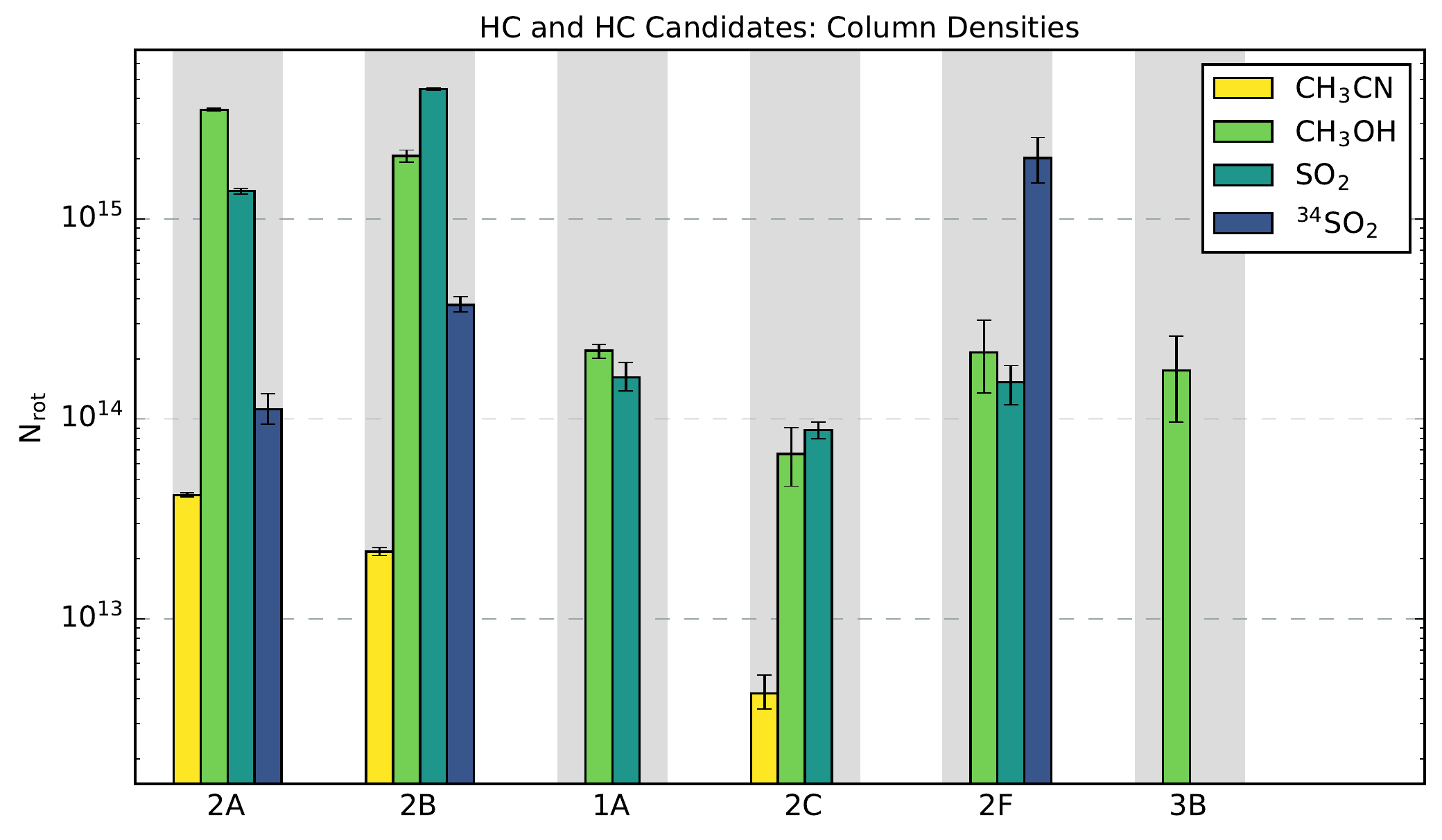}
\caption{Rotational temperatures ({\it left}) and column densities ({\it right}) of hot cores (2\,A and 2\,B) and hot core candidates (1\,A, 2\,C, 2\,F, and 3\,B) in N\,105 determined by spectral modeling for molecular species with the detection of multiple transitions---CH$_3$CN, CH$_3$OH, SO$_2$, and $^{34}$SO$_2$ (where available; see Section 4.3 and Table 6).  To calculate column densities for hot core candidates for species with no independent temperature determination, the CH$_3$OH temperature was assumed. \label{f:histHCsHCcand}}  
\end{figure*}

\section{Discussion}
\label{s:discussion}

\subsection{Hot Cores and Hot Core Candidates} 
\label{s:hotcores}

The physical and chemical properties of the continuum sources N\,105--2\,A and 2\,B indicate that they are bona fide hot cores.  Their rotational temperatures determined based on multiple tracers exceed 100 K (see Table 5) and their effective FWHM sizes are $\sim$0.13 pc (Table 4), consistent with the  definition of hot cores ($\lesssim$0.1 pc; e.g., \citealt{kurtz2000}). Slightly larger sizes of the LMC hot cores compared to Galactic hot cores are not surprising due to less dust and as a result, the radiation reaching further distances from the protostars (see also \citealt{sewilo2018}).  Both sources 2\,A and 2\,B show emission from COMs and are associated with the H$_2$O/OH masers, as typically observed toward Galactic hot cores.

Figure~\ref{f:histHCsHCcand} shows the bar histogram of rotational temperatures and column densities for bona fide hot cores 2\,A and 2\,B and several other sources from our sample exhibiting high temperature in at least one of the four tracers (CH$_3$CN, CH$_3$OH, SO$_2$, and $^{34}$SO$_2$): 2\,C, 2\,F, 3\,B, and 1\,A. We consider these sources as ``hot core candidates.''

Source 2\,C with the high CH$_3$OH rotational temperature of $\sim$95 K with an uncertainty of $\sim$25\% is likely a hot core; however, the SO$_2$ temperature is warm at $\sim$32 K. If 2\,C is indeed a hot core, the relatively low SO$_2$ temperature could be the result of blended hot and cold SO$_2$ components that could not be separated at the signal-to-noise of our observations. Alternatively, the derived SO$_2$ temperature could be suppressed by non-LTE effects; however, it is unclear whether these non-LTE effects alone would explain the large difference in the rotational temperature between CH$_3$OH and SO$_2$.  CH$_3$CN is present toward 2\,C, but only the $K=0$ component of the 14$_K$--13$_K$ ladder has been reliably detected. The 2\,F continuum peak located $\sim$1$''$ to the west from source 2\,B is likely a separate source and a hot core with the CH$_3$OH temperature of $127\pm33$ K.  The CH$_3$OH temperature determination for 3\,B has the highest uncertainty: 159$^{+41}_{-61}$ K, but the source is associated with a CH$_3$OH maser.  Multiple transitions for other species are not available to obtain an independent temperature measurement to confirm the results for 2\,F and 3\,B.  The sizes of 2\,C ($\sim$0.15 pc) and 3\,B ($\sim$0.12 pc) are consistent with them being hot cores; 2\,F is a compact source as well, but it is blended with 2\,B. 

Source 1\,A is an embedded YSO that has started ionizing its immediate surroundings and it is associated with an outflow (see Section~\ref{s:IR}). N\,105--1\,A is associated with cold CH$_3$OH ($\sim$12~K), but SO$_2$ has a rotational temperature of $\sim$$96\pm20$ K.  The SO$_2$ peak is offset from the 1\,A 1.2 mm continuum peak by $\sim$0$\rlap.{''}$6 (see Fig.~\ref{f:N105int1d}). The analysis of the spectrum extracted as a mean at half-peak of the SO$_2$ emission provided a similar result for the SO$_2$ temperature ($98\pm20$ K) and slightly warmer CH$_3$OH ($22\pm4$ K), which may be indicative of a more compact, hotter CH$_3$OH component centered on the hot SO$_2$ source. The SO$_2$ lines in the 1\,A  spectrum we analyzed may (at least partially) originate in the area offset from the continuum source and/or CH$_3$OH may be sub-thermally excited. The region of the hot SO$_2$ emission may be a separate source---externally illuminated since no infrared source has been detected at this position, possibly associated with shocks from the 1\,A outflow.

\subsection{Hot and Cold CH$_3$OH  \& SO$_2$}
\label{s:methanol}

Methanol is detected in all 12 continuum sources identified in the N\,105 region. The bona fide hot cores, 2\,A and 2\,B, and the hot core candidates 2\,C and 2\,F,  all contain both hot and cold CH$_3$OH components.  Only hot CH$_3$OH is detected in the remaining candidate, 3\,B, and only cold methanol is detected in the other 7 sources. SO$_2$ also exhibits both hot and cold components in 2\,A, 2\,B, and 2\,F. 

The formation of CH$_3$OH by solid phase hydrogenation of CO ice is the only viable formation pathway in interstellar chemistry (e.g., \citealt{herbst2009}), thus all the detected CH$_3$OH originates from a prior cold phase when hydrogenation of CO molecules occurred on grain ice mantles. These molecular ices, containing primarily H$_2$O, CO, CO$_2$, CH$_4$, NH$_3$, and CH$_3$OH (e.g., \citealt{boogert2015}), were subsequently released, wholly or partially, into the gas.  For the hot cores, the mantles can be  removed by thermal desorption from dust heated by the protostar through, and/or sputtering in shock waves associated with outflows (e.g., \citealt{jorgensen2020}).  Either mechanism could account for the presence of the hot CH$_3$OH component but the presence of SiO emission, resulting from the sputtering of refractory dust material (e.g., \citealt{schilke1997}),  confirms the role of shock waves in producing the  hot CH$_3$OH component in 2\,A--C and 2\,F.  

In hot cores where the chemistry is initiated by shock waves, cooling of the postshock gas is rapid,  the temperature approaches that expected from thermal balance determined by protostellar radiative heating, and the long-term chemical evolution closely approaches the predictions of pure thermal desorption models \citep{charnley2000}.  The derived temperatures in the N\,105 hot cores, $\sim$95--170 K,  suggest that their observed composition could be explained by simple models of post-evaporation chemistry.  For example,  H$_2$O and H$_2$S injected from grain mantles can drive the production of S-bearing molecules, such as SO and SO$_2$  \citep{charnley1997}.  Hydrogen atom abstraction reactions with H$_2$S are endoergic but can proceed in hot gas to produce SH and atomic S $$\rm H_2S \buildrel H \over \longrightarrow SH \buildrel H \over \longrightarrow S $$   
Protonation of H$_2$O by molecular ions ($\rm  XH^+ = H_3^+, HCO^+$) followed by  electron dissociative recombination reactions can  release OH and then atomic O  $$\rm H_2O \buildrel XH^+ \over \longrightarrow H_3O^+ \buildrel e^-\over \longrightarrow OH \buildrel XH^+ \over \longrightarrow H_2O^+ \buildrel e^-\over \longrightarrow O $$   and lead to SO and SO$_2$ through 
$$\rm SH \buildrel O \over \longrightarrow SO \buildrel OH \over \longrightarrow SO_2 $$  with an additional contribution to SO formation from reaction of atomic S with OH. Reformation of H$_2$O and H$_2$S in reactions of H$_2$,  with each of O, OH, S and SH, is  inhibited below about 250 K  \citep{charnley1997}. 

In 2\,D, SiO is also detected but the cold CH$_3$OH gas is accompanied by a warm component ($\sim$31 K), significantly less abundant and much cooler than in higher temperature cores. However,  the SiO abundance is $\sim$8--30 times less abundant than in the hot CH$_3$OH cores, perhaps indicating the presence of lower sputtering yields in weaker shocks. 
Less efficient sputtering of ices means less H$_2$S and H$_2$O injected into the gas, lower abundances of SH and OH,  and hence may also account for the non-detection of  SO$_2$ in 2\,D. 

\def\ltsim{\lower 0.7ex\hbox{$\buildrel < \over \sim\ $}}
\def\gtsim{\lower 0.7ex\hbox{$\buildrel > \over \sim\ $}} 

Molecular desorption from ice mantles is most probably also the origin of the cold gas detected in CH$_3$OH and, indirectly, in SO$_2$. However, it is less clear exactly how these ice mantles were deposited into the gas, considering that the inferred dust temperatures are far too low ($\sim$10--17 K) to allow  thermal desorption.  The same unresolved issue arises in Galactic dark clouds, where emission from water, methanol, and other complex molecules is detected in dense clumps at locations far from protostars or any outflows  (e.g., \citealt{bacmann2012}; \citealt{cernicharo2012}; \citealt{vastel2014}; \citealt{wirstrom2014}; \citealt{taquet2017}; \citealt{soma2018}; \citealt{agundez2021}). 

Several desorption mechanisms have been proposed and include photodesorption, explosion of UV-irradiated ices, grain heating following cosmic-ray impact (\citealt{leger1985}), cosmic-ray sputtering (\citealt{wakelam2021}), as well as  ``reactive desorption'' in which the energy released in exothermic grain-surface reactions is sufficient to overcome the physisorption binding energy (\citealt{minissale2016}; \citealt{chuang2018}).  Alternatively, a localized kinematic origin has been proposed for the origin of methanol and other putative ice-mantle molecules in Galactic dark clouds. This involves   transient heating of dust grains in low-velocity grain-grain collisions.  The center-of-mass kinetic energy resulting from  drift velocities of $\ltsim$1 km s$^{-1}$ heats the grains,  which then cool by evaporation of surface molecules. Relative grain-grain streaming could occur through wave motions (\citealt{markwick2001}),  in merging collisions between small clumps or filaments (\citealt{dickens2001}; \citealt{buckle2006}; Taquet, in prep.), or in fluid dynamical instabilities \citep{harju2020}.
   
 These mechanisms could also lead indirectly to SO and SO$_2$ in cold gas via the neutral processes described above:  if H$_2$S is also desorbed from the ices, SH and S can still be produced through protonation and dissociative electron recombination  
 $$\rm H_2S \buildrel XH^+ \over \longrightarrow H_3S^+ \buildrel e^-\over \longrightarrow SH \buildrel XH^+ \over \longrightarrow SH^+\buildrel e^-\over \longrightarrow S $$  
even though H$_2$S destruction by H atoms is inefficient.  Model calculations based on the grain-streaming  picture are able to explain the  observed close spatial correspondence between CH$_3$OH and SO in maps of several dark clouds \citep{buckle2006}. 

\begin{figure*}[ht!]
\centering
\includegraphics[width=\textwidth]{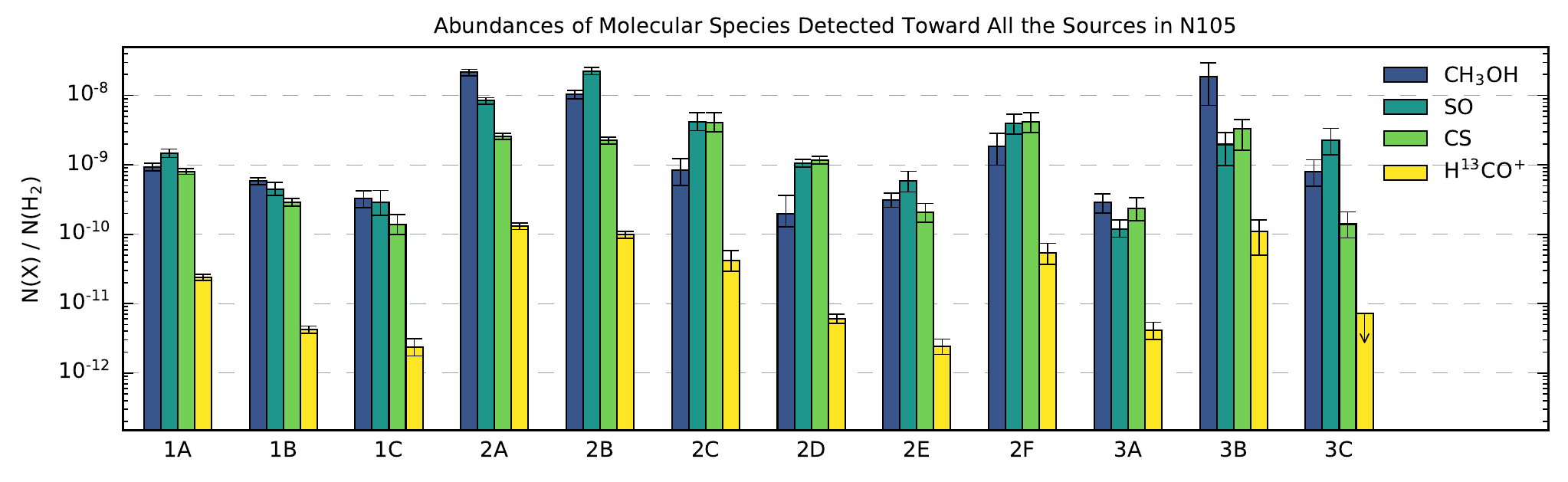}
\caption{Comparison of the CH$_3$OH, SO, CS, and H$^{13}$CO$^{+}$ abundances between all the continuum sources in N\,105.  CH$_3$OH, SO, CS, and H$^{13}$CO$^{+}$ are the only molecules detected toward all the sources. \label{f:histabundall}}
\end{figure*}

Thus, the origin of the cold methanol, as well as of SO and SO$_2$, could in principle be due to several processes. For the cold CH$_3$OH found in N\,105, photodesorption is very unlikely due to the high visual extinctions and because ice photolysis experiments indicate that CH$_3$OH dissociates on desorption (\citealt{bertin2016}; \citealt{martind2016}). Processes involving cosmic-rays or reactive desorption may be expected to produce extended, almost homogenous, CH$_3$OH distributions.  We find that cold CH$_3$OH emission is widespread in N\,105--1\,A relative to other sources and this may reflect a higher cosmic-ray flux in this field. It is difficult to estimate the cosmic-ray flux necessary to explain the cold CH$_3$OH emission as this requires knowledge of the cosmic-ray composition, the manner in which the grain is heated (whole-grain vs. spot heating), as well as whether a radical explosion can be initiated (e.g. \citealt{leger1985}).  A markedly higher cosmic-ray flux in N\,105--1 should have consequences for other molecules. In N\,105--1, we only detect simpler molecules that most likely originate in gas-phase reactions. Of these, H$^{13}$CO$^{+}$ would be most sensitive to the cosmic-ray ionization rate. The fact that the derived H$^{13}$CO$^{+}$ abundances in N105-1 are lower than toward most of the sources in N\,105--2 and N\,105--3 probably rules out the cosmic-ray desorption mechanism as the important contributor to the production of the cold CH$_3$OH emission.  

If reactive desorption of CH$_3$OH upon surface formation produced the widespread cold CH$_3$OH emission in N\,105--1\,A, it is difficult to understand why it is also not evident in N\,105--2 and N\,105--3.  For cores where the cold CH$_3$OH emission is more compact, grain collisions could be responsible for mantle desorption;  evaluating this contribution  will require a detailed understanding of the kinematics and higher angular resolution observations to probe structures (clumps, filaments) not resolved in our ALMA observations.

\subsection{Molecular Abundances: N\,105}
\label{s:abundancesN105}

\begin{figure*}[ht!]
\centering
\includegraphics[width=0.8\textwidth]{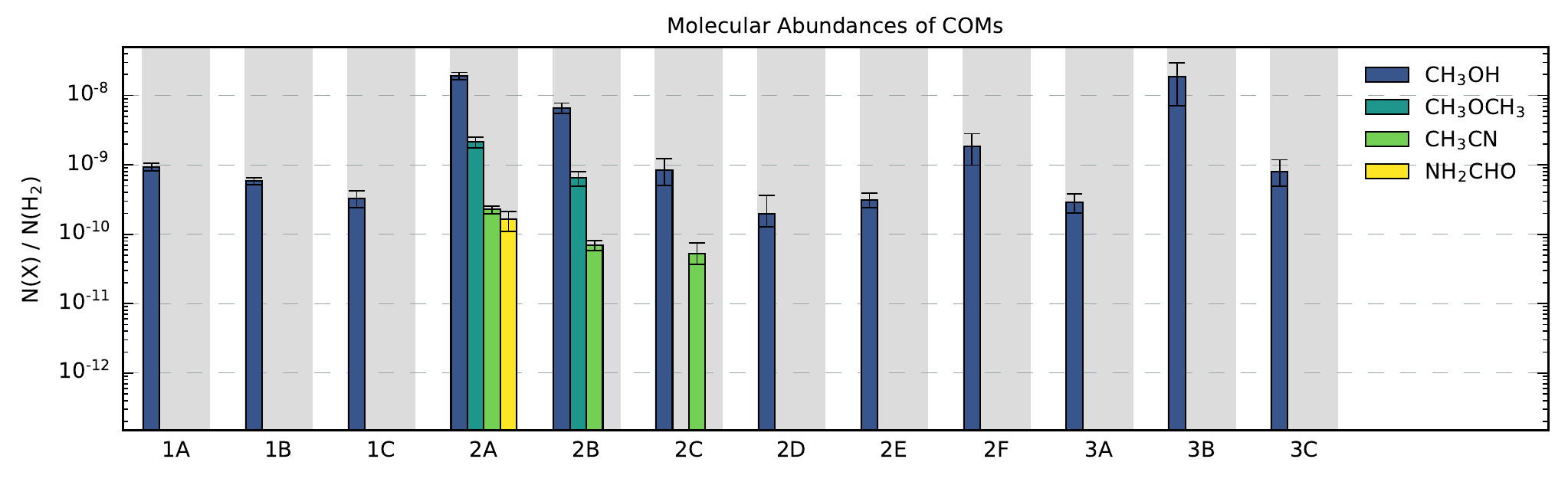}
\includegraphics[width=0.8\textwidth]{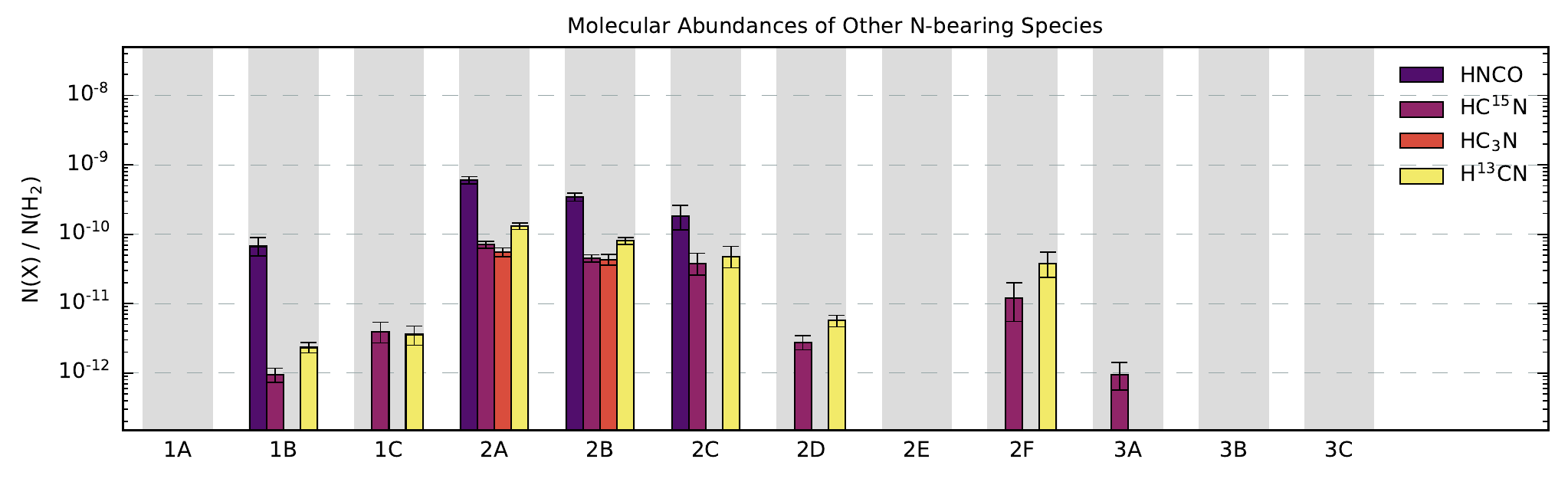}
\includegraphics[width=0.8\textwidth]{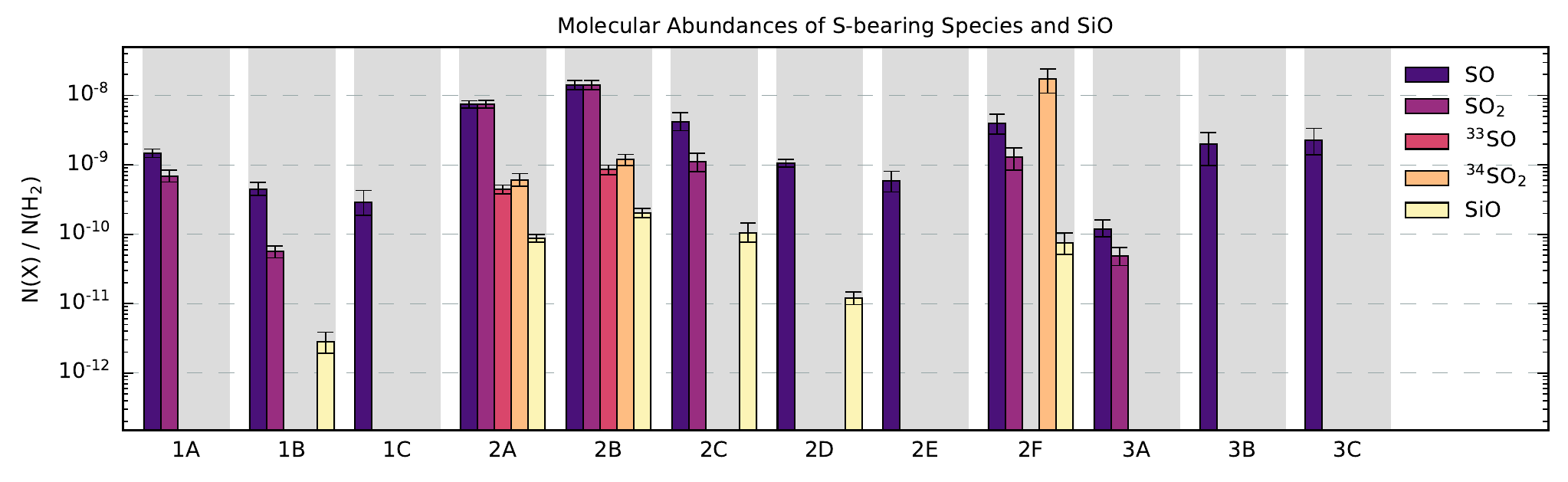}
\includegraphics[width=0.8\textwidth]{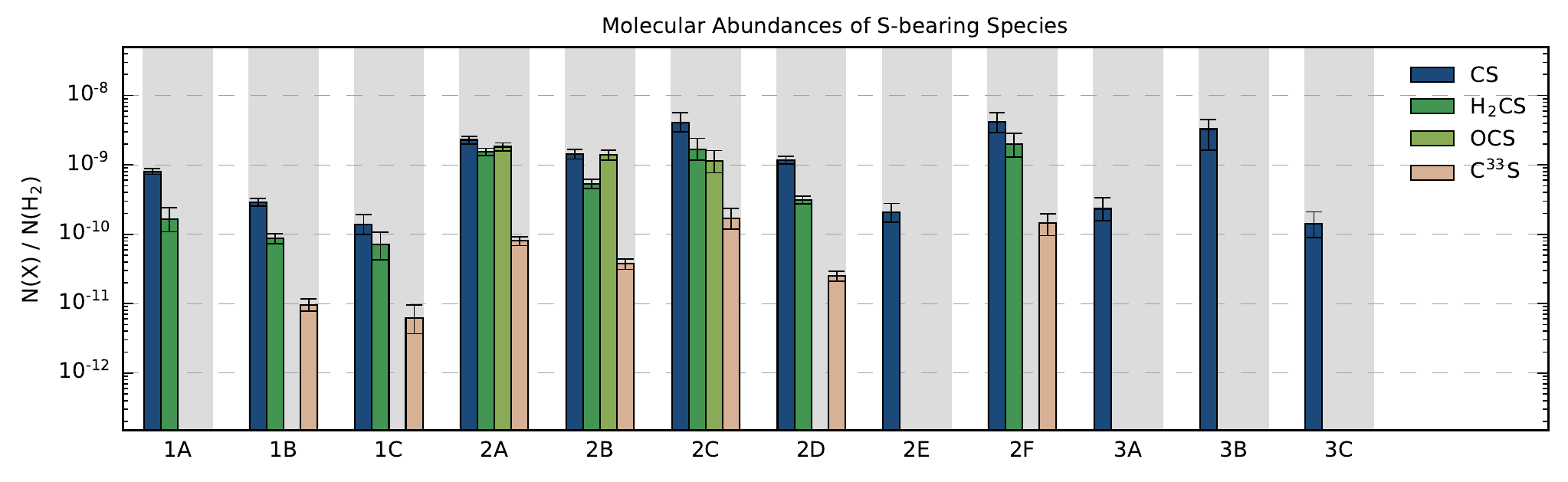}
\caption{Comparison of molecular abundances between all the 1.2 mm continuum sources detected in N\,105 with ALMA (from top to bottom): COMs (CH$_3$OH, CH$_3$OCH$_3$, CH$_3$CN, and NH$_2$CHO); simple N-bearing species (HNCO, HC$^{15}$N, HC$_3$N, and H$^{13}$CN); S-bearing species (SO, SO$_2$, $^{33}$SO, and $^{34}$SO$_2$) and SiO; other S-bearing species (CS, H$_2$CS, OCS, C$^{33}$S). In terms of the number of detected species and molecular abundances, N\,105--2 is the most chemically rich ALMA field in N\,105. \label{f:histabundNS}}
\end{figure*}

In Figs.~\ref{f:histabundall} and \ref{f:histabundNS}, we compare the fractional abundances with respect to H$_2$ for all the 1.2 mm continuum sources in N\,105, calculated as described in Section~\ref{s:nh2}.  Figure~\ref{f:histabundall} shows the bar histogram for the species detected toward all the sources: CH$_3$OH, SO, CS, and H$^{13}$CO$^{+}$.  The largest differences between the sources are observed in H$^{13}$CO$^{+}$. 

In Fig.~\ref{f:histabundNS}, we separately compare fractional abundances of COMs, N-bearing species, S-/O-bearing species and SiO, and C-/S-bearing species. In general, N\,105--2 is the most chemically-rich region with the detection of COMs other than CH$_3$OH and the highest fractional abundances of other species; the only exception is SO and CS which have comparable abundances toward the hot core candidate 3\,B.  The fewest number of species has been detected toward the continuum sources in N\,105--3, but those detected have fractional abundances comparable to those observed toward other fields, indicating a smaller size or lower density for these sources, resulting in reduced emission line strengths.  N-bearing species have lower abundances than S-bearing species in all three ALMA fields. 

Sources 2\,A, 2\,B, and 2\,C are the only sources in the N\,105 ALMA fields with the detection of CH$_3$CN.  Recently, \citet{mininni2021} analyzed the CH$_3$CN data for a sample of high-mass star-forming regions at different evolutionary stages (high-mass starless cores, high-mass protostellar objects, and UC H\,{\sc ii} regions) and concluded that the mean abundance of CH$_3$CN is a good tracer of the early stages of high-mass star formation; it shows an order of magnitude increase from starless cores to later evolutionary stages. In the LMC, we can use this result to investigate the relative ages of sources with the detection of CH$_3$CN. The lowest CH$_3$CN abundance was measured toward 2\,C: two times lower than that observed toward 2\,B and five times lower than for 2\,A.  Thus, 2\,C appears to be the youngest of the three sources, which is consistent with its location in the optical dark lane and a lack of an IR match in the available data.  

The continuum source A in N\,105--1 is the only source in our ALMA fields with the detection of H recombination lines, indicating the presence of ionized gas. It is still an embedded protostar, but it has already started ionizing its surroundings.  N\,105--2 appears to be the site of the most vigorous on-going star formation with multiple maser sites, a YSO with the detection of the CO$_2$ ice band, and the presence of COMs, hot cores, and deuterated species. In the south, hot core candidate N\,105--3\,B is associated with 6.7 GHz and 12.2 GHz CH$_3$OH masers. These are radiatively excited Class\,{\sc ii} masers known to be tracers of a very early phase of massive star formation (e.g., \citealt{cragg1992}; \citealt{ellingsen2006}). It is plausible that the source is at the early hot core phase and thus not all hot core tracers (such as SO$_2$) have achieved detectable levels.

\subsection{Tentative Detection of Formamide in the Low Metallicity Environment}
\label{s:formamide}

Formamide (NH$_2$CHO) is the simplest naturally occurring amide and has been proposed as a precursor of prebiotic molecules with a key role in the emergence of life on Earth (see \citealt{lopez2019} for a review). We detected the NH$_2$CHO 12$_{2,10}$--11$_{2,9}$ transition at 260.189 GHz (the strongest NH$_2$CHO line in the observed spectral range) at a 3.2$\sigma$ level toward hot core N\,105--2\,A. The statistical uncertainty of our NH$_2$CHO detection takes into account the uncertainty in the emission due to the contribution of the overlapping CH$_2$CO transition at 260.192 GHz to the detected line flux. Since our identification of NH$_2$CHO is based on a single, low signal-to-noise transition which is blended with another line, we can only consider this detection as tentative. If the presence of NH$_2$CHO in the LMC is confirmed, it will constitute the first detection of this astrobiologically relevant molecule in the extragalactic low-metallicity environment.

Based on the CH$_3$OH rotational temperature, N\,105--2\,A is the hottest of the sources observed in N\,105 which is likely the reason for the detection of gas-phase NH$_2$CHO only in this core. NH$_2$CHO has one of the highest physisorption binding energies of the most common interstellar COMs (including CH$_3$OH; \citealt{penteado2017}).  Lower maximum dust temperatures could account for its non-detection in the other sources in N\,105. 

\begin{figure*}[ht!]
\centering
\includegraphics[width=0.7\textwidth]{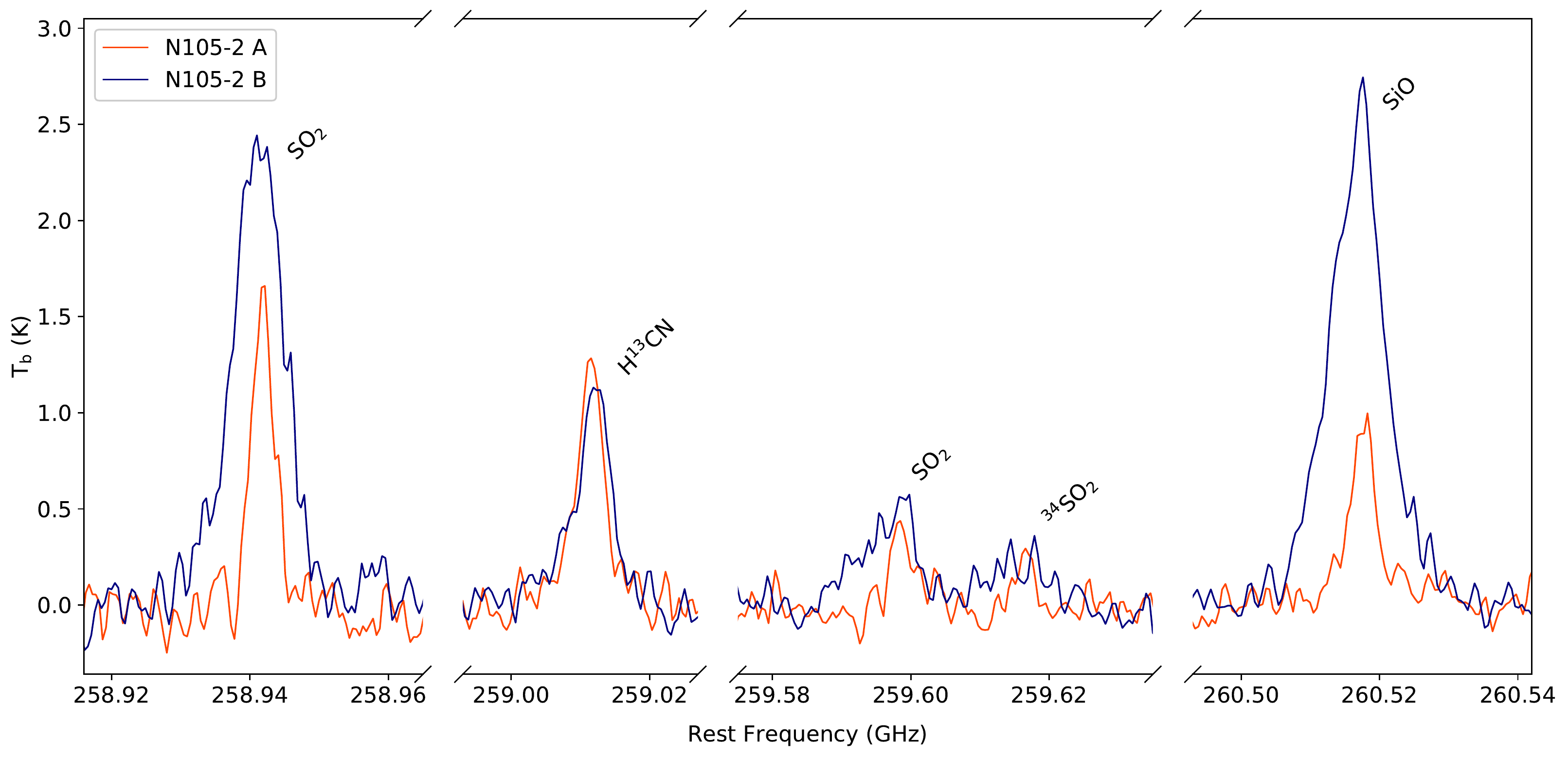}
\caption{A comparison between the line profiles of selected S- and N-bearing species, as well as a shock tracer SiO detected toward the N\,105--2 A and 2\,B hot cores. \label{f:linewidths}}
\end{figure*}

The formation routes of NH$_2$CHO are still debated. It may proceed via gas phase pathways involving H$_2$CO and NH$_2$ (e.g., \citealt{barone2015}; \citealt{skouteris2017}) and/or grain surface reactions involving hydrogenation of HNCO  (e.g., \citealt{charnley2008}); although laboratory studies indicate that the latter  route may not be viable (e.g., \citealt{noble2015}; \citealt{lopez2019} and references therein). Observations show that NH$_2$CHO is most abundant in the hot parts of protostellar envelopes ($T>100$ K; hot cores: e.g., \citealt{bisschop2007}; \citealt{allen2017}; hot corinos, similar to hot cores but formed around low- and intermediate-mass protostars: e.g., \citealt{lopez2015}; \citealt{marcelino2018}; \citealt{bianchi2019}) and regions dominated by shocks (e.g., in protostellar outflows; e.g., \citealt{yamaguchi2012}; \citealt{mendoza2014}; \citealt{codella2017}; \citealt{ceccarelli2017}). It has not been established yet whether the formation route of NH$_2$CHO depends on the environment, although gas-phase reactions seem to be the dominant pathway leading to NH$_2$CHO in the protostellar outflows (e.g., \citealt{codella2017}). Observations of NH$_2$CHO in the LMC offer unique tests of its formation models in the metal-poor environments with lower dust content and higher UV radiation fields.  While our observations do not provide enough information to draw reliable conclusions, we have attempted a preliminary investigation of the NH$_2$CHO formation routes in the LMC. 

Our observations did not cover any H$_2$CO lines, but we detected HNCO toward 2\,A. The NH$_2$CHO and HNCO fractional abundances with respect to H$_2$ for 2\,A are $X({\rm NH_{2}CHO}) = (1.7^{+0.5}_{-0.6})\times10^{-10}$  and  $X({\rm HNCO}) =  (5.4\pm0.7)\times10^{-10}$, respectively (see Table 6). One of the methods used to test the dominant formation route of NH$_2$CHO is to investigate its abundance correlation with other molecules that are thought to be chemically linked (i.e., forming from a common precursor or one forming from the other). Almost a linear correlation over several orders of magnitude in the fractional abundances was found for NH$_2$CHO and HNCO in the Milky Way (e.g., \citealt{lopez2019}).  We can test whether the fractional abundances observed toward 2\,A follow the $X({\rm NH_{2}CHO})-X({\rm HNCO})$ correlation found for Galactic  sources by estimating the expected $X({\rm NH_{2}CHO})$ from the observed  $X({\rm HNCO})$ and comparing it to the observed value (after correcting both observed values for a difference in metallicity, assuming $Z_{\rm LMC} = 0.5\,Z_{\odot}$).  The observational correlation (power-law fit) that holds for sources with the NH$_2$CHO detection (no upper limits) is available in literature: $X({\rm NH_{2}CHO}) = 0.04 X({\rm HNCO})^{0.93}$ \citep{lopezsepulcre2015}, resulting in an expected $X({\rm NH_{2}CHO})$ for 2\,A of $\sim$$1.8\times10^{-10}$ based on the observed, metallicity-corrected $X({\rm HNCO})$. The observed, metallicity-corrected  $X({\rm NH_{2}CHO})$ ($\sim$$3.4\times10^{-10}$) is a factor of $\sim$1.9 higher than the expected value, but still within the scatter of the $X({\rm NH_{2}CHO})-X({\rm HNCO})$ relation. The $X({\rm NH_{2}CHO})-X({\rm HNCO})$ correlation observed toward Galactic sources may not come from a direct chemical link between HNCO and NH$_2$CHO, but rather be the result of their similar response to the temperature of their environment (e.g., \citealt{quenard2018}). 

Using the ALMA Band 6 and Band 7 observations, \citet{shimonishi2021} have recently detected multiple transitions of NH$_2$CHO toward the submillimeter continuum source SMM1 in the WB89--789 star-forming region in the extreme outer Galaxy.  WB89--789 SMM1 is located at the galactocentric distance of 19 kpc where metallicity (traced by the oxygen abundance) is expected to be a factor of four lower than in the solar neighborhood (e.g., \citealt{fernandez2017}; see \citealt{shimonishi2021} and references therein).  \citet{shimonishi2021} found that even though WB89--789 SMM1 and hot cores in the LMC represent low-metallicity environments, there is no resemblance between the extreme outer Galaxy and the LMC sources. The authors suggest that the dissemblance might be a result of differences in the environments such as the strength of the interstellar radiation field which is significantly higher in the LMC (see Section~\ref{s:intro}). 

\citet{shimonishi2021} determined $X({\rm NH_{2}CHO})$ of $(1.8\pm0.1)\times10^{-11}$ and $X({\rm HNCO})$ of $(2.7\pm0.8)\times10^{-10}$ for WB89--789 SMM1 for physical scales corresponding to those probed by the ALMA observations of the LMC hot cores (0.1 pc), which is an order of magnitude lower than $X({\rm NH_{2}CHO})$ estimated for 2\,A.  The expected value of $X({\rm NH_{2}CHO})$ based on the $X({\rm NH_{2}CHO})-X({\rm HNCO})$ correlation of \citet{lopezsepulcre2015} for metallicity-corrected $X({\rm HNCO})$ measured toward WB89--789 SMM1 is $\sim$$5.3\times10^{-11}$. This value is a factor of $\sim$1.4 lower than the metallicity-corrected observed $X({\rm NH_{2}CHO})$ ($\sim$$7.2\times10^{-11}$) -- a result similar to that we obtained for 2\,A. 

Figure~\ref{f:histgal} shows that the abundance of NH$_2$CHO in 2\,A is higher than those measured toward the Orion Hot Core and Sgr\,B2(N) in single-dish observations (see a discussion in Section~\ref{s:galcompare}). There are, however, single-dish measurements of $X({\rm NH_{2}CHO})$ toward Galactic hot cores which are an order of magnitude higher than $X({\rm NH_{2}CHO})$ in 2\,A. For example,  \citet{bisschop2007} measured $X({\rm NH_{2}CHO})$ of a few times $10^{-9}$ for six hot cores: G24.78$+$0.08, G75.78$+$0.34, NGC\,6334\,IRS1, NGC\,7538\,IRS1, W3(H$_2$O), and W33\,A; these values are one to two orders of magnitude higher than $X({\rm NH_{2}CHO})$ reported in the literature for the same sources.

The NH$_2$CHO transition in the spectrum of 2\,A is blended with a CH$_2$CO 13$_{1,13}$--12$_{1,12}$ line at 260.192 GHz (a frequency/velocity shift of $\sim$2.9 MHz/$\sim$3.3 km s$^{-1}$; see Table~\ref{t:detections}), making the measurements of the NH$_2$CHO column density and abundance less reliable. We have investigated the NH$_2$CHO--CH$_2$CO line blending issue in detail based on other CH$_2$CO transitions detected in our observations: the CH$_2$CO line at 244.712 GHz (Table~\ref{t:detections}) and two weak lines at 242.375 GHz and 242.398 GHz ($<$3$\sigma$ detections) which helped constraining the fit. 

While the formation on grains is the most likely scenario for CH$_2$CO, a gas-phase formation route is also possible (e.g., \citealt{bisschop2007}). The NH$_2$CHO emission peak is offset from the HNCO peak in 2\,A ($\sim$0$\rlap.{''}$14 or $\sim$0.034 pc / $\sim$7000 AU at 50 kpc), but it is also not coincident with the CH$_2$CO peak. Observations of additional transitions of NH$_2$CHO are needed to confirm our tentative detection and to conduct a more reliable investigation of the spatial correlation between the NH$_2$CHO and HNCO emission that would allow us to test the NH$_2$CHO grain-surface formation scenario in the significantly different chemical laboratory of the LMC.  Observations of H$_2$CO and NH$_2$, when compared to the NH$_2$CHO data, would help test the gas-phase NH$_2$CHO formation route proposed by \citet{barone2015} in which H$_2$CO and NH$_2$ are its precursors. Interstellar NH$_2$ was detected from the ground in Sgr B2 (\citealt{vandishoeck1993}) and could be observed with ALMA. In addition, the abundance ratio of the deuterated forms of NH$_2$CHO can provide a strong constraint on its formation route (e.g., \citealt{coutens2016}; \citealt{lopez2019}).

\begin{figure*}[ht!]
\centering
\includegraphics[width=0.8\textwidth]{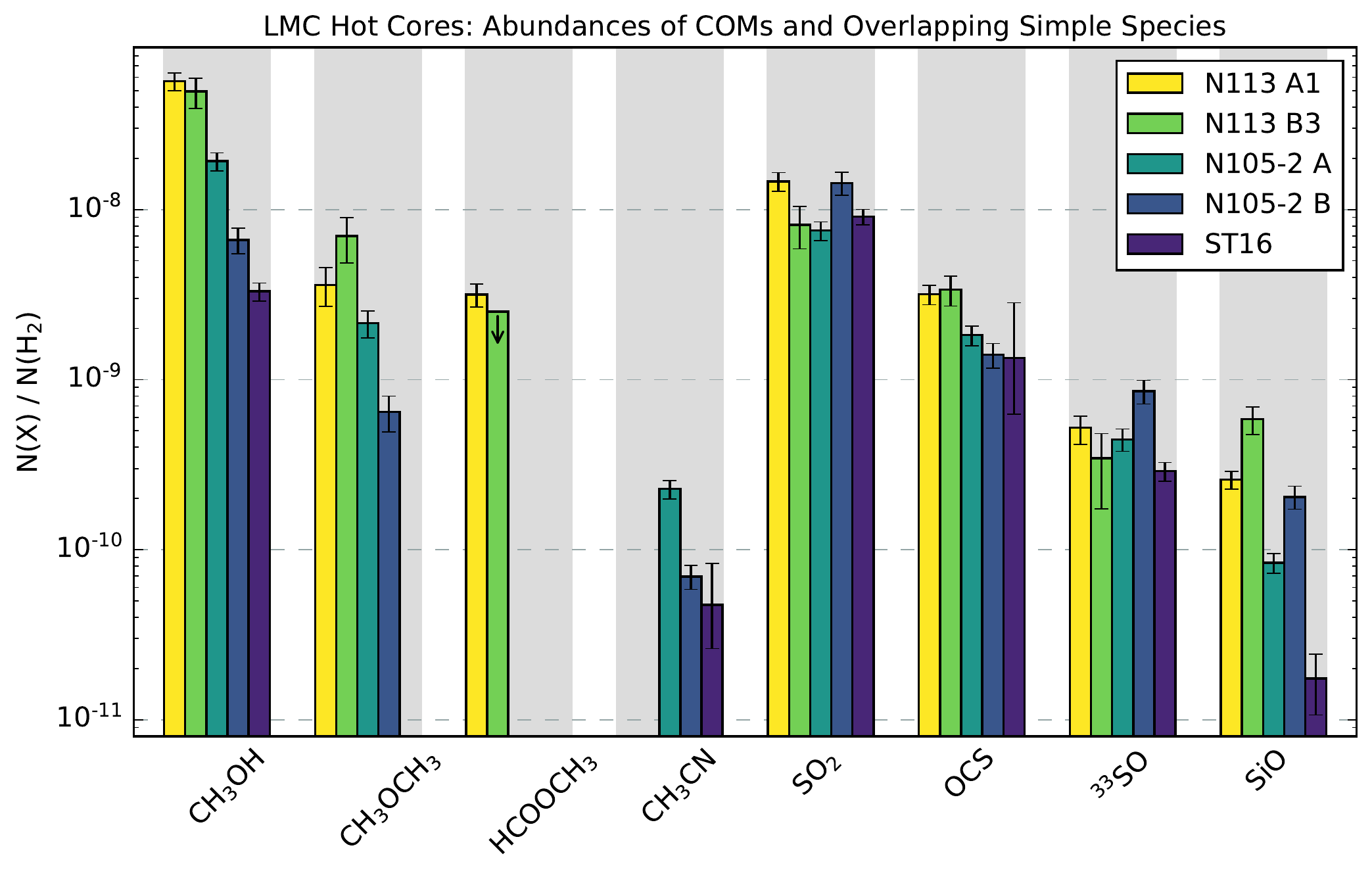}
\caption{Comparison of COM abundances, as well as abundances of simple species detected toward hot cores N\,113\,A1 and B3, N\,105--2\,A and 2\,B, and ST\,16. ST\,16 represents `organic-poor' while the remaining sources represent the `organic-rich' hot cores (see text for details).  CH$_3$CN lines were not covered by ALMA observations of N\,113; the program was designed for a different science goal and the detection of COMs was serendipitous. \label{f:histabundN113}}
\end{figure*}

\subsection{Chemical Differences Between Hot Cores N\,105--2 A and B}
\label{s:chemdiff2AB}

The integrated intensity images of N\,105--2 and the spectra of individual continuum components reveal differences in the chemical make-up of hot cores 2\,A and 2\,B, which are separated by $\sim$3$\rlap.{''}$2 ($\sim$0.78 pc / $\sim$160,000 au).  N-bearing species and deuterated species have higher abundances toward 2\,A (up to a factor of $\sim$2 for  CH$_3$CN and HDO), while the S-bearing species and SiO have higher abundances toward source 2\,B (by a factor of $\sim$3--4). The hot SO$_2$ abundance is larger in 2\,B than any other LMC hot core observed to date.  NH$_2$CHO and HDCO have only been detected toward source 2\,A. Finally, the kinematic structure observed toward 2\,B is much more complex than in 2\,A, indicating the presence of at least two velocity structures.  

In general, spectral lines for all the species are broader toward 2\,B than 2\,A, indicating the presence of more significant large scale motions in this region. The difference in line widths, as well as line intensities, is particularly striking for SO, $^{33}$SO, SO$_2$,  $^{34}$SO$_2$, and SiO (see Fig.~\ref{f:linewidths}). The enhanced abundance of S-bearing species and SiO toward 2\,B indicates the shock origin of some of the emission, consistent with the observed greater kinematic complexity, which is reflected in the broad line profiles. The enhanced production of the S- and Si-bearing molecules in shocks is a result of sputtering or destruction of refractory grain cores that release the Si and S atoms to the gas, making them available for chemical reactions (e.g., \citealt{schilke1997}; \citealt{gusdorf2008}; \citealt{vandishoeck2018} and references therein; see also a discussion in Section\ref{s:methanol}). 

Other species that are often detected in low-velocity outflows have broader lines toward 2\,B than 2\,A; these include the ice chemistry products such as CH$_3$OH, CH$_3$CN, HNCO, and HDO that can be be released to the gas by shock-driven sublimation in addition to thermal sublimation in the hot core region (e.g., \citealt{vandishoeck2018}; \citealt{oberg2021}), as well as H$^{13}$CN that can be produced via the hot gas phase chemistry in the cavity walls (e.g., \citealt{bruderer2009}). Rotation of the envelope can also broaden spectral lines, but we are not able to distinguish between the outflow and rotation based on our relatively low spatial resolution data. Molecular species with narrower lines mostly originated in the more quiescent hot core region. All these processes most likely take place in 2\,A as well, but are not as much affected by complex kinematics as in 2\,B; since the temperature is higher in the 2\,A hot core (two times for CH$_3$CN), the thermal evaporation of grain ice mantles is more efficient in this region, resulting in higher or comparable fractional abundances for all molecules expect the S- and Si-bearing species.

\subsection{Molecular Abundances: N\,105 vs. Other LMC Hot Cores with COMs}
\label{s:abundancesHC}

In Fig.~\ref{f:histabundN113}, we compare the fractional abundances with respect to H$_2$ for the LMC hot cores with COMs from literature (N\,113 A1 and B3, \citealt{sewilo2018}; ST16, \citealt{shimonishi2020}) and the newly identified hot cores in N\,105: 2\,A and 2\,B.  We re-analyzed the spectra of N\,113 A1 and B3 using the same spectral extraction and modeling techniques and the subsequent analysis for N\,113 A1/B3 as for hot cores and other sources in N\,105 and obtained results consistent with those reported in \citet{sewilo2018}; the results are listed in Table~\ref{t:tempdens}. 

As described in Section~\ref{s:nh2}, we adopted $T(\rm CH{_3}CN)$ for hot cores N\,105--2\,A and 2\,B to calculate $N(\rm H_2)$. For ST16, we recalculated the molecular abundances from \citet{shimonishi2020} by estimating $N(\rm H_2)$ using the dust temperature of 60 K provided in the paper (consistent with $T(\rm CH{_3}CN)$ of 53$^{+10}_{-7}$ K) and assuming the same LMC dust--to--gas ratio as for N\,105 and N\,113 (Section~\ref{s:nh2}). $T(\rm CH{_3}OH)$ is the only temperature determination available for hot cores A1 and B3 in N\,113 and it was used for the analysis.

\begin{figure*}[ht!]
\centering
\includegraphics[width=\textwidth]{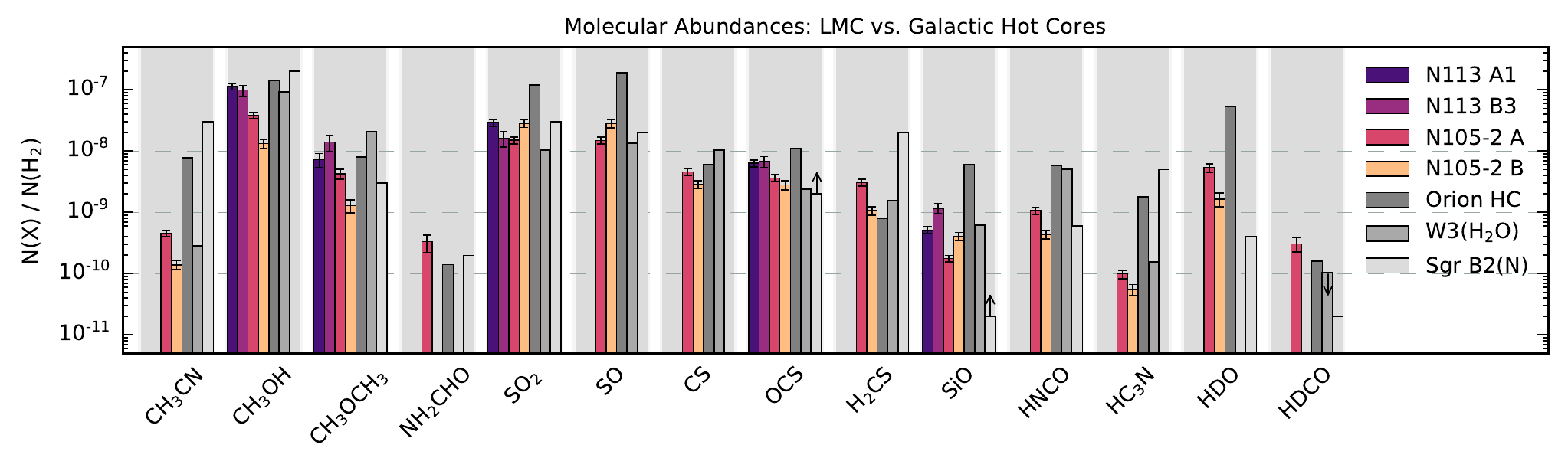} 
\caption{Comparison of the molecular abundances observed toward the organic-rich LMC (metallicity-scaled assuming $Z_{\rm LMC}$=0.5 $Z_{\odot}$) and selected Galactic hot cores. Upper and lower limits are indicated with arrows. The data for Galactic regions come from \citet{blake1987} and \citet{sutton1995}  for Orion Hot Core (HC),  from \citet{helmich1997} for Sgr\,B2(N), and from \citet{nummelin2000} for W3(H$_2$O). The largest differences in molecular abundances between the LMC and Galactic hot cores are seen for N-bearing species. \label{f:histgal}}
\end{figure*}

Figure~\ref{f:histabundN113} compares fractional abundances of COMs detected toward at least two LMC hot cores (CH$_3$OH, CH$_3$OCH$_3$, HCOOCH$_3$, and CH$_3$CN) and simple molecules that were observed toward all of the sources (SO$_2$, OCS, $^{33}$SO, and SiO).  N\,113\,A1 and B3 hot cores represent a class of ``organic-rich'' hot cores, while ST16 is an example of an ``organic-poor'' hot core as defined by \citet{shimonishi2020}; see Section~\ref{s:intro}.  

It is evident from Fig.~\ref{f:histabundN113} that N\,113\,A1 and B3 remain the LMC hot cores with the highest abundance of CH$_3$OH with respect to H$_2$, the only detection of HCOOCH$_3$, and the most reliable detection of CH$_3$OCH$_3$. The CH$_3$OCH$_3$ lines detected toward N\,105--2\,A and 2\,B have low signal-to-noise ratios and thus the molecular abundance calculations are less reliable. As discussed above, the enhanced abundances of SiO and SO$_2$ toward N\,105--2\,B are likely the results of the strong shock activity in the region. The SiO abundance toward N\,113\,B3 is even higher; similarly to 2\,B, the environment of this source is highly dynamic. There are no large variations in the OCS abundance between the known LMC hot cores.  The local environment appears to have a significant impact on the observed molecular abundances. 

Figure~\ref{f:histabundN113} demonstrates that N\,105--2\,A and 2\,B are chemically more similar to the organic-rich hot cores A1 and B3 in N\,113 than to ST16 and the other organic-poor hot core ST11 where no COMs were detected \citep{shimonishi2016b}. In the following section, we compare the fractional abundances of the species detected toward N\,105--2\,A and 2\,B to those detected toward several representative Galactic hot cores to confirm their classification as organic-rich hot cores.  It is expected that in general the fractional abundances of the molecular species observed toward the organic-rich hot cores scale with metallicity.

\subsection{Comparison to Galactic Hot Cores} 
\label{s:galcompare}

To compare the fractional molecular abundances derived for the LMC hot cores to those observed toward the Galactic hot cores, we first correct them for a difference in metallicity between these two galaxies. For this analysis, we adopt a mean metallicity of $-0.30\pm0.08$ dex (i.e., half of that observed in the solar neighborhood; based on the abundance of O and Si) estimated by \citet{rolleston2002} based on the analysis of all the classes of objects used to trace the chemical composition of the ISM in the LMC (e.g., H\,{\sc ii} regions, F-type supergiants, Cepheids, and B-type giants/bright giants) which is in agreement with a metallicity of $-0.31\pm0.04$ they determined based on the young B-type dwarfs.  This is also consistent with the metallicity determination based on the Fe abundance measured toward F-type supergiants (e.g., \citealt{russell1992} and references therein) and Cepheids (e.g., \citealt{luck1998}). The overall conclusions presented in this section do not change for a lower value of $Z_{\rm LMC}$ of 0.4 $Z_{\odot}$ found in some studies (see e.g., \citealt{maeder1999} and references therein). 
 
To correct the fractional abundances measured toward the LMC hot cores for a difference in metallicity between the LMC and the Galaxy, we have multiplied them by a factor of $1/Z_{\rm LMC}$. This simple scaling does not take into account individual elemental abundances; in the LMC, the C, O, and N elemental abundances are lower than in the Galaxy by a factor of 2.45, 2.19, and 4.83, respectively (\citealt{russell1992}). 
While the individual elemental abundances matter, it is not clear without detailed physio-chemical modelling (out of scope of the present paper) how multi-metal species should be scaled with the metallicity. The scaling of these species by the product of their metal abundances may be correct for the molecules formed in the gas phase; however, for those species that form on dust grains it only depends on the grain abundance (a total surface area available for chemistry) and the abundance of the least abundant atom. 

 The metallicity-corrected molecular abundances for N\,105--2\,A and 2\,B, as well as hot cores in N\,113, are compared to the abundances observed toward selected Galactic hot cores in Fig.~\ref{f:histgal}. We compare our observations to the single-dish observations of the Orion Hot Core, W3(H$_2$O), and Sgr\,B2(N) that trace similar physical scales---within a factor of a few of the physical scales probed by our LMC ALMA observations. The data for the Orion Hot Core come from \citet{sutton1995} where available, otherwise the fractional abundances from \citet{blake1987} are plotted (CH$_3$CN, HNCO, and HDO).  The data for W3(H$_2$O) and Sgr\,B2(N) come from \citet{helmich1997} and \citet{nummelin2000}, respectively.  The W3(H$_2$O) observations are the most similar to our ALMA observations of N\,105 in terms of the probed physical scales ($\sim$0.16 pc vs. $\sim$0.12 pc in N\,105).

Metallicity-scaled CH$_3$OH abundances with respect to H$_2$ observed toward N\,113 A1/B3 and N\,105--2\,A are the most similar to those observed toward Galactic hot cores, as can been seen in Fig.~\ref{f:histgal}; they are at the lower end of the range of those observed toward a larger Galactic hot cores sample (e.g., \citealt{mookerjea2007}).  The CH$_3$OH abundance estimated for N\,105--2\,B is lower than for N\,105--2\,A  and two hot cores in N\,113 and is more similar to that of ST16, indicating that it may be an intermediate case between the organic--rich and organic--poor types of the LMC hot cores proposed by \citet{shimonishi2020}. \citet{gerner2014} reported a low median CH$_3$OH abundance of $2.6\times10^{-8}$ (for a typical hot core temperature of 100~K),  comparable (within a factor of 2) to the 2\,B's metallicity-scaled CH$_3$OH abundance of $1.3\times10^{-8}$, estimated based on the IRAM 30m telescope observations tracing the 0.02--0.4 pc spatial scales of eleven Galactic hot molecular cores (G9.62$+$0.19, G10.47$+$0.03, G29.96$-$0.02, G31.41$+$0.31, G34.26$+$0.15, G45.47$+$0.05, G75.78$+$0.34, NGC\,7538B, Orion--KL, W31\,IRS5, and W3(H$_2$O));  the authors acknowledge that their calculations provide CH$_3$OH abundances up to a factor of ten lower than reported in literature for the same objects. 

For all hot cores in N\,105 and N\,113,  the metallicity-corrected CH$_3$OCH$_3$ (N\,105 and N\,113) and HCOOCH$_3$ (N\,113) abundances are at the lower end of the range of those observed toward Galactic hot cores.  The difference between the fractional abundances of the N-bearing species between the LMC (metallicity-scaled) and Galactic hot cores appears to be larger than for COMs (e.g., CH$_3$CN and HC$_3$N; see Fig.~\ref{f:histgal}); however, they are still consistent with Galactic values;  in Fig.~\ref{f:histgal}, similarly low abundances of CH$_3$CN, HNCO, and HC$_3$N as for 2\,A and 2\,B have been reported for the Galactic hot core W3(H$_2$O). The interesting result highlighted in Fig.~\ref{f:histgal} is that the metallicity-corrected abundance of NH$_2$CHO observed toward 2\,A is higher than for the Orion and Sgr\,B2(N) hot cores. White intriguing, we stress that more reliable observations of NH$_2$CHO toward 2\,A (higher sensitivity and more than one transition) are needed to draw reliable conclusions based on this result. No observations of N-bearing species exist for N\,113\,A1/B3 hot cores \citep{sewilo2018}. 

Molecular abundances reported in the literature for a single region can vary significantly depending on the spatial scales probed and differences in the method for determining the column densities and abundances; e.g., for W3(H$_2$O) there is a difference of about two orders of magnitude between the measurements reported in \citet{helmich1997} and \citet{bisschop2007}. The measurements from  \citet{helmich1997}  that match the metallicity-scaled LMC data well (Fig.~\ref{f:histgal}) is based on the James Clerk Maxwell Telescope (JCMT) observations, which at the distance of W3 trace similar spatial scales as our ALMA observations in the LMC; also, the data analysis is similar.  These examples highlight the importance of comparing the abundances observed toward the LMC hot cores with those of a carefully selected Galactic hot cores measured based on the observations sampling similar physical scales and using a similar data analysis methods. However, a comparison to a broader population of Galactic hot cores does not change our overall conclusion that hot cores 2\,A and 2\,B are not significantly different than Galactic hot cores. 

Figure~\ref{f:histgal} shows that hot cores 2\,A and 2\,B indeed belong to the organic-rich category proposed by \citet{shimonishi2020} (yet to be confirmed as a class based on a larger sample of hot cores) since the molecular abundances roughly scale with metallicity and larger COMs are detected.

Chemically distinct hot core types observed in the LMC may be explained if there are metallicity inhomogeneities in the LMC disk.  The chemical composition and dynamics of part of the LMC might have been altered by a close encounter between the LMC and SMC ($Z_{\rm SMC} = 0.1-0.2~Z_{\odot}$) about 0.2 Gyr ago as suggested by the hydrodynamical simulations (e.g., \citealt{fujimoto1990}; \citealt{bekki2007a}; \citealt{yozin2014}) and the observational evidence including a highly asymmetric distribution of the H\,{\sc i} and CO gas in the LMC and the presence of two H\,{\sc i } velocity components separated by $\sim$50 km s$^{-1}$ in velocity, spatially connected by bridge H\,{\sc i} features, and showing complementary spatial distributions on a kpc scale (e.g., \citealt{luks1992}). One of the H\,{\sc i } velocity components corresponds to the gas extending over the entire LMC disk and is dubbed the ``D-component.'' The lower velocity gas (``L-component'') is more spatially confined and introduces asymmetry in the distribution of the H\,{\sc i} gas in the LMC. 

According to the model, during the tidal interaction between the LMC and SMC the H\,{\sc i} gas is stripped from both galaxies and the remnant gas is falling down on each galaxy. In the LMC, this infalling H\,{\sc i} gas is observed as the H\,{\sc i} L-component and has a relative velocity of $\sim$50 km s$^{-1}$. A collision with the H\,{\sc i} gas in the LMC disk triggered the formation of the young massive cluster R136 in 30\,Dor and massive stars in the region extending south of 30\,Dor--H\,{\sc i} Ridge; it includes two major elongated CO clouds in the south-east of the LMC: the Molecular Ridge and the CO Arc (\citealt{fukui2017}). A similar scenario of massive star formation triggered by the colliding H\,{\sc i} flows as a result of the SMC--LMC interaction was recently proposed for the star-forming region N\,44 (\citealt{tsuge2019}; \citealt{tokuda2019}; \citealt{fukui2019}). 

Numerical simulations show that the gas infalling on the LMC disk contains a large amount of metal-poor gas from the SMC (\citealt{bekki2007b}), replenishing the higher metallicity material in the LMC with metal-poor gas and causing the metallicity inhomogeneities. \citet{fukui2017} and \citet{tsuge2019} found that the H\,{\sc i} L-component is metal-poor in the H\,{\sc i} Ridge and N\,44, respectively, confirming this model prediction. \citet{tsuge2019} estimated the fraction of the SMC gas in the H\,{\sc i} Ridge (including R136) and N\,44 of 0.5 and 0.3, respectively. The metal content was estimated based on the correlation between the dust optical depth at 353 GHz ($\tau_{353}$; measured using the combined {\it Planck} and IRAS data) and the H\,{\sc i} intensity (gas/dust $\sim$ $W$(H\,{\sc i})/$\tau_{353}$). 

The differences in the physical and chemical conditions in each galaxy are likely significant enough for gas mixing to result in a range of environments that would lead to variations in the hot core COM abundances. Regions affected by tidal interactions between the LMC and SMC would be characterized by a lower dust-to-gas ratio, stronger UV radiation fields, and consequently higher dust temperatures than in the unaffected LMC gas (e.g., \citealt{vanloon2010b}; \citealt{duval2014}). The lower dust content means that there would be less dust grains for surface chemistry.  Higher dust temperatures may result in less efficient formation of COMs as predicted by the warm ice chemistry model and  astrochemical simulations (e.g., \citealt{shimonishi2016a}; \citealt{acharyya2018}). The number of heavy elements available for chemistry would also be lower because in the SMC the elemental abundances of gaseous (N, C, O) atoms are $\sim$(3, 2, 2) lower than in the LMC (e.g., \citealt{russell1992}). The cosmic-ray density in the SMC is only $\sim$15\% of that observed in the solar neighborhood (compared to $\sim$25\% in the LMC; \citealt{abdo2010smc}), making the cosmic-ray-induced UV radiation less effective than in the LMC.  

The hot core ST11 with a non-detection of COMs is located near the region thought to be affected by the interaction between the LMC and SMC, possibly in a lower-metallicity region, while all known hot cores with COMs are associated with the unaffected areas in the disk. This distribution of hot cores with and without COMs may be the result of a source selection effect; a larger sample of hot cores is needed to investigate the connection between the metallicity inhomogeneities and the properties of hot cores.

\section{Summary and Conclusions}
\label{s:summary} 

To increase a sample of hot cores in the LMC, we conducted ALMA observations toward seven fields in the LMC having common characteristics with two fields in the star-forming region N\,113 hosting the only hot cores with COMs known prior to our observations (association with YSOs, OH/H$_2$O masers, and/or the SO emission).  Our ALMA observations covered four 1875-MHz spectral windows between $\sim$241 GHz and $\sim$261 GHz.  Here, we present the analysis of the three ALMA fields located in the star-forming region N\,105 at the western edge of the LMC bar.  

We performed the spectral analysis of twelve 1.2 mm continuum sources in N\,105.  We identified S-bearing species: SO, $^{33}$SO, SO$_2$, $^{34}$SO$_2$, CS, C$^{33}$S, OCS, H$_2$CS; N-bearing species: HNCO, HC$_3$N, HC$^{15}$N, H$^{13}$CN; three deuterated molecules: HDO, HDCO, and HDS, as well as SiO, H$^{13}$CO$^{+}$, and CH$_2$CO.  We detected COMs CH$_3$OH, CH$_3$CN, CH$_3$OCH$_3$, and tentatively detected NH$_2$CHO (a 3.2$\sigma$ detection of the strongest transition in the observed frequency range). If the presence of NH$_2$CHO in the LMC is confirmed, it will constitute the first detection of this astrobiologically relevant molecule in the extragalactic sub-solar environment, providing us an insight into the metal-poor systems from the earlier cosmological epochs. 

Methanol has been detected toward all the sources and shows both the extended and compact emission, while other COMs are mainly associated with the 1.2 mm continuum sources N\,105--2\,A and 2\,B. Based on the spectral line modeling, we estimated rotational temperatures and column densities, as well as the fractional molecular abundances for all the continuum sources.  The physical and chemical properties of 2\,A and 2\,B indicate that these sources are bone fide hot cores.  We also identified sources 2\,C, 2\,F, 3\,B, and 1\,A as hot core candidates; they have high temperatures in at least one of the four species with multiple transitions. 

We compared the fractional molecular abundances of hot cores 2\,A and 2\,B to those observed toward other known LMC hot cores  and toward representative Galactic hot cores. We concluded that hot cores 2\,A and 2\,B are ``organic-rich'' as defined by \citet{shimonishi2020} because they are associated with COMs more complex than six atoms and the observed molecular abundances roughly scale with metallicity. Chemically distinct hot core types observed in the LMC may be explained if there are metallicity inhomogeneities in the LMC disk. Such metallicity inhomogeneities may be the result of the tidal interactions between the LMC and the SMC. 

We report the detection of the CO$_2$ ice band at 15.2 $\mu$m in the {\it Spitzer}/IRS spectrum of the mid-IR source likely associated with both 2\,A and 2\,B, which was missed in previous studies, indicating that the source is at the early stage of the protostellar evolution, consistent with our ALMA observations. The near-IR {\it VLT}/KMOS spectroscopic observations provided us an insight into the nature of three near-IR sources in N\,105, two found to be embedded and associated with continuum sources 1\,A and 2\,E.  The third KMOS source located nearby 2\,B is unlikely to be associated with this ALMA source;  however, the KMOS data show evidence for the presence of the extended shocked emission in the region, consistent with the ALMA data. 

Our observations highlight the need for higher sensitivity observations that would allow for a more reliable detection of larger molecules (higher signal-to-noise and multiple transitions). 

Finally, our study further confirms that larger COMs can be formed in low-metallicity galaxies, thus a possibility of the emergence of life as it happened on Earth is open in these systems.

\acknowledgements 
We thank the anonymous referee for insightful comments that helped us improve the manuscript. The material is based upon work supported by NASA under award number 80GSFC21M0002 (M. S.). A. S. M. carried out this research within the Collaborative Research Centre 956 (subproject A6), funded by the Deutsche Forschungsgemeinschaft (DFG) - project ID 184018867. A. K. acknowledges support from the First TEAM grant of the Foundation for Polish Science No. POIR.04.04.00-00-5D21/18-00. This article has been supported by the Polish National Agency for Academic Exchange under Grant No. PPI/APM/2018/1/00036/U/001 (A. K.).  We thank Dr. J{\"u}rgen Ott for making an unpublished list of water masers in the LMC available to us.  The National Radio Astronomy Observatory is a facility of the National Science Foundation operated under cooperative agreement by Associated Universities, Inc. This paper makes use of the following ALMA data: ADS/JAO.ALMA\#2019.1.01720.S and \#2017.1.00093.S.  ALMA is a partnership of ESO (representing its member states), NSF (USA) and NINS (Japan), together with NRC (Canada), NSC and ASIAA (Taiwan), and KASI (Republic of Korea), in cooperation with the Republic of Chile. The Joint ALMA Observatory is operated by ESO, AUI/NRAO and NAOJ.  The SAGE and HERITAGE datasets are made available by the Infrared Science Archive (IRSA) at IPAC, which is operated by the California Institute of Technology under contract with the National Aeronautics and Space Administration.  This research made use of APLpy, an open-source plotting package for Python \citep{robitaille2012}. 

\facilities{ALMA, Spitzer, Herschel, VLT:Antu (KMOS)}

\clearpage

\bibliographystyle{aasjournal}
\bibliography{refs.bib}

\clearpage

\appendix 
\counterwithin{figure}{section}

\section{The Results of Spectral Modeling and Fractional Abundances for Continuum Sources in N\,105}
\label{s:fittingresultstab}

Table~\ref{t:tempdens} provides the results of the LTE spectral modeling described in Section~\ref{s:modeling} for all the ALMA 1.2 mm continuum sources in N\,105--1, N\,105--2, and N\,105--3. It also includes fractional abundances with respect to both H$_2$ and CH$_3$OH. Table~\ref{t:SO2fits} provides the results of the additional XCLASS LTE  modeling of SO$_2$ for hot cores N\,105--2\,A and 2\,B; the XCLASS analysis is described in Section~\ref{s:xclass}. 

\startlongtable
\begin{deluxetable*}{llcrccll}
\centering
\tablecaption{The Results of Spectral Modeling and Fractional Abundances with Respect to H$_2$ and CH$_3$OH \label{t:tempdens}}
\tablewidth{0pt}
\tablehead{
\multicolumn{1}{c}{Source} &
\multicolumn{1}{c}{Species, ${\rm X}$} &
\colhead{$T_{\rm rot} ({\rm X})$} &
\colhead{$N({\rm X})$}&
\colhead{$v_{\rm LSR}$} &    
\colhead{$\Delta v_{\rm FWHM}$} &
\colhead{$N(\rm X)/N({\rm H_2})$}  &
\colhead{$N(\rm X)/N({\rm CH_{3}OH})$} \\
\colhead{} &
\colhead{} &
\colhead{(K)} &
\colhead{(cm$^{-2}$)} &
\colhead{(km s$^{-1}$)} &
\colhead{(km s$^{-1}$)} &
\colhead{} & 
\colhead{}
}
\startdata
N\,105--1\,A & CH$_3$OH & 12.1$^{+1.0}_{-0.6}$ & (2.2$^{+0.2}_{-0.2})\times10^{14}$ & 236.5$^{+0.1}_{-0.1}$ & 3.2$^{+0.3}_{-0.3}$ & (9.4$^{+1.1}_{-1.2})\times10^{-10}$ & \nodata \\   
                   & H$^{13}$CO$^{+}$ & \ditto & (5.6$^{+0.2}_{-0.2})\times10^{12}$ & 237.2$^{+0.1}_{-0.1}$ & 4.5$^{+0.2}_{-0.2}$ & (2.4$^{+0.2}_{-0.2})\times10^{-11}$ & (2.6$^{+0.2}_{-0.2})\times10^{-2}$ \\ 
                   & H$^{13}$CN           & \ditto  & (1.3$^{+0.2}_{-0.2})\times10^{12}$ & 234.2$^{+0.3}_{-0.3}$ & 2.5$^{+0.7}_{-0.6}$ & (5.6$^{+1.1}_{-1.1})\times10^{-12}$ & (6.0$^{+1.2}_{-1.1})\times10^{-3}$ \\ 
                   & CS           & \ditto & (1.90$^{+0.03}_{-0.03})\times10^{14}$ & 236.72$^{+0.02}_{-0.02}$ & 3.03$^{+0.04}_{-0.04}$ & (8.1$^{+0.8}_{-0.8})\times10^{-10}$ & (8.6$^{+0.7}_{-0.7})\times10^{-1}$ \\ 
                   & H$_2$CS & \ditto & $<$$3.9\times10^{13}$ & 237.2$^{+0.5}_{-0.5}$ & 2.2$^{+1.0}_{-0.7}$ & $<$1.7$\times$10$^{-10}$ & $<$1.8$\times10^{-1}$ \\ 
                   & SO$_2$ & 95.5$^{+19.9}_{-20.3}$ & (1.6$^{+0.3}_{-0.2})\times10^{14}$ & 237.7$^{+0.2}_{-0.2}$ & 4.9$^{+0.5}_{-0.4}$ & (6.9$^{+1.5}_{-1.2})\times10^{-10}$ & (7.4$^{+1.5}_{-1.3})\times10^{-1}$ \\ 
                   & SO         & \ditto & (3.5$^{+0.4}_{-0.3})\times10^{14}$ & 237.46$^{+0.04}_{-0.04}$ & 4.5$^{+0.1}_{-0.1}$ & (1.5$^{+0.2}_{-0.2})\times10^{-9}$ & 1.6$^{+0.2}_{-0.2}$ \\ 
 N\,105--1\,B & CH$_3$OH & 16.1$^{+0.3}_{-0.3}$ & (5.0$^{+0.2}_{-0.2})\times10^{14}$ & 242.00$^{+0.03}_{-0.04}$ & 2.6$^{+0.1}_{-0.1}$ & (5.9$^{+0.7}_{-0.7})\times10^{-10}$ & \nodata \\ 
                   & H$^{13}$CO$^{+}$ &  \ditto & (3.7$^{+0.2}_{-0.2})\times10^{12}$ & 242.3$^{+0.1}_{-0.1}$ & 2.8$^{+0.2}_{-0.1}$ & (4.3$^{+0.5}_{-0.5})\times10^{-12}$ & (7.3$^{+0.4}_{-0.4})\times10^{-3}$ \\ 
                   & H$^{13}$CN &  \ditto & (2.0$^{+0.3}_{-0.2})\times10^{12}$ & 242.2$^{+0.1}_{-0.1}$ & 2.3$^{+0.3}_{-0.3}$ & (2.3$^{+0.4}_{-0.4})\times10^{-12}$ & (4.0$^{+0.5}_{-0.5})\times10^{-3}$ \\ 
                   & HC$^{15}$N &  \ditto & (8.2$^{+1.7}_{-1.7})\times10^{11}$ & 242.2\tablenotemark{\footnotesize a} & 1.9$^{+0.6}_{-0.4}$ & (9.5$^{+2.3}_{-2.2})\times10^{-13}$ & (1.6$^{+0.3}_{-0.3})\times10^{-3}$ \\ 
                   & CS &  \ditto & (2.5$^{+0.1}_{-0.1})\times10^{14}$ & 242.18$^{+0.01}_{-0.01}$ & 2.43$^{+0.04}_{-0.03}$ & (2.9$^{+0.4}_{-0.4})\times10^{-10}$ & (4.9$^{+0.3}_{-0.3})\times10^{-1}$ \\ 
                   & C$^{33}$S &  \ditto & (8.3$^{+1.6}_{-1.3})\times10^{12}$ & 242.9$^{+0.4}_{-0.4}$ & 4.6$^{+0.8}_{-0.9}$ & (9.6$^{+2.1}_{-1.9})\times10^{-12}$ & (1.7$^{+0.3}_{-0.3})\times10^{-2}$ \\ 
                   & HNCO &  \ditto & (5.8$^{+1.9}_{-1.5})\times10^{13}$ & 242.1\tablenotemark{\footnotesize a} & 3.0$^{+1.3}_{-0.9}$ & (6.7$^{+2.3}_{-1.9})\times10^{-11}$ & ($1.2^{+0.4}_{-0.3})\times10^{-1}$ \\ 
                   & H$_{2}$CS &  \ditto & (7.5$^{+0.9}_{-0.9})\times10^{13}$ & 242.4$^{+0.1}_{-0.1}$ & 2.3$^{+0.3}_{-0.2}$ & (8.8$^{+1.4}_{-1.4})\times10^{-11}$ & (1.5$^{+0.2}_{-0.2})\times10^{-1}$ \\ 
                   & SiO &  \ditto & $<$2.4 $\times 10^{12}$ & 242.9$^{+0.5}_{-0.5}$ & 2.9$^{+1.4}_{-1.0}$ & $<$2.8$\times$10$^{-12}$ & $<$4.8$\times10^{-3}$ \\ 
                   & SO$_2$ & 24.9$^{+3.7}_{-3.5}$ & (4.8$^{+0.8}_{-0.7})\times10^{13}$ & 242.9$^{+0.2}_{-0.2}$ & 2.1$^{+0.4}_{-0.3}$ & (5.6$^{+1.1}_{-1.1})\times10^{-11}$ & (9.6$^{+1.5}_{-1.5})\times10^{-2}$ \\ 
                   & SO &  \ditto & (3.8$^{+0.9}_{-0.5})\times10^{14}$ & 242.09$^{+0.02}_{-0.02}$ & 2.62$^{+0.05}_{-0.05}$ & (4.4$^{+1.2}_{-0.8})\times10^{-10}$ & (7.6$^{+1.8}_{-1.1})\times10^{-1}$ \\  
N\,105--1\,C & CH$_3$OH           & 17.3$^{+2.8}_{-2.5}$ & (1.6$^{+0.2}_{-0.2})\times10^{14}$ & 239.3$^{+0.1}_{-0.1}$ & 2.1$^{+0.3}_{-0.2}$ & (3.3$^{+0.9}_{-0.9})\times10^{-10}$ & \nodata \\ 
                   & H$^{13}$CO$^{+}$ &  \ditto & (1.2$^{+0.2}_{-0.1})\times10^{12}$ & 239.2$^{+0.1}_{-0.1}$ & 1.6$^{+0.3}_{-0.2}$ & (2.4$^{+0.7}_{-0.6})\times10^{-12}$ & (7.2$^{+1.6}_{-1.3})\times10^{-3}$ \\            
                   & H$^{13}$CN           &  \ditto & (1.8$^{+0.4}_{-0.3})\times10^{12}$ & 239.6$^{+0.2}_{-0.2}$ & 2.1$^{+0.5}_{-0.4}$ & (3.6$^{+1.2}_{-1.1})\times10^{-12}$ & (1.1$^{+0.3}_{-0.2})\times10^{-2}$ \\                 
                   & HC$^{15}$N           &  \ditto & (1.9$^{+0.6}_{-0.4})\times10^{12}$ & 239.4$^{+0.5}_{-0.5}$ & 5.1$^{+1.2}_{-1.3}$ & (3.9$^{+1.5}_{-1.2})\times10^{-12}$ & (1.2$^{+0.4}_{-0.3})\times10^{-2}$ \\                         
                   & CS           &  \ditto & (6.9$^{+2.1}_{-1.1})\times10^{13}$ & 239.41$^{+0.02}_{-0.03}$ & 2.2$^{+0.1}_{-0.1}$ & (1.4$^{+0.5}_{-0.4})\times10^{-10}$ & (4.2$^{+1.4}_{-0.9})\times10^{-1}$ \\ 
                   & C$^{33}$S   &  \ditto & $<$3.1 $\times 10^{12}$ & 239.6$^{+0.3}_{-0.3}$ & 1.4$^{+0.9}_{-0.5}$ & $<$6.2$\times$10$^{-12}$ & $<$1.9$\times10^{-2}$ \\ 
                   & SO         &  \ditto & (1.4$^{+0.6}_{-0.4})\times10^{14}$ & 239.3$^{+0.1}_{-0.1}$ & 2.1$^{+0.2}_{-0.2}$ & (2.9$^{+1.4}_{-1.0})\times10^{-10}$ & (8.8$^{+3.7}_{-2.6})\times10^{-1}$ \\                  
                   & H$_2$CS &  \ditto & (3.6$^{+1.5}_{-1.2})\times10^{13}$ & 240.1$^{+0.3}_{-0.3}$ & 2.1$^{+0.6}_{-0.6}$ & (7.2$^{+3.5}_{-2.9})\times10^{-11}$ & (2.2$^{+0.9}_{-0.8})\times10^{-1}$ \\ 
N\,105--2\,A  &  CH$_3$CN & 152.3$^{+9.7}_{-10.5}$ & (4.2$^{+0.1}_{-0.1})\times10^{13}$ & 243.0$^{+0.1}_{-0.1}$ & 4.2$^{+0.1}_{-0.1}$ &  (2.3$^{+0.3}_{-0.3})\times10^{-10}$ & (11.8$^{+0.4}_{-0.4})\times10^{-3}$ \\   
                    & H$^{13}$CO$^{+}$ & \ditto & (2.14$^{+0.03}_{-0.03})\times10^{13}$ & 241.90$^{+0.02}_{-0.02}$ & 3.13$^{+0.04}_{-0.04}$ &  (1.2$^{+0.1}_{-0.1})\times10^{-10}$ & (6.1$^{+0.1}_{-0.1})\times10^{-3}$ \\ 
                    & H$^{13}$CN           &  \ditto & (2.1$^{+0.1}_{-0.1})\times10^{13}$ & 242.4$^{+0.1}_{-0.1}$ & 5.5$^{+0.2}_{-0.2}$ &  (1.2$^{+0.1}_{-0.1})\times10^{-10}$ & (6.1$^{+0.2}_{-0.2})\times10^{-3}$ \\ 
                    & HC$^{15}$N           &  \ditto & (1.2$^{+0.1}_{-0.1})\times10^{13}$ & 242.4$^{+0.1}_{-0.1}$ & 4.6$^{+0.2}_{-0.2}$ &  (6.3$^{+0.8}_{-0.8})\times10^{-11}$ & (3.3$^{+0.2}_{-0.2})\times10^{-3}$ \\                         
                    & HC$_{3}$N             &  \ditto & (9.1$^{+0.8}_{-0.9})\times10^{12}$ & 242.5$^{+0.2}_{-0.2}$ & 4.3$^{+0.5}_{-0.5}$ &  (5.0$^{+0.7}_{-0.8})\times10^{-11}$ & (2.6$^{+0.2}_{-0.3})\times10^{-3}$ \\                         
                    & CS            &  \ditto & (4.21$^{+0.01}_{-0.01})\times10^{14}$ & 241.55$^{+0.01}_{-0.01}$ & 3.88$^{+0.01}_{-0.01}$ &  (2.3$^{+0.3}_{-0.3})\times10^{-9}$ & (12.0$^{+0.2}_{-0.2})\times10^{-2}$ \\ 
                    & C$^{33}$S  &  \ditto & (1.5$^{+0.1}_{-0.1})\times10^{13}$ & 241.2$^{+0.1}_{-0.1}$ & 3.0$^{+0.3}_{-0.3}$ &  (8.0$^{+1.2}_{-1.2})\times10^{-11}$ & (4.2$^{+0.4}_{-0.3})\times10^{-3}$ \\ 
                    & H$_2$CS &  \ditto & (2.8$^{+0.1}_{-0.1})\times10^{14}$ & 241.9$^{+0.1}_{-0.1}$ & 3.6$^{+0.1}_{-0.1}$ &  (1.6$^{+0.2}_{-0.2})\times10^{-9}$ & (8.1$^{+0.3}_{-0.3})\times10^{-2}$ \\ 
                    & OCS         &  \ditto & (3.4$^{+0.2}_{-0.2})\times10^{14}$ & 242.9$^{+0.1}_{-0.1}$ & 4.0$^{+0.3}_{-0.3}$ &  (1.8$^{+0.2}_{-0.2})\times10^{-9}$ & (9.5$^{+0.5}_{-0.6})\times10^{-2}$ \\ 
                    & SiO             &  \ditto & (1.6$^{+0.1}_{-0.1})\times10^{13}$ & 242.5$^{+0.1}_{-0.1}$ & 5.5$^{+0.3}_{-0.2}$ &  (8.8$^{+1.1}_{-1.1})\times10^{-11}$ & (4.6$^{+0.2}_{-0.2})\times10^{-3}$ \\                  
                    & NH$_{2}$CHO  &  \ditto & (3.0$^{+0.8}_{-0.9})\times10^{13}$ & 243.3$^{+0.5}_{-0.6}$ & 4.0$^{+1.3}_{-1.0}$ & (1.7$^{+0.5}_{-0.6})\times10^{-10}$ & (8.6$^{+2.3}_{-2.7})\times10^{-3}$ \\    
                    & HNCO  &  \ditto & (9.9$^{+0.6}_{-0.6})\times10^{13}$ & 243.8\tablenotemark{\footnotesize a} & 4.2\tablenotemark{\footnotesize a} &  (5.4$^{+0.7}_{-0.7})\times10^{-10}$ & (2.8$^{+0.2}_{-0.2})\times10^{-2}$ \\                
                    & HDCO         &  \ditto & (2.8$^{+0.7}_{-0.6})\times10^{13}$ & 242.2$^{+0.4}_{-0.4}$ & 3.4$^{+1.0}_{-0.8}$ &  (1.5$^{+0.4}_{-0.4})\times10^{-10}$ & (8.0$^{+1.9}_{-1.8})\times10^{-3}$ \\        
                    & CH$_3$OCH$_3$  &  \ditto & (3.9$^{+0.5}_{-0.5})\times10^{14}$ & 242.9$^{+0.3}_{-0.2}$ & 3.4$^{+0.8}_{-0.5}$ &  (2.1$^{+0.4}_{-0.4})\times10^{-9}$ & (1.1$^{+0.2}_{-0.1})\times10^{-1}$ \\         
                    & CH$_2$CO  &  \ditto & (4.8$^{+0.9}_{-0.9})\times10^{13}$ & 242.7$^{+0.3}_{-0.3}$ & 3.4$^{+0.7}_{-0.6}$ &  (2.6$^{+0.6}_{-0.6})\times10^{-10}$ & (1.4$^{+0.2}_{-0.3})\times10^{-2}$ \\      
                    & HDO  &  \ditto & (4.9$^{+0.5}_{-0.4})\times10^{14}$ & 242.7$^{+0.2}_{-0.2}$ & 4.2$^{+0.4}_{-0.4}$ &  (2.7$^{+0.4}_{-0.4})\times10^{-9}$ & (1.4$^{+0.2}_{-0.1})\times10^{-1}$ \\     
                    & CH$_3$OH (hot) & 170.9$^{+4.6}_{-4.2}$ & (3.5$^{+0.1}_{-0.1})\times10^{15}$ & 242.83$^{+0.04}_{-0.05}$ & 4.3$^{+0.1}_{-0.1}$ &  (1.9$^{+0.2}_{-0.2})\times10^{-8}$ & \nodata \\ 
                    & CH$_3$OH (cold) & 14.2$^{+1.3}_{-1.0}$ & (3.9$^{+0.2}_{-0.2})\times10^{14}$ & 241.3$^{+0.1}_{-0.1}$ & 3.1$^{+0.2}_{-0.1}$ &  (2.1$^{+0.3}_{-0.3})\times10^{-9}$ & (11.2$^{+0.6}_{-0.6})\times10^{-2}$ \\ 
                    & SO$_2$ (hot) & 176.0$^{+5.6}_{-4.9}$ & (1.38$^{+0.04}_{-0.04})\times10^{15}$ & 242.82$^{+0.05}_{-0.04}$ & 4.7$^{+0.1}_{-0.1}$ &  (7.5$^{+0.9}_{-1.0})\times10^{-9}$ & (3.9$^{+0.1}_{-0.1})\times10^{-1}$ \\ 
                    & SO              &  \ditto & (1.37$^{+0.03}_{-0.02})\times10^{15}$ & 242.07$^{+0.01}_{-0.01}$ & 4.24$^{+0.03}_{-0.03}$ &  (7.5$^{+0.9}_{-0.9})\times10^{-9}$ &(3.9$^{+0.1}_{-0.1})\times10^{-1}$  \\                
                    & $^{33}$SO  &  \ditto & (8.2$^{+0.8}_{-0.7})\times10^{13}$ & 243.3$^{+0.2}_{-0.2}$ & 3.2$^{+0.6}_{-0.4}$ &  (4.5$^{+0.7}_{-0.7})\times10^{-10}$ & (2.3$^{+0.2}_{-0.2})\times10^{-2}$ \\ 
                    & SO$_2$ (cold) & 24.2$^{+1.5}_{-1.7}$ & (3.1$^{+0.2}_{-0.2})\times10^{14}$ & 242.4$^{+0.1}_{-0.1}$ & 3.8$^{+0.1}_{-0.1}$ &  (1.7$^{+0.2}_{-0.2})\times10^{-9}$ &(8.8$^{+0.5}_{-0.5})\times10^{-2}$  \\ 
                    & $^{34}$SO$_2$ & 114.8$^{+36.0}_{-21.5}$ & (1.1$^{+0.2}_{-0.2})\times10^{14}$ & 243.1$^{+0.1}_{-0.1}$ & 2.6$^{+0.3}_{-0.3}$ &  (6.1$^{+1.4}_{-1.2})\times10^{-10}$ & (3.2$^{+0.6}_{-0.5})\times10^{-2}$ \\ 
N\,105--2\,B  &  CH$_3$CN & 88.2$^{+10.1}_{-9.4}$ & (2.2$^{+0.1}_{-0.1})\times10^{13}$ & 243.2$^{+0.2}_{-0.2}$ & 6.7$^{+0.5}_{-0.4}$ & (6.9$^{+1.2}_{-1.1})\times10^{-11}$ & (10.5$^{+0.9}_{-0.8})\times10^{-3}$ \\ 
                   & H$^{13}$CO$^{+}$ & \ditto & (1.96$^{+0.02}_{-0.02})\times10^{13}$ & 243.32$^{+0.02}_{-0.02}$ & 3.49$^{+0.04}_{-0.04}$ & (6.3$^{+1.0}_{-1.0})\times10^{-11}$ & (9.5$^{+0.7}_{-0.7})\times10^{-3}$ \\ 
                    & H$^{13}$CN           &  \ditto & (1.59$^{+0.04}_{-0.04})\times10^{13}$ & 243.3$^{+0.1}_{-0.1}$ & 7.1$^{+0.3}_{-0.2}$ & (5.1$^{+0.8}_{-0.8})\times10^{-11}$ & (7.7$^{+0.6}_{-0.6})\times10^{-3}$ \\ 
                    & HC$^{15}$N           &  \ditto & (8.9$^{+0.4}_{-0.4})\times10^{12}$ & 243.3$^{+0.1}_{-0.1}$ & 5.9$^{+0.3}_{-0.2}$ & (2.9$^{+0.5}_{-0.4})\times10^{-11}$ & (4.3$^{+0.3}_{-0.3})\times10^{-3}$ \\                         
                    & HC$_{3}$N             &  \ditto & (8.5$^{+1.3}_{-1.0})\times10^{12}$ & 242.9$^{+0.4}_{-0.3}$ & 4.8$^{+0.8}_{-0.6}$ & (2.7$^{+0.6}_{-0.5})\times10^{-11}$ & (4.1$^{+0.7}_{-0.5})\times10^{-3}$ \\                         
                    & CS            &  \ditto & (4.48$^{+0.01}_{-0.01})\times10^{14}$ & 242.8\tablenotemark{\footnotesize a} & 4.09$^{+0.01}_{-0.01}$ & (1.4$^{+0.2}_{-0.2})\times10^{-9}$ & (2.2$^{+0.2}_{-0.1})\times10^{-1}$ \\ 
                    & C$^{33}$S  &  \ditto & (1.2$^{+0.1}_{-0.1})\times10^{13}$ & 242.8$^{+0.1}_{-0.1}$ & 3.3$^{+0.3}_{-0.3}$ & (3.7$^{+0.7}_{-0.6})\times10^{-11}$ & (5.7$^{+0.7}_{-0.6})\times10^{-3}$ \\ 
                    & H$_2$CS &  \ditto & (1.7$^{+0.1}_{-0.1})\times10^{14}$ & 242.8$^{+0.1}_{-0.1}$ & 3.6$^{+0.1}_{-0.1}$ & (5.4$^{+0.9}_{-0.8})\times10^{-10}$ & (8.1$^{+0.6}_{-0.6})\times10^{-2}$ \\ 
                    & OCS         &  \ditto & (4.4$^{+0.2}_{-0.3})\times10^{14}$ & 243.0$^{+0.1}_{-0.2}$ & 5.8$^{+0.4}_{-0.4}$ & (1.4$^{+0.2}_{-0.2})\times10^{-9}$ & (2.1$^{+0.2}_{-0.2})\times10^{-1}$ \\ 
                    & SiO             &  \ditto & (6.4$^{+0.1}_{-0.1})\times10^{13}$ & 244.9$^{+0.1}_{-0.1}$ & 11.1$^{+0.1}_{-0.1}$ & (2.0$^{+0.3}_{-0.3})\times10^{-10}$ & (3.1$^{+0.2}_{-0.2})\times10^{-2}$ \\                  
                    & HNCO  &  \ditto & (6.8$^{+0.4}_{-0.5})\times10^{13}$ & 242.9\tablenotemark{\footnotesize a} & 4.9\tablenotemark{\footnotesize a} & (2.2$^{+0.4}_{-0.4})\times10^{-10}$ & (3.3$^{+0.3}_{-0.3})\times10^{-2}$ \\                
                    & HDO  &  \ditto & (2.6$^{+0.5}_{-0.5})\times10^{14}$ &245.8$^{+0.8}_{-0.7}$ & 8.2$^{+2.3}_{-1.8}$ & (8.2$^{+2.1}_{-2.0})\times10^{-10}$ & (1.2$^{+0.3}_{-0.3})\times10^{-1}$ \\     
                    & CH$_3$OCH$_3$  &  \ditto & (2.0$^{+0.4}_{-0.4})\times10^{14}$ & 243.1$^{+0.6}_{-0.5}$ & 4.9$^{+1.3}_{-1.1}$ & (6.5$^{+1.6}_{-1.5})\times10^{-10}$ & (9.8$^{+1.9}_{-1.9})\times10^{-2}$ \\       
                    & CH$_3$OH (hot) & 136.1$^{+8.2}_{-7.6}$ & (2.1$^{+0.1}_{-0.1})\times10^{15}$ & 243.1$^{+0.1}_{-0.1}$ & 6.2$^{+0.3}_{-0.2}$ & (6.6$^{+1.1}_{-1.1})\times10^{-9}$ & \nodata \\ 
                    & CH$_3$OH (cold) & 15.6$^{+1.1}_{-1.0}$ & (7.1$^{+0.3}_{-0.3})\times10^{14}$ & 242.9$^{+0.1}_{-0.1}$ & 3.7$^{+0.1}_{-0.1}$ & (2.3$^{+0.4}_{-0.4})\times10^{-9}$ & (3.4$^{+0.3}_{-0.3})\times10^{-1}$ \\ 
                    &  SO$_2$ (hot) & 161.5$^{+2.4}_{-2.4}$ & (4.5$^{+0.1}_{-0.1})\times10^{15}$ & 245.2$^{+0.1}_{-0.1}$ & 11.6$^{+0.1}_{-0.1}$ & (1.4$^{+0.2}_{-0.2})\times10^{-8}$ &2.2$^{+0.2}_{-0.2}$  \\ 
                    & SO            &  \ditto & (4.46$^{+0.04}_{-0.05})\times10^{15}$ & 243.84$^{+0.01}_{-0.01}$ & 7.96$^{+0.02}_{-0.02}$ & (1.4$^{+0.2}_{-0.2})\times10^{-8}$ & 2.2$^{+0.2}_{-0.1}$ \\           
                     & $^{33}$SO  &  \ditto & (2.7$^{+0.1}_{-0.1})\times10^{14}$ & 244.4$^{+0.2}_{-0.2}$ & 8.5$^{+0.4}_{-0.4}$ & (8.5$^{+1.4}_{-1.3})\times10^{-10}$ & (1.3$^{+0.1}_{-0.1})\times10^{-1}$ \\        
                    &  SO$_2$ (cold) & 28.2$^{+0.9}_{-0.8}$ & (1.08$^{+0.04}_{-0.03})\times10^{15}$ & 243.9$^{+0.1}_{-0.1}$ & 7.7$^{+0.1}_{-0.1}$ & (3.5$^{+0.6}_{-0.5})\times10^{-9}$ & (5.2$^{+0.4}_{-0.4})\times10^{-1}$ \\ 
                    & $^{34}$SO$_2$ & 113.6$^{+12.2}_{-12.2}$ & (3.7$^{+0.4}_{-0.3})\times10^{14}$ & 245.2$^{+0.2}_{-0.2}$ & 9.3$^{+0.5}_{-0.5}$ & (1.2$^{+0.2}_{-0.2})\times10^{-9}$ & (1.8$^{+0.2}_{-0.2})\times10^{-1}$ \\ 
 N\,105--2\,C & CH$_3$OH (hot) & 95.3$^{+28.3}_{-20.4}$ & (6.7$^{+2.4}_{-2.1})\times10^{13}$ & 242.5$^{+0.1}_{-0.1}$ & 1.0$^{+0.2}_{-0.2}$  & (8.3$^{+4.1}_{-3.3})\times10^{-10}$ &\nodata  \\ 
                    & CH$_3$CN & \ditto & (4.3$^{+1.0}_{-0.7})\times10^{12}$ & 242.8\tablenotemark{\footnotesize a} & 3.9$^{+1.0}_{-0.7}$  & (5.3$^{+2.2}_{-1.6})\times10^{-11}$ & (6.4$^{+2.7}_{-2.2})\times10^{-2}$ \\    
                    & H$^{13}$CO$^{+}$ & \ditto & (3.3$^{+0.7}_{-0.5})\times10^{12}$ & 242.3$^{+0.1}_{-0.1}$  & 3.1$^{+0.2}_{-0.2}$  & (4.2$^{+1.7}_{-1.2})\times10^{-11}$ & (5.0$^{+2.1}_{-1.7})\times10^{-2}$ \\ 
                    & H$^{13}$CN           & \ditto  & (3.8$^{+0.9}_{-0.7})\times10^{12}$ & 243.1$^{+0.3}_{-0.2}$  & 4.6$^{+0.5}_{-0.5}$  & (4.8$^{+1.9}_{-1.5})\times10^{-11}$ & (5.7$^{+2.4}_{-2.1})\times10^{-2}$ \\ 
                    & HC$^{15}$N           & \ditto  & (3.0$^{+0.7}_{-0.6})\times10^{12}$ & 241.8$^{+0.3}_{-0.3}$  & 5.5$^{+0.8}_{-0.7}$  & (3.8$^{+1.5}_{-1.2})\times10^{-11}$ & (4.5$^{+1.9}_{-1.7})\times10^{-2}$ \\                        
                    & CS            & \ditto  & (3.3$^{+0.6}_{-0.4})\times10^{14}$ & 242.47$^{+0.01}_{-0.01}$  & 4.52$^{+0.03}_{-0.04}$  & (4.1$^{+1.5}_{-1.1})\times10^{-9}$ & 4.9$^{+1.9}_{-1.6}$ \\ 
                    & C$^{33}$S  & \ditto  & (1.3$^{+0.3}_{-0.2})\times10^{13}$ & 242.3$^{+0.1}_{-0.1}$  & 2.6$^{+0.2}_{-0.2}$  & (1.7$^{+0.7}_{-0.5})\times10^{-10}$ & (2.0$^{+0.8}_{-0.7})\times10^{-1}$ \\ 
                    & H$_{2}$CS  & \ditto  & (1.4$^{+0.4}_{-0.2})\times10^{14}$ & 242.33$^{+0.03}_{-0.05}$  & 2.8$^{+0.1}_{-0.1}$  & (1.7$^{+0.7}_{-0.5})\times10^{-9}$ & 2.0$^{+0.9}_{-0.7}$ \\ 
                    & OCS  & \ditto & (9.2$^{+2.1}_{-1.9})\times10^{13}$ & 243.2$^{+0.3}_{-0.3}$ & 3.0$^{+0.8}_{-0.6}$  & (1.1$^{+0.5}_{-0.4})\times10^{-9}$ & 1.4$^{+0.6}_{-0.5}$ \\ 
                    & SiO  & \ditto  & (8.5$^{+1.4}_{-1.0})\times10^{12}$ & 242.1$^{+0.2}_{-0.2}$  & 6.4$^{+0.4}_{-0.3}$  & (1.1$^{+0.4}_{-0.3})\times10^{-10}$ & (1.3$^{+0.5}_{-0.4})\times10^{-1}$ \\ 
                    & HNCO  & \ditto  & (1.5$^{+0.4}_{-0.4})\times10^{13}$ & 243.7\tablenotemark{\footnotesize a} & 2.6\tablenotemark{\footnotesize a} & (1.8$^{+0.8}_{-0.7})\times10^{-10}$ & (2.2$^{+0.9}_{-0.9})\times10^{-1}$ \\ 
                    &  CH$_3$OH (cold) & 10.4$^{+0.1}_{-0.1}$ & (4.2$^{+0.1}_{-0.1})\times10^{15}$ & 242.39$^{+0.02}_{-0.02}$  & 3.53$^{+0.03}_{-0.03}$  & (5.3$^{+1.8}_{-1.3})\times10^{-8}$ & 63.3$^{+22.6}_{-19.6}$ \\ 
                    & HDS  & \ditto & (4.4$^{+0.5}_{-0.6})\times10^{13}$ & 242.0$^{+0.1}_{-0.1}$  & 2.4$^{+0.4}_{-0.3}$  & (5.5$^{+1.9}_{-1.6})\times10^{-10}$ & (6.6$^{+2.5}_{-2.2})\times10^{-1}$ \\ 
                    & SO$_2$ & 32.0$^{+3.7}_{-3.2}$ & (8.8$^{+0.8}_{-0.8})\times10^{13}$ & 241.7$^{+0.3}_{-0.3}$  & 6.0$^{+0.7}_{-0.6}$  & (1.1$^{+0.4}_{-0.3})\times10^{-9}$ & 1.3$^{+0.5}_{-0.4}$ \\ 
                    & SO         & \ditto & (3.4$^{+0.3}_{-0.2})\times10^{14}$ & 242.24$^{+0.03}_{-0.02}$  & 3.7$^{+0.1}_{-0.1}$  & (4.2$^{+1.4}_{-1.1})\times10^{-9}$ & 5.0$^{+1.8}_{-1.6}$ \\ 
N\,105--2\,D & CH$_3$OH (cold) & 10.0$^{+0.1}_{-0.1}$ & (3.0$^{+0.1}_{-0.1})\times10^{15}$ & 243.38$^{+0.02}_{-0.01}$ & 2.31$^{+0.03}_{-0.03}$ & (1.5$^{+0.2}_{-0.2})\times10^{-8}$ & 71.1$^{+63.7}_{-23.5}$ \\ 
                    & CH$_3$OH (warm) & 31.3$^{+1.4}_{-1.1}$ & (4.1$^{+3.5}_{-1.3})\times10^{13}$ & 240.6$^{+2.4}_{-0.6}$ & 3.2$^{+6.1}_{-1.4}$ & (2.0$^{+1.7}_{-0.7})\times10^{-10}$ & \nodata \\ 
                    & H$^{13}$CO$^{+}$ & \ditto & (1.3$^{+0.1}_{-0.1})\times10^{12}$ & 243.7$^{+0.1}_{-0.2}$ & 3.5$^{+0.3}_{-0.3}$ & (6.1$^{+0.9}_{-0.9})\times10^{-12}$ & (3.1$^{+2.7}_{-1.0})\times10^{-2}$ \\ 
                    & H$^{13}$CN  & \ditto & (1.2$^{+0.2}_{-0.2})\times10^{12}$ & 242.9$^{+0.3}_{-0.2}$ & 3.3$^{+0.6}_{-0.6}$ & (5.7$^{+1.1}_{-1.1})\times10^{-12}$ & 2.9$^{+2.5}_{-1.0}$ \\ 
                    & HC$^{15}$N  & \ditto & (5.7$^{+1.3}_{-1.0})\times10^{11}$ & 243.3$^{+0.2}_{-0.2}$ & 2.0$^{+0.6}_{-0.4}$ & (2.8$^{+0.7}_{-0.6})\times10^{-12}$ & 1.4$^{+1.3}_{-0.5})\times10^{-2}$ \\ 
                    & CS  & \ditto & (2.4$^{+0.1}_{-0.1})\times10^{14}$ & 243.2\tablenotemark{\footnotesize a} & 2.96$^{+0.03}_{-0.03}$ & (1.2$^{+0.2}_{-0.1})\times10^{-9}$ & 6.0$^{+5.1}_{-1.9}$ \\ 
                    & C$^{33}$S  & \ditto & (5.2$^{+0.5}_{-0.6})\times10^{12}$ & 243.3$^{+0.2}_{-0.1}$ & 2.6$^{+0.3}_{-0.3}$ & (2.5$^{+0.4}_{-0.4})\times10^{-11}$ & (1.3$^{+1.1}_{-0.4})\times10^{-1}$ \\ 
                    & H$_2$CS  & \ditto & (6.5$^{+0.3}_{-0.3})\times10^{13}$ & 243.3$^{+0.1}_{-0.1}$ & 2.5$^{+0.1}_{-0.1}$ & (3.2$^{+0.4}_{-0.4})\times10^{-10}$ & 1.6$^{+1.4}_{-0.5}$ \\ 
                    & SO  & \ditto & (2.2$^{+0.1}_{-0.1})\times10^{14}$ & 243.40$^{+0.02}_{-0.02}$ & 2.8$^{+0.1}_{-0.1}$ & (1.1$^{+0.1}_{-0.1})\times10^{-9}$ & 5.5$^{+4.7}_{-1.7}$ \\ 
                    & SiO  & \ditto & (2.5$^{+0.4}_{-0.4})\times10^{12}$ & 244.1$^{+0.5}_{-0.5}$ & 7.0$^{+1.5}_{-1.1}$ & (1.2$^{+0.3}_{-0.2})\times10^{-11}$ & (6.2$^{+5.4}_{-2.2})\times10^{-2}$ \\ 
N\,105--2\,E & CH$_3$OH & 13.5$^{+1.5}_{-1.1}$ & (1.9$^{+0.3}_{-0.3})\times10^{14}$ & 241.4$^{+0.1}_{-0.1}$ & 2.1$^{+0.3}_{-0.3}$ & (3.1$^{+0.8}_{-0.7})\times10^{-10}$ & \nodata \\ 
                   & H$^{13}$CO$^{+}$ & \ditto & (1.5$^{+0.3}_{-0.2})\times10^{12}$ & 241.2$^{+0.1}_{-0.1}$ & 1.5$^{+0.3}_{-0.2}$ & (2.4$^{+0.7}_{-0.6})\times10^{-12}$ & (7.8$^{+2.0}_{-1.7})\times10^{-3}$ \\ 
                   & CS & \ditto & (1.3$^{+0.4}_{-0.3})\times10^{14}$ & 241.27$^{+0.02}_{-0.02}$ & 1.9$^{+0.1}_{-0.1}$ & (2.1$^{+0.7}_{-0.6})\times10^{-10}$ & (6.6$^{+2.2}_{-1.7})\times10^{-1}$ \\ 
                   & SO & \ditto & (3.7$^{+1.1}_{-1.0})\times10^{14}$ & 241.2$^{+0.1}_{-0.1}$ & 1.8$^{+0.2}_{-0.2}$ & (5.9$^{+2.1}_{-1.9})\times10^{-10}$ & 1.9$^{+0.7}_{-0.6}$ \\ 
N\,105--2 \,F & CH$_3$OH (hot) & 126.0$^{+33.3}_{-30.1}$ & (2.2$^{+1.0}_{-0.8})\times10^{14}$ & 242.8$^{+0.6}_{-0.6}$ & 2.8$^{+1.0}_{-1.0}$ & (1.8$^{+0.9}_{-0.8})\times10^{-9}$ & \nodata \\ 
                     & SO$_2$ (hot) & \ditto  & (1.5$^{+0.3}_{-0.4})\times10^{14}$ & 243.5$^{+0.7}_{-0.6}$ & 6.4$^{+1.7}_{-1.6}$ & (1.3$^{+0.5}_{-0.5})\times10^{-9}$ &(7.1$^{+3.5}_{-3.1})\times10^{-1}$ \\ 
                    & H$^{13}$CO$^{+}$ & \ditto  & (6.3$^{+1.5}_{-1.0})\times10^{12}$ & 242.0$^{+0.1}_{-0.1}$ & 2.8$^{+0.3}_{-0.2}$ & (5.4$^{+2.0}_{-1.7})\times10^{-11}$ & (2.9$^{+1.5}_{-1.2})\times10^{-2}$\\ 
                    & H$^{13}$CN & \ditto  & (4.5$^{+1.5}_{-1.1})\times10^{12}$ & 242.0$^{+0.2}_{-0.2}$ & 2.3$^{+0.5}_{-0.4}$ & (3.8$^{+1.7}_{-1.4})\times10^{-11}$ & (2.1$^{+1.1}_{-0.9})\times10^{-2}$\\ 
                    & HC$^{15}$N & \ditto & $<$1.4 $\times 10^{12}$ & 241.5$^{+0.7}_{-0.6}$ & 2.4$^{+1.9}_{-1.0}$ & $<$1.2$\times10^{-11}$ & (6.6$^{+4.9}_{-4.0})\times10^{-3}$\\ 
                    & CS & \ditto  & (4.9$^{+0.9}_{-0.7})\times10^{14}$ & 241.9$^{+0.1}_{-0.1}$ & 3.6$^{+0.1}_{-0.1}$ & (4.2$^{+1.4}_{-1.3})\times10^{-9}$ & 2.3$^{+1.1}_{-0.9}$\\  
                    & C$^{33}$S & \ditto  & (1.7$^{+0.4}_{-0.3})\times10^{13}$ & 241.9$^{+0.3}_{-0.3}$ & 4.2$^{+0.7}_{-0.7}$ & (1.5$^{+0.5}_{-0.5})\times10^{-10}$ & (7.9$^{+3.9}_{-3.4})\times10^{-2}$\\ 
                    & H$_{2}$CS & \ditto  & (2.3$^{+0.7}_{-0.5})\times10^{14}$ & 242.0$^{+0.1}_{-0.1}$ & 3.1$^{+0.2}_{-0.2}$ & (2.0$^{+0.9}_{-0.7})\times10^{-9}$ & 1.1$^{+0.6}_{-0.5}$\\ 
                    & SO & \ditto  & (4.7$^{+0.8}_{-0.6})\times10^{14}$ & 241.9$^{+0.1}_{-0.1}$ & 3.3$^{+0.1}_{-0.1}$ & (4.0$^{+1.4}_{-1.2})\times10^{-9}$ & 2.2$^{+1.0}_{-0.9}$\\ 
                    & SiO & \ditto  & (8.8$^{+2.2}_{-1.5})\times10^{12}$ & 242.5$^{+0.5}_{-0.6}$ & 7.6$^{+1.1}_{-1.2}$ & (7.5$^{+2.9}_{-2.4})\times10^{-11}$ & (4.1$^{+2.1}_{-1.7})\times10^{-2}$\\ 
                    & CH$_3$OH (cold) & 6.9$^{+0.2}_{-0.2}$ & (6.7$^{+1.1}_{-1.3})\times10^{15}$ & 241.8$^{+0.1}_{-0.1}$ & 2.4$^{+0.1}_{-0.1}$ & (5.7$^{+1.9}_{-1.9})\times10^{-8}$& 30.9$^{+14.7}_{-13.0}$\\ 
                    & SO$_2$ (cold) & \ditto  & (3.9$^{+0.8}_{-0.9})\times10^{14}$ & 241.9$^{+0.4}_{-0.4}$ & 3.6$^{+1.1}_{-0.8}$ & (3.3$^{+1.2}_{-1.2})\times10^{-9}$ & 1.8$^{+0.9}_{-0.8}$\\ 
                    & $^{34}$SO$_2$ & \ditto & (2.0$^{+0.5}_{-0.5})\times10^{15}$ & 244.0$^{+0.2}_{-0.3}$ & 1.7$^{+0.7}_{-0.5}$ & (1.7$^{+0.7}_{-0.6})\times10^{-8}$ & 9.4$^{+4.9}_{-4.2}$ \\ 
N\,105--3\,A & CH$_3$OH & 10.6$^{+1.3}_{-1.0}$ & (1.2$^{+0.3}_{-0.3})\times10^{14}$ & 238.4$^{+0.1}_{-0.1}$ & 2.0$^{+0.3}_{-0.3}$ & (2.9$^{+1.0}_{-0.9})\times10^{-10}$ & \nodata \\ 
                   & H$^{13}$CO$^{+}$ & \ditto & (1.7$^{+0.4}_{-0.3})\times10^{12}$ & 238.3$^{+0.1}_{-0.1}$ & 1.9$^{+0.2}_{-0.2}$ & (4.1$^{+1.3}_{-1.1})\times10^{-12}$ & (1.4$^{+0.4}_{-0.4})\times10^{-2}$ \\ 
                   & HC$^{15}$N & \ditto & (4.0$^{+1.7}_{-1.4})\times10^{11}$ & 238.5\tablenotemark{\footnotesize a}& 1.3\tablenotemark{\footnotesize a} & (9.5$^{+4.7}_{-3.9})\times10^{-13}$ & (3.3$^{+1.6}_{-1.4})\times10^{-3}$ \\ 
                   & CS &  \ditto & (9.7$^{+3.7}_{-2.6})\times10^{13}$ & 238.43$^{+0.02}_{-0.02}$ & 2.0$^{+0.1}_{-0.1}$ & (2.3$^{+1.0}_{-0.8})\times10^{-10}$ & (8.0$^{+3.6}_{-2.8})\times10^{-1}$ \\ 
                   & SO$_2$ & 30.5$^{+7.1}_{-7.0}$ & (2.0$^{+0.5}_{-0.4})\times10^{13}$ & 238.8$^{+0.2}_{-0.2}$ & 1.8$^{+0.6}_{-0.5}$ & (4.9$^{+1.6}_{-1.3})\times10^{-11}$ & (1.7$^{+0.6}_{-0.5})\times10^{-1}$ \\ 
                   & SO & \ditto & (5.0$^{+1.3}_{-0.7})\times10^{13}$ & 238.5$^{+0.1}_{-0.1}$ & 2.3$^{+0.2}_{-0.1}$ & (1.2$^{+0.4}_{-0.3})\times10^{-10}$ & (4.1$^{+1.5}_{-1.1})\times10^{-1}$ \\  
N\,105--3\,B & CH$_3$OH & 159.1$^{+41.3}_{-60.6}$ & (1.8$^{+0.9}_{-0.8})\times10^{14}$ & 238.1$^{+0.3}_{-0.2}$ & 2.3$^{+0.6}_{-0.5}$ & (1.9$^{+1.1}_{-1.2})\times10^{-8}$ & \nodata \\ 
                   & H$^{13}$CO$^{+}$ & \ditto & (1.0$^{+0.3}_{-0.4})\times10^{12}$ & 237.4$^{+0.2}_{-0.2}$ & 1.8$^{+0.5}_{-0.3}$ & (1.1$^{+0.5}_{-0.6})\times10^{-10}$& (5.9$^{+3.5}_{-3.3})\times10^{-3}$\\ 
                   & CS & \ditto & (3.1$^{+0.7}_{-0.9})\times10^{13}$ & 237.46$^{+0.02}_{-0.02}$ & 1.61$^{+0.05}_{-0.04}$ &  (3.3$^{+1.2}_{-1.7})\times10^{-9}$& (1.8$^{+0.9}_{-0.9})\times10^{-1}$\\ 
                   & SO & \ditto & (1.9$^{+0.7}_{-0.5})\times10^{13}$ & 237.8\tablenotemark{\footnotesize a} & 1.8$^{+0.5}_{-0.4}$ & (2.0$^{+0.9}_{-1.0})\times10^{-9}$ & (1.1$^{+0.6}_{-0.6})\times10^{-1}$ \\ 
N\,105--3\,C & CH$_3$OH & 10.8$^{+3.1}_{-2.1}$ & (1.7$^{+0.2}_{-0.2})\times10^{14}$ & 238.9$^{+0.2}_{-0.1}$ & 1.8$^{+0.3}_{-0.3}$ & (7.9$^{+4.0}_{-3.0})\times10^{-10}$& \nodata \\ 
                   & H$^{13}$CO$^{+}$ & \ditto & $<$1.6 $\times 10^{12}$ & 239.2$^{+0.5}_{-0.4}$ & 3.7$^{+1.5}_{-0.9}$ & $<$$7.2\times10^{-12}$& (9.1$^{+1.9}_{-2.2})\times10^{-3}$ \\ 
                   & CS & \ditto & (3.1$^{+0.2}_{-0.2})\times10^{13}$ & 239.0$^{+0.1}_{-0.1}$ & 1.6$^{+0.2}_{-0.2}$ & (1.4$^{+0.7}_{-0.5})\times10^{-10}$& (1.8$^{+0.2}_{-0.3})\times10^{-1}$ \\ 
                   & SO & \ditto & (5.0$^{+0.5}_{-0.6})\times10^{14}$ & 238.5$^{+0.2}_{-0.2}$ & 3.6$^{+0.6}_{-0.5}$ & (2.3$^{+1.1}_{-0.9})\times10^{-9}$& 2.8$^{+0.4}_{-0.5}$ \\
                   \hline
N\,113\,A1 & CH$_3$OH & 133.2$^{+6.6}_{-6.8}$ & (1.5$^{+0.1}_{-0.1})\times10^{16}$ & 239.4$^{+0.1}_{-0.1}$ & 4.9$^{+0.3}_{-0.2}$ & (5.7$^{+0.7}_{-0.7})\times10^{-8}$&  \nodata \\                  
                  & HCOOCH$_3$ & \ditto & (8.5$^{+0.8}_{-1.0})\times10^{14}$ & 239.4\tablenotemark{\footnotesize a}  & 4.9\tablenotemark{\footnotesize a} & (3.2$^{+0.5}_{-0.5})\times10^{-9}$& (5.6$^{+0.6}_{-0.7})\times10^{-2}$  \\      
                  & CH$_3$OCH$_3$ & \ditto & (9.7$^{+2.4}_{-2.1})\times10^{14}$ & 239.4\tablenotemark{\footnotesize a}  & 4.9\tablenotemark{\footnotesize a} & (3.6$^{+1.0}_{-0.9})\times10^{-9}$& (6.3$^{+1.6}_{-1.4})\times10^{-2}$ \\                                   
                  & $^{33}$SO & \ditto & (1.4$^{+0.2}_{-0.2})\times10^{14}$ & 239.4\tablenotemark{\footnotesize a} & 4.9\tablenotemark{\footnotesize a} & (5.2$^{+0.9}_{-1.0})\times10^{-10}$& (9.2$^{+1.3}_{-1.5})\times10^{-3}$ \\
                  & SO$_2$ & \ditto & (3.9$^{+0.3}_{-0.2})\times10^{15}$ & 238.9$^{+0.2}_{-0.2}$ & 8.1$^{+0.6}_{-0.4}$ & (1.5$^{+0.2}_{-0.2})\times10^{-8}$& (2.6$^{+0.2}_{-0.2})\times10^{-1}$ \\
                  & H$_2$S & \ditto & (1.2$^{+0.1}_{-0.1})\times10^{15}$ & 238.7$^{+0.2}_{-0.2}$ & 5.7$^{+0.5}_{-0.4}$ & (4.5$^{+0.6}_{-0.6})\times10^{-9}$&(7.8$^{+0.7}_{-0.7})\times10^{-2}$  \\
                  & SiO & \ditto & (7.0$^{+0.3}_{-0.3})\times10^{13}$ & 238.3$^{+0.1}_{-0.1}$ & 6.5$^{+0.2}_{-0.2}$ & (2.6$^{+0.3}_{-0.3})\times10^{-10}$&(4.6$^{+0.3}_{-0.3})\times10^{-3}$  \\
                  & DCN & \ditto & (1.1$^{+0.2}_{-0.2})\times10^{13}$ & 238.8$^{+0.4}_{-0.4}$ & 4.4$^{+1.0}_{-0.8}$ & (3.9$^{+0.8}_{-0.8})\times10^{-11}$& (6.9$^{+1.3}_{-1.2})\times10^{-4}$ \\
                  & OCS & \ditto & (8.5$^{+0.6}_{-0.5})\times10^{14}$ & 239.2$^{+0.2}_{-0.2}$ & 5.2$^{+0.4}_{-0.3}$ & (3.2$^{+0.4}_{-0.4})\times10^{-9}$& (5.6$^{+0.5}_{-0.4})\times10^{-2}$ \\
                  & $^{13}$CS & \ditto & (6.4$^{+0.5}_{-0.4})\times10^{13}$ & 238.9$^{+0.2}_{-0.1}$ & 4.4$^{+0.3}_{-0.3}$ & (2.4$^{+0.3}_{-0.3})\times10^{-10}$& (4.2$^{+0.4}_{-0.3})\times10^{-3}$ \\
                  & c--C$_3$H$_2$ & 12.5$^{+3.1}_{-2.1}$ & (2.0$^{+0.7}_{-0.5})\times10^{13}$ & 237.6$^{+0.3}_{-0.3}$ & 3.7$^{+0.7}_{-0.8}$ & (7.4$^{+2.9}_{-2.0})\times10^{-11}$& (1.3$^{+0.5}_{-0.3})\times10^{-3}$ \\
N\,113\,B3 & CH$_3$OH & 132.9$^{+15.1}_{-16.4}$ & (6.2$^{+0.7}_{-0.8})\times10^{15}$ & 231.8$^{+0.5}_{-0.5}$ & 7.9$^{+0.2}_{-0.2}$ & (5.0$^{+1.0}_{-1.0})\times10^{-8}$& \nodata \\               
                  & HCOOCH$_3$ & \ditto & $<$3.1$\times10^{14}$ & 231.9\tablenotemark{\footnotesize a}  & 7.9\tablenotemark{\footnotesize a} & $<$2.5$\times10^{-9}$& $<$5.1$\times10^{-2}$ \\                        
                  & CH$_3$OCH$_3$ & \ditto & (8.7$^{+2.1}_{-2.2})\times10^{14}$ & 231.9\tablenotemark{\footnotesize a}  & 7.9\tablenotemark{\footnotesize a} & (7.0$^{+2.0}_{-2.1})\times10^{-9}$&  (1.4$^{+0.4}_{-0.4})\times10^{-1}$\\                                   
                  & $^{33}$SO & \ditto & (4.3$^{+1.6}_{-2.0})\times10^{13}$ & 231.9\tablenotemark{\footnotesize a} & 7.9\tablenotemark{\footnotesize a} & (3.5$^{+1.4}_{-1.7})\times10^{-10}$&  (7.0$^{+2.7}_{-3.4})\times10^{-3}$\\
                  & SO$_2$ & \ditto & (1.0$^{+0.2}_{-0.2})\times10^{15}$ & 232.1$^{+1.4}_{-1.1}$ & 11.3$^{+3.2}_{-2.7}$ & (8.2$^{+2.3}_{-2.3})\times10^{-9}$&  (1.6$^{+0.4}_{-0.4})\times10^{-1}$\\ 
                  & H$_2$S & \ditto & (5.7$^{+0.9}_{-0.7})\times10^{14}$ & 233.3$^{+0.2}_{-0.2}$ & 3.8$^{+0.5}_{-0.4}$ & (4.6$^{+1.0}_{-0.9})\times10^{-9}$& (9.3$^{+1.9}_{-1.6})\times10^{-2}$ \\
                  & SiO & \ditto & (7.3$^{+0.7}_{-0.6})\times10^{13}$ & 234.0$^{+0.2}_{-0.2}$ & 14.4$^{+0.5}_{-0.4}$ & (5.9$^{+1.1}_{-1.1})\times10^{-10}$&  (1.2$^{+0.2}_{-0.2})\times10^{-2}$\\
                  & DCN & \ditto & (1.8$^{+0.3}_{-0.2})\times10^{13}$ & 233.0$^{+0.2}_{-0.2}$ & 5.0$^{+0.5}_{-0.5}$ & (1.5$^{+0.3}_{-0.3})\times10^{-10}$& (3.0$^{+0.6}_{-0.5})\times10^{-3}$ \\
                  & OCS & \ditto & (4.2$^{+0.5}_{-0.5})\times10^{14}$ & 232.6$^{+0.4}_{-0.5}$ & 6.8$^{+1.2}_{-0.8}$ & (3.4$^{+0.7}_{-0.7})\times10^{-9}$& (6.8$^{+1.2}_{-1.2})\times10^{-2}$ \\
                  & $^{13}$CS & \ditto & (4.2$^{+0.6}_{-0.4})\times10^{13}$ & 233.4$^{+0.2}_{-0.2}$ & 4.6$^{+0.4}_{-0.4}$ & (3.4$^{+0.7}_{-0.6})\times10^{-10}$&  (6.8$^{+1.3}_{-1.1})\times10^{-3}$\\
                  & c--C$_3$H$_2$ & 26.3$^{+18.6}_{-8.0}$ & (1.9$^{+0.4}_{-0.2})\times10^{13}$ & 234.3$^{+0.3}_{-0.3}$ & 4.8$^{+0.8}_{-0.6}$ & (1.5$^{+0.4}_{-0.3})\times10^{-10}$&  (3.1$^{+0.7}_{-0.5})\times10^{-3}$\\                 
\enddata
\vspace*{0.2cm}
\tablenotemark{\footnotesize a}{For some molecules with noisy or tentative line detections, reliable parameter error estimates could not be obtained using the Monte Carlo resampling method. In these cases, $v_{\rm LSR}$ or both $v_{\rm LSR}$ and $\Delta v_{\rm FWHM}$ were held fixed at the value given by the initial least-squares fit, with the other parameters allowed to vary freely (see Section~\ref{s:modeling} for details). }
\end{deluxetable*}

\begin{deluxetable*}{cccccc}
\centering
\tablecaption{The Results of XCLASS LTE SO$_2$ Fitting for Hot Cores N\,105--2\,A and 2\,B\tablenotemark{\footnotesize a,b}}
\label{t:SO2fits}
\tablewidth{0pt}
\tablehead{
\multicolumn{1}{c}{Component} &
\multicolumn{1}{c}{Source Size} &
\colhead{$T_{\rm rot} ({\rm SO_2})$} &
\colhead{$N({\rm SO_2})$}&
\colhead{$v_{\rm LSR}$} &    
\colhead{$\Delta v_{\rm FWHM}$} \\
\colhead{} &
\colhead{($''$)} &
\colhead{(K)} &
\colhead{(cm$^{-2}$)} &
\colhead{(km s$^{-1}$)} &
\colhead{(km s$^{-1}$)} 
}
\startdata
\multicolumn{6}{c}{N105--2\,A} \\
 1 & 1.00 [0.58, 1.70] & 47 [83, 207] &  $(3.0\, [0.6, 9.4])\times10^{14}$ & 242.2 [241.7, 243.6] & 5.6 [2.3, 4.5]  \\
2 & 0.05 [0.04, 0.15] & 102 [40, 178] & $(2.0\, [0.1, 2.0])\times10^{17}$ & 242.6 [241.9, 243.4] & 2.1 [2.6, 5.1]  \\ 
        	 \hline
\multicolumn{6}{c}{N105--2\,B} \\
1 & 1.24 [0.34, 1.68] & 46 [43, 76] & $(8.4\, [9.7, 21.3])\times10^{14}$ & 241.7 [242.9, 245.8] & 8.1 [8.7, 10.0]  \\ 
2 & 0.05 [0.07, 0.84] & 139 [105, 275] & $(5.8\, [0.01, 2.5])\times10^{17}$ & 244.8 [241.2, 247.7] & 10.5 [8.2, 12.3]  \\ 
\enddata
\tablenotetext{\footnotesize a}{For each parameter, the best $\chi^2$ value is given, as well as the Bayesian credibility intervals calculated by an MCMC error estimate (orange and black dashed lines in Figs.~\ref{f:corner1}--\ref{f:corner2}, respectively).}
\tablenotetext{\footnotesize b}{For comparison, we provide the results of the SO$_2$ single-component rotational diagram analysis for N105--2\,A and 2\,B: $[T_{\rm rot}, N]$ = $[134.1^{+2.2}_{-2.1}, (1.61\pm0.05)\times10^{15}]$ for 2\,A and $[122.9\pm1.0, (3.53\pm0.05)\times10^{15}]$ for 2B (see Section\ref{s:modeling}).}
\end{deluxetable*}

\begin{figure*}
\centering
\includegraphics[width=0.8\textwidth]{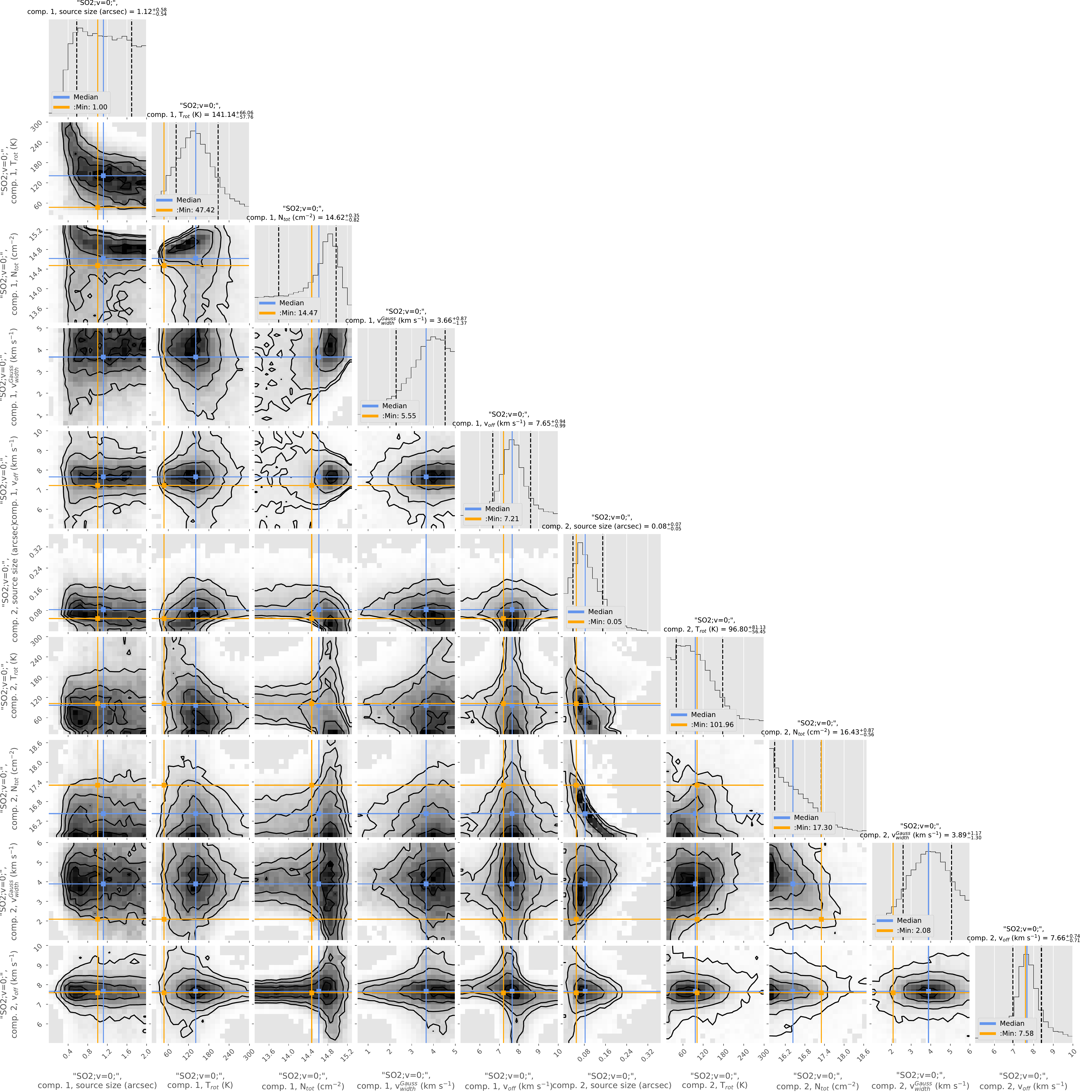}
\caption{Corner plot of the MCMC error estimate for the XCLASS LTE SO$_2$ fit for N\,105--2\,A.  The 16\% and 84\% quantiles are plotted as black dashed lines, the 50\% quantile (median) is shown in blue, while the lowest $\chi^2$ value is shown in orange.  For very asymmetric distributions, the lowest $\chi^2$ value is at quite some distance from the median, at or even beyond the limits of the credibility interval.} \label{f:corner1}
\end{figure*}

\begin{figure*}
\centering
\includegraphics[width=0.8\textwidth]{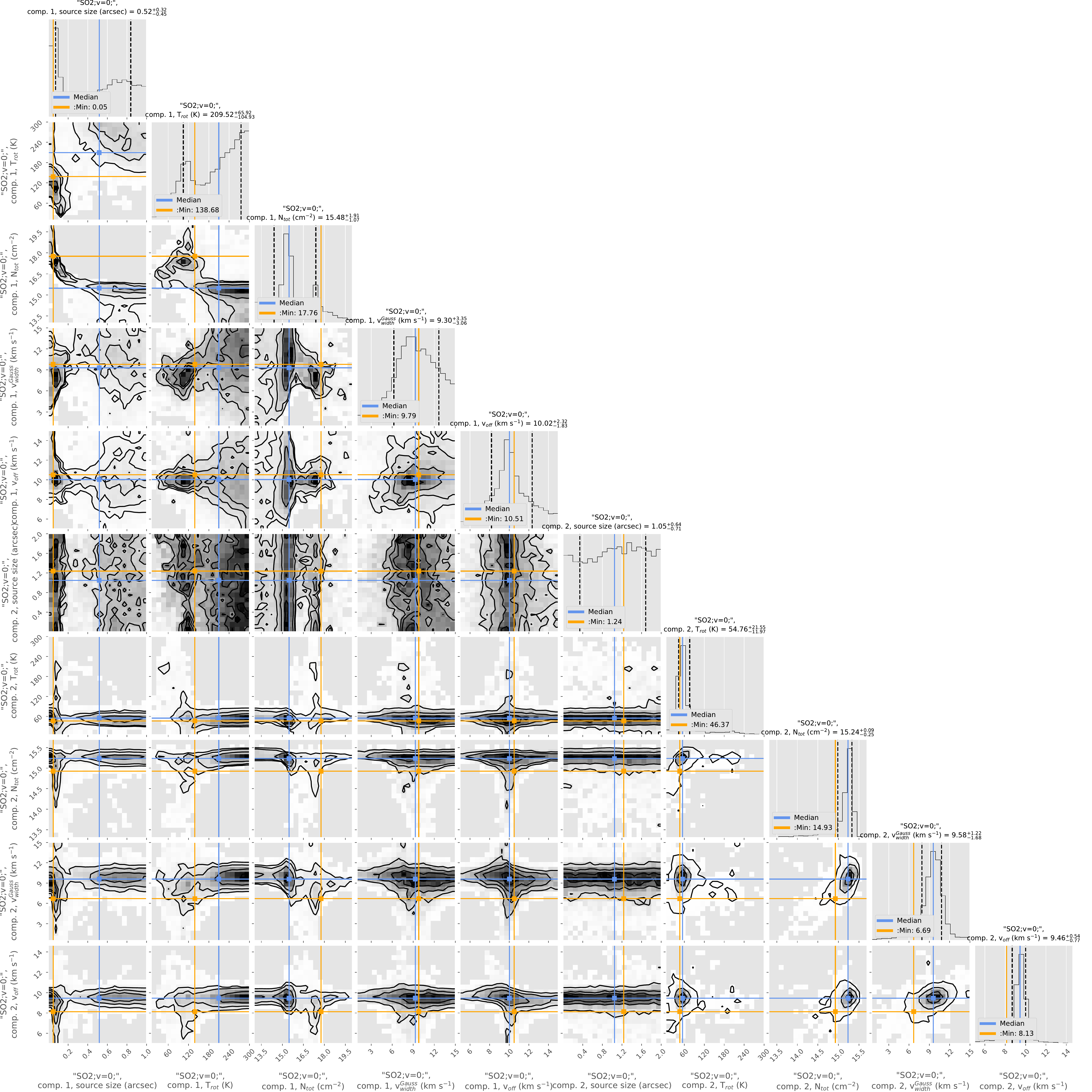}
\caption{The same as Fig.~\ref{f:corner1}, but for source N\,105--2\,B.} \label{f:corner2}
\end{figure*}

\section{Observed Spectra and Model Fits} 
\label{s:appspectra}

\begin{figure*}[h!]
\includegraphics[width=0.33\textwidth]{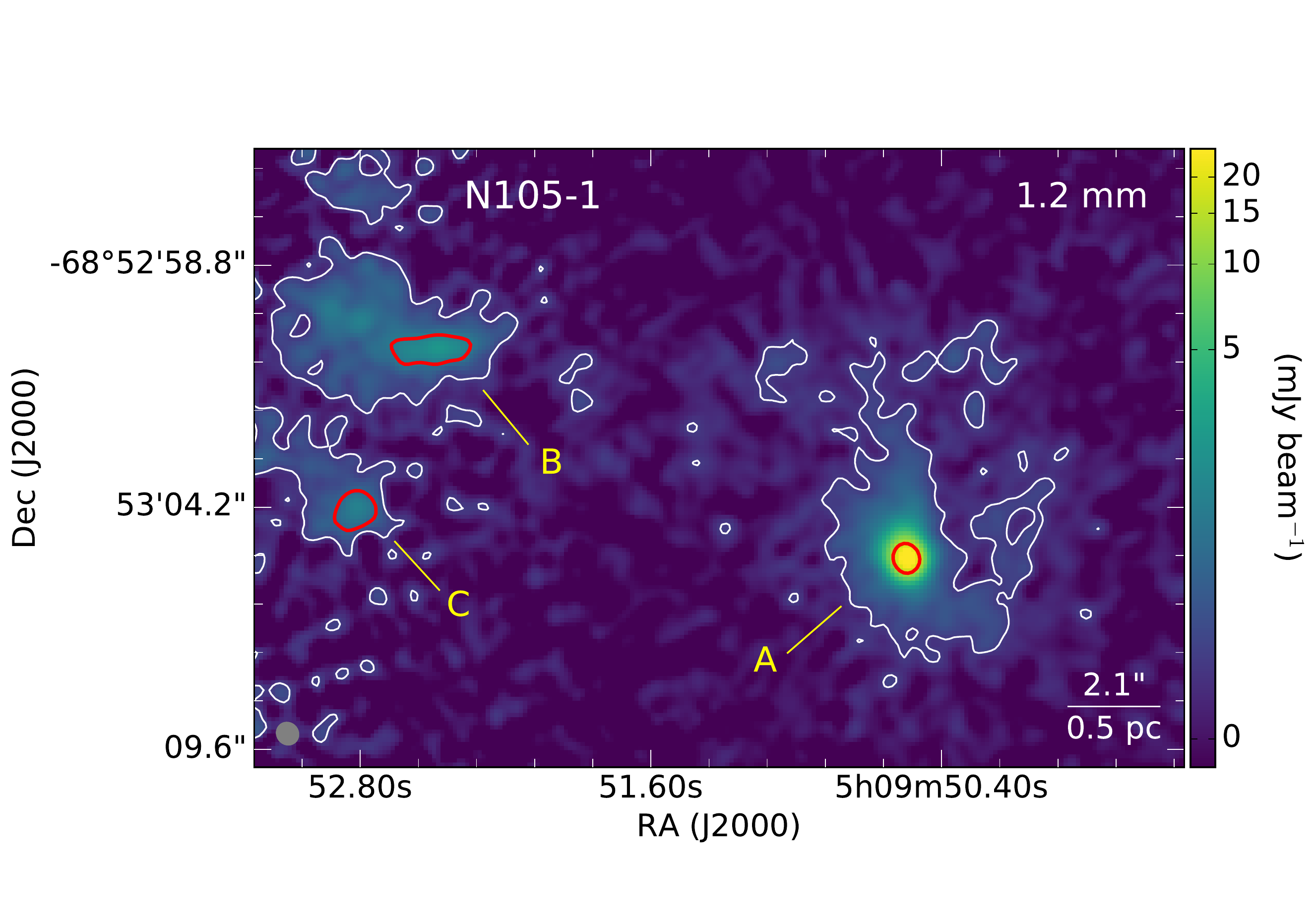}
\includegraphics[width=0.33\textwidth]{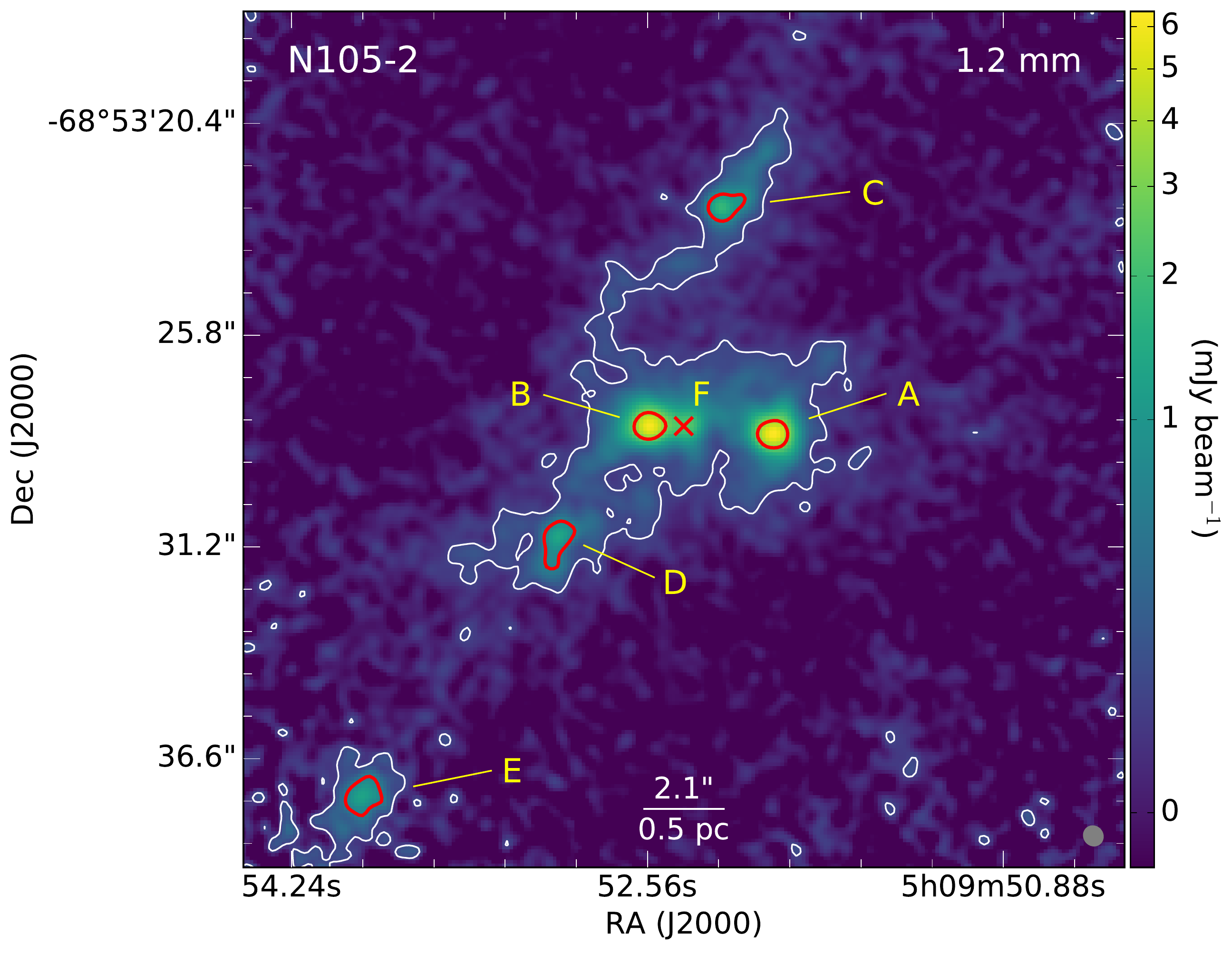}
\includegraphics[width=0.33\textwidth]{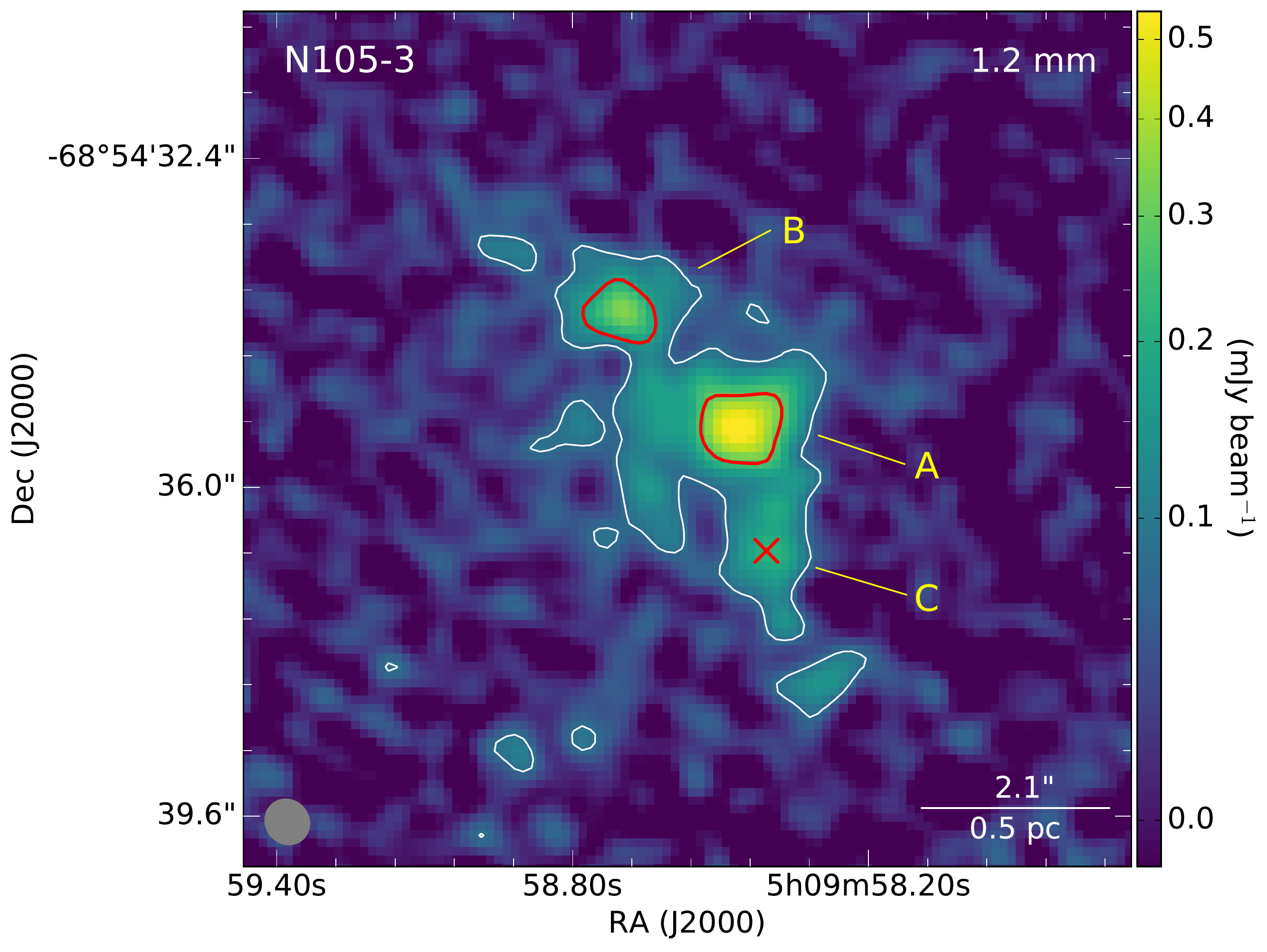}
\caption{The 1.2 mm continuum images of N\,105--1 ({\it left}), N\,105--2 ({\it center}), and N\,105--3 ({\it right}) with spectral extraction regions indicated in red; they are contours corresponding to 50\% of the 1.2 mm continuum peak of individual sources (see Section~\ref{s:spectral} for details). White contours represent the 3$\sigma$-level continuum emission (as in Fig.~\ref{f:N105B6cont}). \label{f:extraction}}
\end{figure*}

\begin{figure*}
\centering
\includegraphics[width=\textwidth]{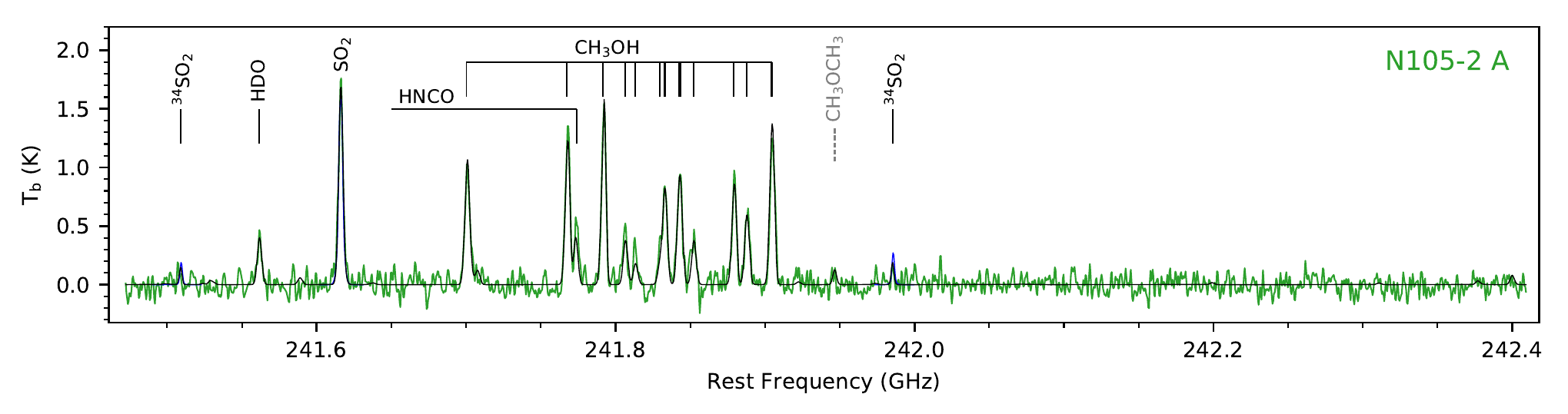}
\includegraphics[width=\textwidth]{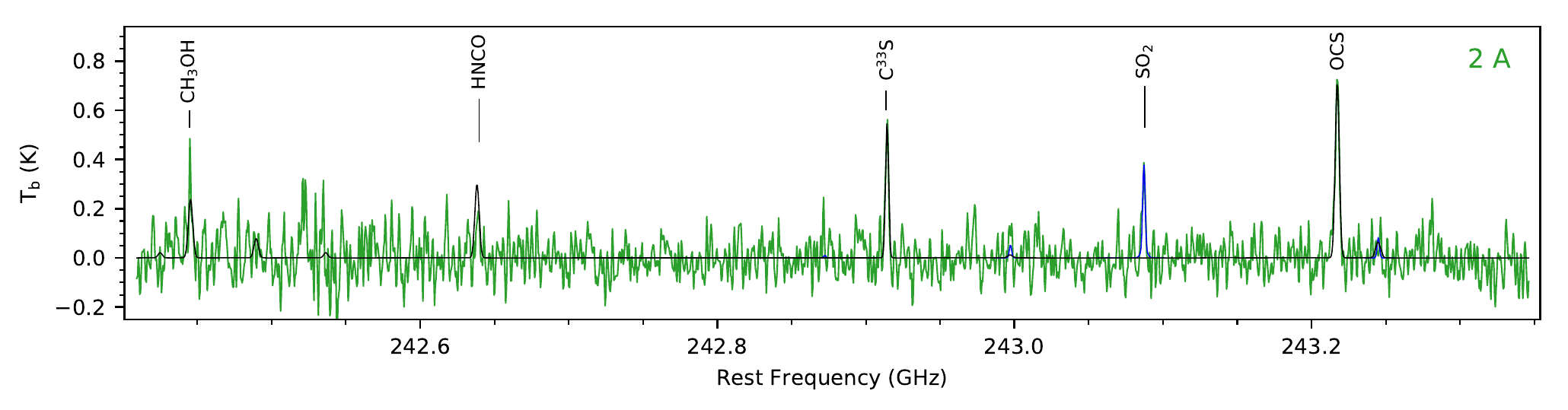}
\includegraphics[width=\textwidth]{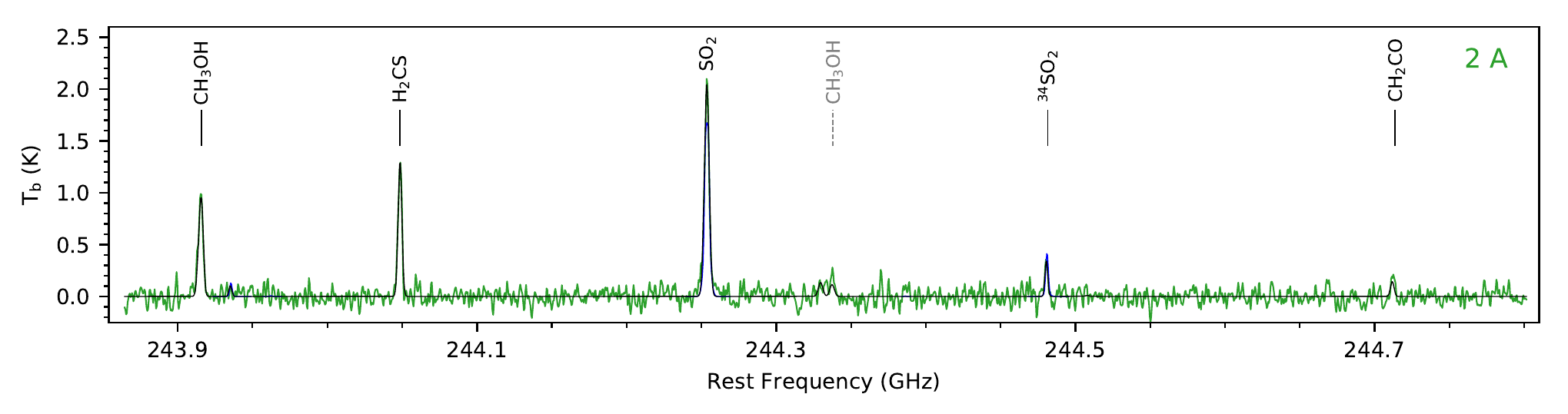}
\includegraphics[width=\textwidth]{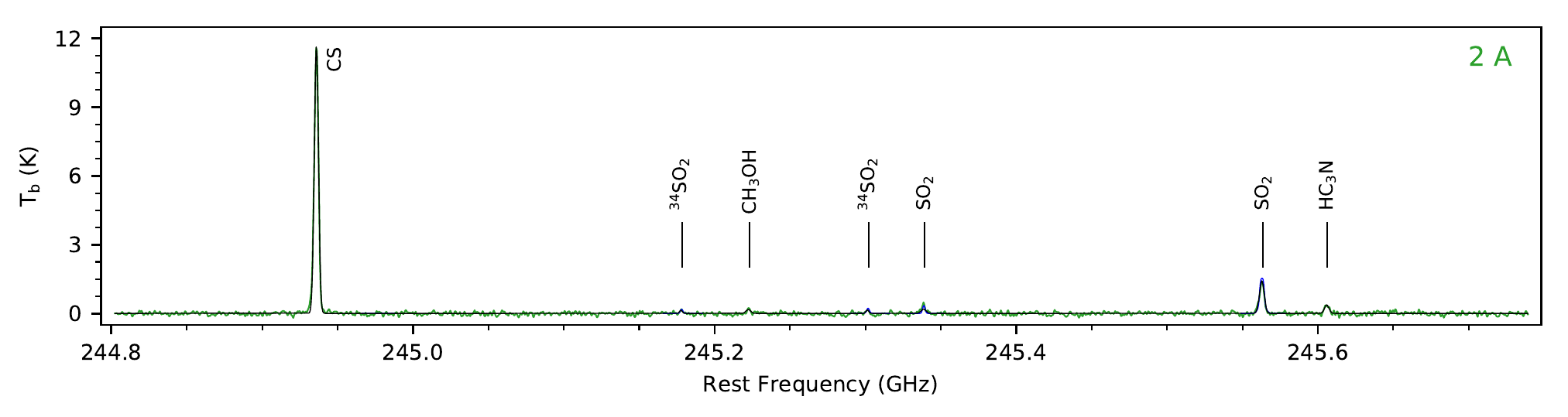}
\caption{ALMA Band 6 spectra of the N\,105--2 A hot core in the $\sim$242~GHz ({\it top two panels}) and $\sim$245~GHz ({\it bottom two panels}) spectral windows. The detected ({\it black}) and tentatively detected ({\it gray}) spectral lines are labeled (see also Table~\ref{t:detections}).  The LTE synthetic spectra including all species and described in Section~\ref{s:modeling} are shown in black. The SO$_2$ transition at $\sim$243.09 GHz is not present in these models because it was not possible to achieve a satisfactory fit to this line; it was excluded from the analysis to improve the overall fit. The XCLASS LTE synthetic spectra of SO$_2$ and $^{34}$SO$_2$ are shown in blue and include the SO$_2$ transition at $\sim$243.09 GHz (see Section~\ref{s:xclass}). \label{f:spec2A_1_app}}
\end{figure*}

\begin{figure*}
\centering
\includegraphics[width=\textwidth]{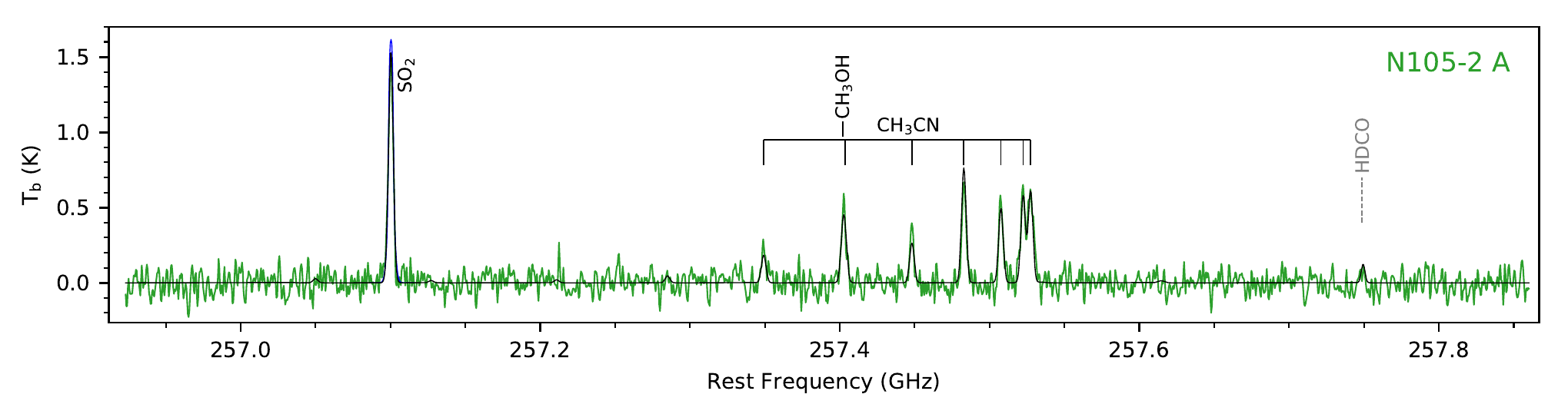}
\includegraphics[width=\textwidth]{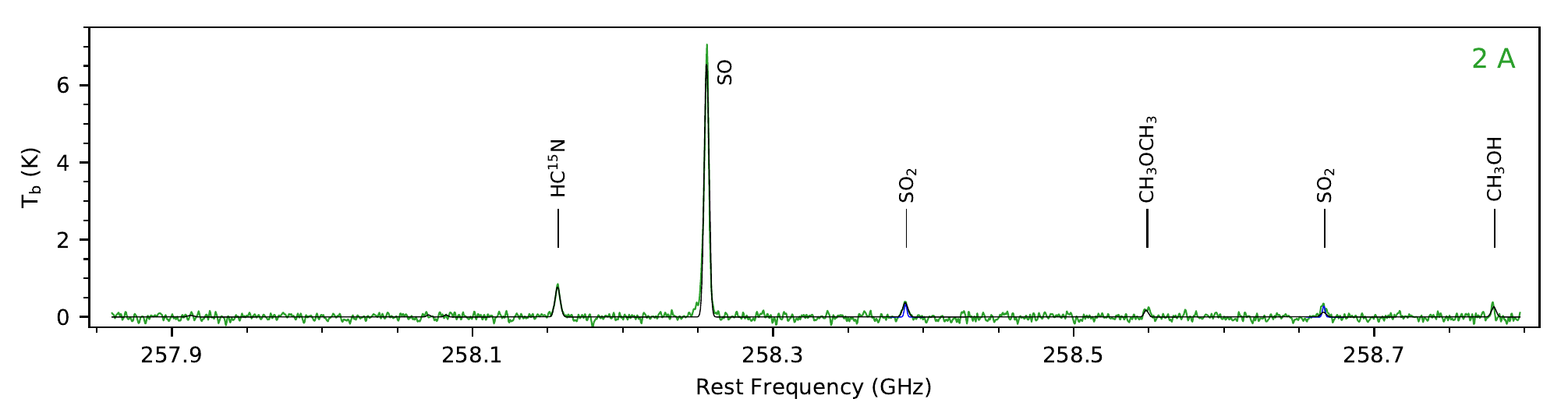}
\includegraphics[width=\textwidth]{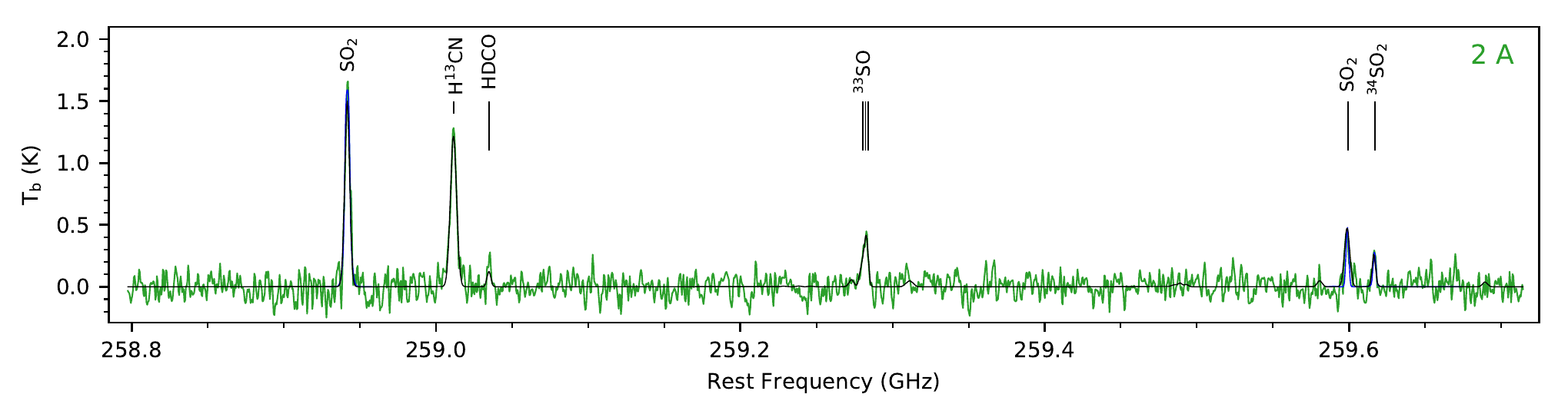}
\includegraphics[width=\textwidth]{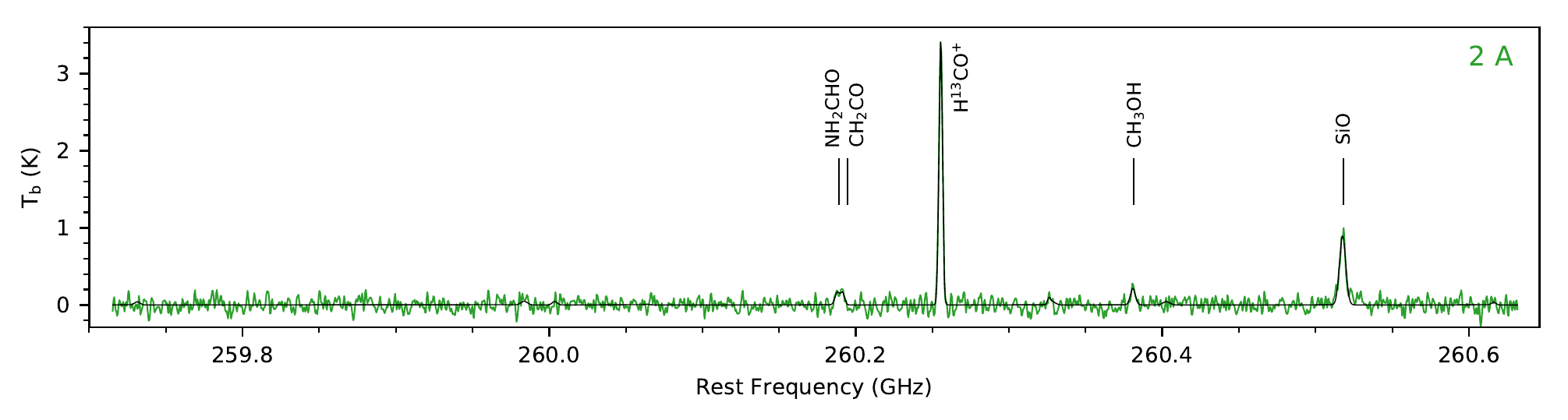}
\caption{The same as Fig.~\ref{f:spec2A_1_app}, but for the $\sim$258~GHz ({\it top two panels}) and $\sim$260~GHz ({\it bottom two panels}) spectral windows.  \label{f:spec2A_2_app}}
\end{figure*}

\begin{figure*}
\centering
\includegraphics[width=\textwidth]{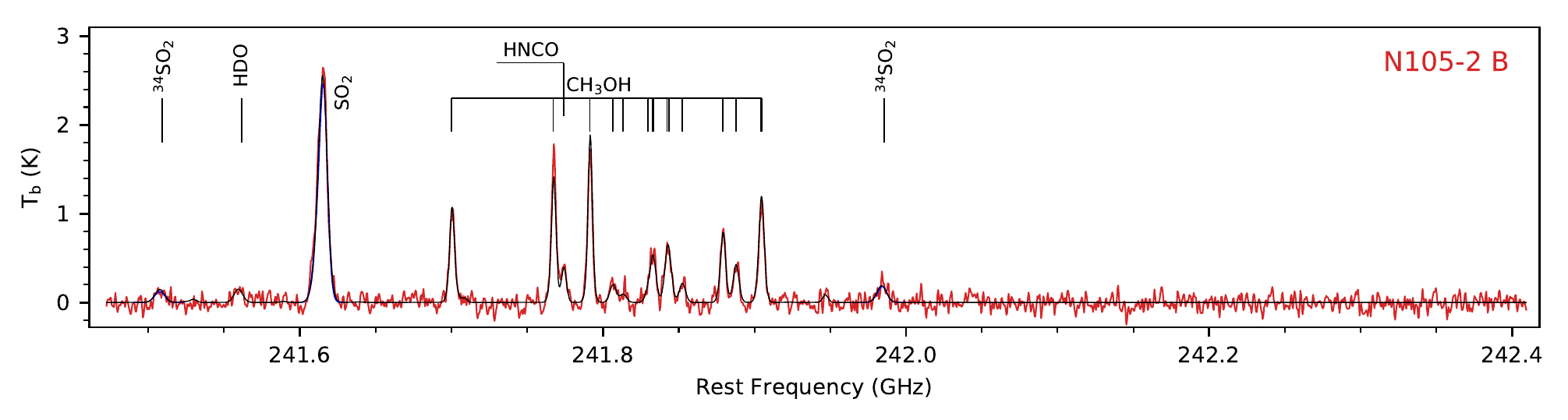} 
\includegraphics[width=\textwidth]{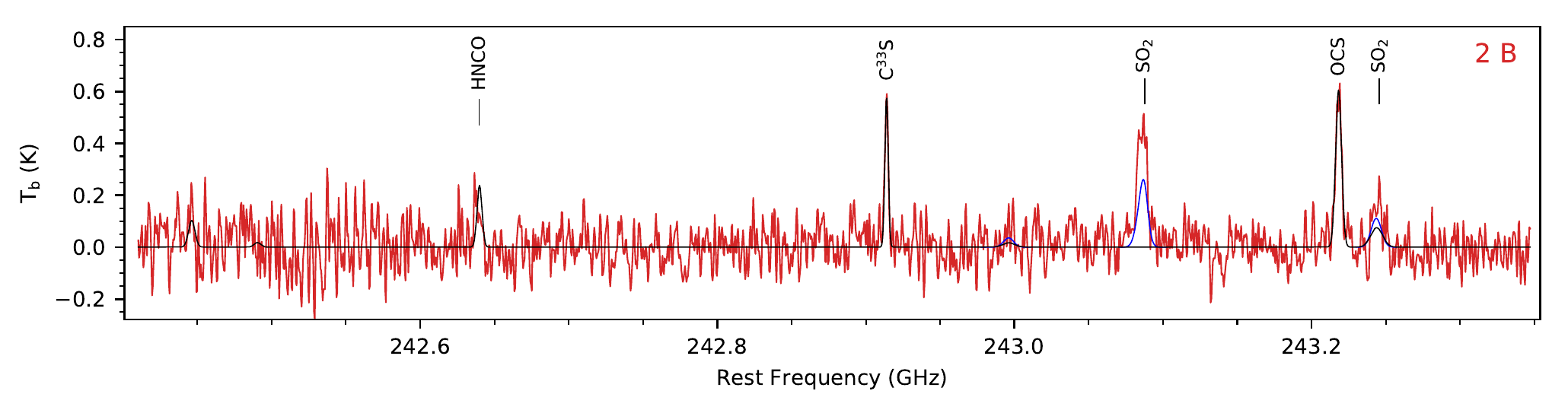}
\includegraphics[width=\textwidth]{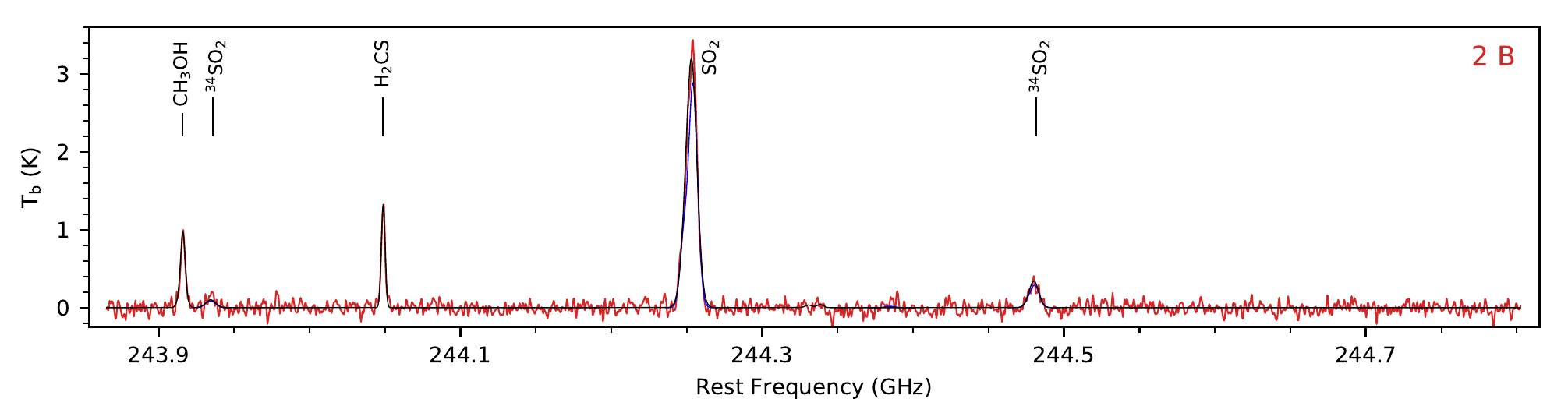}
\includegraphics[width=\textwidth]{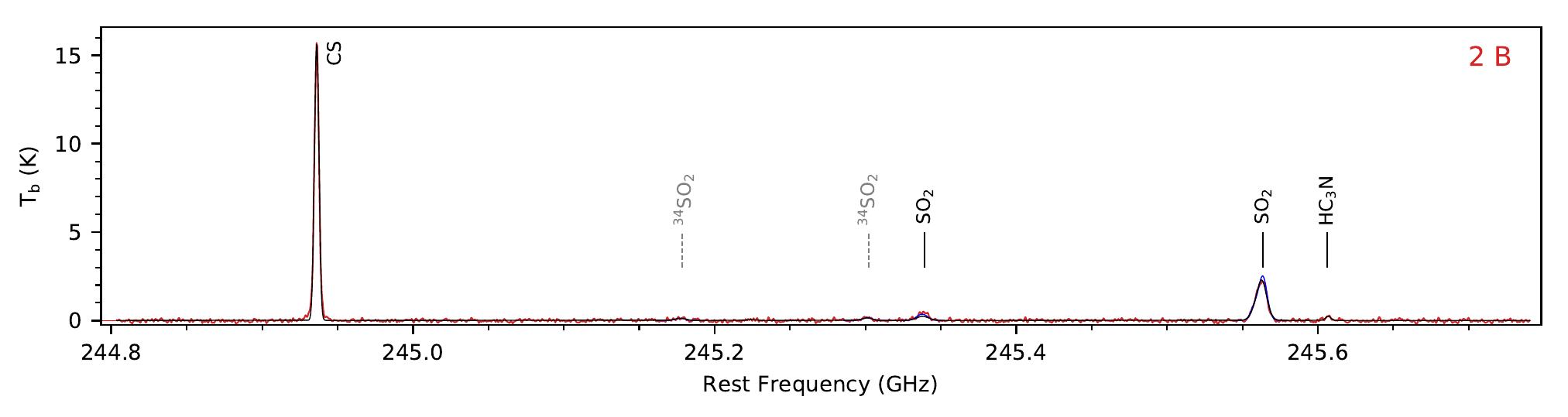}
\caption{The same as Fig.~\ref{f:spec2A_1_app}, but for the N\,105--2\,B hot core. \label{f:spec2B_1_app}}
\end{figure*}

\begin{figure*}
\centering
\includegraphics[width=\textwidth]{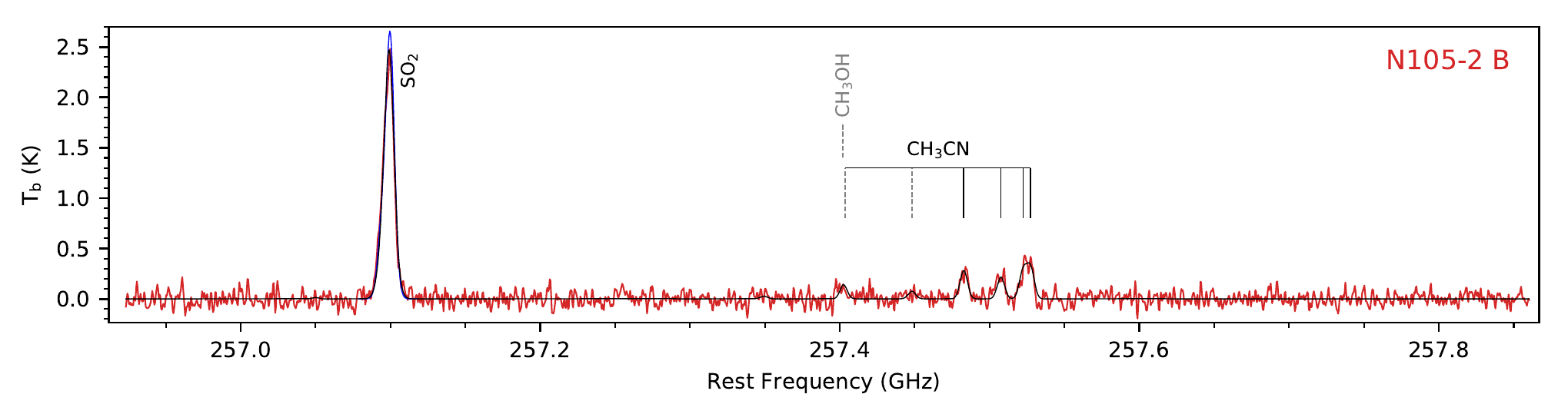}
\includegraphics[width=\textwidth]{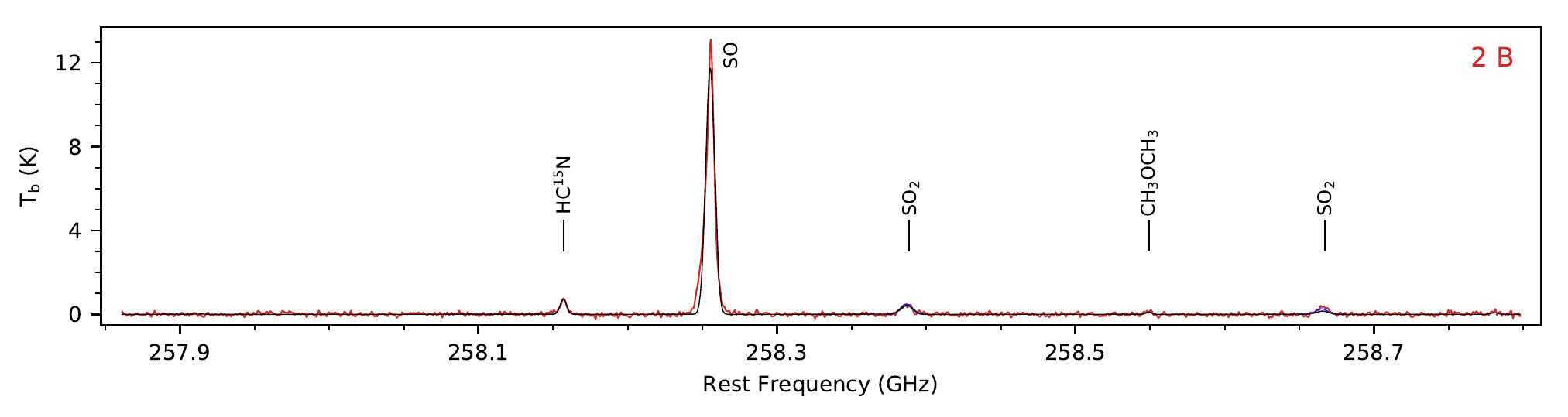}
\includegraphics[width=\textwidth]{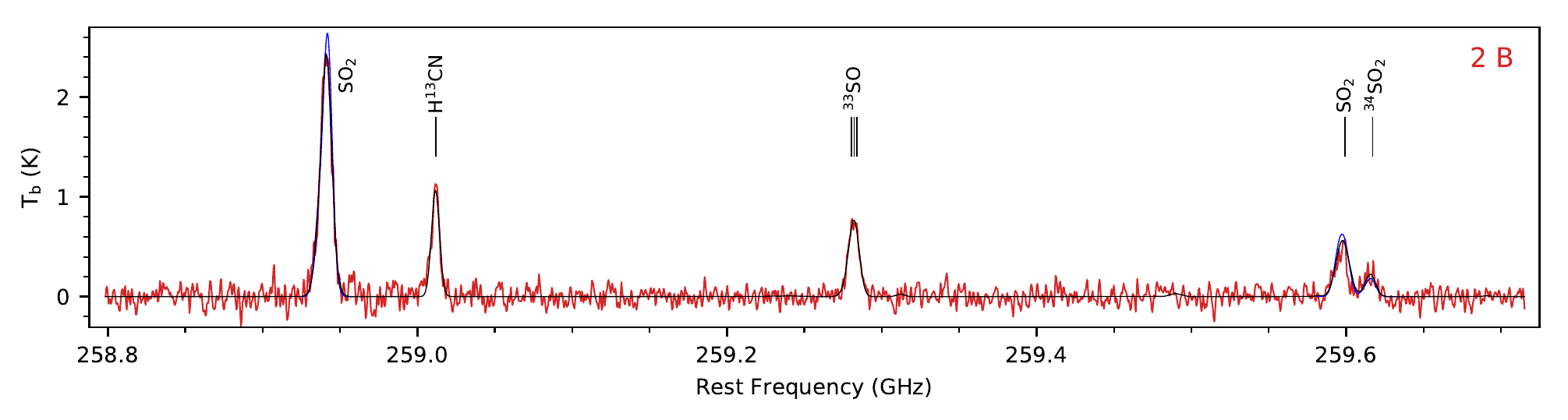}
\includegraphics[width=\textwidth]{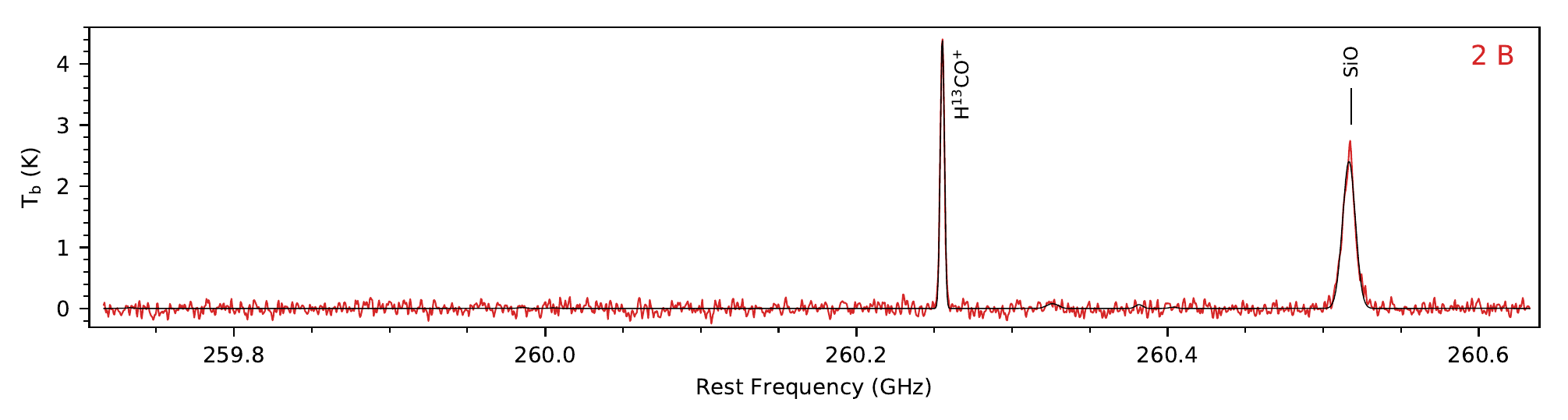}
\caption{The same as Fig.~\ref{f:spec2A_2_app}, but for the N\,105--2\,B hot core.  \label{f:spec2B_2_app}}
\end{figure*}

\begin{figure*}
\centering
\includegraphics[width=\textwidth]{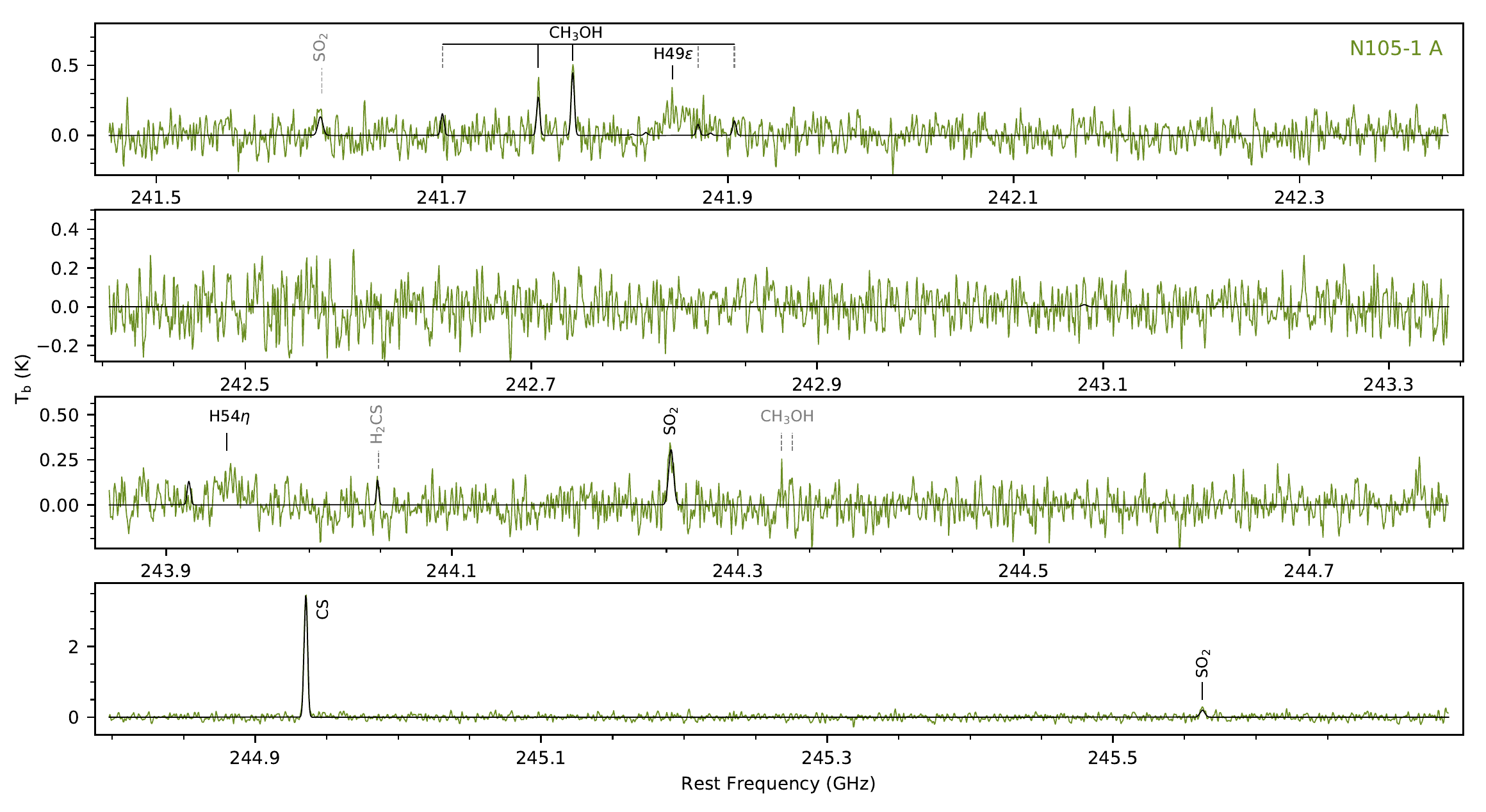}
\includegraphics[width=\textwidth]{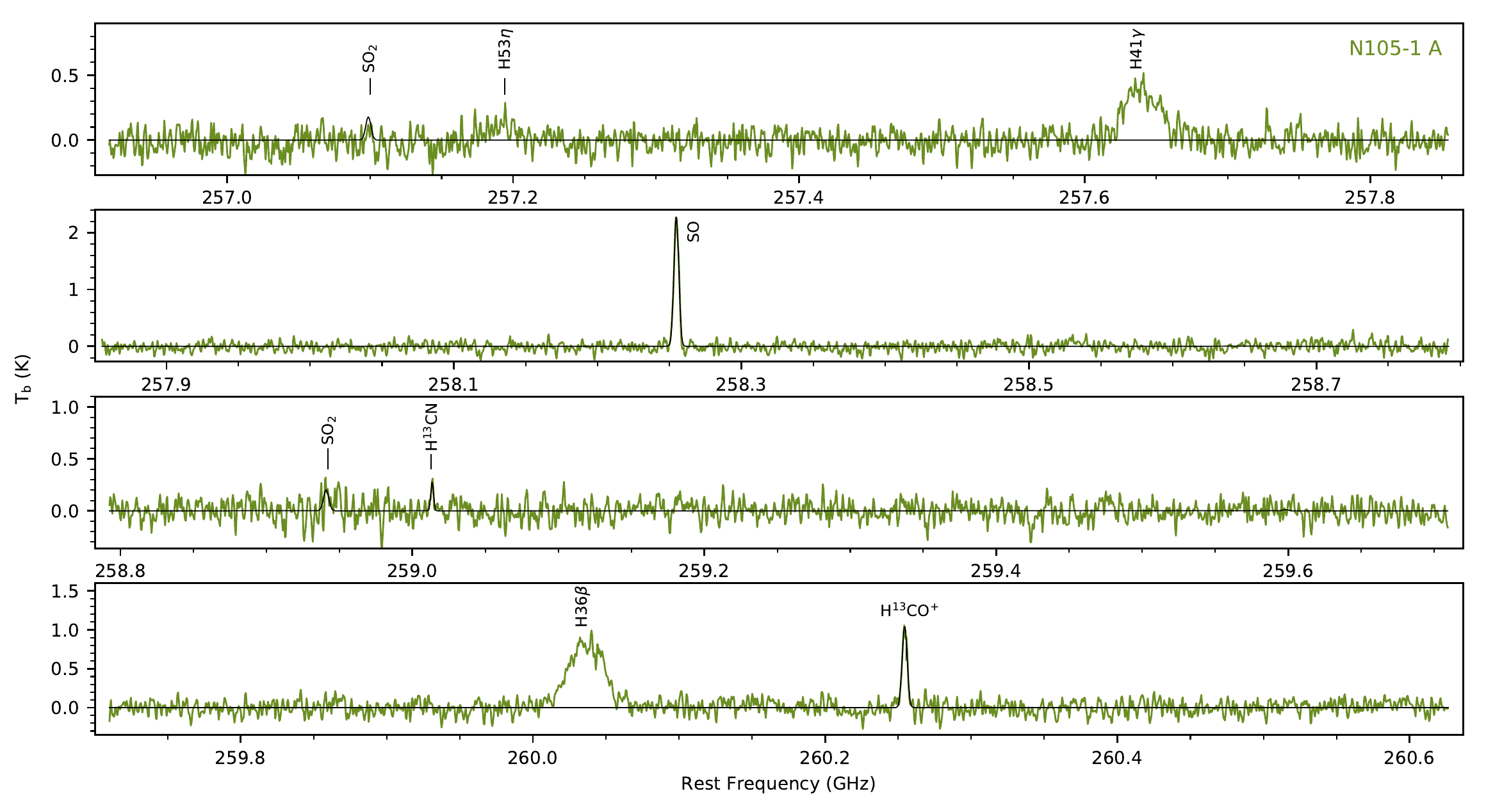}
\caption{The same as Fig.~\ref{f:spec2A_1_app} ({\it top}) and Fig.~\ref{f:spec2A_2_app} ({\it bottom}), but for source N\,105--1\,A.  \label{f:spec1A_app}}
\end{figure*}

\begin{figure*}
\centering
\includegraphics[width=\textwidth]{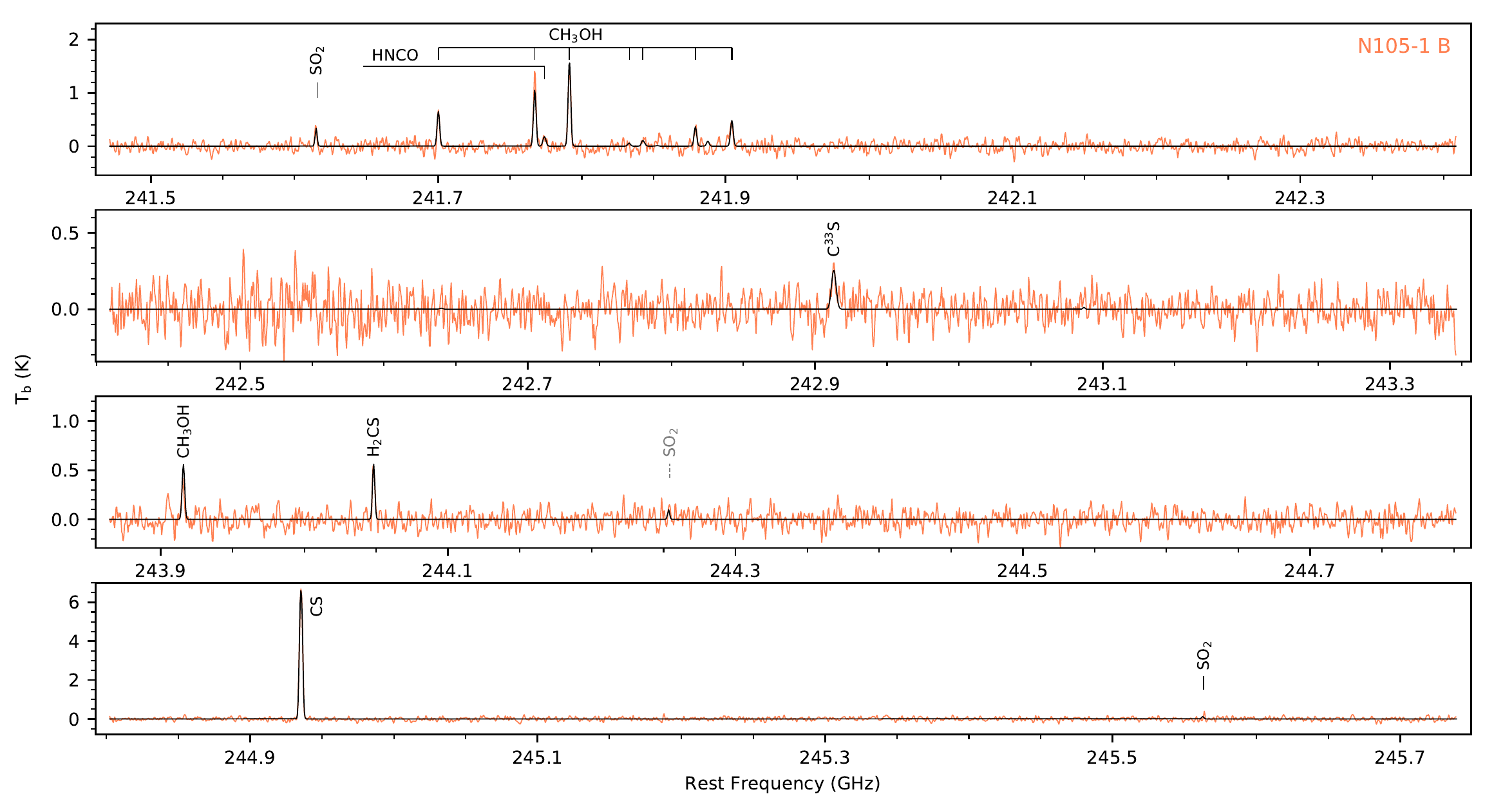}
\includegraphics[width=\textwidth]{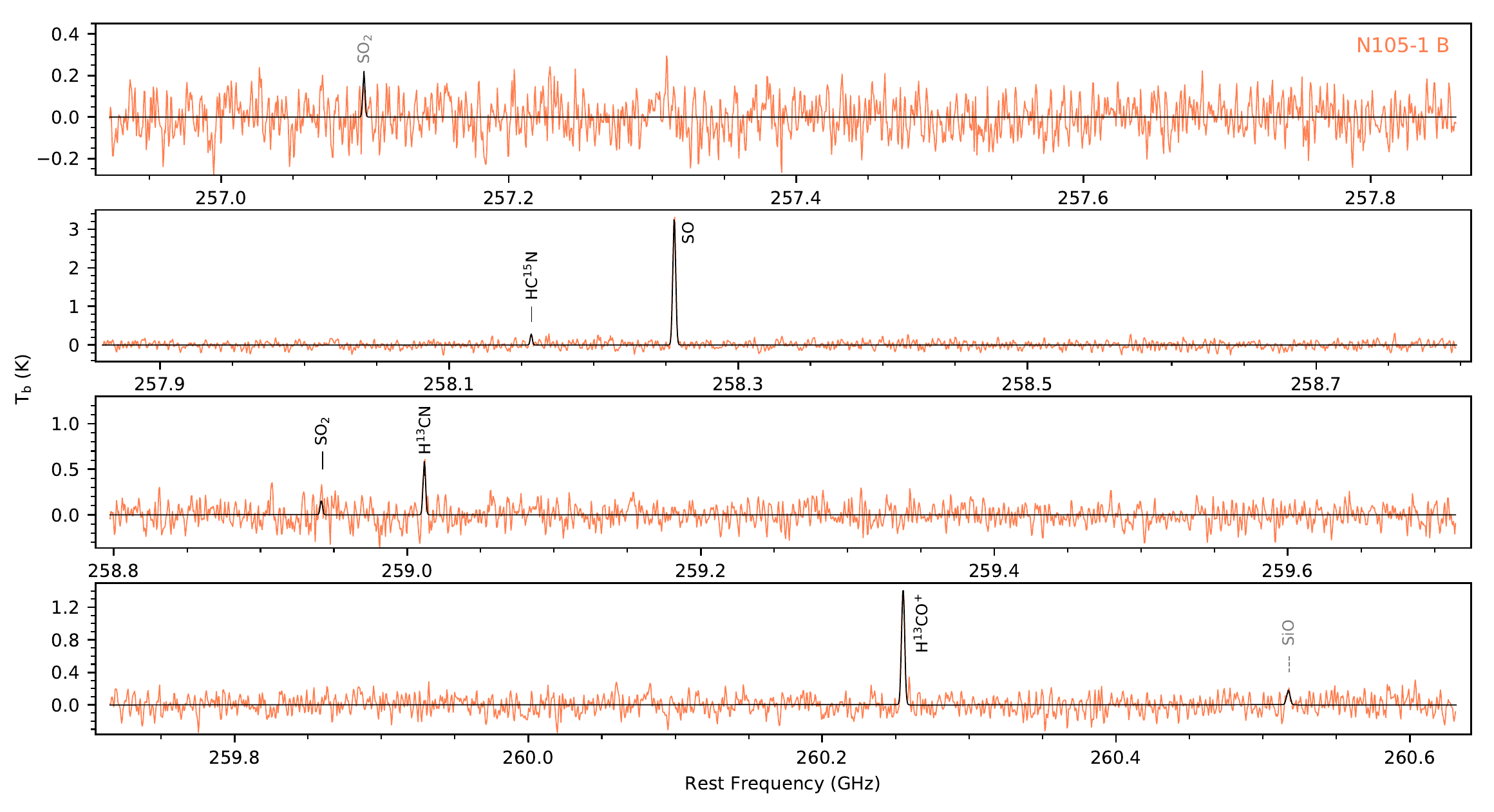}
\caption{The same as Fig.~\ref{f:spec2A_1_app} ({\it top}) and Fig.~\ref{f:spec2A_2_app} ({\it bottom}), but for source N\,105--1\,B.  \label{f:spec1B_app}}
\end{figure*}

\begin{figure*}
\centering
\includegraphics[width=\textwidth]{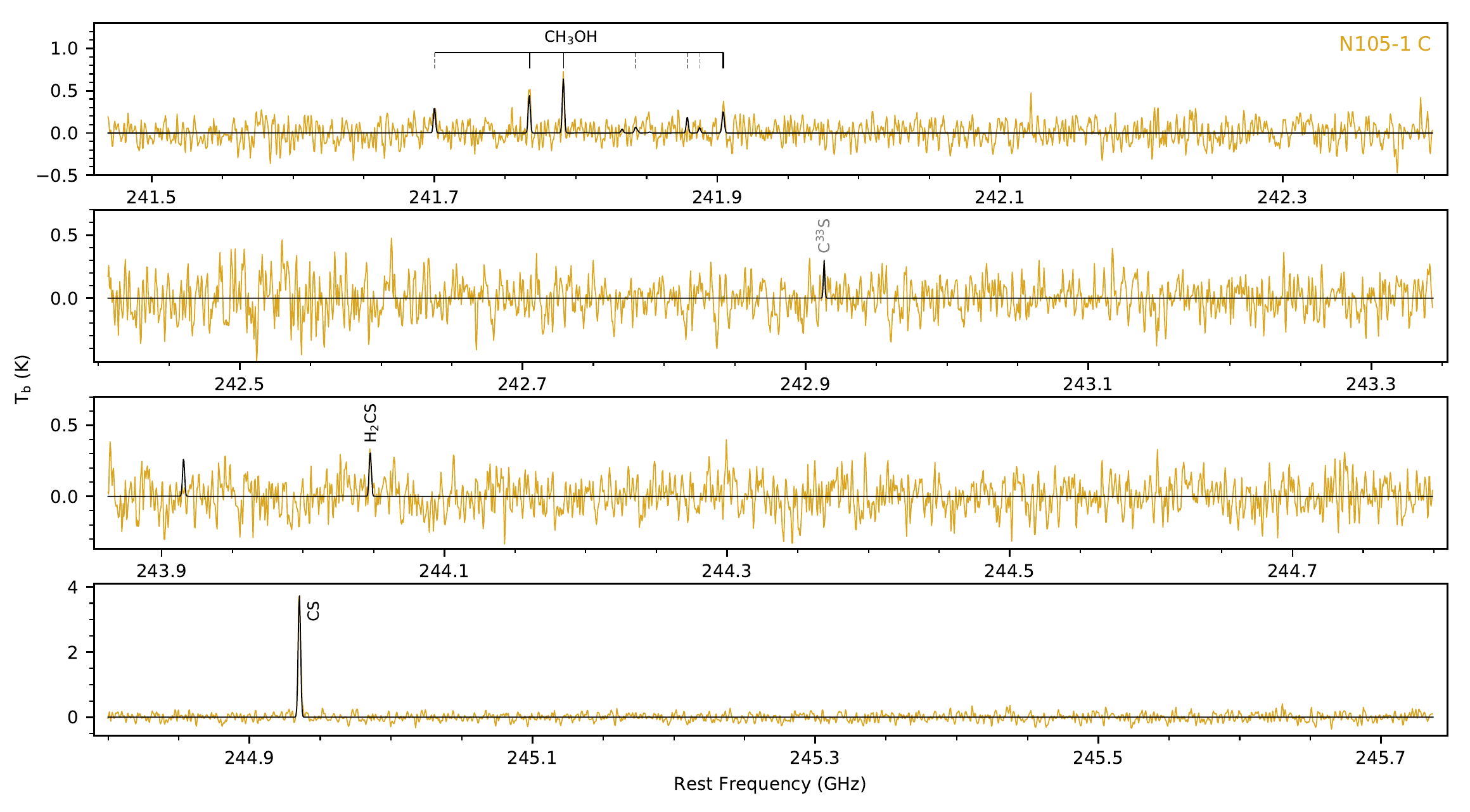}
\includegraphics[width=\textwidth]{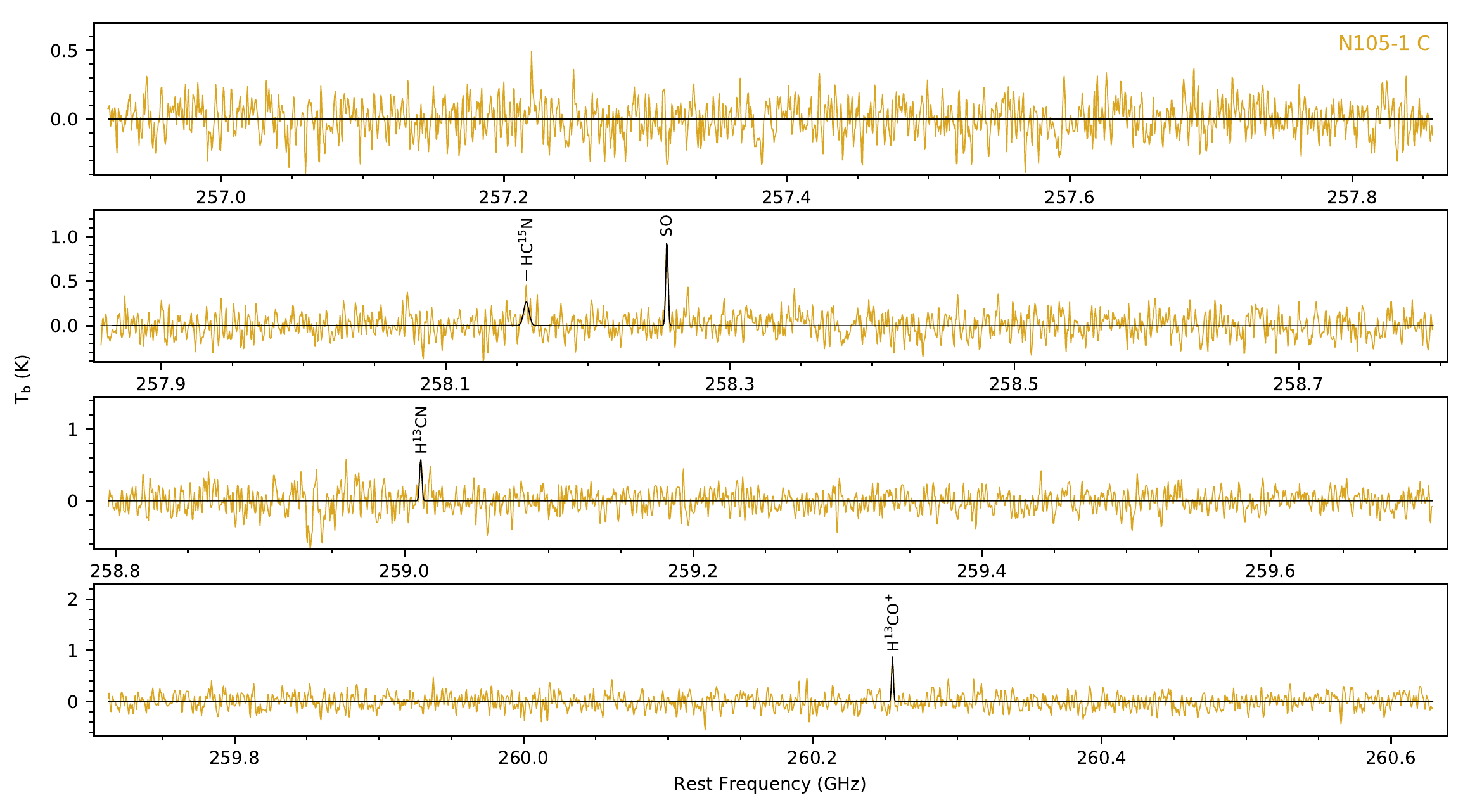}
\caption{The same as Fig.~\ref{f:spec2A_1_app} ({\it top}) and Fig.~\ref{f:spec2A_2_app} ({\it bottom}), but for source N\,105--1\,C. \label{f:spec1C_app}}
\end{figure*}

\begin{figure*}
\centering
\includegraphics[width=\textwidth]{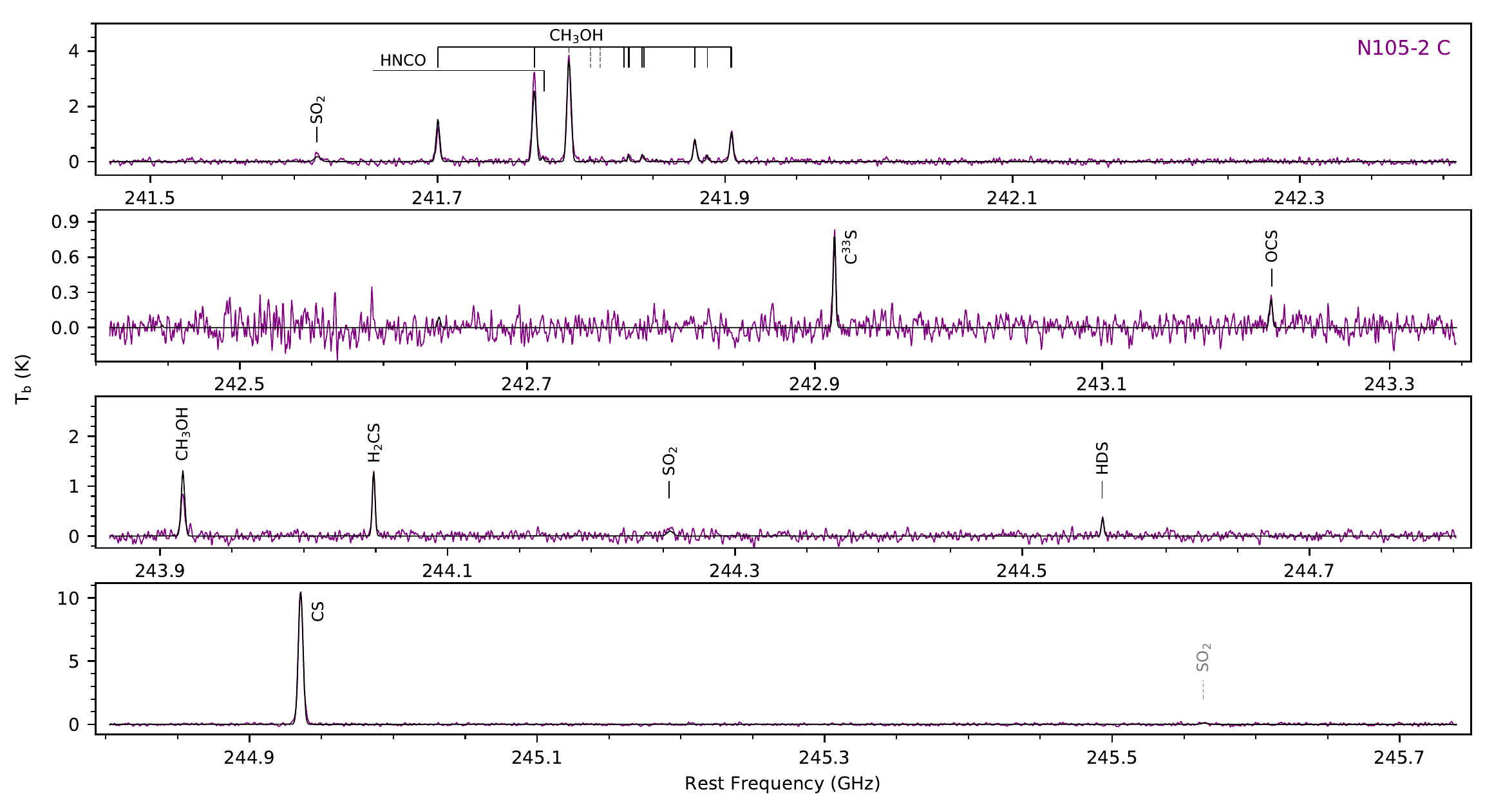}
\includegraphics[width=\textwidth]{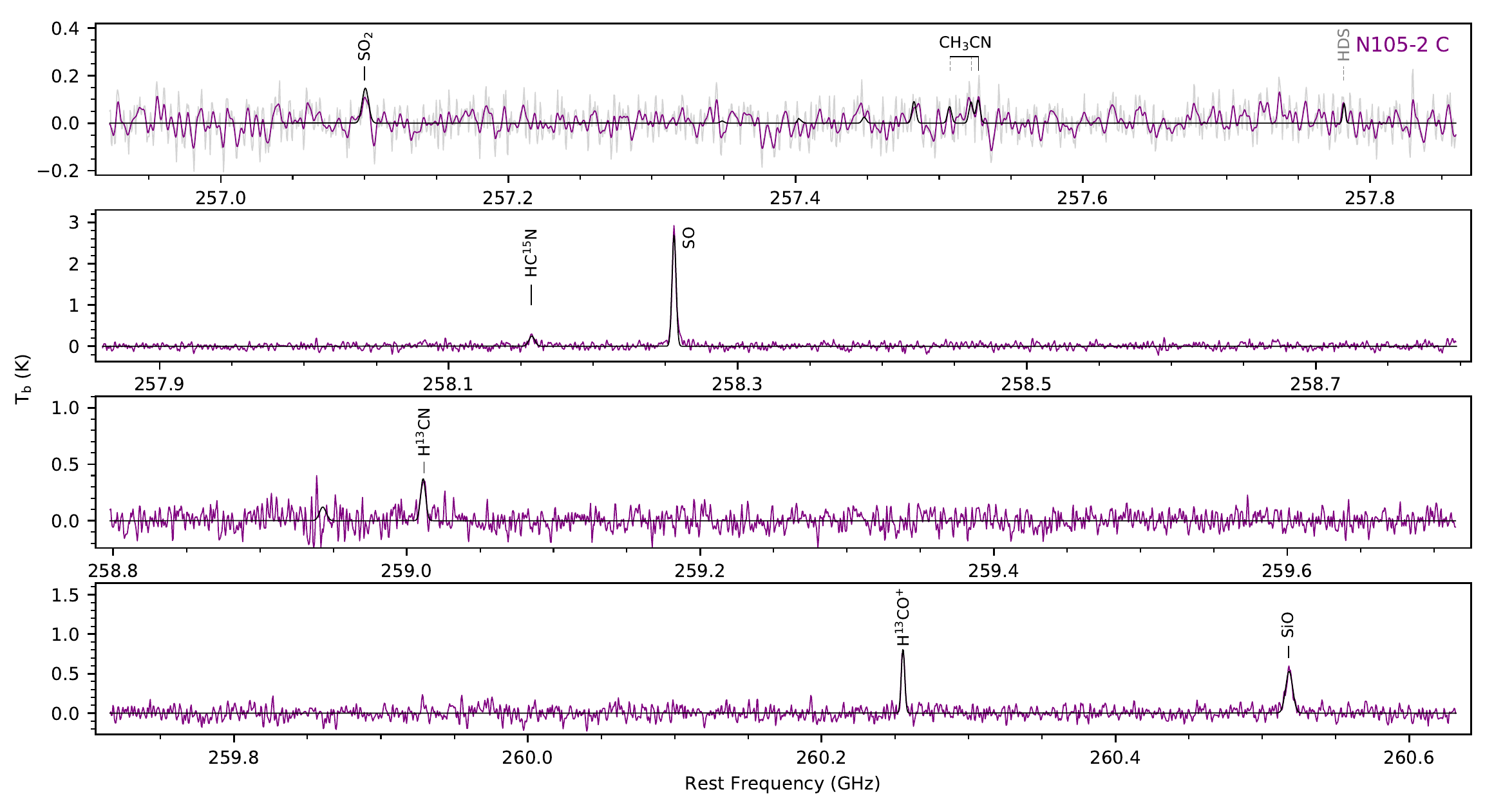}
\caption{The same as Fig.~\ref{f:spec2A_1_app} ({\it top}) and Fig.~\ref{f:spec2A_2_app} ({\it bottom}), but for source N\,105--2\,C.  In the plot at the top of the lower panel, the observed spectrum is shown in gray, while the Hanning-smoothed spectrum is shown in purple to highlight the detection of the CH$_3$CN lines.  The observed spectrum is shown in purple in all the other plots. \label{f:spec2C_app}}
\end{figure*}

\begin{figure*}
\centering
\includegraphics[width=\textwidth]{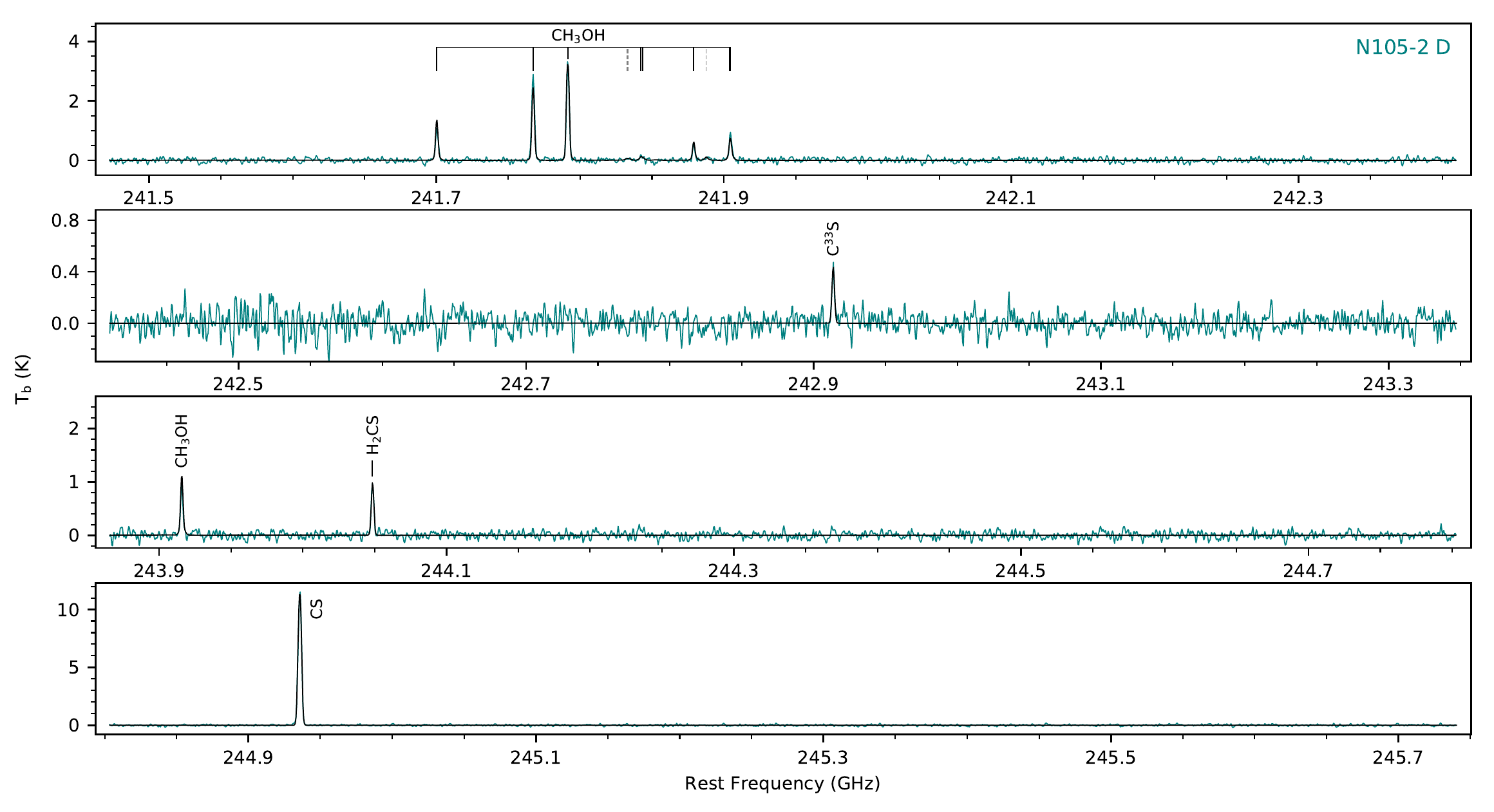}
\includegraphics[width=\textwidth]{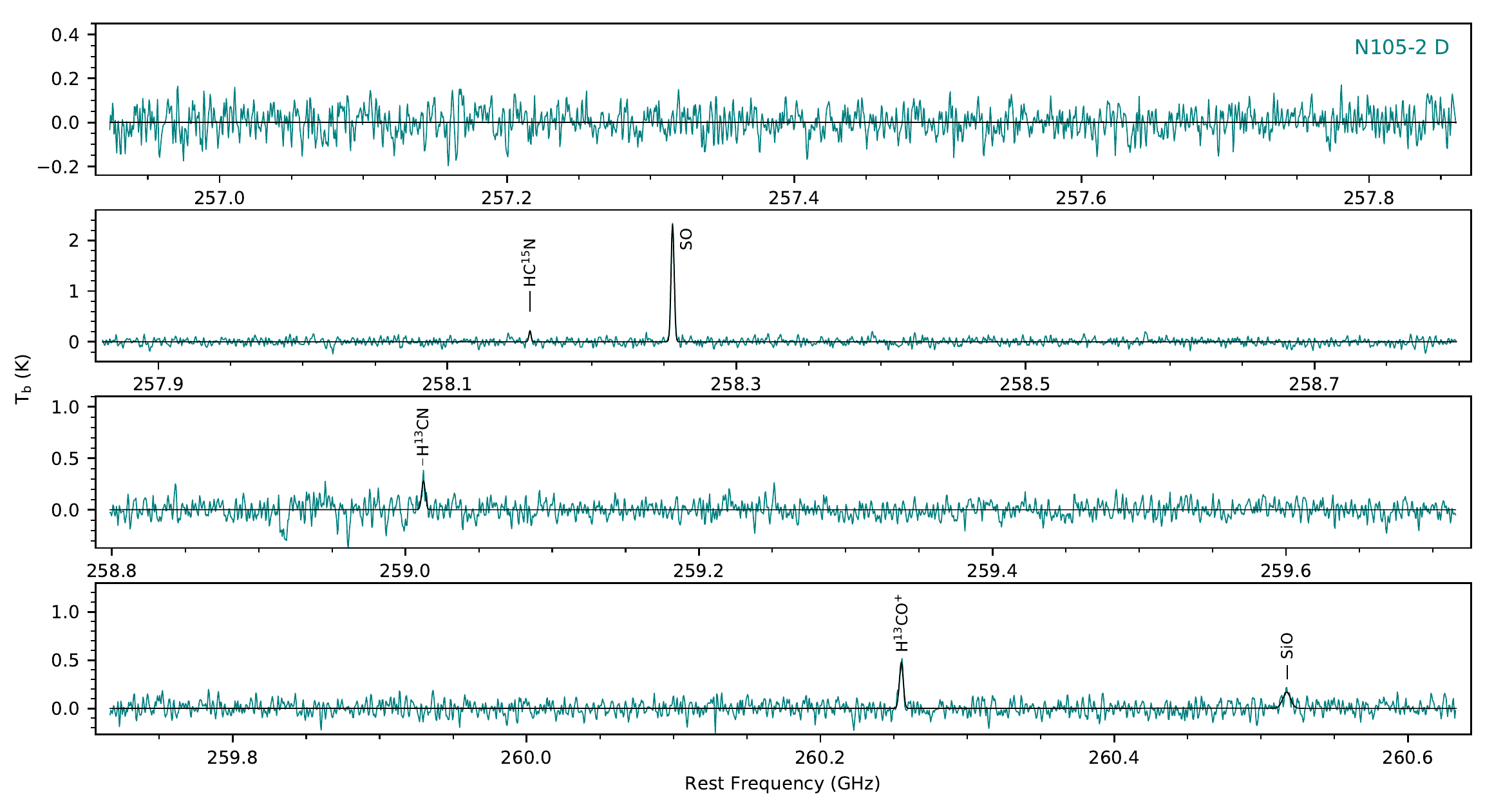}
\caption{The same as Fig.~\ref{f:spec2A_1_app} ({\it top}) and Fig.~\ref{f:spec2A_2_app} ({\it bottom}), but for source N\,105--2\,D. \label{f:spec2D_app}}
\end{figure*}

\begin{figure*}
\centering
\includegraphics[width=\textwidth]{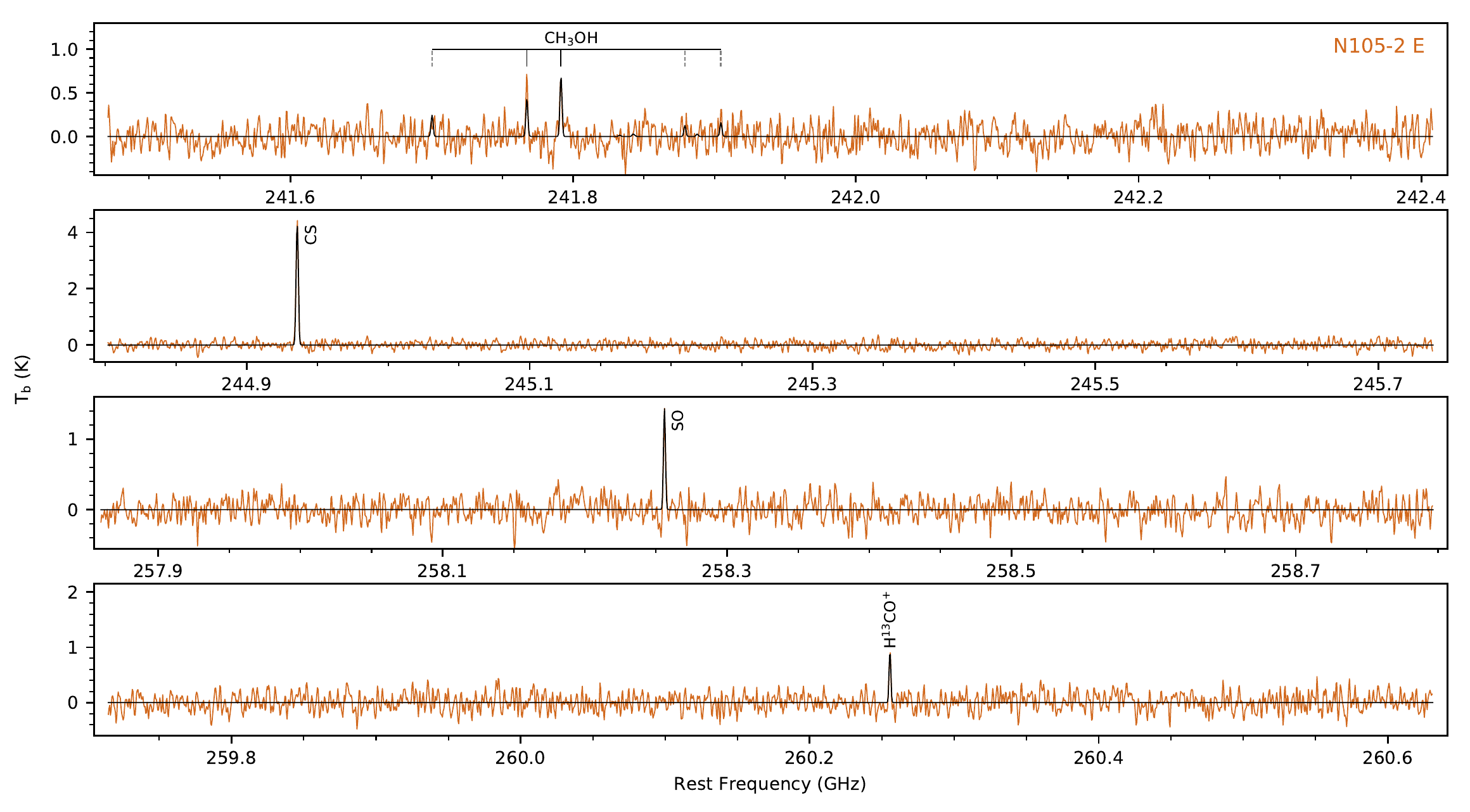}
\caption{A subset of plots shown in Figs.~\ref{f:spec2A_1_app} and \ref{f:spec2A_2_app} for 2\,A, but for source N\,105--2\,E; only spectral ranges with line detections are shown. \label{f:spec2E_app}}
\end{figure*}

\begin{figure*}
\centering
\includegraphics[width=\textwidth]{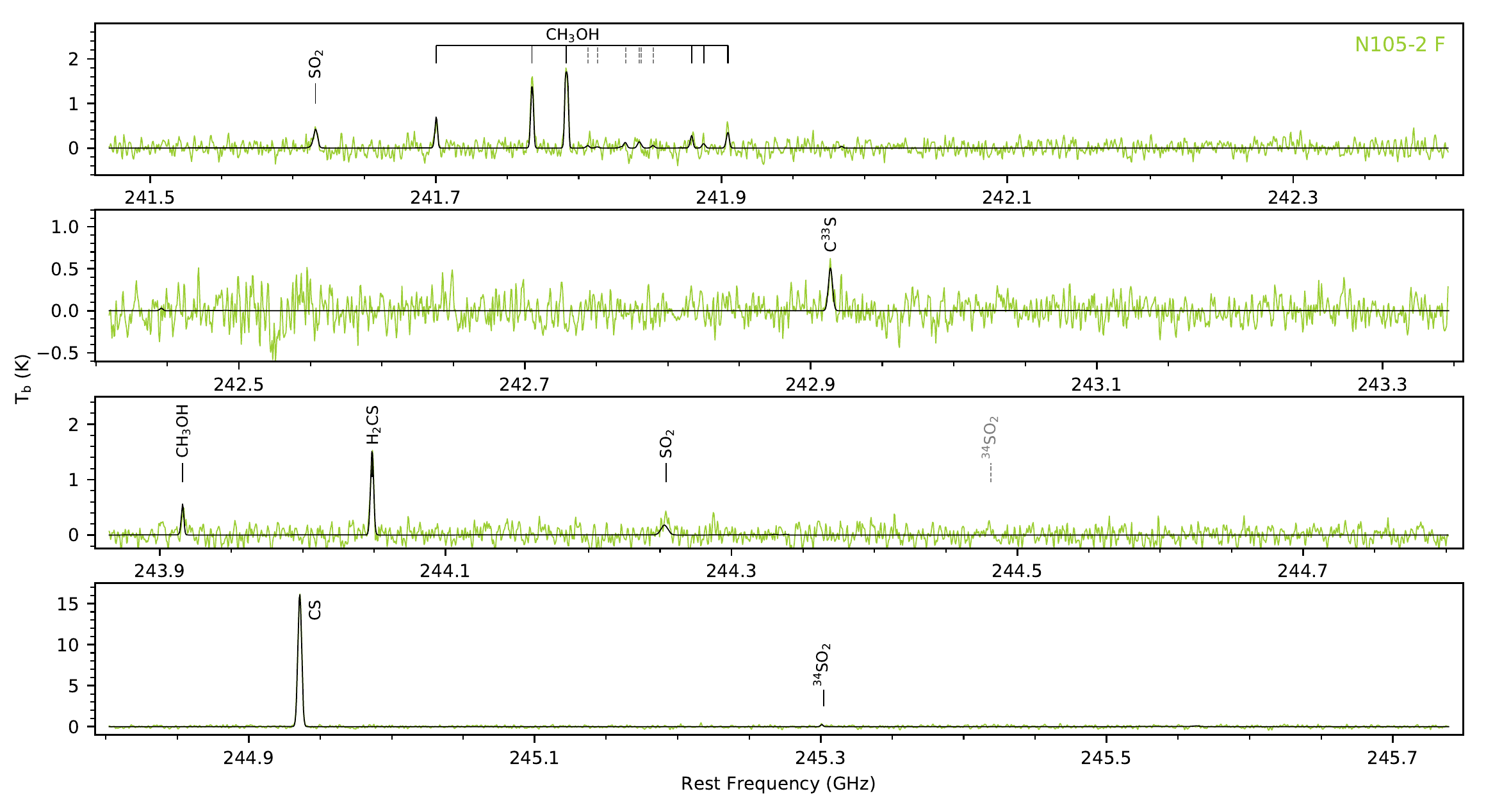}
\includegraphics[width=\textwidth]{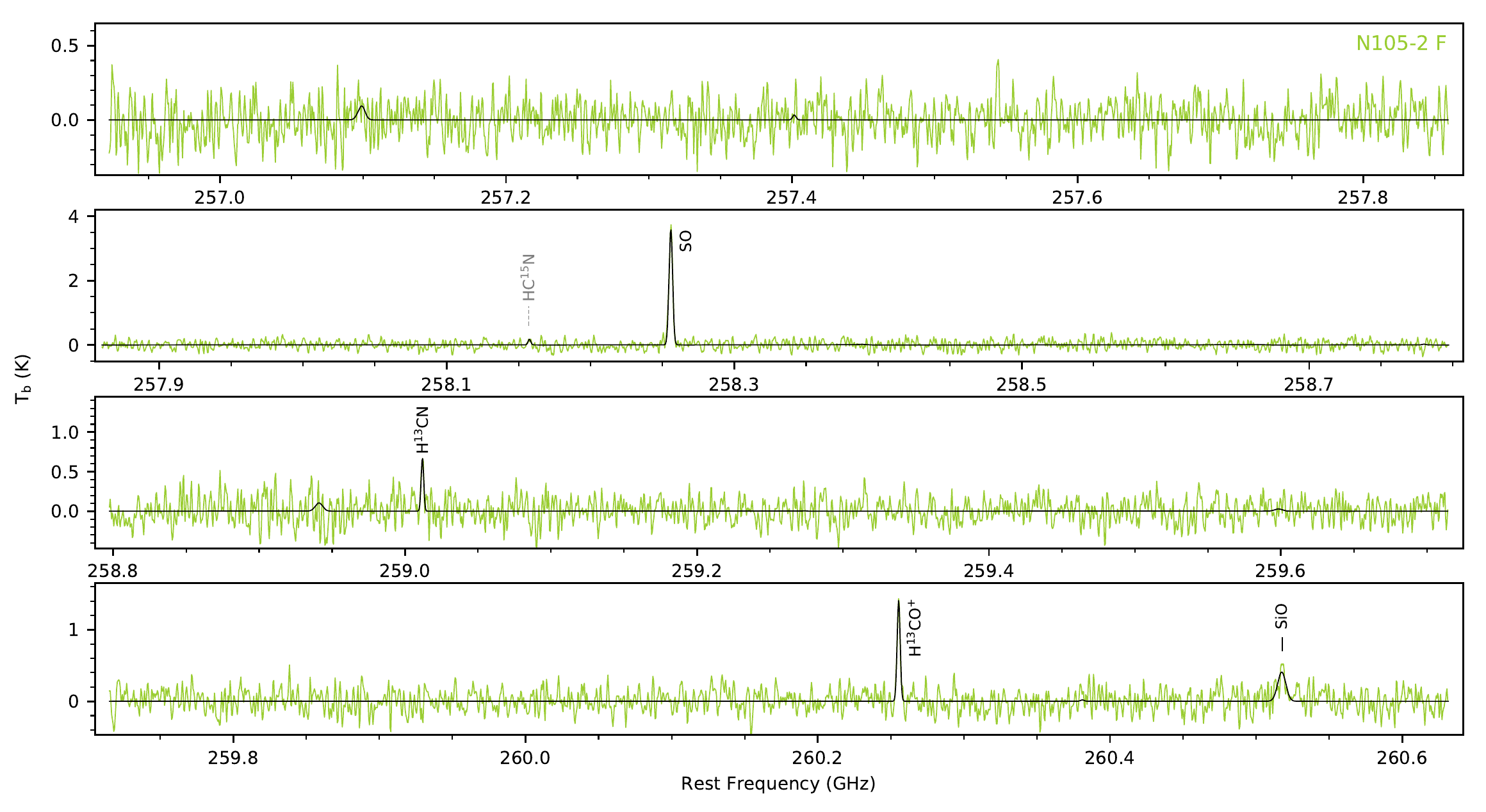}
\caption{The same as Fig.~\ref{f:spec2A_1_app} ({\it top}) and Fig.~\ref{f:spec2A_2_app} ({\it bottom}), but for source N\,105--2\,F. \label{f:spec2F_app}}
\end{figure*}

\begin{figure*}
\centering
\includegraphics[width=\textwidth]{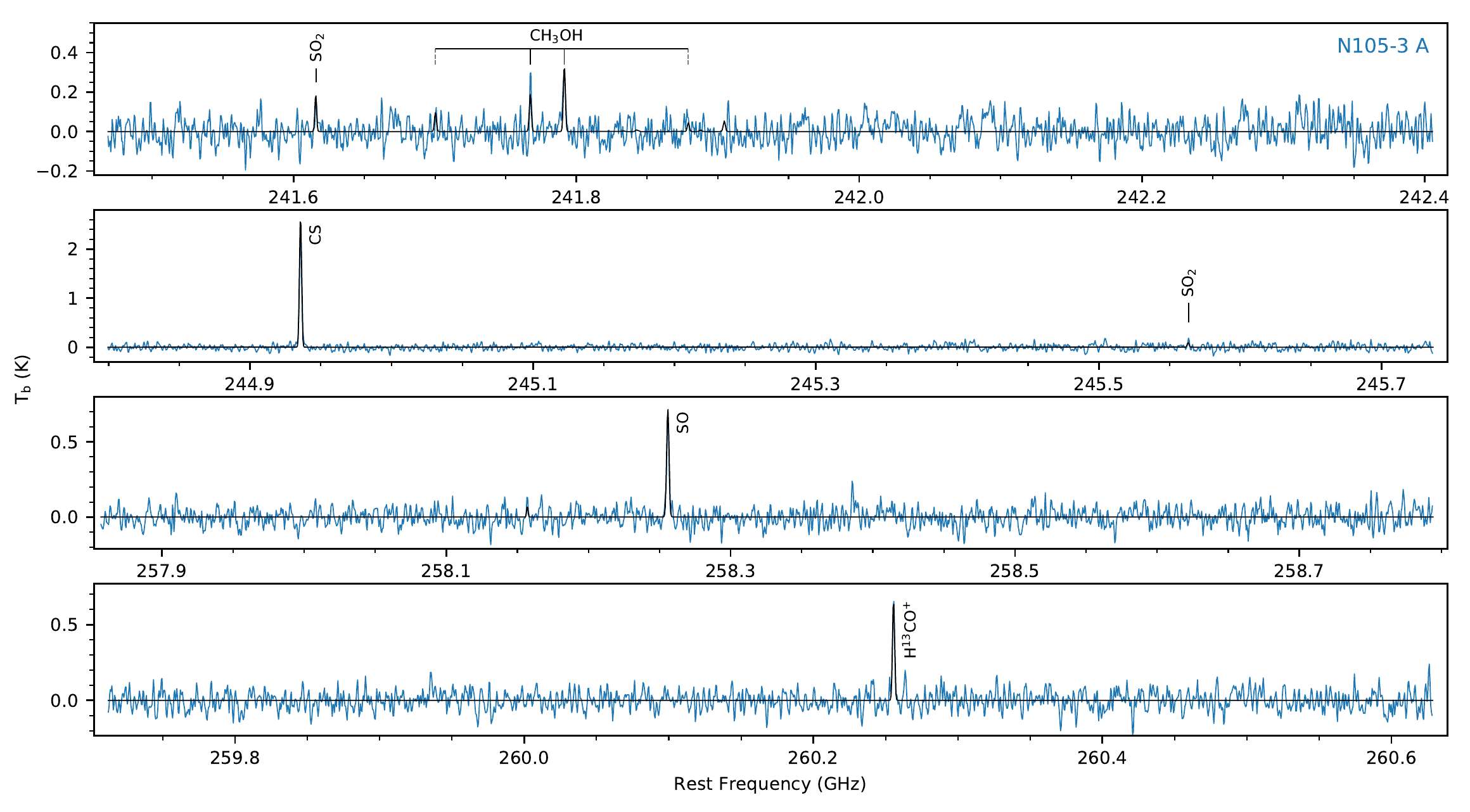}
\caption{The same as Fig.~\ref{f:spec2E_app}, but for source N\,105--3\,A. \label{f:spec3A_app}}
\end{figure*}

\begin{figure*}
\centering
\includegraphics[width=\textwidth]{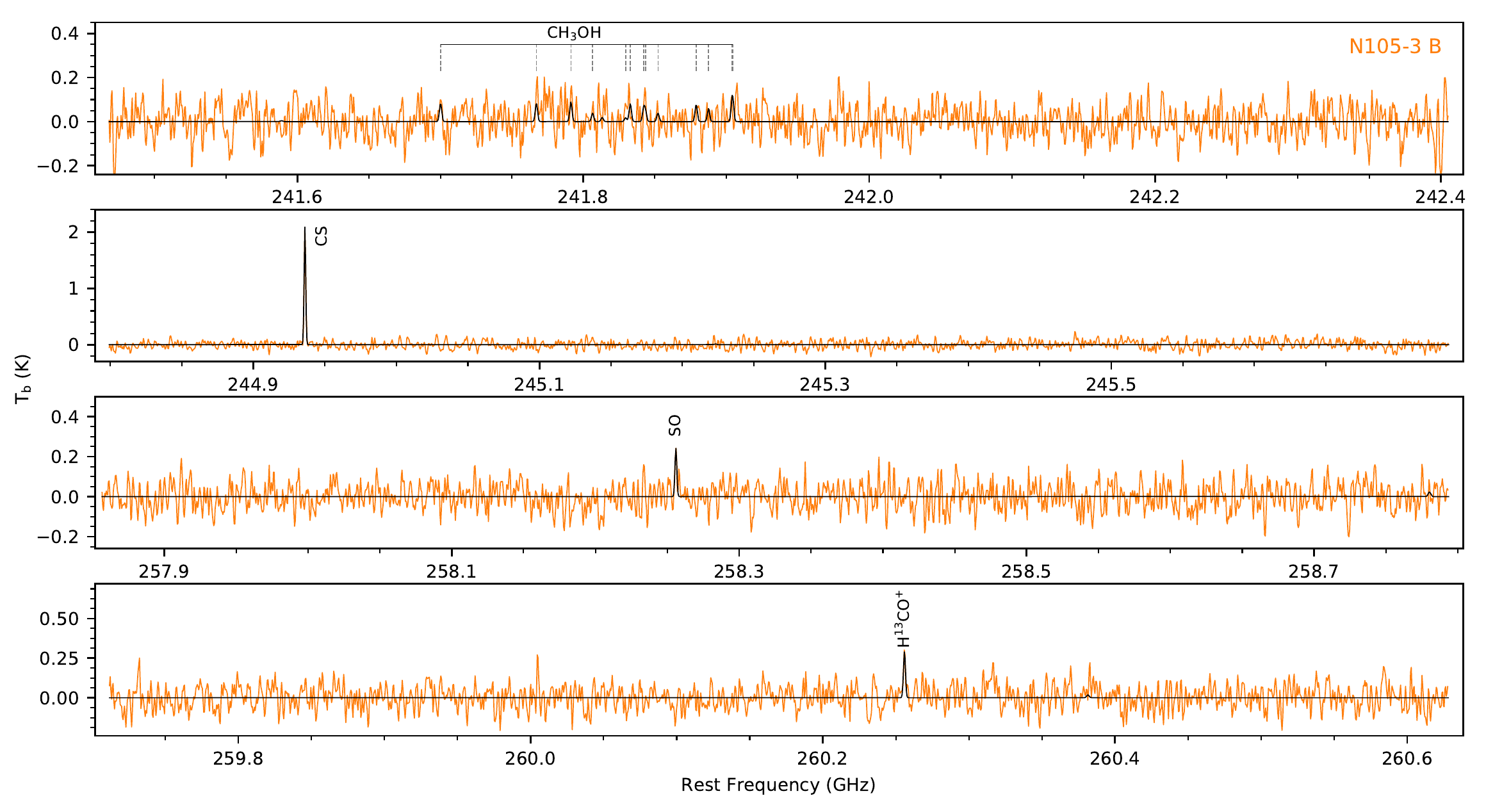}
\caption{The same as Fig.~\ref{f:spec2E_app}, but for source N\,105--3\,B. \label{f:spec3B_app}}
\end{figure*}

\begin{figure*}
\centering
\includegraphics[width=\textwidth]{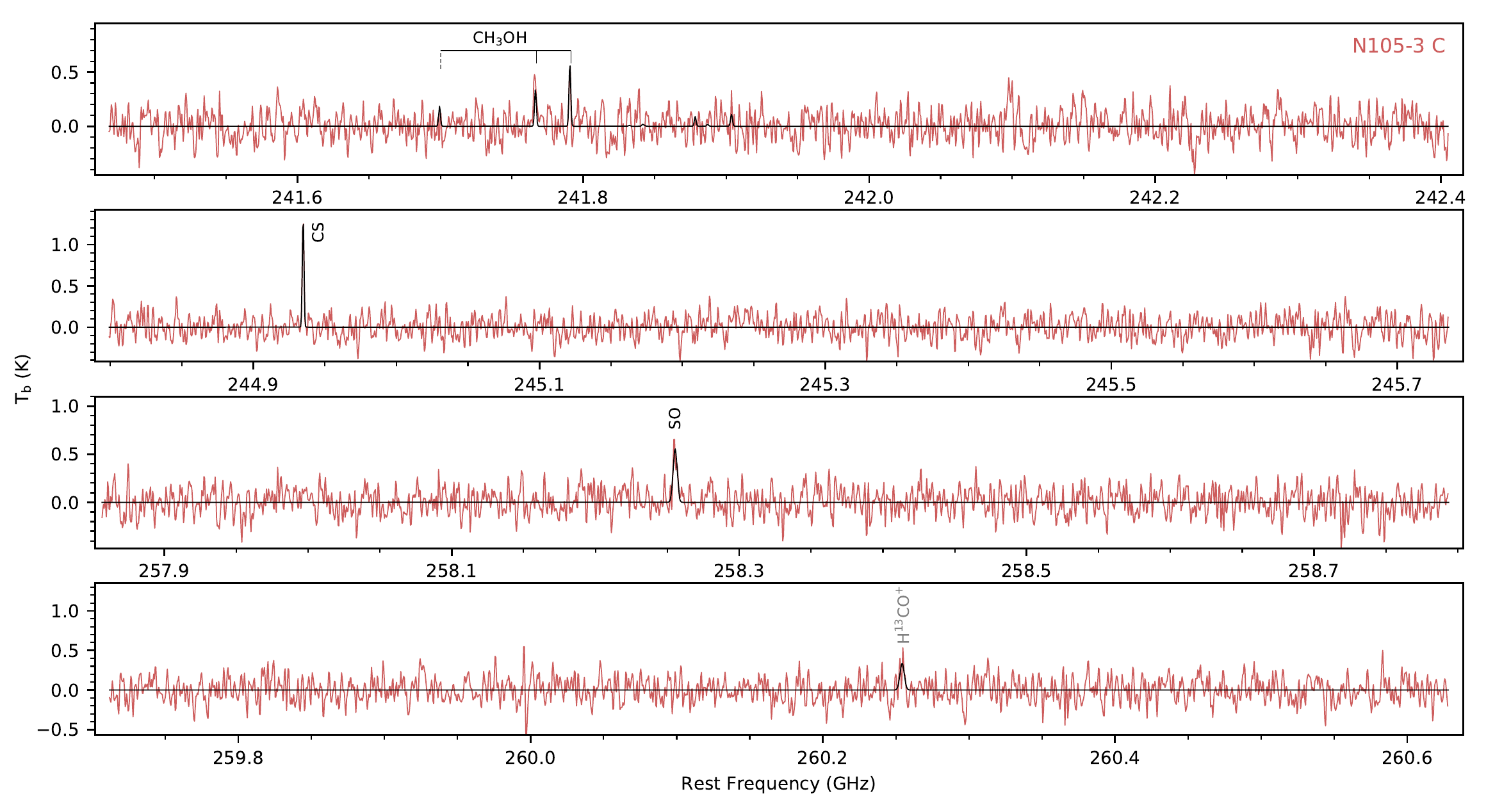}
\caption{The same as Fig.~\ref{f:spec2E_app}, but for source N\,105--3\,C.  \label{f:spec3C_app}}
\end{figure*}

In Figs.~\ref{f:spec2A_1_app}--\ref{f:spec3C_app}, we present the ALMA Band 6 spectra from all four spectral windows covered by our observations (1\,A--1\,C, 2\,A--2\,D, 2\,F) or selected sub-windows with line detections (2\,E, 3\,A--3\,C) for each continuum source analyzed in this paper.  The spectra of hot cores 2\,A and 2\,B are shown first (Figs.~\ref{f:spec2A_1_app}--\ref{f:spec2B_2_app}). The spectral extraction method is described in Section~\ref{s:spectral} and Fig.~\ref{f:extraction} shows the spectral extraction regions overlaid on the 1.2 mm continuum images. The synthetic spectra are overlaid on the observed spectra in Figs.~\ref{f:spec2A_1_app}--\ref{f:spec3C_app}. The spectral line modeling method and the fitting results are discussed in Section~\ref{s:modeling}.

\begin{figure*}
\centering
\includegraphics[width=\textwidth]{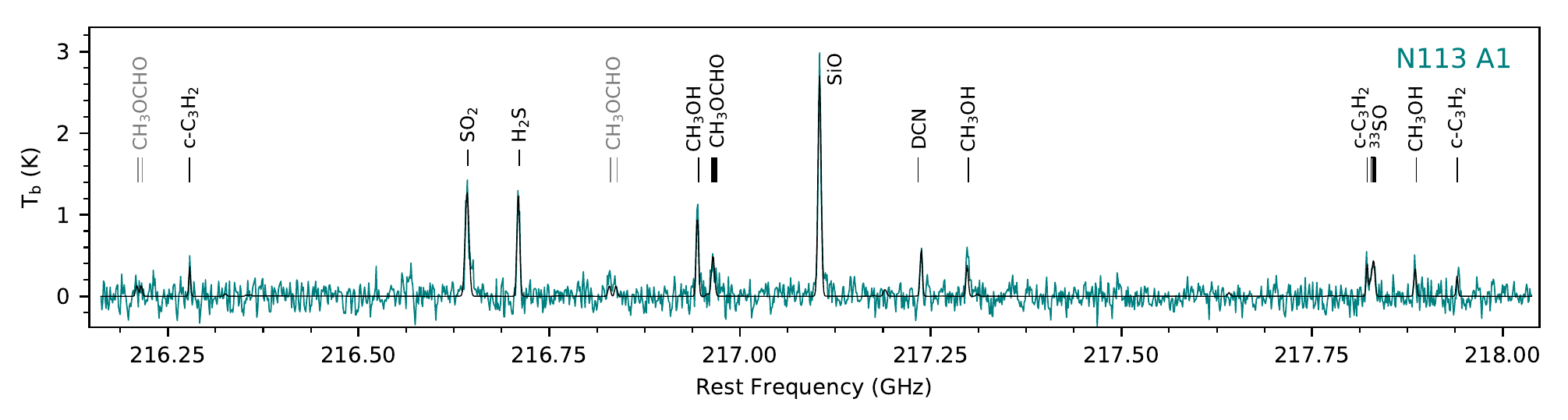}
\includegraphics[width=\textwidth]{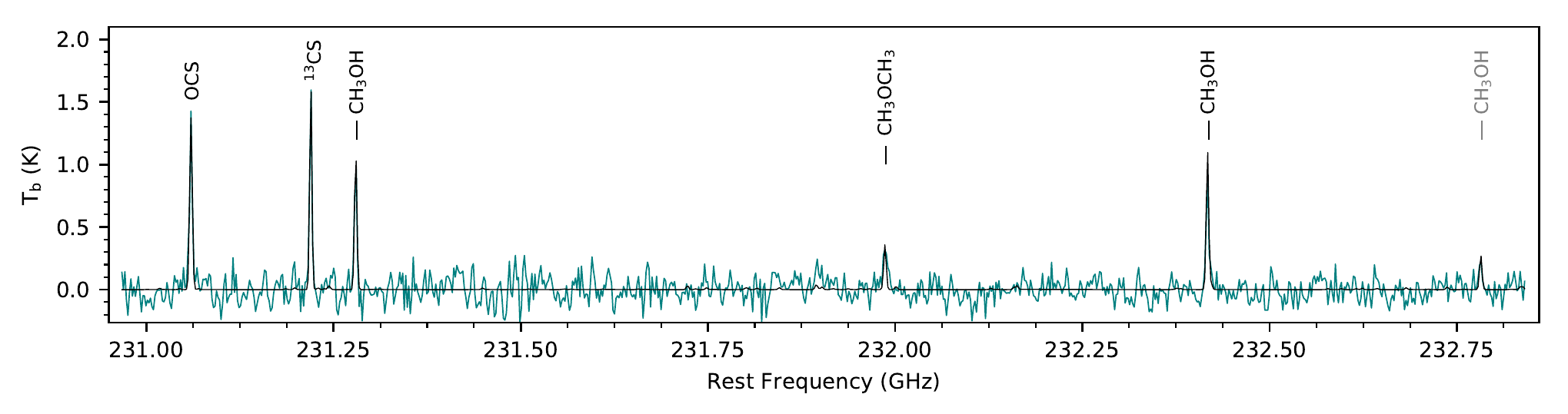}
\includegraphics[width=\textwidth]{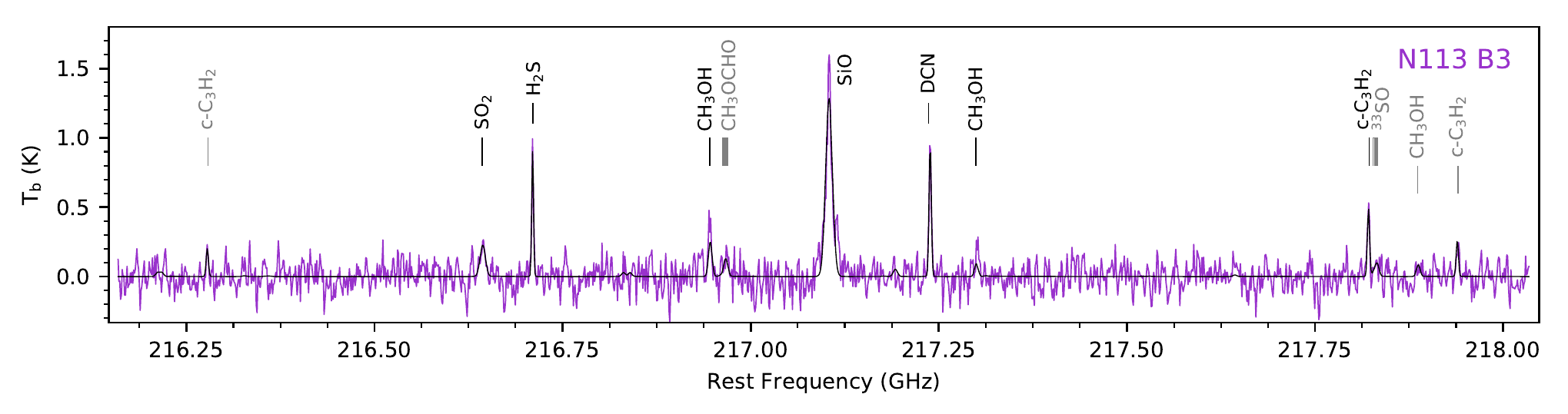}
\includegraphics[width=\textwidth]{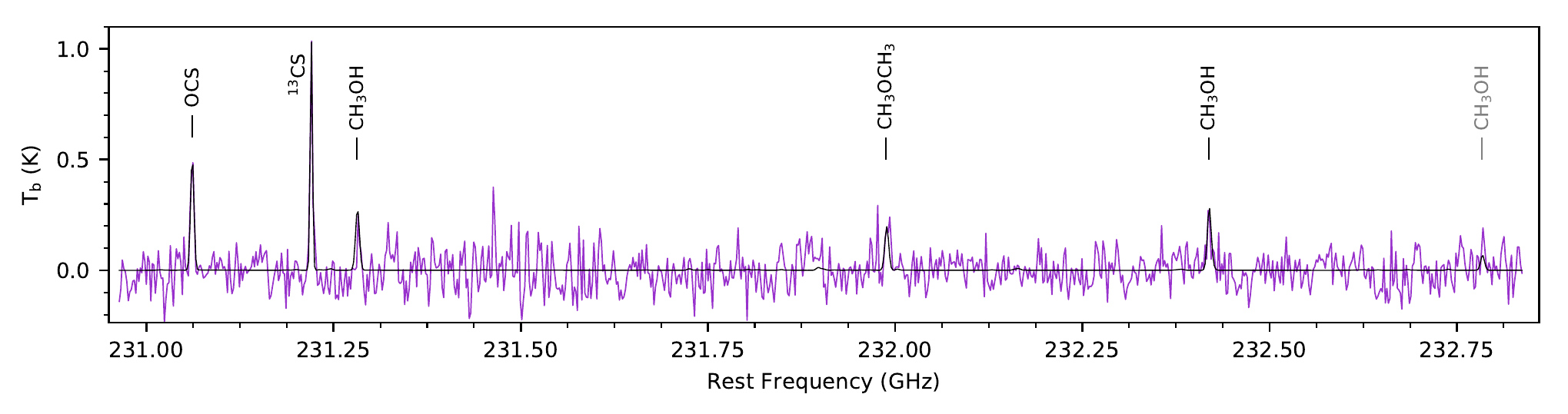}
\caption{ALMA spectra of the N\,113\,A1 ({\it top two panels}) and B3 ({\it bottom two panels}) hot cores for two spectral windows in Band 6. The spectra were first reported in \citet{sewilo2018}.  Here, we present the synthetic spectra (shown in black) obtained in the re-analysis of the data following the methods used for N\,105 (see Section~\ref{s:modeling}).  The detected (black) and tentatively detected (gray) spectral lines are labeled (see also Table~1 in \citealt{sewilo2018}).  \label{f:specN113_app}}
\end{figure*}

\section{N\,105 Star-Forming Region at Far-IR and Radio Wavelengths}

\begin{figure*}
\centering
\includegraphics[width=\textwidth]{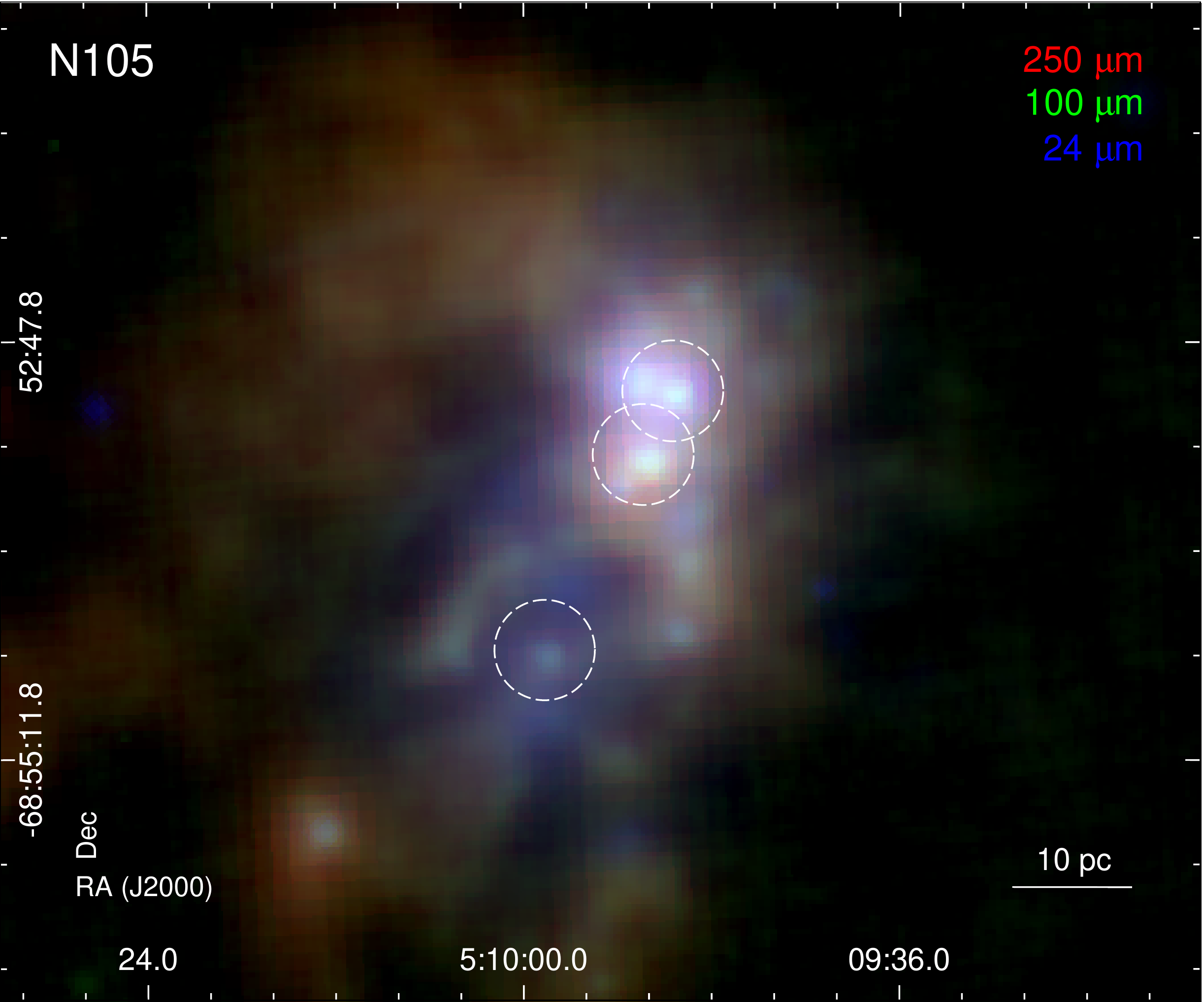}
\caption{Three-color mosaic of N\,105 combining the {\it Spitzer}/SAGE MIPS 24 $\mu$m ({\it red}), {\it Herschel}/HERITAGE PACS  100 $\mu$m ({\it green}), and SPIRE 250 $\mu$m ({\it blue}) images.  Three ALMA pointings are indicated (from north-west to south-east): N\,105--1, N\,105--2, and N\,105--3. \label{f:N105herschel3col}}
\end{figure*}

\begin{figure*}
\centering
\includegraphics[width=\textwidth]{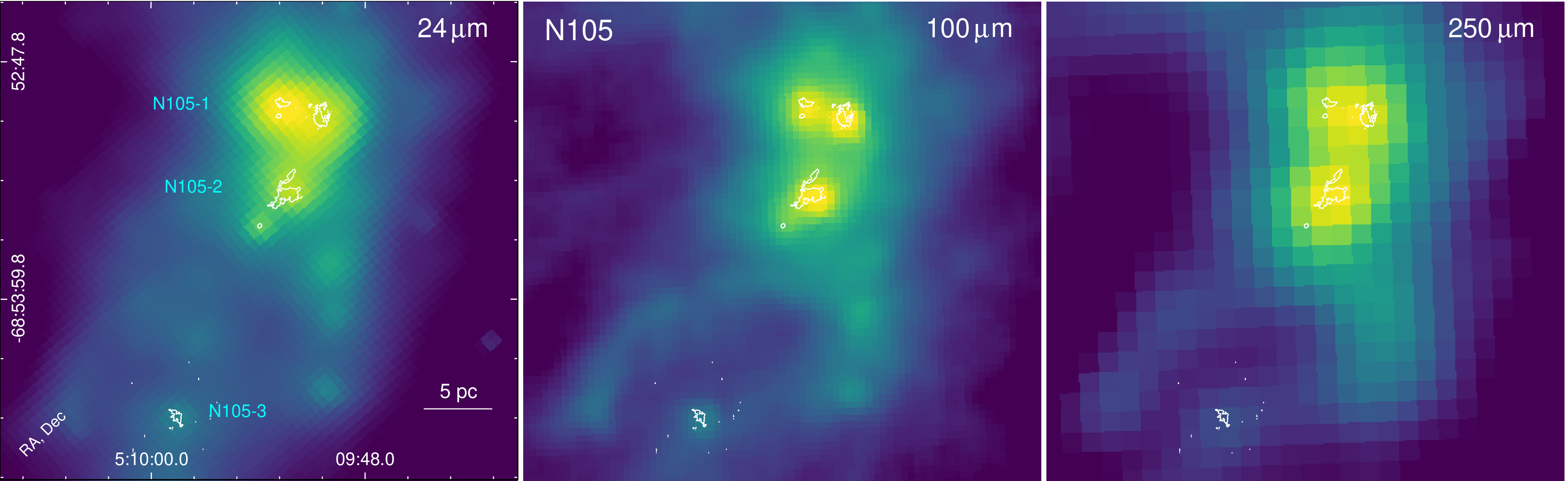}
\caption{The {\it Spitzer}/SAGE MIPS 24 $\mu$m ({\it left}), {\it Herschel}/HERITAGE PACS 100 $\mu$m ({\it center}), and SPIRE 250 $\mu$m ({\it right}) images of the N\,105 star-forming region.  The 3$\sigma$ ALMA contours are overlaid for reference. \label{f:N105herschel3im}}
\end{figure*}

\begin{figure*}
\centering
\includegraphics[width=0.495\textwidth]{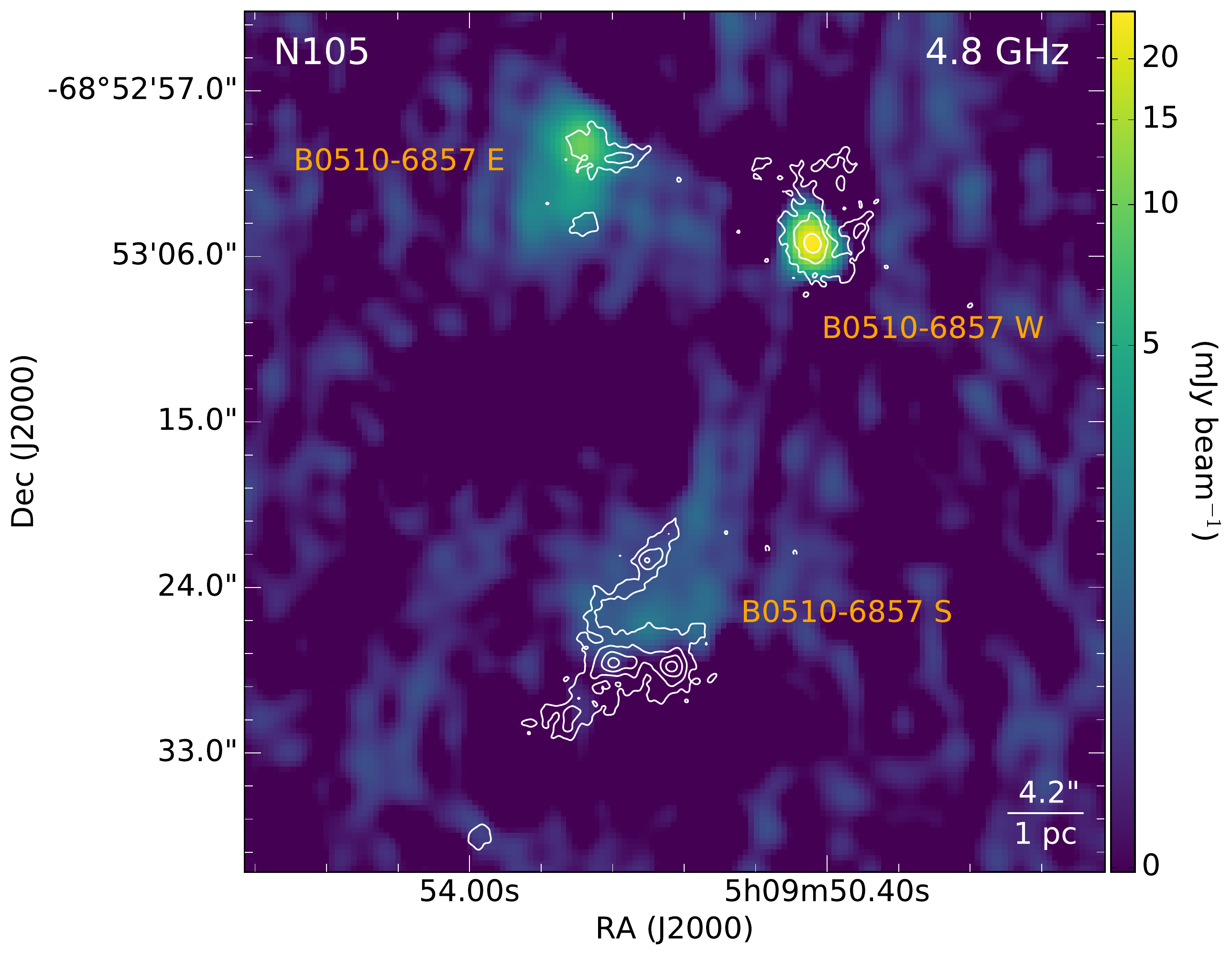}
\includegraphics[width=0.495\textwidth]{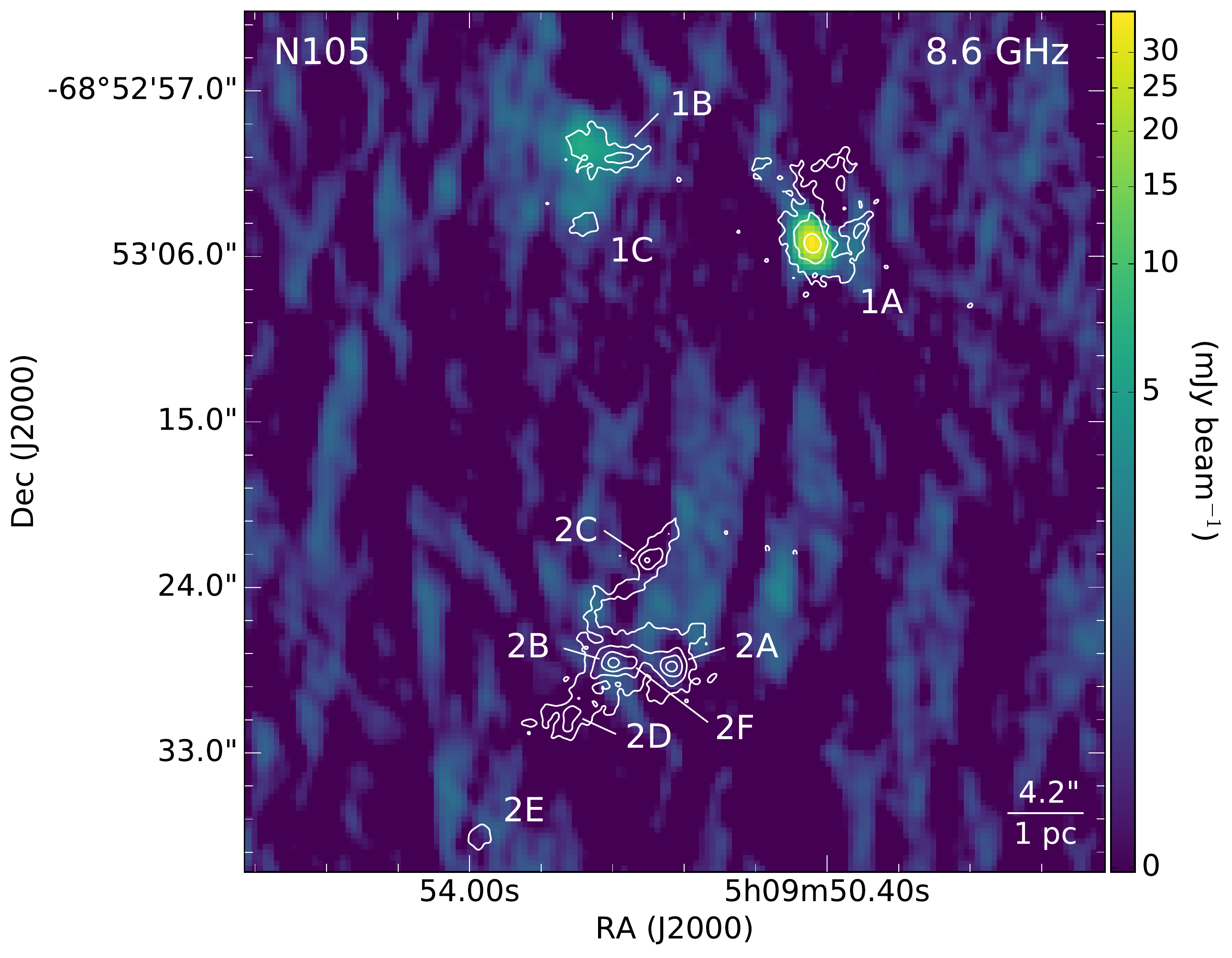}
\caption{The ATCA 4.8 GHz (6 cm; {\it left}) and 8.6 GHz (3 cm; {\it right}) images of N\,105 \citep{indebetouw2004} covering the ALMA N\,105--1 (sources 1\,A--1\,C) and N\,105--2 (sources 2\,A--2\,F) fields. The 1.2 mm continuum contours are overlaid with contour levels of (3, 10, 100)$\sigma_1$ with $\sigma_1$ = $6.8\times10^{-5}$ Jy beam$^{-1}$ for N\,105--1 and (3, 10, 30, 80)$\sigma_2$ with $\sigma_2$ = $5.0\times10^{-5}$ Jy beam$^{-1}$ for N\,105--2.  The ATCA 4.8 GHz/ 8.6 GHz radio continuum sources and ALMA 1.2 mm continuum sources are labeled in the left and right panel, respectively. The synthesized beam sizes are $2\rlap.{''}19\times1\rlap.{''}70$ and $1\rlap.{''}82\times1\rlap.{''}24$  for 4.8 GHz and 8.6 GHz images, respectively. \label{f:radio}}
\end{figure*}

In Fig.~\ref{f:N105herschel3col}, we present a three-color mosaic combining the longer wavelength images of N\,105: {\it Spitzer}/SAGE MIPS 24 $\mu$m (\citealt{meixner2006}; \citealt{sage}),  {\it Herschel}/HERITAGE PACS 100 $\mu$m, and SPIRE 250 $\mu$m images (\citealt{meixner2013}; \citealt{heritage}); the observed ALMA fields N\,105--1, N\,105--2, and N\,105--3 are indicated. Figure~\ref{f:N105herschel3im} shows each image separately with the 3$\sigma$ 1.2 mm continuum contour overlaid.  The ATCA 4.8 GHz and 8.6 GHz radio images from \citet{indebetouw2004} covering the ALMA N\,105--1 and N\,105--2 fields are presented in Fig.~\ref{f:radio}.

\end{document}